\documentclass[a4paper, 10pt, 
headings=small,
headings=twolinechapter,
headings=optiontotocandhead,
a4paper,oneside,openright,
toc=flat,
headsepline,
titlepage=false,
bibliography=totoc, 
abstract]{scrreprt}
\usepackage[top=2cm,left=2cm,right=2cm,bottom=2cm]{geometry}

\KOMAoption{chapterprefix}{false}
\addtokomafont{part}{\raggedright}

\usepackage[dvipsnames]{xcolor}
\usepackage[utf8]{inputenc}
\usepackage[T1]{fontenc}  

\usepackage{lmodern}
\usepackage{relsize}
\usepackage{scrhack}
\usepackage{enumitem}
\usepackage{dsfont} 
\usepackage{amsthm}
\usepackage{amsmath}
\usepackage{amssymb}
\usepackage{siunitx}
\usepackage{caption}
\setkomafont{caption}{\small\rmfamily}
\setkomafont{captionlabel}{\upshape\bfseries\rmfamily}
\usepackage{subfig}
\setcapindent{0pt}
\addtokomafont{disposition}{\rmfamily}
\usepackage{setspace}
\usepackage[affil-it]{authblk}

\numberwithin{equation}{section}
\usepackage{mathtools}
\usepackage{physics}
\ExplSyntaxOn
\msg_redirect_name:nnn { siunitx } { physics-pkg } { none }
\ExplSyntaxOff
\usepackage[backend=biber, doi=false, isbn=false, eprint=false, url=false, maxcitenames=2, sorting=none, date=year, style=numeric-comp, maxbibnames=10, biblabel=brackets, style=phys, backref=true]{biblatex}
\usepackage[breaklinks=true,colorlinks=true,linkcolor=blue,urlcolor=blue,citecolor=green]{hyperref}
\usepackage{booktabs} 
\usepackage{longtable}
\usepackage{tikz}
\usepackage{float}
\usepackage{xurl}
\usepackage{qcircuit}

\usepackage{tcolorbox}
\usepackage{adjustbox}
\usepackage{bm}

\usepackage{chngcntr}
\counterwithin{figure}{section}

\newcommand{\qvdots}{
  \raisebox{0.3em}{\ensuremath{\vdots}}%
}

\usepackage{changepage}
\newenvironment{indentmetric}{\begin{adjustwidth}{0.5cm}{}}{\end{adjustwidth}}

\DeclareFieldFormat{title}{#1}                      
\DeclareFieldFormat{citetitle}{#1} 
\renewbibmacro{in:}{}
\DeclareFieldFormat{pages}{#1}
\DeclareFieldFormat[article,periodical]{volume}{\mkbibbold{#1}}
\DeclareFieldFormat{journaltitle}{#1\isdot}

\providecommand{\sourceurl}[1]{the QCMet software repository under:\newline\href{https://gitlab.npl.co.uk/qc-metrics-and-benchmarks/qcmet/-/tree/main/tutorials/#1}{\texttt{\detokenize{#1}}}}

\providecommand{\sourceurltwo}[1]{\newline\href{https://gitlab.npl.co.uk/qc-metrics-and-benchmarks/qcmet/-/tree/main/tutorials/#1}{\texttt{\detokenize{#1}}}}

\newcommand*{\fullsechyperref}[1]{\hyperref[{#1}]{\ref*{#1}: \nameref*{#1}}}
\newcommand{\metrichref}[2]{\hyperref[#1]{\color{Black}{#2}}}%

\defbibheading{subbibliography}[\refname]{\subsection*{#1}}

\newcommand{\printsectionbib}[0]{\begingroup
\let\clearpage\relax
\printbibliography[heading=subbibliography]
\endgroup}

\DeclareMathOperator*{\argmax}{argmax}

\makeatletter
\newcommand*{\shifttext}[2]{%
  \settowidth{\@tempdima}{#2}%
  \makebox[\@tempdima]{\hspace*{#1}#2}%
}
\makeatother

\newif\ifshowmetricfig
\showmetricfigtrue
\providecommand{\metricfig}[1]{\ifshowmetricfig #1 \fi}

\RedeclareSectionCommand[tocnumwidth=2.em]{chapter}

\newbool{showpart1references}
\boolfalse{showpart1references}

\newcommand{\metricbibliography}{
\begingroup
\renewbibmacro{pageref}{}
\printbibliography[heading=subbibliography,segment=\therefsegment]
\endgroup
}

\DefineBibliographyStrings{english}{
  backrefpage={Cited on page},
  backrefpages={Cited on pages}
}

\newbool{anchorsegments}
\boolfalse{anchorsegments}

\makeatletter

\AtBeginDocument{%
  \ifbool{anchorsegments}
    {\long\def\blx@bibhyperref[#1]#2{%
       \blx@sfsave\hyper@natlinkstart{\the\c@refsection @\the\c@refsegment @#1}%
       \blx@sfrest
       #2%
       \blx@sfsave\hyper@natlinkend\blx@sfrest}%
     \protected\long\def\blx@imc@bibhyperlink#1#2{%
       \blx@sfsave\hyper@natlinkstart{\the\c@refsection:\the\c@refsegment:#1}%
       \blx@sfrest
       #2%
       \blx@sfsave\hyper@natlinkend\blx@sfrest}%
     \protected\long\def\blx@imc@bibhypertarget#1#2{%
       \blx@sfsave\hyper@natanchorstart{\the\c@refsection:\the\c@refsegment:#1}%
       \blx@sfrest
       #2%
       \blx@sfsave\hyper@natanchorend\blx@sfrest}%
     \protected\def\blx@anchor{%
       \xifinlist
         {\the\c@refsection @\the\c@refsegment @\abx@field@entrykey}
         {\blx@anchors}
         {}
         {\listxadd
            {\blx@anchors}
            {\the\c@refsection @\the\c@refsegment @\abx@field@entrykey}%
          \hyper@natanchorstart{%
            \the\c@refsection @\the\c@refsegment @\abx@field@entrykey}%
          \hyper@natanchorend}}%
     \defbibheading{subbibliography}{\subsection*{References}}}
    {\defbibheading{subbibliography}{%
       \AtNextBibliography{\let\blx@anchor\relax}%
       \subsection*{Metric references}}}}

\makeatother

\BiblatexSplitbibDefernumbersWarningOff

\usepackage[autooneside=false]{scrlayer-scrpage}
\leftmark
\automark[chapter]{chapter}
\clearpairofpagestyles 
\ihead[]{\leftmark}

\cehead{title}         
\setkomafont{pageheadfoot}{\normalcolor\bfseries}

\usepackage{etoolbox}

\makeatletter
\def\@endpart{} 

\let\nonewpagechapter\chapter 
\patchcmd\nonewpagechapter{\if@openright\cleardoublepage\else\clearpage\fi}{}{}{}

\patchcmd{\part}{\null\vfil}{}{}{}
\makeatletter
\def\@endchapter{} 

\let\nonewpagechapter\chapter 
\patchcmd\nonewpagechapter{\if@openright\cleardoublepage\else\clearpage\fi}{}{}{}

\patchcmd{\part}{\null\vfil}{}{}{}

\makeatletter
\patchcmd{\scr@startchapter}{\if@openright\cleardoublepage\else\clearpage\fi}{}{}{}
\makeatother

\makeatletter
\newcommand{\unchapter}[1]{%
  \begingroup
  \let\@makechapterhead\@gobble 
  \chapter[tocentry={}]{#1}
  \endgroup
}
\makeatother

\RedeclareSectionCommand[style=chapter,beforeskip=\baselineskip, afterskip=0.1\baselineskip]{part}

\newcommand*{\fullref}[1]{\hyperref[{#1}]{\nameref*{#1}}}

\titlehead{%
\vspace*{-\headsep}\vspace{-\headheight}\vspace{-\topskip}\usekomafont{pageheadfoot}{}}

\title{\vspace{-5ex}\textbf{A Review and Collection of Metrics and Benchmarks for Quantum Computers:\\
definitions, methodologies and software}}

\author[1]{Deep Lall\thanks{These authors contributed equally}}
\author[1]{Abhishek Agarwal$^*$}
\author[1,2]{Weixi Zhang$^*$}
\author[1]{Lachlan Lindoy}
\author[1]{Tobias Lindstr\"om}
\author[1]{Stephanie Webster}
\author[1]{Simon Hall}
\author[3,4]{Nicholas Chancellor}
\author[2]{Petros Wallden}
\author[2,5]{Raul Garcia-Patron}
\author[2,6]{Elham Kashefi}
\author[7]{Viv Kendon}
\author[7]{Jonathan Pritchard}
\author[1,7]{Alessandro Rossi}
\author[8]{Animesh Datta}
\author[9]{Theodoros Kapourniotis}
\author[9]{Konstantinos Georgopoulos}
\author[1,10]{\\{Ivan Rungger}\thanks{\href{mailto:ivan.rungger@npl.co.uk}{ivan.rungger@npl.co.uk}}}

\affil[1]{National Physical Laboratory, Hampton Road, Teddington TW11 0LW, United Kingdom}
\affil[2]{School of Informatics, QSL, University of Edinburgh, 10 Crichton Street, Edinburgh EH8 9AB, United Kingdom}
\affil[3]{School of Computing, Newcastle University, 1 Science Square, Newcastle upon Tyne NE4 5TG, United Kingdom}
\affil[4]{Department of Physics, Joint Quantum Centre, Durham University, South Road, Durham DH1 3LE, United Kingdom}
\affil[5]{Phasecraft Ltd., London, United Kingdom}
\affil[6]{Sorbonne Université, CNRS, LIP6, 75005 Paris, France}
\affil[7]{Department of Physics, University of Strathclyde, Glasgow G4 0NG, United Kingdom}
\affil[8]{Department of Physics, University of Warwick, Coventry CV4 7AL, United Kingdom}
\affil[9]{National Quantum Computing Centre, Didcot OX11 0QX, United Kingdom}
\affil[10]{Department of Computer Science, Royal Holloway, University of London, Egham, TW20 0EX, United Kingdom}

\date{\vspace{-5ex}}
\begin{document}
\maketitle
\begin{abstract}
Quantum computers have the potential to provide an advantage over classical computers in a number of areas.
The quest towards such quantum advantage has required the development of methods to benchmark the performance of quantum computers against classical computers as well as the relative performance of different quantum computing platforms. This is a challenging task, firstly due to the large diversity of hardware platforms, and secondly due to the emergence of two different approaches to quantum computing, one being a gate-based approach for universal quantum computation, and the other an analogue approach tailored to outperforming classical computers for specific tasks. 
Numerous metrics to benchmark the performance of quantum computers, ranging from their individual hardware components to entire applications, have been proposed over the years. Navigating the resulting extensive literature can be overwhelming. Objective comparisons are further hampered in practice as different variations of the same metric are used, and the data disclosed together with a reported metric value is often not sufficient to reproduce the measurements. This article addresses these challenges by providing a review of metrics and benchmarks for quantum computers and 1) a comprehensive collection of benchmarks allowing holistic comparisons of quantum computers, 2) a consistent format of the definitions across all metrics including a transparent description of the methodology and of the main assumptions and limitations, and 3) a reproducible approach by linking the metrics to open-source software used to evaluate them. The benchmarks span aspects of a device such as architectural properties, quality, speed, and stability, and also include metrics tailored to several non-gate-based architectures.

Quantum computing technology has reached the stage where a number of methods for performance characterization are backed by a large body of real-world implementation and use, as well as by theoretical proofs. These mature benchmarking methods will benefit from commonly agreed-upon approaches. Indeed, such agreement is the only way to fairly, unambiguously, and objectively benchmark quantum computers across manufacturers. However, there are also benchmarking techniques that are still significantly evolving and for which standardization would be premature. We identify five areas where international standardization working groups could be established, namely: i) the identification and agreement on the categories of metrics that comprehensively benchmark device performance; ii) the identification and agreement on a set of well-established metrics that together comprehensively benchmark performance; iii) the identification of metrics specific to hardware platforms, including non-gate-based quantum computers; iv) inter-laboratory comparison studies to develop best practice guides for measurement methodology; and v) agreement on what data and software should be reported together with a metric value to ensure trust, transparency and reproducibility. We provide potential routes to advancing these areas, thereby contributing to accelerate the progress of quantum computing hardware towards quantum advantage.
\end{abstract}
\newpage

\setcounter{tocdepth}{1}
\tableofcontents
\vspace{1em}
\part{Introduction}
\label{part:introduction}
\addtocontents{toc}{\protect\setcounter{tocdepth}{2}}
\begin{refsegment}
\renewcommand{\thechapter}{\Roman{chapter}}
\renewcommand{\theHchapter}{R\Roman{chapter}}
\setcounter{chapter}{0}
\unchapter{Introduction}
\renewcommand{\thechapter}{\arabic{chapter}}
\setcounter{chapter}{0}
The field of quantum computation has experienced significant advances in the last decades. The elementary units of quantum computation -- qubits, have been implemented on a wide range of platforms such as superconducting circuits~\cite{kjaergaard2020superconducting,krantz_quantum_2019,siddiqi2021engineering,devoret2004superconducting,wendin2017quantum,clarke2008superconducting,devoret2013superconducting,you2005superconducting}, trapped ions~\cite{ion:progress-review,haffner2008quantum, romaszko2020engineering,ion:cirac-zoller,benhelm2008towards,monroe2013scaling, debnath2016demonstration, wrightBenchmarking11qubitQuantum2019}, neutral atoms~\cite{saffman2019quantum, henriet2020quantum,bluvstein22,wurtz2023aquila, wintersperger2023neutral, graham2022multi, bluvstein2024logical}, as well as photonic~\cite{o2009photonic,Kok2007Linear,wang2020integrated,barz2015quantum,flamini2018photonic,slussarenko2019photonic,knill2001scheme}, and semiconductor devices~\cite{RevModPhys.95.025003,chatterjee2021semiconductor,De_Michielis_2023,maurand2016cmos,awschalom2013quantum,vandersypen2019quantum,kloeffel2013prospects}. The number of qubits implemented in a single device has grown significantly, and devices with hundreds of qubits have now been demonstrated on various platforms~\cite{bluvstein2024logical,kim2023evidence,patra2024efficient,Xanadu22}. Improvements in the hardware and qubit control have allowed significant reductions in the error rates~\cite{ion:progress-review,kjaergaard2020superconducting,wintersperger2023neutral,RevModPhys.95.025003,flamini2018photonic}, and research on reducing the noise induced errors in quantum computers is ongoing.  An outstanding challenge for all hardware platforms towards making quantum computers practically useful is to reduce error rates further. In addition to developments in the hardware, there have also been advancements in the design of quantum algorithms and in their resilience to noise~\cite{motta2022emerging,kim2023evidence,google2020hartree}, which have led to reduced requirements for the implementation of algorithms on hardware~\cite{dalzell2023quantum,childs2018toward, motta2022emerging}.

The most prevalent model of quantum computation is denoted as gate-based (or circuit-based) quantum computing, where a discrete set of unitary operations, denoted as gates, is used to perform the computation. Non-gate-based (or non-circuit-based) quantum computing approaches on the other hand are typically designed to capitalize on specific strengths of each hardware platform to provide advantage over classical computing, with the drawback that the computations are usually not universal. Such non-gate-based approaches include quantum annealing~\cite{johnson11aManufacturedSpins,hauke2020perspectives,finilla1994quantumannealing}, boson sampling~\cite{GBS2017,Aaronson2010,ScattershotBS,Gard_2015}, and analogue quantum simulation~\cite{daley2022practical,georgescu2014quantum,buluta2009quantum,ebadi2021quantum}. Gate-based and non-gate-based approaches can also be combined in an algorithm, where each approach is used for those parts of the algorithm where it provides an advantage over the other.
Although claims of quantum computers outperforming classical computers at certain tasks have been made for both gate-based and non-gate-based quantum computers~\cite{Arute2019, Xanadu22,wu2021strongquantumcomputationaladvantage, zhu2022quantum, daley2022practical, Jiuzhang3}, there has yet to be a clear demonstration of quantum advantage for practically useful tasks, which is typically denoted as practically useful quantum advantage. 

Furthermore, quantum computers are now being made available to academic and industrial users in the cloud, and it is important that such users have a transparent way to evaluate the capability of the services they access. Therefore, the need to benchmark these devices using suitable metrics for objective comparisons arises, both to evaluate the status of a hardware platform and to help guide it towards achieving practically useful quantum advantage. To this aim the benchmarks need to holistically characterize all aspects of the performance of a quantum computer that are relevant for the achievement of practically useful quantum advantage. Such benchmarks make it possible to track the progress of the field over time, and also to guide the development of the hardware and algorithms to eventually outperform classical computers for certain problems~\cite{proctor2024benchmarkingquantumcomputers,herrmann2023quantum}. 

The development and study of metrics and benchmarks for quantum computers have proliferated immensely over the past decades, where improvements both in  the hardware and in the performance metrics have led to mutual benefits and faster development. 
Since its inception, errors in quantum computation due to interaction of a qubit with its environment were identified among the main difficulties in constructing quantum computers~\cite{divincenzo1995two,divincenzo1995quantum}. The ratio between the time taken for individual qubit operations and the time over which qubits can avoid being perturbed by the environment was used to estimate the number of operations that the devices can be expected to run successfully~\cite{divincenzo1995two}. These timescales and their ratio are among the first performance metrics proposed for quantum computers.
In 1995, the first demonstration of a two-qubit gate was performed on trapped-ions~\cite{monroe1995demonstration}. The state of the qubits with and without the application of the gate on different initial states was measured and compared to theoretically ideal outcomes. The difference between the observed and ideal outcomes in the absence of the quantum gate was attributed to errors in state preparation, measurement, laser cooling, and decoherence. The addition of the gate led to further errors and a larger difference compared to the ideal outcomes. This method for characterization constituted an early version of tomography of the quantum state and process, and the evaluation of a fidelity between quantum computing output and the ideal theoretical target is the foundation of a large part of metrics and benchmarks used today. In the mid-1990s the development and implementation of tomographic methods for reconstructing quantum states and processes from experimental data was developed~\cite{raymer1994complex, poyatos1997complete,chuang1997prescription,nielsen1998complete, laflamme1998nmr,hradil1997quantum}, and related theoretical work introduced metrics such as entanglement fidelity and diamond norm~\cite{schumacher1996sending,jozsa1994fidelity, nielsen1996entanglement,aharonov1998quantum}. The first quantum algorithms were run in 1998~\cite{jones1998implementation,chuang1998experimental}. In the latter, the quality of the algorithm output was evaluated by reconstructing the quantum state and quantifying its difference with the ideal outcomes. Variants of this methodology are still used today to evaluate the quality of algorithm execution on a quantum computer. The fundamental challenge for such methods is that they are difficult to scale to large number of qubits, because of the exponentially increasing complexity of carrying out full state or process tomography.

In the late 1990s and early 2000s, research in the metrics for quantum processes led to relations being found between metrics such as average gate fidelity and entanglement fidelity, generalizations of the metrics, and methods for experimental determination of the metrics~\cite{NIELSEN2002249,PhysRevA.60.1888,fortunato2002implementation,gilchrist2005distance}. The evaluation of these metrics required the full characterization of the quantum process, thus limiting their use to small number of qubits. An alternate approach for characterizing the quality of quantum processes, randomized benchmarking (RB), was proposed in 2005~\cite{emerson2005scalable}. RB uses random gate sequences to obtain averaged fidelities for a set of gates. Subsequently, significant research effort, still ongoing today, has been dedicated to developing RB, both theoretically and practically~\cite{knillRandomizedBenchmarkingQuantum2008,emerson2007symmetrized,levi2007efficient,dankert2009exact,magesanCharacterizingQuantumGates2012,proctorDirectRandomizedBenchmarking2019,PhysRevLett.109.240504,PhysRevA.89.062321,PhysRevLett.108.260503,magesanEfficientMeasurementQuantum2012,SheldonIterativeRandomizedBenchmarking,PhysRevLett.122.200502,proctorScalableRandomizedBenchmarking2021,onorati2019randomized,morvan2021qutrit,goviaRandomizedBenchmarkingSuite2022,helsen2022general}. This has led to the development of multiple RB variants that allow characterizing different kinds of processes, or different kinds of noise, and the creation of a framework encompassing the many kinds of RB has been proposed~\cite{helsen2022general}. Today, RB based methods are widely used to obtain metrics for the quality of quantum gates, and often reported by the hardware manufacturers~\cite{tomeshSupermarQScalableQuantum2022}. It is important to note that the experimentally often-measured average fidelities do not directly correspond to theoretically relevant quantities such as fault tolerance thresholds, which are quantified by the diamond norm~\cite{gilchrist2005distance,PhysRevLett.117.170502}. Another important step in the development of characterization techniques was the proposal of gate set tomography (GST) in the early 2010s~\cite{merkel_self-consistent_2013,blume2013robust}. GST circumvents various issues faced by prior methods such as quantum process tomography~\cite{nielsenGateSetTomography2021,greenbaum_introduction_2015}, at the cost of increased complexity and runtime, limiting its use to only few-qubit systems. Making GST more efficient and scalable to larger systems is an active research endeavor~\cite{nielsenGateSetTomography2021,brieger2023compressive,ostrove2023near,cao2022efficient}.

As the number of qubits in a device grew over the years, with devices having tens of qubits being accessible to the public by the late 2010s~\cite{mooney2019entanglement}, so did the need for the characterization of the performance of entire devices, rather than just individual operations. One class of approaches towards full device characterization utilizes randomized circuits. Such approaches include quantum volume~\cite{crossValidatingQuantumComputers2019,jurcevic2021demonstration,pelofske2022quantum}, mirrored circuits average polarization~\cite{proctorScalableRandomizedBenchmarking2021, mayerTheoryMirrorBenchmarking2021, proctorMeasuringCapabilitiesQuantum2022, amico2023defining,moses2023race}, and upper bound of variation distance~\cite{Ferracin2019,Ferracin2021}. Another approach characterizes the performance of a device by evaluating the quality of outputs for chosen sets of algorithms and applications. Various benchmarking suites, with different sets of algorithms and different metrics to quantify the quality of a device in executing those algorithms, have been proposed in recent years~\cite{lubinskiApplicationOrientedPerformanceBenchmarks2021,liQASMBenchLowlevelQASM2022a,patelExperimentalEvaluationNISQ2020,kochDemonstratingNISQEra2020,reschBenchmarkingQuantumComputers2021,georgopoulosQuantumComputerBenchmarking2021,quetschlichMQTBenchBenchmarking2022,10.1145/3307650.3322273,dallaire-demersApplicationBenchmarkFermionic2020, mccaskeyQuantumChemistryBenchmark2019,tomeshSupermarQScalableQuantum2022, donkersQPackScoresQuantitative2022,linkeExperimentalComparisonTwo2017,wrightBenchmarking11qubitQuantum2019,martielBenchmarkingQuantumCoprocessors2021,liaoBenchmarkingQuantumProtocols2022,lubinski2023optimization, lubinski2024quantum}. 
With the growth of  cloud-based services, a third approach that introduces notions such as blindness to certify quantum computation in a distributed setting could become especially appealing~\cite{barz2013experimental, gheorghiu2019verification, eisertQuantumCertificationBenchmarking2020}. Metrics that quantify the speed of devices, rather than just quality, have also been proposed~\cite{wackQualitySpeedScale2021}. 

In recent years increasing consideration has been devoted to metrics of relevance for quantum error correction (QEC), since QEC promises a solution to the large error susceptibility of quantum computers~\cite{steane1998quantum, horowitz2019quantum, A_Yu_Kitaev_1997, doi:10.1080/00107514.2019.1667078, devitt2013quantum, lidar2013quantum, preskill1998reliable, knill1997theory, gottesman1997stabilizer, gottesman2010introduction}. In QEC, the information is processed in so called logical qubits, which are abstract qubits that are encoded on the states of a larger number of physical qubits in hardware. Certain errors on the logical qubits can be detected and corrected, and QEC involves the repeated detection and correction of errors during quantum computations. When error levels on physical qubits are below a threshold that depends on the specific QEC framework, the logical error rates can be suppressed to arbitrarily low levels and the device is called fault-tolerant. QEC requires a large number of physical qubits with sufficiently low error-rates, as well as fast classical processing and feedback to the quantum processor. Today, the improvements in quantum computing capabilities have allowed for various proof-of-principle demonstrations of QEC~\cite{google2023suppressing, bluvstein2024logical,da2024demonstration, sivak2023real,postler2022demonstration, erhard2021entangling,PhysRevX.11.041058, egan2021fault, acharya2024quantum, brock2024quantum, reichardt2024demonstration, putterman2024hardware}. Various metrics that evaluate how well the error-correction performs have been used in the recent literature, including the logical coherence times~\cite{egan2021fault, sivak2023real, PhysRevX.11.041058}, the ratio of the logical error rate to the physical error rate~\cite{sivak2023real, da2024demonstration, brock2024quantum,acharya2024quantum}, and error per QEC cycle~\cite{PhysRevX.11.041058, sivak2023real, da2024demonstration,google2023suppressing, reichardt2024demonstration, brock2024quantum, acharya2024quantum}. As QEC development is fast progressing, the development of QEC specific metrics is ongoing.

Despite the resulting wide range of available metrics, only a few metrics are widely used in practice. These widely used metrics are usually more mature in terms of their development, and typically include the qubit relaxation and dephasing times, typically denoted as $T_1$ and $T_2$, respectively, or RB-based gate error metrics. The lack of widespread adoption of a number of other more recently proposed metrics reflects the lack of consensus on these metrics compared to more mature metrics. Given the significant advancements in the field, and the increasing need for unambiguous and objective metrics to compare devices, a number of articles reviewing metrics and proposing various levels of standardization have been compiled recently~\cite{amico2023defining,acuaviva2024benchmarking,proctor2024benchmarkingquantumcomputers,hashim2024practical}.  
Standardization of metrics and benchmarks is as important as it is challenging, given the large number of metrics available and the potential influence of a standard set of benchmarks on commercial success of a platform. Some of the benefits of a standardized approach are that it avoids the use of different variants for evaluating a metric, which all ultimately give the same information with some quantitative deviations. There can be significant overlaps between the performance indicators of a device that different metrics aim to quantify, so that the identification of a subset of benchmarks that holistically measures device performance would be useful. A challenge to this aim is the need of ensuring that no particular hardware platform is unfairly disadvantaged nor unduly advantaged. The set of benchmarks thus needs to be discussed in view of all the hardware platforms. 

For well established metrics, a standardized approach to evaluate them will be beneficial. There are various subtleties in the evaluation of the metrics, such as the number of measurements taken, averaging procedures, classical circuit optimizations, and many more. Standardizing these aspects and the information that is reported together with a metric value is important to ensure transparency and reproducibility~\cite{amico2023defining,pelofske2022quantum,acuaviva2024benchmarking}. 
Another challenge is that, although significant research has been devoted to developing metrics for gate-based quantum computers, much more limited research exists for non-gate-based devices such as quantum annealers, boson sampling devices, or analog quantum simulators. It is important not just to devise metrics applicable to these devices, but also devise metrics that enable comparisons between these devices and traditional gate-based quantum computers. We identify this as a key area where improvements are needed. For example, there are very few metrics for quantum annealers which can be applied across different hardware implementations, and even those which do exist need to be modified and improved as the technology progresses and the devices move into a more coherent operating regime.

The intent of this article is to cater to the need for objective performance comparisons by defining metrics and the associated methodologies by which the performance of emerging quantum processors can be holistically measured and compared. We review the literature and use it to identify a collection of metrics that forms a comprehensive benchmarking suite covering the relevant aspects of device performance such as speed, quality, and stability, as well as performs different levels of benchmarking: from component-level benchmarking to application-level benchmarking. 
Along with short definitions and longer descriptions of the metrics, we provide detailed methodologies to evaluate them, their assumptions and limitations, relevant references, and finally demonstrator software that implements the methodology to measure the metric. In the first part of the document we also give a general introduction to quantum computing and of some of the widely used hardware platforms, introducing all the required concepts also for the non-expert readers. 
The provided collection of metrics and the associated software package can contribute to the development of a standardized approach for measuring the performance of quantum computers. By design we restrict the selected metrics in the collection to a limited number. Considering the fast advancement of the field, the document and software is set up as dynamic online resource that is regularly updated with new metrics and improved insights on existing ones. As metrics develop, these can be added in future versions. The methodologies presented in the document for obtaining metrics are based on what is commonly employed in practice, and as research on metrics and benchmarking progresses, these methodologies may also be updated.

The linked software package implements the methodologies to evaluate the metrics and is designed to be used in conjunction with this document, both as tutorial to provide all information on the implementation details of the described methodology, as well as a practical tool to evaluate the collection of metrics on a quantum computing emulator or on a hardware platform. The software either implements the methodology directly or links to further open source libraries such as Qiskit~\cite{qiskit2024} and PyGSTi~\cite{nielsenPyGSTioPyGSTiVersion2022}. It provides a platform that allows one to evaluate all the metrics in the collection in a consistent manner, for example by using a single consistent noise model for the evaluation of all metrics. The effects of changing the noise in the model on all metrics can then be evaluated and potential correlations across metrics investigated. The software directly implements the methodology described in this document, focusing on clarity and transparency and hence avoiding optimizations that may obscure the relation between the described methodology and the software. Having such open-source software enables reproducibility of the metrics by making visible any parameters used to obtain them, which may not always be reported otherwise, as well as providing enough output data to be able to others to reproduce the calculations.

\textit{Structure of document:} The remainder of this document is divided into four parts: Part \ref{chapter:overviewmetrics} provides an overview of the chosen metrics across all categories, Part \ref{part:background} provides the background, and Part \ref{part:metrics} presents the database of metrics, and finally Part \ref{part:conclusions} provides the conclusions. 

In Part \ref{chapter:overviewmetrics} we both review the relevant literature on performance metrics and list the collection of metrics, alongside with the motivation of the choice of each metric and of their categorization. We provide a short summary for each of the metrics included in the collection and what performance related information each metric provides. At the end of this part we provide a discussion and an outlook on future research in the field as well as the proposed steps for standardization of metrics and benchmarks.

Part \ref{part:background} provides the background with all the required information for the more detailed definitions of the metrics and methodologies given in Part \ref{part:metrics}. In Sec. \ref{chapter:qc_approaches} we provide a brief overview of the different quantum computing approaches and their performance indicators. In Sec. \ref{chapter:hardwarePlatforms} we provide a description of some of the most widely used hardware platforms including the hardware-specific performance aspects. In Sec. \ref{chapter:methodologies} we give a description of the concepts and methodologies that are used to obtain various metrics presented in the document. The concepts in this part are presented in the context of quantum computing performance, and also accessible to the non-expert to equip those new to the field with the knowledge needed for understanding the technical details of the metrics presented in the database. 

In Part \ref{part:metrics} the detailed definitions of the metrics are presented. Here, the individual sections correspond to the different categories and metrics presented in part \ref{chapter:overviewmetrics}. This part contains, for each metric, a definition, a description, the measurement methodology listed as a step by step guide, assumptions and limitations, a link to the associated software, and references. This part goes together with the software provided at the \href{https://qcmet.npl.co.uk/}{QCMet software repository (qcmet.npl.co.uk)}.  

\ifbool{showpart1references}{\printbibliography[heading=subbibliography,segment=\therefsegment]}{}
\end{refsegment}

\part{Overview and review of metrics and benchmarks}
\label{chapter:overviewmetrics}
\addtocontents{toc}{\protect\setcounter{tocdepth}{1}}
\begin{refsegment}

\chapter{Motivation of metrics selection and categorization}
\label{chapter:metrics_categories}

The drive to develop quantum computers stems from their potential to be able to perform computations significantly faster than classical computers, or even to perform computations that are impossible to perform on classical computers. A number of performance metrics used to benchmark classical computers can therefore also directly be applied to quantum computers, such as the speed of operations, the number of processing units, as well as the probability of errors to occur in the computation. This last aspect is particularly important in emerging quantum computing hardware, where the error probability is still too large to perform very large computations for real-world applications. 

The collection of metrics selected in this document is designed to comprehensively benchmark quantum computers across all relevant performance criteria. The metrics are grouped in five categories, each of which aims to answer questions related to the performance, which combined allow obtaining a comprehensive evaluation of the performance of a device:  
\begin{itemize}
    \item {\bfseries M1. \nameref*{chapter:hardware_architecture_properties}:} What are the device's architectural properties and what operations is it able to perform?
    \item {\bfseries M2-5. Quality metrics:} How accurate are the outputs of computations on the device?
    \item {\bfseries M6. \nameref*{chapter:speed_metrics}:} How fast are operations and algorithms executed on the device?
    \item {\bfseries M7. \nameref*{chapter:stability_metrics}:} How stable and reliable is the device over time?
    \item {\bfseries M8-10. Non-gate-based quantum computer metrics}: What algorithms are non-gate-based quantum computers targeting and what is their performance?
\end{itemize}
Reporting the values of metrics in categories M1-7 allows obtaining a comprehensive performance evaluation of a device for general quantum computers; the categories M8-10 on the other hand only apply to specific non-gate-based quantum computing architectures.

A large number of metrics are dedicated to measuring the quality of the device. This is because measuring quality is a complex task which can be performed at different levels of abstraction, ranging from the intrinsic quality of the qubits as components of the quantum hardware to the quality of output of entire applications. The quality metrics are therefore divided into the following set of categories:
\begin{itemize}
    \item {\bfseries M2.~\nameref*{chapter:qubit_quality_metrics}:} How much are idle qubits affected by errors?
    \item {\bfseries M3.~\nameref*{chapter:gate_quality_metrics}:} How accurately can operations on qubits, denoted as gates, be performed?
    \item {\bfseries M4.~\nameref*{chapter:circuit_quality_metrics}:} How accurately can general quantum circuits be executed?
    \item {\bfseries M5.~\nameref*{chapter:well_studied_tasks}:} How accurately can real-world applications be executed?
\end{itemize}

There is a variety of different non-gate-based quantum computing technologies. In this article performance metrics for three widely studied types of non-gate-based quantum computing approaches are provided:
\begin{itemize}
    \item {\bfseries M8.~\nameref*{chapter:annealers}:} How can one characterize the performance of quantum annealers?
    \item {\bfseries M9.~\nameref*{chapter:boson_sampling}:} How can one characterize the performance of boson sampling devices?
     \item {\bfseries M10.~\nameref*{chapter:neutral_atoms}:} How can one characterize the performance of neutral atom devices for analogue simulation?
\end{itemize}
    
\begin{figure*}[!htb]
    \centering
    \includegraphics[width=\textwidth]{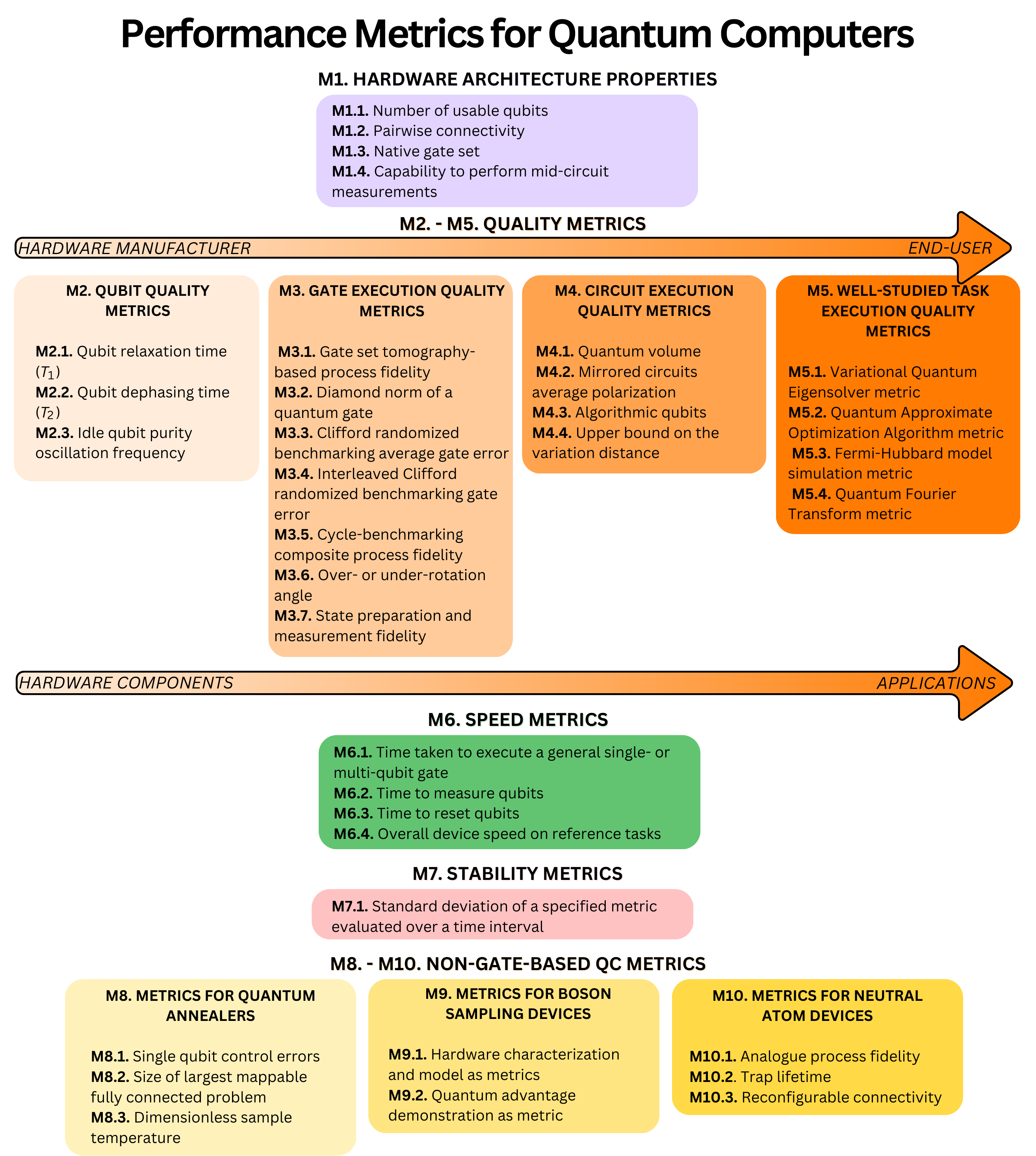}
    \caption{Schematic of the collection of metrics presented in this document, with each category illustrated in a different color.}
    \label{fig:metrics_schematic}
\end{figure*}

The full set of metrics is schematically illustrated in Fig.~\ref{fig:metrics_schematic}. In the remainder of this section we provide a summary and overview of the literature for each metric in a category, motivating the relevance of a metric and explaining which performance criteria it addresses. While the metrics are intended to benchmark physical qubits, when error-corrected quantum computers are available~\cite{steane1998quantum, horowitz2019quantum, A_Yu_Kitaev_1997, doi:10.1080/00107514.2019.1667078, devitt2013quantum, lidar2013quantum, preskill1998reliable, knill1997theory, gottesman1997stabilizer, gottesman2010introduction}, most of the metrics are directly applicable to the logical qubits generated within quantum error correcting codes as well. For example, the fidelity of a qubit operation is applicable to both a physical and a logical qubit.

\chapter{Short descriptions of the comprehensive set of metrics and benchmarks}
\label{chapter:metrics_summary}

\let\theoldsection\thesection
\def\thesection{\ref*{chapter:hardware_architecture_properties}}
\section*{M1. \nameref*{chapter:hardware_architecture_properties}}
The hardware architecture properties determine general capabilities of the hardware. In this document a set of properties is specified that determines general device performance across hardware platforms. Individual hardware platforms may specify additional architectural device specific architectural properties.\\ 

\noindent\fullsechyperref{sec:number_of_usable_qubits}
\begin{adjustwidth}{0.5cm}{}
One of the key metrics that determines whether or not a given quantum computer can be used for a particular task is the number of qubits available in the device~\cite{nielsen2002quantum, ladd2010quantum,wurtz2023aquila,Preskill2018QuantumCI,ladd2010quantum, steane1998quantum}. Thus, the number of usable qubits is a key metric that quantifies the capabilities of the hardware~\cite{horowitz2019quantum, corcolesChallengesOpportunitiesNearTerm2020, Arute2019, debnath2016demonstration, wrightBenchmarking11qubitQuantum2019,IBMUnveils400, kielpinski2002architecture, google2020hartree,cirac2000scalable,wackQualitySpeedScale2021,gambetta2020ibm, preskill2023quantum}. \\
\end{adjustwidth}

\noindent\fullsechyperref{sec:connectivity}
\begin{indentmetric}
To be able to perform complex computations across the whole quantum computer, one needs to ensure that the states of individual qubits can be entangled across different qubits~\cite{divincenzo2000physical}. Entanglement across different qubits is the property of a quantum system, where the state of the full quantum system across multiple qubits cannot be described by the states of individual qubits alone. Entanglement across qubits is generated by applying multi-qubit gates, typically two-qubit gates. 
The pairwise connectivity of qubits is specified by a list of pairs of qubits between which entangling operations can be applied directly. Qubits can also be entangled indirectly by swapping the states of qubits, then applying the direct entangling operation between a pair of directly connected qubits, and then swapping the states of qubits back. The swapping operations can be done either by physically moving around the qubits~\cite{ion:progress-review, moses2023race, bluvstein22,bluvstein2024logical} or by applying entangling operations that swap the information in the qubits~\cite{o2019generalized,gokhale2021faster,wille2014optimal}. The overhead of swapping qubits can be large~\cite{linkeExperimentalComparisonTwo2017, Holmes_2020, herbert2018depth, almudever2020realizing}, leading to longer execution times and worse performance due to noise. Thus, a high degree of connectivity can be very advantageous. Knowledge of the pairwise qubit connectivity can be used to optimize quantum circuits in order to improve performance~\cite{cowtan2019qubit,lin2022domain, Zulehner2019, li2019tackling, huang2023near, serrano2022quantum, siraichi2018qubit, sivarajah2020t}. Thus, the pairwise connectivity of a device is an important architectural property~\cite{10.1145/3307650.3322273,linkeExperimentalComparisonTwo2017, wrightBenchmarking11qubitQuantum2019, crainHighspeedLowcrosstalkDetection2019,kim2023evidence,Arute2019,crossValidatingQuantumComputers2019}.\\
\end{indentmetric}

\noindent\fullsechyperref{sec:native_gate_set}
\begin{indentmetric}
For a quantum computer to be universal it needs to be able to perform arbitrary unitary operations~\cite{divincenzo2000physical, nielsen2002quantum}. Any multi-qubit unitary operation can be decomposed into a circuit consisting of general single-qubit and two-qubit gates~\cite{shende2005synthesis, nielsen2002quantum}. It is important to determine the gates that the hardware can perform natively~\cite{10.1145/3307650.3322273,amy2013meet}, and whether these can be combined to generate general single-qubit gates and a two-qubit gate allowing for universal computation. The set of these  native gates is denoted as the native gate set. Quantum circuits for a specific hardware can be designed to predominantly use the native gates, thereby avoiding more general gates that can require expensive decompositions in terms of number of native gates~\cite{sivarajah2020t,khatri2019quantum,shi2020resource,alam2020,mccaskey2018hybrid,miller2011elementary}. Decomposing general operations into native gates can introduce a significant overhead~\cite{10.1145/3307650.3322273,khatri2019quantum, haner2018software}. \\
\end{indentmetric}

\noindent\fullsechyperref{sec:capability_mid_circuit_measurements}
\begin{indentmetric}
Quantum algorithms can utilize qubit re-use in order to improve the resource requirements of the algorithms~\cite{yirka2021qubit, yalovetzky2021hybrid,chertkov2022holographic, chertkov2023characterizing,foss2021holographic,PhysRevX.13.041057}. Qubit re-use involves measuring and then resetting a qubit multiple times within a single quantum circuit execution. Thus, the capability to perform mid-circuit measurements determines whether such quantum algorithms can be run on a device. This capability is also important because it is required to perform quantum error correction \cite{devitt2013quantum} and for applying gates conditioned on previous measurement results~\cite{kitaev1995quantum,monz2016realization,griffiths1996semiclassical,parker2000efficient,mosca1998hidden}. \\
\end{indentmetric}

\def\thesection{M2-5}
\section*{M2-5. Quality metrics}
Compared to existing classical computers the current generation of quantum computers is more error-prone, limiting the usefulness of the devices. The possibility to apply quantum error correction also requires error rates to be below some code dependent thresholds. The quality metrics are designed to quantify the error rates of quantum computing hardware, which is achieved by quantifying various differences of the outputs of a device when compared to the outputs expected for an ideal noise-free quantum computer. In evaluating the performance of a quantum computer, quantifying the quality is one of the most challenging aspects~\cite{Preskill2018QuantumCI}, since there are various different sources of errors, and these sources can be different for different hardware platforms.

\subsection*{\ref*{chapter:qubit_quality_metrics}. \nameref*{chapter:qubit_quality_metrics}}
\label{sec:overview_qubit_quality}
Qubits are not isolated from their environment, and their interaction with the environment can significantly perturb their quantum state, potentially leading to a full loss of information on the state of the qubit~\cite{zurek1991decoherence, schlosshauer2019quantum,bennett2000quantum, bouwmeester2000physics,brandt1999qubit, siddiqi2021engineering, nielsen2002quantum,preskill1998lecture}. The effects of the environment are generally described by various types of noise acting on the qubit.\\

\noindent\fullsechyperref{sec:t1} and \fullsechyperref{sec:t2}
\begin{indentmetric}
Two dominant sources of noise due to interactions with the environment are amplitude damping (also denoted as relaxation), and dephasing: amplitude damping occurs due to the qubit losing energy to the environment and changing its state, while dephasing predominantly occurs due to fast fluctuations of the qubit frequency due to the environment interaction. The qubit relaxation time, $T_1$, and the qubit dephasing time, $T_2$, quantify the effects of amplitude damping and dephasing, respectively~\cite{nielsen2002quantum, preskill1998lecture,krantz_quantum_2019, burnett2019decoherence,mcdermott2009materials,chirolli2008decoherence,Klimov_2018,carroll2022dynamics,ion:two-qubit1,plenio1997decoherence,kielpinski2001entanglement,ion:progress-review,graham2022multi,}. 
These metrics quantify the major noise contributions that typically determine the duration over which a qubit preserves its state and stays coherent, which is denoted as the coherence time. They are therefore central metrics that quantify the quality of single qubits.
Longer coherence times allow running more operations on the qubit before the results are significantly affected by decoherent noise such as amplitude damping and dephasing, which reduce the qubit coherence. 

The $T_1$ and $T_2$ taken in isolation provide only a partial picture for qubit performance in quantum circuits, since what determines the number of gates that can be successfully applied is the ratio between these times and the duration of a quantum gate. Different hardware platforms can have gate durations that differ by orders of magnitude~\cite{linkeExperimentalComparisonTwo2017, bluvstein2024logical,moses2023race,kim2023evidence, google2023suppressing}. For example, ion-trap based devices typically have orders of magnitude longer coherence times than devices such as superconducting qubits, but also typically have much longer gate durations~\cite{linkeExperimentalComparisonTwo2017}.\\ 
\end{indentmetric}

\noindent\fullsechyperref{sec:amount_of_1q_nonmarkovian_noise}
\phantomsection
\label{sec:overview_idle_qubit_purity_oscillation}
\begin{indentmetric}
In quantum computing hardware, the qubit-environment interaction typically changes over time, both over fast and slow time-scales. If the time-scale of such variations is comparable to the circuit execution time, then the noise affecting the qubits can vary significantly during one execution of a quantum circuit. In such cases the noise due to the environment cannot be accurately modelled as Markovian noise, which assumes time-independence of the noise. Instead, one must account for the non-Markovianity due to the changing environment~\cite{Rivas_2014,breuer2002theory}. The idle qubit purity oscillation frequency is a metric that indicates the presence of non-Markovian effects in the time evolution of a qubit~\cite{agarwal2023modelling, gulacsi2023smoking,maniscalco2006non}. Various other methodologies for characterizing more general non-Markovian noise have been proposed~\cite{white2020demonstration,white2022non, su2023characterizing,luchnikov2020machine,li2024non,PRXQuantum.2.040351}. However, they typically require running a large number of circuits as part of the characterization.\\
\end{indentmetric}

\subsection*{\ref*{chapter:gate_quality_metrics}. \nameref*{chapter:gate_quality_metrics}}

The quality of individual gates that are applied is determined by different factors such as qubit decoherence~\cite{schlosshauer2019quantum,burnett2019decoherence,plenio1997decoherence, reschBenchmarkingQuantumComputers2021}, miscalibrations~\cite{PhysRevA.82.042339,PhysRevA.96.022330,wittler2021integrated,SheldonIterativeRandomizedBenchmarking,wallman2015estimating,shapira2018robust,PhysRevLett.113.220501}, or errors such as leakage~\cite{werninghaus2021leakage,motzoi2009simple, chen2016measuring,wood2018quantification,hayes2020eliminating,negnevitsky2018repeated,levine2019parallel,kjaergaard2020superconducting,xia2015randomized, ion:progress-review} and crosstalk~\cite{krinner2020,sarovar2020detecting,mundada2019,heinz2021,piltz2014trapped,mehta2016integrated,pogorelov2021compact,zhukas2021high,gaebler2021suppression,tripathi2022suppression,sheldon2016procedure,xia2015randomized,saffman2019quantum,wintersperger2023neutral,zhao2022quantum,zhou2023quantum,parrado2021crosstalk,ion:osc-grad,urban2009observation,levine2018high}. The metrics included in category \fullref{sec:overview_qubit_quality} also quantify some of the dominant sources of errors that affect the quality of the gates \cite{corcoles2014processverificationoftwoqubit,huang2019fidelity,Marxer2023longdistancetransmoncoupler,google2023suppressing}. Often, there is a trade-off between the gate control parameters: longer gate durations can allow reducing the effects of leakage but come at the cost of increased qubit decoherence during gate execution~\cite{PhysRevA.96.022330,motzoi2009simple}.

The effects of these errors are that the state resulting from the application of a gate is different from the ideal expected state in the absence of errors. The metrics included in this category are designed to quantify the overall difference between the ideal noise-free gate and the gate executed on hardware. The metrics do not directly relate gate execution qualities to the individual noise sources, which would require accurate modelling of the noise in the devices and its environment, which is a challenging task. Instead, they provide averaged effects on the gate operations generated by such noise. Since the level and type of averaging is not unique, there are many metrics for gate quality.

The noisy gates are not described by fully unitary quantum operations due to the effects of decoherence.
Characterizing these effective non-unitary operations can be a challenging task~\cite{nielsenGateSetTomography2021}. One method which allows obtaining such descriptions of the effective quantum processes is known as quantum process tomography~\cite{chuang1997prescription, dariano2001quantum,poyatos1997complete,altepeter2003ancilla,nielsenGateSetTomography2021,childs2001realization,bialczak2010quantum,bendersky2008selective}. Process tomography consists of determining the quantum states resulting from the application of the target gate on a set of input states that spans the space of all possible input states. The quantum states are measured using quantum state tomography~\cite{mohseni2008quantum,leibfried1996experimental,banaszek2013focus,blume2010optimal,christandl2012reliable,granade2016practical,haah2016sample,hradil1997quantum,james2001measurement,smolin2012efficient,nielsenGateSetTomography2021,}, which involves measurements of the quantum state in different bases, such that the measurements can be used to reconstruct the quantum state. 
This method is sensitive to errors such as state preparation and measurement errors, limiting its usefulness~\cite{merkel_self-consistent_2013, mohseni2008quantum,nielsenGateSetTomography2021}. A variant of process tomography that takes into account state preparation and measurement (SPAM) errors is
Gate Set Tomography (GST)~\cite{merkel_self-consistent_2013,greenbaum_introduction_2015,nielsenGateSetTomography2021,blume2017demonstration,blume2013robust,dehollain2016optimization,Nielsen_2020,cao2022efficient}. It involves solving an optimization problem that characterizes the quantum gates, state preparation errors, and readout errors simultaneously. 
GST requires running a large number of circuits and is time-consuming, and so is only feasible for few-qubit operations. Once one has obtained a description of the quantum process using GST, one can quantify the difference with the ideal process. There are various metrics that can be used to quantify the difference~\cite{gilchrist2005distance,white2007measuring,NIELSEN2002249,PhysRevA.60.1888}.
A fundamental assumption in GST is that the noise is constant over the measurement acquisition time, or more generally that the noise is Markovian. This is often not the case in quantum computing hardware.
The assumption of Markovian noise can be a significant shortcoming in devices with a large amount of non-Markovian noise, and other methods which can account for the temporal or system-environment correlations causing the non-Markovianity must be used in these cases. Some examples include process tensor tomography~\cite{white2020demonstration}, non-Markovian quantum process tomography~\cite{white2022non}, non-Markovian gate set tomography~\cite{li2024non}, fast Bayesian tomography~\cite{su2023characterizing}, and machine learning based methods~\cite{guo2020tensor,luchnikov2020machine}. The drawback of these methods is that they typically require running a larger number of circuits and solving more challenging classical optimisation problems.

In what follows we describe the metrics included in this category, some of which are based on GST.\\

\noindent\fullsechyperref{sec:process_fidelity}
\begin{indentmetric}
 The process fidelity, also known as entanglement fidelity, quantifies the deviation of the gate executed on hardware from the ideal noiseless gate, and can be calculated using GST~\cite{NIELSEN2002249,proctorMeasuringCapabilitiesQuantum2022,gilchrist2005distance, schumacher1996sending}.
 It is related to the average fidelity of a quantum gate~\cite{NIELSEN2002249,proctorMeasuringCapabilitiesQuantum2022,PhysRevA.60.1888,fortunato2002implementation}, which quantifies the average closeness, over all possible pure input states, between the states resulting from the application of the actual and the ideal process~\cite{bowdrey2002fidelity,NIELSEN2002249, gilchrist2005distance}.
\end{indentmetric}

\noindent\fullsechyperref{sec:diamond_norm}
\begin{indentmetric}
In contrast to the average fidelity, which averages over all initial states, the diamond norm~\cite{aharonov1998quantum,blume2017demonstration,gilchrist2005distance} is the minimum trace distance, over all possible initial states, of the states resulting from the application of the actual and ideal process. Thus, this metric measures the maximum distinguishability of the ideal and actual process. Such worst-case measures are especially relevant when comparing fidelities against fault-tolerant thresholds~\cite{PhysRevLett.117.170502,blume2017demonstration,aharonov1997fault,aharonov1998quantum}.\\
\end{indentmetric}

\noindent\fullsechyperref{sec:RB}
\begin{indentmetric}
The process fidelity, average fidelity, and the diamond norm can all quantify the quality of a specific gate. However, in a quantum circuit, one typically uses a wide set of gates, so that for a performance indicator for a quantum computer to execute general quantum circuits one may consider an averaged fidelity over a set of gates included in typical circuits. Clifford Randomized Benchmarking (RB) is a method that allows calculating such an average error rate, where the considered set of operations are the $N_\mathrm{q}-$qubit Clifford operations~\cite{emerson2005scalable,kimmel2015robust,SheldonIterativeRandomizedBenchmarking,emerson2007symmetrized,knillRandomizedBenchmarkingQuantum2008,magesanScalableRobustRandomized2011,proctor2017randomized,magesanEfficientMeasurementQuantum2012,magesanCharacterizingQuantumGates2012, huangFidelityBenchmarksTwoqubit2019}. The advantage of RB is that it does not require expensive tomographic methods to find the effective operators. It instead runs random circuits and evaluates the probability of obtaining the correct output. The error rate is estimated from the exponential decay of this success probability with the circuit length. This error rate corresponds to the gate fidelity averaged over all possible input states and all $N_\mathrm{q}-$qubit Clifford gates~\cite{helsen2022general}. Note that this only holds if the noise in the quantum computer is completely Markovian and gate independent\cite{helsen2022general,proctor2017randomized}. It is therefore important to also quantify the amount of non-Markovianity of the noise, for example using the the metric \hyperref[sec:overview_idle_qubit_purity_oscillation]{M2.3 Idle qubit purity oscillation frequency}.\\
\end{indentmetric}

\noindent\fullsechyperref{sec:1qinterleaved_rb}
\begin{indentmetric}
Variants of RB have been proposed and used to quantify the quality of individual operations. Interleaved randomized benchmarking (IRB)~\cite{magesanEfficientMeasurementQuantum2012, PhysRevLett.108.260503,helsen2022general,SheldonIterativeRandomizedBenchmarking,onorati2019randomized,harper2017estimating} is one such protocol that can be utilized to estimate the average error of individual Clifford gates, or even individual non-Clifford gates with some variants~\cite{harper2017estimating,helsen2022general}. In this approach one interleaves multiple repetitions of the gate of interest within the RB circuits, so that the difference between the fidelity of the RB circuits and the interleaved circuits allows estimating the fidelity of the interleaved gate. Other methods that have been proposed to characterize the quality of individual operations include random circuit sampling~\cite{liuBenchmarkingNeartermQuantum2022} and direct randomized benchmarking~\cite{proctorDirectRandomizedBenchmarking2019}. Methods based on RB can also used to evaluate the quality of mid-circuit measurements~\cite{goviaRandomizedBenchmarkingSuite2022}. For an overview of different variants of RB, see Ref.~\cite{helsen2022general}.\\
\end{indentmetric}

\noindent\fullsechyperref{sec:cycle_benchmarking}
\begin{indentmetric}
Running GST to obtain the process fidelity of a quantum gate can be costly in terms of the number of circuits required to run, and the amount of time spent on classical optimization. Cycle benchmarking~\cite{erhardCharacterizingLargescaleQuantum2019a,morvan2021qutrit, zhang2024generalized, hashim2023benchmarking, mitchell2021hardware, Hashim2021PRX} has been proposed as a way to more efficiently measure the process fidelity of a quantum gate, or a layer of quantum gates, by using randomized compiling~\cite{Wallman-EmersonPRA2016, Hashim2021PRX}. Cycle benchmarking has also been proposed to benchmark multi-qubit operations while also benchmarking their effects on other qubits. This allows for the quantification of an important kind of noise that affects multiple qubits simultaneously called crosstalk. Crosstalk describes how qubits affect each other, both in the idle state and also when gates are applied. This noise can take various forms, and various metrics have been proposed to quantify the different effects~\cite{sarovar2020detecting}. There are metrics that quantify the averaged effect of the crosstalk errors on quantum gates rather than describing the individual sources of crosstalk noise. Some examples of this are simultaneous randomized benchmarking~\cite{PhysRevLett.109.240504,magesanScalableRobustRandomized2011} or extensions of it, such as correlated randomized benchmarking~\cite{mckayCorrelatedRandomizedBenchmarking2020}.
These are variants of randomized benchmarking that can be used to characterize locality and weight of crosstalk errors. Methods such as cycle benchmarking~\cite{erhardCharacterizingLargescaleQuantum2019a,morvan2021qutrit, zhang2024generalized, hashim2023benchmarking, mitchell2021hardware, Hashim2021PRX} have been proposed to benchmark multi-qubit operations while taking into account correlated noise sources such as crosstalk.\\
\end{indentmetric}

\noindent\fullsechyperref{sec:amount_of_over_under_rotation}
\begin{indentmetric}
Limited gate calibration precision, drifts on device properties, or undesired interactions that cannot be turned off can lead to systematic or coherent errors in the device~\cite{kimmel2015robust,SheldonIterativeRandomizedBenchmarking,lazuar2023calibration,wittler2021integrated,PhysRevA.82.042339,thomas2011robustness,blume2017demonstration,berberich2024robustness,trout2018simulating,vepsalainen2022improving}. Single-qubit unitary gate operations can be represented as rotation gates on the Bloch sphere. One measure of the quality of such gates is how accurately these rotations are implemented. Control errors can lead to over-rotation or under-rotation errors, where the actual rotation amount corresponding to the operation differs from the desired rotation. Characterizing these kinds of errors often involves repeated application of these gates in order to amplify the errors before measuring their effects~\cite{SheldonIterativeRandomizedBenchmarking, lazuar2023calibration}.\\
\end{indentmetric}

\noindent\fullsechyperref{sec:spam_fidelity}
\begin{indentmetric}
Aside from the typical unitary operations applied to the qubits, it is also important to quantify the quality of the state preparation and measurement  (SPAM)~\cite{chen2019detector,geller2021toward,bravyi2021mitigating,fiuravsek2001maximum,keith2018joint,sun2018efficient,chow2010detecting,dicarlo2010preparation,PhysRevLett.113.220501,ion:readout,landa2022experimental}. 
Similar to how quantum state tomography can be used to characterize quantum states, quantum measurement tomography~\cite{fiuravsek2001maximum,lundeen2009tomography,maciejewski2020mitigation,luis1999complete,blumoff2016implementing} can be used to characterize the native measurement operations. However, state tomography assumes an ideal readout, and measurement tomography assumes ideal state preparation. Thus, when both state preparation and readout errors are non-negligible, it is challenging to directly characterize the two processes independently without other assumptions, such as ideal qubit gates~\cite{lin2021independent}. Since GST involves the characterization of the qubit gates together with SPAM errors, it circumvents the issue of characterizing operations independently by jointly characterizing the operations during the optimization procedure~\cite{nielsenGateSetTomography2021}. The SPAM fidelity can therefore be obtained using the GST results.\\
\end{indentmetric}

\noindent The methods discussed above can differ significantly in the required execution time. For example, RB based methods involve running random operations such as the application of random Clifford gates. On the other hand, evaluating the diamond norm typically requires a complete description of the process, which can be obtained by performing GST. Typically, GST requires running more circuits than RB methods, and involves a solving a much more difficult classical optimization problem, so that it takes significantly longer to run. The longer execution times of GST also reduce its ability to provide time-resolved fluctuations of metrics. On the other hand, the advantage of GST is that it provides a much more detailed characterization the quantum processes, so that the results of GST can be used to evaluate multiple metrics unlike RB based methods. Thus, when choosing which metrics to measure, one must consider factors such as the number of quantum circuits that need to be executed and the classical computational time overhead.

\subsection*{\ref*{chapter:circuit_quality_metrics}. \nameref*{chapter:circuit_quality_metrics}}

The effect of individual gate errors on the full circuit execution result is difficult to estimate, since each quantum circuit amplifies individual gate errors in a different way. The quality estimates of individual quantum gates obtained with the metrics in category \ref{chapter:qubit_quality_metrics} can therefore only give a very approximate indication of how well a device executes entire quantum circuits. Instead, to evaluate the quality of circuit execution one can use metrics that characterize the quality of execution of entire circuits directly~\cite{proctorScalableRandomizedBenchmarking2021,crossValidatingQuantumComputers2019, blume-kohoutVolumetricFrameworkQuantum2020, millsApplicationMotivatedHolisticBenchmarking2021}.\\

\noindent\fullsechyperref{sec:quantum_volume}
\begin{indentmetric}
A single-number metric that aims to quantify the performance of a device in running certain kinds of quantum circuits is the quantum volume~\cite{crossValidatingQuantumComputers2019,baldwin2022re,wackQualitySpeedScale2021,pelofske2022quantum,jurcevic2021demonstration}. The metric is calculated by running randomized square circuits, where the number of circuit layers are equal to the number of qubits, for increasing numbers of qubits. The quantum volume is evaluated by finding the largest square circuit that can be run on the device and that also passes an acceptance criterion that quantifies the quality of the output. Importantly, this requires classical simulation of the quantum circuits, which limits scalability. A further drawback of quantum volume is that it is not necessarily indicative of the performance of a device in running circuits which are not square, such as circuits which have depths that scale non-linearly with the number of qubits. Extensions to quantum volume that generalize from square circuits to rectangular circuits with different scaling of widths and depths have been proposed~\cite{blume-kohoutVolumetricFrameworkQuantum2020, millerImprovedVolumetricMetric2022}.\\
\end{indentmetric}

\noindent\fullsechyperref{sec:mirror_circuits}
\begin{indentmetric}
A method that can be used to perform characterization of entire circuits without needing to perform expensive classical simulations is mirror benchmarking~\cite{proctorScalableRandomizedBenchmarking2021, mayerTheoryMirrorBenchmarking2021,proctorMeasuringCapabilitiesQuantum2022,amico2023defining,moses2023race}. In order to avoid the classical simulations, mirror benchmarking transforms the circuits so that their outputs are easy-to-predict and have simple measurement outcomes. The main idea of mirror benchmarking involves appending the circuit of interest with its inverse mirror copy, with a layer of random single qubit Pauli operations in between. Specific noise contributions can cancel out when the inverse mirror copy is applied, thus a set of circuits with different random Pauli operations must be used to ensure that these noise contributions are not cancelled out. The known ideal output of the circuits involved in these benchmarking methods makes the results easily verifiable, hence making the benchmarking methods scalable to large numbers of qubits. The quality metric is the average polarization after circuit execution, which is a rescaled form of the success probability of circuit execution. This method can be used with Clifford circuits as well as general non-Clifford circuits, although in the latter case the mirrored circuit needs to modified in order to account for the layer of random Pauli operations~\cite{proctorMeasuringCapabilitiesQuantum2022,crossValidatingQuantumComputers2019, millsApplicationMotivatedHolisticBenchmarking2021}.\\
\end{indentmetric}

\noindent\fullsechyperref{sec:algorithmic_qubits}
\begin{indentmetric}
Ref~\cite{blume-kohoutVolumetricFrameworkQuantum2020} generalizes quantum volume and provides a framework for quantifying how well a device executes circuits with different depths and widths. This is denoted as volumetric benchmarking. The algorithmic qubits~\cite{AlgorithmicQubitsBetter,chen2023benchmarking} benchmarking method uses circuits corresponding to six chosen quantum algorithms and combines the volumetric benchmarking results for those algorithms and outputs a single-number metric that is obtained by finding the number of qubits at which all 6 benchmarks pass a specific success criteria.\\
\end{indentmetric}

\noindent\fullsechyperref{sec:accreditation}
\begin{indentmetric}
To evaluate the quality of noisy quantum circuit execution one may use quantum accreditation protocols~\cite{Ferracin2019,Ferracin2021}, which do not involve computationally expensive classical simulation of quantum circuits. The upper bound on the variation distance is a metric within the accreditation protocol that bounds the distance between the probability distributions corresponding to a quantum circuit run on a noisy quantum computer and its ideal noiseless counterpart.\\
\end{indentmetric}

\noindent 
Given the many possible types of circuits, there are a number of other potential circuit execution quality metrics. One such example is cross-entropy benchmarking (XEB), which measures circuit execution quality by comparing the outputs of random quantum circuits and then calculating how likely the measurement outcomes would be if measured on an ideal noiseless quantum computer~\cite{boixo2018characterizing}. 
One key limitation of circuit execution benchmarks such as quantum volume \cite{crossValidatingQuantumComputers2019} and XEB \cite{Arute2019,boixo2018characterizing,helsen2022general} is the requirement of expensive classical simulations that scale exponentially with the number of qubits. Thus, once quantum computers reach the threshold where the classical simulations become infeasible, other methods, which do not rely on exponentially-scaling simulations, need to be used.\\ 

\subsection*{\ref*{chapter:well_studied_tasks}. \nameref*{chapter:well_studied_tasks}}

Metrics that quantify the performance of a device in executing specific well-studied tasks can provide a view of the overall capabilities of a device. Such tasks can include entire applications, algorithms, and subroutines of applications. There are a number of articles that propose algorithm or application level benchmarking suites for gate-based quantum computers~\cite{lubinskiApplicationOrientedPerformanceBenchmarks2021,liQASMBenchLowlevelQASM2022a,patelExperimentalEvaluationNISQ2020,kochDemonstratingNISQEra2020,reschBenchmarkingQuantumComputers2021,georgopoulosQuantumComputerBenchmarking2021,quetschlichMQTBenchBenchmarking2022,10.1145/3307650.3322273,dallaire-demersApplicationBenchmarkFermionic2020,mccaskeyQuantumChemistryBenchmark2019,tomeshSupermarQScalableQuantum2022,donkersQPackScoresQuantitative2022,linkeExperimentalComparisonTwo2017,wrightBenchmarking11qubitQuantum2019,martielBenchmarkingQuantumCoprocessors2021,liaoBenchmarkingQuantumProtocols2022,}. Most such proposed benchmarking suites quantify the performance of a quantum computer in solving specific algorithms, with the inputs being derived from target applications. In Ref.~\cite{sawaya2023hamlib} the authors present a database of qubit-based quantum Hamiltonians which can be used to benchmark quantum computers. The main differences between the various proposed benchmarking suites are both the choice of algorithms and of the metrics used to quantify the quality of outputs. The algorithms that appear commonly in these suites include the Quantum Fourier Transform~\cite{shor1997polynomial,nielsen2002quantum, lubinskiApplicationOrientedPerformanceBenchmarks2021, liQASMBenchLowlevelQASM2022a, patelExperimentalEvaluationNISQ2020, kochDemonstratingNISQEra2020, reschBenchmarkingQuantumComputers2021}, Grover's search~\cite{grover1996, georgopoulosQuantumComputerBenchmarking2021, lubinskiApplicationOrientedPerformanceBenchmarks2021, quetschlichMQTBenchBenchmarking2022, 10.1145/3307650.3322273, liQASMBenchLowlevelQASM2022a, patelExperimentalEvaluationNISQ2020}, the variational quantum algorithm (VQE)~\cite{TILLY20221, dallaire-demersApplicationBenchmarkFermionic2020, mccaskeyQuantumChemistryBenchmark2019, tomeshSupermarQScalableQuantum2022, donkersQPackScoresQuantitative2022, lubinskiApplicationOrientedPerformanceBenchmarks2021, quetschlichMQTBenchBenchmarking2022, liQASMBenchLowlevelQASM2022a, patelExperimentalEvaluationNISQ2020}, the Bernstein-Vazirani algorithm~\cite{Bernstein1997, linkeExperimentalComparisonTwo2017, lubinskiApplicationOrientedPerformanceBenchmarks2021, lubinskiApplicationOrientedPerformanceBenchmarks2021, 10.1145/3307650.3322273, liQASMBenchLowlevelQASM2022a, patelExperimentalEvaluationNISQ2020, wrightBenchmarking11qubitQuantum2019}, the Hidden-shift problem~\cite{vanDamQuantumAlgorithmsHiddenShiftProblems, linkeExperimentalComparisonTwo2017, lubinskiApplicationOrientedPerformanceBenchmarks2021, 10.1145/3307650.3322273, liQASMBenchLowlevelQASM2022a, wrightBenchmarking11qubitQuantum2019}, the Deutsch-Jozsa algorithm~\cite{lubinskiApplicationOrientedPerformanceBenchmarks2021, quetschlichMQTBenchBenchmarking2022, liQASMBenchLowlevelQASM2022a, patelExperimentalEvaluationNISQ2020}, Shor's period finding algorithm\cite{lubinskiApplicationOrientedPerformanceBenchmarks2021, quetschlichMQTBenchBenchmarking2022, liQASMBenchLowlevelQASM2022a}, Quantum Phase Estimation~\cite{kitaev1995quantum, georgopoulosQuantumComputerBenchmarking2021, lubinskiApplicationOrientedPerformanceBenchmarks2021, quetschlichMQTBenchBenchmarking2022, liQASMBenchLowlevelQASM2022a, patelExperimentalEvaluationNISQ2020}, the Quantum Approximate Optimization Algorithm (QAOA)~\cite{finzgarQUARKFrameworkQuantum2022, martielBenchmarkingQuantumCoprocessors2021, tomeshSupermarQScalableQuantum2022, quetschlichMQTBenchBenchmarking2022, liQASMBenchLowlevelQASM2022a, patelExperimentalEvaluationNISQ2020, reschBenchmarkingQuantumComputers2021}, amplitude estimation~\cite{lubinskiApplicationOrientedPerformanceBenchmarks2021, quetschlichMQTBenchBenchmarking2022, liQASMBenchLowlevelQASM2022a}, and quantum communications~\cite{liaoBenchmarkingQuantumProtocols2022, eisertQuantumCertificationBenchmarking2020, zhukovQuantumCommunicationProtocols2019}. Other algorithms and applications that have been considered include quantum singular value transform~\cite{dongRandomCircuitBlockencoded2021}, error correcting circuits~\cite{michielsenBenchmarkingGatebasedQuantum2017}, encoding and decoding of cat states~\cite{knillAlgorithmicBenchmarkQuantum2000}, and the Harrow–Hassidim–Loyd (HHL) algorithm~\cite{cornelissenScalableBenchmarksGateBased2021}. One may also consider general variational quantum machine learning algorithms, which may be somewhat tolerant to noise in the hardware.

The fidelity between the output on an ideal noise-free quantum computer and the output on hardware is commonly used as the metric to quantify performance, typically averaged over a set of chosen circuits. It requires knowing the exact solution, and thus limits the scalability of these benchmarking methods~\cite{lubinskiApplicationOrientedPerformanceBenchmarks2021}. One may also restrict the tasks to problems that can be solved efficiently classically also for large system sizes. However, performance of a device on simple problems or circuits may not be indicative of the performance of the device for complicated problems or circuits~\cite{lubinskiApplicationOrientedPerformanceBenchmarks2021}.
Other kinds of metrics that have been considered include runtime~\cite{donkersQPackScoresQuantitative2022}, largest circuit that can be run successfully~\cite{AlgorithmicQubitsBetter, dallaire-demersApplicationBenchmarkFermionic2020, donkersQPackScoresQuantitative2022}, and application specific metrics~\cite{finzgarQUARKFrameworkQuantum2022, mccaskeyQuantumChemistryBenchmark2019, martielBenchmarkingQuantumCoprocessors2021}. In Ref.~\cite{tomeshSupermarQScalableQuantum2022, liQASMBenchLowlevelQASM2022a} the authors compare the different algorithms used in benchmarking methods and evaluate what aspects of device performance the methods measure. \\

Four of the well-studied tasks are considered as metrics in this category, and are described in what follows.\\

\noindent\fullsechyperref{sec:vqe}
\begin{indentmetric}
On near term devices the rather large error rates only allow short quantum circuits to run successfully. A representative class of such circuits are those used within the VQE to prepare a wavefunction on the quantum computer, which allows obtaining the ground state energy of a given Hamiltonian, and which is a key task in many types of applications~\cite{TILLY20221}. VQE is therefore widely used on current noisy devices, typically in chemistry and materials science applications~\cite{TILLY20221, peruzzo2014variational,kandala2017hardware,google2020hartree,OMalley2016Scalable,nam2020ground,Hempel2018Quantum,Shen2017Quantum}.\\
\end{indentmetric}

\noindent\fullsechyperref{sec:Q-Score}
\phantomsection
\label{sec:overview_qscore}
\begin{indentmetric}
Another important type of applications for quantum computers is in the area of optimization.
Representative examples are combinatorial optimization problems, where one aims to find an optimal object out of a set of objects. The quantum approximate optimization algorithm (QAOA)~\cite{farhi2014quantum,zhou2020qaoa,blekos2024review} targets such problems. It can also be executed at low circuit depths, which makes it suitable for noisy devices, similarly to VQE. An advantage of using QAOA for benchmarking is that it can be run using both gate-based and non-gate-based quantum computing approaches, such as quantum annealing devices~\cite{PhysRevX.10.021067,chancellor2019domain, blekos2024review}. The so-called Q-Score~\cite{martielBenchmarkingQuantumCoprocessors2021} is a representative example for quantifying the performance of a device in executing QAOA, and is used as metric in this document.\\
\end{indentmetric}

\noindent\fullsechyperref{sec:hubbard_model}
\begin{indentmetric}
The 1-D Fermi-Hubbard model is exactly solvable using the Bethe ansatz, and the outputs of methods involving the simulation of that system can be verified~\cite{dallaire-demersApplicationBenchmarkFermionic2020}. The quality of the computation of this task on a quantum computer therefore provides a scalable benchmark. Solution of such Hamiltonian simulation problems is part of the set of applications where quantum advantage may be found. It is challenging to run on noisy devices, and fault tolerance may be required to achieve quantum advantage for Hamiltonian simulation.\\
\end{indentmetric}

\noindent\fullsechyperref{sec:qft}
\begin{indentmetric}
The quantum Fourier transform (QFT) is an algorithm that forms a central role in many applications that are expected to provide quantum advantage, for example for the factorization of integers within Shor's algorithm. The quality of the a QFT output on a device is an indicator for the performance of the device for such applications. It typically requires larger circuit depths for its execution, so that it is challenging to run successfully on noisy devices, and hence representative of the device performance towards quantum advantage.
\end{indentmetric}

\def\thesection{\ref*{chapter:speed_metrics}}
\section*{M6. \nameref*{chapter:speed_metrics}}
\label{sec:generalSpeed}

Device speeds can vary over orders of magnitude, depending on the device hardware platform~\cite{linkeExperimentalComparisonTwo2017}.  
Variations in the speed over orders of magnitude can therefore make the difference between an algorithm finishing in a day or within many years of runtime. Thus, the speed of operations is an important consideration for the achievement of practical quantum advantage.

There are many contributions to the total time taken to run a particular algorithm on a quantum computer, and the magnitude of the times taken for the individual contributions vary across hardware platforms. These contributions include not only the actual time to run the quantum circuits, but also delay times between circuit execution, compilation and other pre-processing times, data transfer times between control equipment, and other similar times~\cite{wackQualitySpeedScale2021, mesman2021qpack}.\\

One of these contributions is the time taken to run the quantum circuit itself, which can also be further decomposed into the individual times taken to run specific operations that constitute the circuit. The provision of these individual times therefore allows estimating runtimes for general circuits on a given hardware. These contributions consist of the following.\\

\noindent\fullsechyperref{sec:1q_gate_speed}
\phantomsection
\label{sec:overview_1q_gate_speed}
\begin{indentmetric}
Typically, the time to execute all quantum gates in a circuit forms the dominant contribution to the total time taken to run the circuit. However, different quantum gates require different execution times, which are set by the duration of a pulse sent to the hardware to execute the gate. Some single-qubit gates do not need an actual pulse to be sent to the hardware, but can be performed virtually by modifying parameters of subsequent gates, so that their execution time is effectively zero~\cite{PhysRevA.96.022330}. It is therefore useful to provide a metric that averages timings over all single qubit gates. Such a general metric is the time taken to perform a general single- or multi-qubit gate on the device, which effectively averages over all possible single- or multi-qubit gates. \\
\end{indentmetric}

\noindent \fullsechyperref{sec:timemeasure}
\label{sec:overview_timemeasure}
\begin{indentmetric}
At the end of a quantum circuit, a subset of the qubits are measured in order to extract the results of the circuit execution. The time taken to perform the measurement on each qubit is thus another contribution to the total time taken to run the quantum circuit.\\
\end{indentmetric}

\noindent\fullsechyperref{sec:timereset}
\label{sec:overview_timereset}
\begin{indentmetric}
After the execution of a quantum circuit, typically the qubits need to be reset before the execution of the next quantum circuit. This time to reset qubits contributes to the total time to run a set of quantum circuits.\\
\end{indentmetric}

\noindent\fullsechyperref{sec:timereferencetasks}
\phantomsection
\label{sec:overview_timereferencetasks}

\begin{indentmetric}
In order to evaluate how fast a quantum computer is at running particular algorithms, one can directly evaluate the time taken to run those algorithms~\cite{georgopoulosQuantumComputerBenchmarking2021,lubinski2024quantum,lubinski2023optimization,donkersQPackScoresQuantitative2022,sankar2024benchmarking}. Thus, the overall device speed on reference tasks can provide estimates of device speeds in real-world usage scenarios.\\
\end{indentmetric}

\noindent By using the gate time metrics \hyperref[sec:overview_1q_gate_speed]{M6.1-3} one can estimate the time taken to run a set of quantum circuits. To account for factors such as pre-processing times and data transfer times, rather than quantifying the contributions individually one can quantify the total time taken to run specific quantum algorithms using \hyperref[sec:overview_timereferencetasks]{M6.4}.
In practice evaluating time metrics and using them for cross-platform comparisons can be challenging~\cite{lubinskiApplicationOrientedPerformanceBenchmarks2021}. This is due to the differences in the way timing information is reported by different hardware vendors, and also due to the fact that often only limited, if any, timing information is provided. \\

One can also evaluate the rate at which a device executes layers of quantum computation~\cite{wackQualitySpeedScale2021,
donkersQPackScoresQuantitative2022}. One proposed metric for this is Circuit Layer Operations per Second (CLOPS)~\cite{wackQualitySpeedScale2021}. The CLOPS metric aims to not only factor in the time taken to perform the individual operations mentioned above, but also other contributions such as delays between circuits and classical processing times. Although such a single-number metric can be useful in comparing the speeds of different devices, it may not always be indicative of the speed of the device in running particular algorithms because different algorithms can have different kinds of circuits or different classical processing and feedback. For example, the VQE algorithm~\cite{TILLY20221} involves a feedback loop between the quantum and classical device, where the results obtained from the QC are processed classically to generate new circuits. Such an algorithm would benefit more from a device with fast communication speeds than algorithms which generate and submit all quantum circuits required for execution at once.\\

\def\thesection{\ref*{chapter:stability_metrics}}
\section*{M7. \nameref*{chapter:stability_metrics}}
\label{sec:stabilityTime}

Qubit and device properties can change significantly over different time-scales~\cite{dasgupta2021stability, proctor2020detecting,dasgupta2020characterizing,yeter2022measuring,van2023evaluating,dasgupta2024stability,alam2019addressing,agarwal2023modelling,carroll2022dynamics, vepsalainen2022improving, Klimov_2018,muller2015interacting, wan2019quantum,merkel2019magnetic, burnett2019decoherence, schlor2019correlating,de2021quantifying,wei2022measurement}. The stability over time of a device is therefore an important metric to quantify the reliability of the device as well as confidence in the obtained metrics. The typically observed changes include random fluctuations around an average value~\cite{vepsalainen2022improving,proctor2020detecting,burnett2019decoherence,schlor2019correlating}, slow drifts of the average value over time~\cite{casparis2016gatemon,yang2019silicon,ruster2016long}, as well as sudden abrupt changes of the values at random times~\cite{Klimov_2018,schlor2019correlating,de2021quantifying,casparis2016gatemon,yang2019silicon,agarwal2023modelling}.
In what follows a number of aspects that affect the methodologies to capture the different aspects of the stability of the device performance are outlined.

The changes of metrics values over time typically exhibit one of the following types of behaviours:
\begin{enumerate}
    \item Randomly distributed fluctuations around an average value~\cite{vepsalainen2022improving,proctor2020detecting,burnett2019decoherence,schlor2019correlating}: the values of a metric fluctuate randomly around the average within a continuous distribution. The standard deviation can provide a good metric for the stability in such cases, especially if the distribution is Gaussian.
    \item Random fluctuations between discrete values~\cite{Klimov_2018,schlor2019correlating,de2021quantifying,casparis2016gatemon,yang2019silicon,agarwal2023modelling}: over a given time interval a metric can fluctuate between a discrete set of values. In this case the resulting distribution over a time interval is not a single continuous distribution, but consists of sharp peaks around the possible values of the metric. While the standard deviation can still provide overall information on the fluctuations, a more detailed description of the specific peaks in the distribution can be beneficial.
    \item Drifts of the metrics over time~\cite{casparis2016gatemon,yang2019silicon,ruster2016long}: the distributions of metrics over a given time-interval obtained in points 1 and 2 above can further exhibit drifts over longer time-scales. While also such drifts can be captured for a given time interval by reporting the standard deviation, it can be beneficial to provide information on the detailed drifting behavior. 
\end{enumerate}

A key aspect in the metric fluctuations described above is the considered time scale. One can mainly distinguish three main types of time-scales:
\begin{enumerate}
    \item Time scales significantly shorter than the circuit execution time: fluctuations of metrics that occur at significantly shorter time scales than the circuit execution time are typically not measurable, and the acquired metric value effectively averages over such fast fluctuations~\cite{schneider1999decoherence,vepsalainen2022improving,muller2015interacting}.
    \item Time scales similar to the circuit execution time: for such time scales, the acquired measurements after each execution of the circuit, denoted as a shot, can vary due to the variation of the metric from one circuit execution to the next. When acquiring a metric over many, often thousands of shots, the resulting value is an average over the underlying fluctuations of the metric between different circuit executions~\cite{vepsalainen2022improving}. In such cases, an evaluation of the metric with a smaller number of shots may be performed to obtain time-resolved information on the fluctuations. Such information can provides insights on the noise sources in the device and on how they affect the execution of quantum circuits. The drawback of a reduced number of shots is a higher measurement uncertainty.
    \item Time scales much longer than the circuit execution time: these can be accurately tracked by systematically evaluating a metric at specific time intervals, which can be minutes, hours, days or months~\cite{carroll2022dynamics,dasgupta2021stability,agarwal2023modelling}.
\end{enumerate}

The type of time interval that needs to be considered varies depending on the aim. For example, for many quantum algorithms it is assumed that the qubits remain stable while the algorithm is executed. Changes in behavior over time can lead to worse algorithm performance. For example, if the over- or under-rotation error changes significantly during execution of the minimization process in a variational quantum eigensolver (VQE), convergence may not be achievable~\cite{TILLY20221}. In this case the time scale to be considered is the duration of the execution of the algorithm. For hardware manufacturers that aim to understand and eventually reduce the noise in the devices, the evaluation of the fluctuations over all the time scales is important.

Since the times taken to evaluate some of the metrics can be large, it is important to report, together with the value of each metric, the start and end times of its measurement, as that allows estimating how much the metrics have fluctuated over time while measuring the metric.\\

\noindent\fullsechyperref{sec:stdTime}
\begin{indentmetric}
A metric to capture the overall stability is the standard deviation of metrics evaluated over a specified time interval. For all types of fluctuations this provides a first estimate of their magnitude. To obtain more detailed insights into the types of instabilities of the metrics and of their origins, the methodologies to characterize the stability of the device over time need to be adapted to the specific type of time-dependence of the fluctuations. \\
\end{indentmetric}

\def\thesection{M8-10}
\section*{M8-10. Non-gate-based quantum computer metrics}
Benchmarking methods and metrics for non-gate-based quantum computers such as annealers, boson sampling devices, and neutral atom quantum simulators are less developed than for gate model quantum computers.
The main reason for this is that each of these approaches has very specific requirements, which may even vary significantly from one manufacturer to the next, so that it is more difficult to find a unique set of metrics that applies to all.
Furthermore, over many years the number of manufacturers for these methodologies was relatively small, so that there was less incentive to develop benchmarks to allow for performance comparisons.
In quantum annealers, for example, for a long time there has been only one manufacturer of large scale quantum annealing systems, D-Wave Systems Inc.\footnote{\url{https://www.dwavesys.com/}}. However, the landscape is now changing, and a number of companies are seeking to provide similar analogue solvers, for example Pasqal\footnote{\url{https://www.pasqal.com/}}, Qilimanjaro\footnote{\url{https://www.qilimanjaro.tech/}}, Orca \footnote{\url{https://orcacomputing.com/}} and Quantum Computing Inc\footnote{\url{https://www.quantumcomputinginc.com/}}. Therefore, there is a clear need for metrics for quantum annealers and other non-gate-based quantum computers. Nevertheless, there is much less literature on metrics for such devices to draw from. Hence, the sections below provide a general overview of the needs for metrics in these areas, and propose a number of metrics based on existing work. The goal of these sections is to provide a starting point for understanding how to assess non-gate-based devices, which can be used in the future development of well defined sets of metrics. We highlight this as an important area for future work. 

\subsection*{\ref*{chapter:annealers}. \nameref*{chapter:annealers}}

While formally adiabatic quantum computation is a type of universal quantum computation~\cite{Aharonov2004adiabaticUniversal}, the annealers that are currently available are non-universal machines. For such non-universal machines, there may not be a single set of metrics that is applicable to all devices. 
As a result, metrics based on demonstrations of quantum effects specifically tailored to the D-Wave devices~\cite{johnson11aManufacturedSpins,lanting14aEntanglement,boixo16aTunneling,Chancellor21SearchRange} are not usually portable to other manufacturers, since it is unlikely that other devices have the controls available to perform comparable experiments.
This category instead includes metrics which assume only basic general annealing protocols. Another option for assessing annealers is to directly measure their performance on optimization problems, such as for example done for the QUARK framework\footnote{\url{https://github.com/QUARK-framework/QUARK}}, and for the Q-Score metric\footnote{\url{https://atos.net/en/solutions/q-score}}. Such measurements also allow for cross-platform comparisons, which is explicitly the goal of the Q-Score~\cite{q-score_annealing,martielBenchmarkingQuantumCoprocessors2021,van-der-Schoot2023Q-score} (metric \hyperref[sec:overview_qscore]{M5.2}). Ref.~\cite{derbyshireRandomizedBenchmarkingAnalogue2020} generalizes randomized benchmarking methods to programmable analogue quantum simulators, and Ref.~\cite{mcgeochPrinciplesGuidelinesQuantum2019} proposes guidelines for reporting and analyzing performance of quantum annealers.

The purpose of this section is to present lower level characterization of the annealer devices. In what follows we describe the metrics for annealers.\\

\noindent\fullsechyperref{sec:annealer_single_qubit_control_errors}
\begin{indentmetric}
These are measures of the level of miss-specification of the problem on each qubit, and can be caused by calibration errors or fundamental lack of precision in the device.\\
\end{indentmetric}
\noindent\fullsechyperref{sec:annealer_largest_mappable_fully_connected_problem}
\begin{indentmetric}
This measures how large of a problem could be solved, assuming it is fully connected. This measure is distinct from number of qubits, since many annealers may not allow arbitrary connectivity and therefore incur overhead when more highly connected problems are mapped.\\
\end{indentmetric}
\noindent\fullsechyperref{sec:annealer_samp_temp}
\begin{indentmetric}
This is a measure of the statistics of the distribution returned by the device. Lower temperature implies lower energy states are found more often. This metric is built on the implicit assumption that the device is operating in a regime where the states returned follow an approximately thermal distribution.\\
\end{indentmetric}

\subsection*{\ref*{chapter:boson_sampling}. \nameref*{chapter:boson_sampling}}

Similar to other quantum computing technologies, boson sampling device performance characterisation can range from the characterization of the hardware, as traditionally done in experimental labs, to quantification of how complex it is to simulate the device operation using a classical computer. There are also metrics for photonic quantum processors analogous to those for gate-based quantum computers, such as quantum volume~\cite{zhang2023quantum}.

These are the considered metrics.\\
\noindent\fullsechyperref{sec:boson_sampling_hardware_characterization_and_model_as_metrics}
\begin{indentmetric}
The characterisation of the individual components of a boson sampling device, such as linear-interferometers and detectors, can allow estimating the overall performance of the device as a whole. These metrics are chosen on a hardware-specific basis.\\
\end{indentmetric}
\noindent\fullsechyperref{sec:boson_sampling_quantum_advantage_demonstration_as_metric}
\begin{indentmetric}
Quantifying how challenging it is to classically simulate sampling from the device provides a metric for the overall capability of the boson sampling device. This is especially relevant in the context of quantum advantage, as this metric effectively quantifies the distance to quantum advantage for the device.\\
\end{indentmetric}

\subsection*{\ref*{chapter:neutral_atoms}. \nameref*{chapter:neutral_atoms}}

Neutral atoms offer the combined modalities of performing both digital gate operations and analogue computation in the form of coherent quantum annealing~\cite{bluvstein22}. When used as a gate-based digital quantum computer, the metrics presented for gate-based quantum computers all apply to the device. However, when used as an analogue quantum simulator~\cite{saffman10,morgado21,browaeys20,semeghini21}, there are other metrics that can be used to quantify the performance of the device.\\ 

\noindent\fullsechyperref{sec:neutral_atoms_analog_benchmark}
\begin{indentmetric}
The analogue process fidelity measures the quality of the analogue computation by running quantum simulation on the device and comparing the obtained results with the ideal results obtained via classical simulation~\cite{mark23}. Note that this metric is not limited to neutral atom devices, but is applicable to a wide class of analogue quantum simulators~\cite{mark23}.\\
\end{indentmetric}

\noindent\fullsechyperref{sec:neutral_atoms_trap_lifetime}
\begin{indentmetric}
Due to the relatively shallow trap depth of neutral atom tweezers, it is possible for atoms to be ejected from the trap during computation. The trap lifetime metric is a measure of the characteristic lifetime of the trap~\cite{schymik2021single}. \\
\end{indentmetric}

\noindent\fullsechyperref{sec:neutral_atoms_reconfigurable_connectivity}
\begin{indentmetric}
Neutral atom quantum computers can also offer the ability to dynamically reconfigure the atom array geometry in order to implement arbitrary couplings. The number of mobile tweezers available and hardware restrictions on the allowed moves limit the degree of reconfigurability. Thus, the reconfigurable connectivity is an important architectural property that characterizes the capability.\\
\end{indentmetric}

\chapter{Discussion and outlook}
\label{chapter:conclusions_outlook}

The aim of this document is to provide a comprehensive collection of metrics allowing the evaluation of all aspects of quantum computing performance relevant on the pathway of achieving quantum advantage. The included metrics measure performance going from the individual quantum computing components, mostly of interest to hardware manufacturers, all the way up to well-studied tasks in applications, more relevant to end users. For thorough benchmarking of a given quantum computing hardware system, it is crucial to evaluate a comprehensive and holistic collection of metrics in order to determine advantages and disadvantages across performance indicators and avoid focusing only on a subset of metrics that may lead to biased conclusions. 

This compendium provides a consistent format for the definitions of all metrics in the collection, including the methodology, assumptions and limitations, and a link to software implementing the methodology. 
The inclusion of methodology and software in the metric description is needed to ensure transparency of the obtained metrics values. When reporting a metric, inclusion of information on how well the assumptions are met by the specific hardware provides trust in the results. The included limitations describe aspects such as how scalable a metric is and where the metric may not be applicable.
The linked open-source software, together with the associated tutorials, is necessary to ensure the reproducibility of reported metrics values, as no parameters of the metrics evaluation are hidden. Each metric in the collection can be evaluated by running its associated software. The software uses or is built on open software libraries with permissive licenses such as Qiskit~\cite{qiskit2024} and PyGSTi~\cite{Nielsen_2020}.
The collection of metrics is aimed at a wide number of hardware platforms. Since the relevance of each metric for a particular hardware platform varies, background information for a number of widely used hardware platforms is provided including a description of the most relevant metrics for each platform. Platform-specific metrics are typically also needed for non-gate-based approaches, and in this document we provide an initial collection of such metrics. In these cases the metric description format is partly adapted to the specific systems, since metrics and benchmarks for these systems are at the early stage of their development and may lead to to more specific metrics in the future. We also suggest that, when reporting metrics, all parameters set in the measurement procedure are reported. Additionally, in the case of metrics containing randomized circuits, one should also report the circuits executed to obtain the metric. If classical optimization is used, the user should then also report the un-optimized and classically optimized circuits, along with all other metric parameters, and the metric value.

Extensive research in the development of metrics and benchmarks has led to a number of well established and mature methodologies. Nevertheless, due to the rapid advancement of quantum hardware and algorithms, research is still needed and is rapidly progressing in a number of areas, such as the following:
\begin{itemize}
\item Significant research is needed to improve efficiency with which a number of metrics can be evaluated. 
As discussed in the Sec.~\ref{chapter:metrics_summary}, some metrics can take significantly longer to evaluate than others. 
Metrics that take very long to evaluate can only measure properties averaged over such long time scales, and are therefore not suitable for the evaluation of potentially occurring fast changes in device properties and performance. Speeding up the evaluation time of metrics is therefore an important area of research, as it will allow gaining significantly more information on the time-dependent fluctuations of hardware performance and with it enable fast on-the-fly re-calibration.
\item Research is also needed to improve the scalability of a number of metrics. As the number of qubits increases, it will become progressively more important to establish metrics that can scale-up to thousands of qubits and beyond. Many existing metrics do not scale to such large number of qubits, often because they require a classical computation of the exact solution, which becomes intractable for large qubit numbers. Ideally, the computational resources demanded by the benchmark should grow at most polynomially with increasing quantum computer size. Furthermore, when scaling up quantum computers to thousands of qubits, some component level metrics may start to affect the performance more than others, which may result in the development of new metrics tailored to such scaled-up systems. 
\item A significant amount of past research was devoted to the development of quality metrics, while speed metrics and stability metrics are much less developed. Therefore, research into these metrics categories will be important. Development of such benchmarks will also guide the development of quantum computer characterization techniques aimed to improve the hardware performance in these benchmarks.
\item It will be important to provide metrics that characterize specific aspects of the occurring errors, such as correlated errors, cross-talk errors, or non-Markovian noise. This will allow optimizing quantum error mitigation and error correction techniques for these hardware errors, as well as guide hardware design to reduce them.
\item It will be important to further develop metrics devoid of any noise model assumptions, accommodating all potential noise disturbances. In many of the currently used metrics there are a number of assumptions regarding the impact of noise on computations. For example, for the evaluation of gate fidelities it may be implicitly assumed that there is no cross-talk, or more generally no non-Markovian noise.
\item Research is required into benchmarks for quantum software, especially low level compilers and optimizers, as well as error correction algorithms
\item Another area for research is the development of metrics for non-gate-based quantum computing approaches, which are typically hardware and application specific and also difficult to directly compare to gate-based approaches, and are hence difficult to communicate to non-experts. A resulting risk is that these may not be considered promising approaches compared to their gate-based counterparts. It is also important to develop metrics for combined gate-based and non-gate based algorithms.
\item It will be important to advance the development of metrics for distributed quantum computers~\cite{buhrman2003distributed, cuomo2020towards, caleffi2024distributed, cacciapuoti2019quantum, beals2013efficient, van2016path, cirac1999distributed}. These offer a solution to the challenge of scaling devices to a large number of qubits by using a network of quantum computers which can communicate with each other, and hence effectively form a larger quantum combined computer. One of the primary challenges with such devices is transmitting entanglement across different quantum computers. Typically, this is performed by the use of entangled Bell pairs~\cite{caleffi2024distributed}. Various metrics that measure the rate and fidelity at which these Bell pairs can be generated and transmitted have been proposed~\cite{dahlberg2019link, stephenson2020high, dupuy2023survey, shi2020concurrent}, as well as other metrics that measure the performance of such quantum networks~\cite{vardoyan2023quantum, lee2024quantum, ferrari2024design, caleffi2024distributed}. Such metrics, in conjunction with the metrics quantifying the performance of individual quantum computers, and with application-level metrics that are agnostic to hardware implementations, can be used for thorough evaluation of the performance of such networked quantum computers.
\item With the aim of developing fault-tolerant (FT) quantum computers~\cite{google2023suppressing, bluvstein2024logical,da2024demonstration, sivak2023real,postler2022demonstration, erhard2021entangling,PhysRevX.11.041058, egan2021fault, acharya2024quantum, brock2024quantum, reichardt2024demonstration, putterman2024hardware}, there is the need to develop metrics that guide and track the progress towards that goal. Most of the metrics presented in this document are largely transferable to logical qubits, for example the various gate fidelities and gate times apply also to logical qubits. New metrics are needed that specifically evaluate metrics relevant to quantum error correction (QEC). This can include, for example, time for each QEC cycle~\cite{google2023suppressing, sivak2023real, acharya2024quantum,putterman2024hardware}, amount of change in error probability per gate provided by QEC when compared to physical error rates~\cite{sivak2023real, da2024demonstration, brock2024quantum,acharya2024quantum}, and error-decoding latencies~\cite{acharya2024quantum}. Given the large scale required for these devices, metrics would also be needed to quantify the performance of error correction implemented on networked quantum computers. The combination of metrics for the physical qubits, QEC capabilities, and logical qubits can then together be used in developing intermediate milestones towards the FT era.
\end{itemize}

Since the end user is typically not familiar with the details of hardware platform and algorithms, from an end user perspective there are a number of properties that should be considered for metrics and benchmarks:
\begin{itemize}
\item Hardware agnostic: For the end user to be able to objectively compare across hardware platforms, the results of benchmarks aimed at end users should remain unaffected by any provider’s specific operations or instructions, and may be framed to ensure this. 
\item Algorithm agnostic and future proof: when providing algorithm level benchmarks it is important to develop performance indicators across a multitude of applications envisaged for the long term future including those targeting quantum advantage.  
\item Predictive and certifiable: quantum benchmarks should also offer insights into the performance of quantum computations and routines that are not explicitly included in the metric evaluation, and hence have a predictive capability. In addition, there should be a level of assurance and a guarantee of accuracy.
\end{itemize}
Such paradigms have been recently introduced in Ref. \cite{2024VerificationFrank}.

As research on metrics and benchmarks progresses, the collection of metrics for performance evaluation is expected to evolve. Hence, this document and its associated software are envisaged at being a living online resource, updated at regular intervals to account for community driven developments in the field.

While research about the above-mentioned aspects is important, many of the metrics, in particular for gate-based approaches, have a high level of maturity and are widely used across platforms. Examples of such metrics are the qubit relaxation time ($T_1$), the qubit dephasing time ($T_2$), or the randomized benchmarking average gate fidelity. Systematic laboratory inter-comparison studies for each of these metrics, where a quantum computing chip is characterized independently by different laboratories either by remote or direct access to the device, can be a useful step in the direction of achieving agreement of the community on the methodologies to evaluate these well established metrics. This requires identification and agreement on a set of metrics that can be considered mature, of relevance for multiple platforms, and widely used. These should be separated from metrics and benchmarks that may still significantly evolve due to future progress of hardware platforms, and where it would be too early to define an agreed methodology to measure them. Analogously, many benchmarks for individual quantum computing components are also still rapidly evolving; this may for example apply to qubit couplers or interconnects. In these cases, it is likely too early to aim for an internationally agreed approach.  

A similar separation of benchmarks can be applied to the end user perspective: many applications where quantum computers promise advantage are well-established and may hence be suitable for selection into an agreed set of application or algorithm level benchmarks. These may include subroutines of full algorithms such as Shor's algorithm or Hamiltonian simulation, such as the quantum Fourier transform, quantum phase estimation, and also the variational quantum eigensolver. One problem with such metrics is that even though a specific small-scale quantum computer may perform well in these metrics, the quantum computer implementation may not be scalable to the large systems needed for quantum advantage. As a result, care must be taken when extrapolating these application level metric results for small or medium scale devices to abilities of future large scale devices to achieve quantum advantage. 
Furthermore, many algorithms are being developed alongside the hardware platforms, both for gate-based and non-gate-based approaches, and it is important not to exclude these in the consideration of the best approach towards quantum advantage. Since algorithms are still heavily evolving, algorithm level benchmarks need to be taken with care to avoid prematurely fixing the design of machines optimized solely for the current known algorithms, potentially overlooking emergent solutions. When benchmarking applications for quantum advantage it is also important to identify and optimize the most efficient classical approaches. 

Independently of the specific metrics and methodologies, an international agreement may be sought to ensure trust in benchmarks by providing transparency and reproducibility. An initial step this direction may involve agreements on which information to disclose together with metrics reports to ensure transparency and reproducibility across different manufacturers, making sure that results can be objectively compared. In this approach each manufacturer can still decide which metric to report, the benefit is that for the reported metrics the values are well defined and trusted. 

The importance of objective and relevant performance benchmarks for quantum computers has led to several international standardization bodies initiating work for the evaluation of which specific areas are now ready for standardization. As this may then inform funding decisions, care must be taken to ensure that standardization is beneficial to the development of the field and facilitates work towards quantum advantage, while still providing end users and investors an informed evaluation of the performance of a specific quantum computing platform. Based on the discussion in the preceding paragraphs, listed below are some items that should be considered by international standardization bodies:
\begin{itemize}
\item identification and agreement on the categories of metrics that comprehensively benchmark device performance;
\item identification and agreement on a set of well-established and mature metrics of relevance across multiple hardware platforms that together benchmark the different aspects of performance such as quality, speed, and stability,;
\item identification of hardware platform specific metrics, including metrics for non-gate-based quantum computers;
\item inter-laboratory comparison studies for a set of such well-established metrics to develop best practice guides on describing the measurement methodology;
\item agreement on what data and software should be reported together with a metric value to ensure trust, transparency and reproducibility.
\end{itemize}
A discussion may hence be held within international standardization bodies to identify the initially restricted set of commonly agreed benchmarks that are expected to remain relevant even as the technology fully matures, while leaving room for further development in metrics for areas that are still rapidly progressing. An agreement may also be sought for metrics categories that need to be considered to ensure a comprehensive evaluation of quantum computing performance.

An open question in the field is whether fault tolerant quantum computing based on quantum error correction will be needed for practically useful quantum advantage. To date a number of different avenues for quantum advantage have been considered: i) quantum advantage for specific tasks using gate-based devices without error correction, ii) quantum advantage using fully error-corrected gate-based quantum computers, iii) quantum advantage for specific tasks using gate-based devices with partial error correction, and iv) quantum advantage on specific problems with non-gate-based quantum computers. While for all considered approaches in order to achieve quantum advantage the noise levels need to be small, and speed of operations and device stability high, the importance of individual metrics for the success the various approaches differs. Hence the device improvements required to achieve quantum advantage depend on the quantum advantage approach that a hardware manufacturer aims for. For example, reducing noise during quantum operations to very low levels is essential if quantum advantage is to be gained without error correction, while increasing the number of low-noise qubits by orders of magnitude alongside finding and implementing the most suitable error-correcting codes for a device are the key priorities on the path to error-corrected quantum computers.
As the goal of the field is to achieve practical quantum advantage using one of the approaches above, one needs to evaluate a collection of metrics to quantify qubit performance across all categories relevant for quantum advantage, such as the ones presented in this document. For example, a commonly reported metric is the noise level when compared to thresholds required for fault tolerance. However, for practical advantage it is also important to evaluate the time take to perform the operations, and it is also important to consider if such a technology architecture is scalable both from a hardware and control perspective, or whether it is stable over long time scales. 
A collection of metrics, such as the one presented in this article, together with the associated open-source software needed to evaluate them, will guide the development of standardized benchmarks for quantum computers for all of these approaches and speed up the progress of the field towards practical quantum advantage.

\let\thesection\theoldsection

\ifbool{showpart1references}{\printbibliography[heading=subbibliography,segment=\therefsegment]}{}
\end{refsegment}

\newpage

\part{Background}
\label{part:background}
\chapter{Quantum computing approaches and their performance indicators}
\label{chapter:qc_approaches}
In this section the gate-based and non-gate-based models of quantum computation are described. The first subsection presents concepts of quantum computing relevant for their performance. These include the types of operations on qubits, the effects of noise in the hardware, quantum error correction, fault tolerance and quantum advantage. This is followed by a detailed description of gate-based quantum computing. In the subsequent subsections a number of non-gate-based approaches are presented: measurement based quantum computation, quantum annealing, boson sampling.

\section{Concepts for quantum computing performance}
\label{sec:definitions}

\begin{refsegment}

\subsection{Universal quantum computer}
The concept of a universal quantum computer builds on its classical counterpart. Computation is the transformation of sequences of symbols by precise rules, and a computer is a physical device that performs these transformations.
Turing defined a class of abstract machines, now called Turing machines, that perform a computation by manipulating symbols on a strip of tape according to a list of rules. A Turing machine that can simulate an arbitrary Turing machine on arbitrary inputs is called a universal Turing machine. 
When the transformations obey the laws of quantum mechanics, the physical device is called a quantum computer. In 1985, Deutsch proposed a quantum Turing machine, which introduces the notion of a universal quantum computer as a universal Turing machine augmented by unitary transformations~\cite{Deutsch1985}. A qubit is the unit of quantum information encoded in a two-level system, and is the quantum analogon to the classical bit of information. In this document both universal and non-universal quantum computing approaches are considered. A non-universal quantum computer can only solve a restricted set of tasks, and it may provide an advantage over classical computers for these tasks.

\subsection{Models of quantum computation}
In 1989, Deustch presented the quantum circuit model, a model of quantum computation equivalent to the quantum Turing machine~\cite{Deutsch1989}. Circuit-based quantum computation is also known as gate-based quantum computation, where a small and finite set of well defined operations, denoted as gates, are performed on the quantum states in a quantum computer. 
Gates are unitary transformations, produced by evolving a Hamiltonian for a certain amount of time.
In addition to the circuit model, several other models of quantum computation have been proposed, such as measurement-based quantum computation, quantum annealing, and boson sampling. These will be discussed in more detail in the next sections. A quantum device that implements a model of quantum computation is called a quantum processing unit (QPU), in analogy to the central processing unit (CPU) in classical computation.
The metrics and benchmarking methods presented in this document are for gate-based quantum computers, as well as for selected non-gate-based models.

\subsection{Noise and errors in the physical realization of quantum computers}
Quantum computing hardware platforms create qubits by manipulating physical systems with at least two discrete energy levels. Operations on the state of such a qubit are physically realized using electromagnetic pulses on the qubit~\cite{krantz_quantum_2019, ion:blueprint}. 
Individual discrete operations on qubits are denoted as gates.
The pulse can be tuned by altering the pulse shape, frequency, amplitude, and duration to implement the gate that is required~\cite{Liang_2024, krantz_quantum_2019}. At the hardware level, a quantum circuit is compiled into a pulse sequence that generates the  evolution of the states of the qubits as described by the quantum circuit~\cite{krantz_quantum_2019}.

Any quantum system seeking to realize a physical quantum computer is subject to disturbances from its surrounding environment.
These result in the quantum computer producing erroneous outputs with some probability.
Errors also result from inaccuracies in the physical implementation of the qubit operations, which are continuous and inherently analogue operations.
Errors from both these sources are generally referred to as noise. Noise can accumulate in the course of a long computation, rendering its outputs unreliable.
The coherence of a qubit refers to the qubit's ability to preserve the quantum information in the face of noise~\cite{konik2021quantum}, and 
the coherence time provides an estimate for how long a qubit can retain phase information~\cite{PhysRevB.86.100506} and hence used to perform computations.

\subsection{Noisy intermediate-scale quantum (NISQ) devices}
Currently available devices are typically referred to as noisy intermediate-scale quantum (NISQ) computers. This term reflects the fact that quantum computers are affected by noise in the hardware, which for example limits the coherence time of the states of the qubits. A number of metrics presented in this document can be used to quantify how strongly the quantum computing hardware is affected by noise. The number of qubits in the intermediate scale lies between very small quantum computers of significantly fewer than ten qubits and large quantum computers with millions of qubits. 

\subsection{Quantum Error Correction (QEC)}
Quantum error correction (QEC) is based on the use of multiple qubits and gates to detect and correct errors that occur due to noise and imperfections in the hardware. This is achieved through redundancy provided by the use of multiple qubits when storing and processing information.

In 1996 Shor proposed a scheme of error correction for the quantum circuit model of quantum computation~\cite{Shor1996}.
Quantum error-correcting codes map so-called logical qubits, which are error corrected, into blocks of physical qubits, such that a small number of errors in the physical qubits of any block has effects that can be detected and corrected. 
This mapping is implemented by quantum circuits, where QEC involves correcting errors in the physical qubits, as well as a means for computing with the encoded logical qubits. If these circuits are implemented with slightly noisy gates, with only a small probability of errors occurring in the physical qubits in each block over short time-scales, the encoded logical qubits are effectively not disturbed by noise.

\subsection{Fault-tolerant quantum computing}

Any physical implementation of QEC is subject to noise. If the noise is smaller than a threshold value, then arbitrarily large quantum computations can be performed reliably by concatenating quantum error-corrected circuits. This is referred to as the quantum threshold theorem, and results in fault-tolerant quantum computing~\cite{knill1998resilient,aharonov1997fault,kitaev2003fault,}. Fault-tolerant quantum computation requires poly-logarithmically more resources. If the noise is larger than the threshold, then errors accumulate at a rate faster than they can be corrected, resulting in an unreliable output, and no fault-tolerance is achieved.

The threshold and resource overhead of fault-tolerant quantum computation depend on the model of quantum computation, the actual computation at hand, the type of noise affecting it, and the quantum error-correcting codes employed to counter them.
Depending on these, estimates for the threshold noise levels range from
values corresponding to fidelities of $99\%$ to $99.9999\%$~\cite{steane2003overhead,stephens2014faulttolerantthresholds,gottesman2010introduction}. 
Those for the overheads range from thousands to millions of qubits and gates~\cite{steane2003overhead,cross2009acomparativecodestudy,knill1998resilient,bravyi2024high,google2023suppressing}.

It is important to benchmark how far a given quantum hardware is away from being able to perform fault-tolerant quantum computation.
Such an ability depends on the fidelity of the gates, on the ability to perform mid-circuit measurements to detect errors, and on the ability to apply measurement-dependent gates to correct potentially found errors, which depends on the speed of measurement and processing of the measurement results. The topology and connectivity of the qubits in the hardware determines which QEC codes can be efficiently executed on such hardware. A number of metrics outlined in this document can be used to quantify these aspects, and the combination of the information provided by such metrics characterizes how far away a specific hardware platform is from being able to implement fault-tolerant quantum computation. 

\subsection{Quantum advantage}

Quantum advantage is achieved when a quantum computer can solve a task faster than large conventional high performance computers. Any limited, for example constant or polynomial speedup, can be considered a quantum advantage. The notion of quantum supremacy is sometimes used either interchangeably for quantum advantage, or to emphasize a large quantum advantage, such as an exponential rather than polynomial speedup.

Any claim of quantum advantage is always based on the comparison with the best classical algorithms and hardware available at the time that claim is made.
Given the always ongoing improvements in classical computing hardware, and the development of novel algorithms implemented on them, the requirements on quantum computers to achieve quantum advantage become more stringent with time.
Two instances of claims of quantum advantage with NISQ devices are those based on random circuit sampling~\cite{Arute2019,wu2021strongquantumcomputationaladvantage,} using the gate-based model of quantum computation, and those based on boson sampling~\cite{Jiuzhang2}. Improvements on classical algorithms have subsequently reduced the resources required to simulate such algorithms on classical computers, and the race between improved classical algorithms and improved quantum computing hardware demonstrations is actively developing~\cite{pednault2019leveraging,zhu2022quantum,liu2021closing,zhou2020whatlimitsthesimulation,zlokapa2023boundaries,morvan2023phase}.

\ifbool{showpart1references}{\printbibliography[heading=subbibliography,segment=\therefsegment]}{}

\end{refsegment}

\section{Gate-based quantum computing}
\label{sec:gate_based_quantum_computing}
\begin{refsegment}
Gate-based quantum computing is also known as circuit-based quantum computing. In this model, qubits are described by complex vectors, and quantum operations on qubits are denoted as quantum logic gates, which correspond to unitary matrices. Quantum circuits are constructed by applying a sequence of gates to qubits, in an analogous way to how classical circuits are constructed by classical logic gates. First proposed by Deustch in 1989~\cite{Deutsch1989}, this model is used in algorithms like the Shor's algorithm for prime factorization~\cite{shor1997polynomial} and in variational quantum eigensolvers~\cite{TILLY20221}.

As opposed to classical bits, which can be in either 0 or 1 but not both at the same time, qubits can be in a quantum superposition of orthonormal basis states $\ket{0}=\begin{bsmallmatrix}  1 \\ 0 \end{bsmallmatrix}$ and $\ket{1}=\begin{bsmallmatrix}  0 \\ 1 \end{bsmallmatrix}$. With regard to the quantized energy levels of a quantum system, it is a convention that $\ket{0}$ represents the lowest-energy (ground) state and  $\ket{1}$ represents the first excited state. An arbitrary state of a qubit is written as $\ket{\psi} = \alpha \ket{0} + \beta \ket{1}$, where $\alpha$ and $\beta$ are complex numbers, and $|\alpha|^2 + |\beta|^2 = 1$. A measurement of the qubit state in the $\ket{0}$ and $\ket{1}$ basis causes the qubit's wavefunction to collapse, outputting $\ket{0}$ with probability $|\alpha|^2$, or outputting $\ket{1}$ with probability $|\beta|^2$. By repeatedly preparing a qubit in the same state and then measuring it, a statistic of measurement outcomes is gathered, and one may estimate the values of $|\alpha|$ and $|\beta|$. The number of circuit executions to prepare the state followed by measurements is usually referred to as shots.

\begin{figure}
    \begin{center} 
    \Qcircuit @C=1.2em @R=.7em {
    &&&&&&&&&& \ket{0} & & \gate{X} & \ctrl{1} & \qw & \targ & \meter \\
    &&&&&&&&&& \ket{0} & & \qw & \targ &  \gate{H} & \qw & \meter\\
    &&&&&&&&&& \ket{0} & & \gate{R_y(\pi)} & \qw & \qw & \ctrl{-2} & \meter
    }
    \vspace{1em}
    \Qcircuit @C=1.2em @R=.7em {
    & & & & & & & & & & & & & & \ctrl{1} & \qw & & & & &\\
    & & \gate{X} & \qw & & \gate{R_y{(\pi)}} & \qw & & & \gate{H} & \qw & & & & \targ & \qw & & & & \meter & \\
    & & \text{$X$ gate} & & & \text{$R_y(\pi)$ gate} & & & & \text{Hadamard gate} & & & & & \text{$\mathrm{C}_X$ gate} & & & & & \text{Measurement}
    }
    \caption{Example circuit diagram for a three-qubit quantum circuit. Each horizontal wire represents a qubit, and the initial state is shown at the left end of the circuit. Symbols and boxes represent $X$, $R_y$, Hadamard and $\mathrm{C}_X$ gates, with measurements at the end of the circuit. The mathematical definitions for these gates and their operation are described in section \ref{sec:methodsmaths}. Multi-qubit gates are drawn vertically across wires, with dots or other symbols placed on the wires that participate in the multi-qubit gate. For controlled gates like the $\mathrm{C}_X$ gate, a solid dot denotes the control qubit and a vertically connected symbol denotes the target qubit ($\oplus$ for $\mathrm{C}_X$).}
    \label{fig:example_circuit}
    \end{center}
\end{figure}
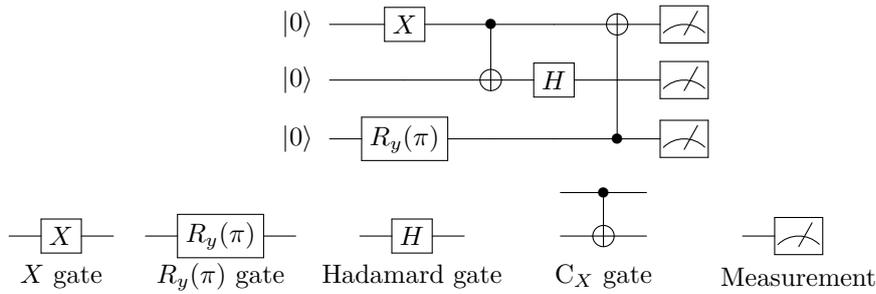

A quantum circuit is a collection of quantum gates applied to an initial quantum state. In a quantum circuit diagram, sequential applications of gates on the same qubit are connected with horizontal wires, and multi-qubit gates are drawn with vertically spanning boxes or with vertical lines. Fig.~\ref{fig:example_circuit} shows an example circuit diagram for a three-qubit quantum circuit containing a number of gates.

If a set of quantum gates can be used to construct an equivalent of any other quantum gate, then this set is called a universal gate set. There are many universal gate sets, examples include \{CNOT and all one-qubit gates\}~\cite{Barenco1995} and \{CNOT, $H$, $\sqrt{S}$\}~\cite{BOYKIN2000101}. The mathematical definitions for these gates and their operation are described in section \ref{sec:methodsmaths}. To achieve universal quantum computation within the gate based model, a quantum computer needs to physically implement a universal gate set. 

There are different technologies of realizing gate based quantum computing, examples including superconducting qubits, trapped ions, neutral atoms and photonic devices. An overview of these technologies is presented in section~\ref{chapter:hardwarePlatforms}. As a result, different quantum hardware may have different gates that are physically implemented, known as native gates. Other quantum gates, not in the set of native gates, need to be decomposed into native gates.
Typically, the native gates are chosen to form a universal gate set, so that universal quantum computation can be achieved.

\ifbool{showpart1references}{\printbibliography[heading=subbibliography,segment=\therefsegment]}{}
\end{refsegment}

\section{Measurement based quantum computation}

\begin{refsegment}
In measurement based quantum computation (MBQC)~\cite{raussendorf2001one},
instead of evolving a quantum state through a circuit, one starts with a large entangled state, denoted as resource state, and performs the computation by adaptively measuring the qubits, one by one with local single-qubit measurements. The main idea is that by measuring one part of an entangled state, one can teleport the quantum information of one layer to the next layer with an added gate applied. Since measurements on quantum states have fundamental randomness, in order to recover the deterministic unitary evolution one needs to adaptively modify the measurements performed, based on the previous measurement outcomes. It is possible to show that MBQC is complexity-wise equivalent to the circuit model, in that one can map a MBQC computation to a circuit one and visa-versa with a constant overhead at most.

MBQC may offer a number of advantages for some applications. It may be easier to implement such a model for some hardware approaches, such as photonic quantum computers~\cite{rudolph2017optimistic}. 
MBQC is also the natural quantum computation model for applications such as blind quantum computing~\cite{broadbent2009universal, barz2012demonstration}
and verifiable blind quantum computing~\cite{barz2013experimental}, which is a secure protocol for quantum computation whereby a server performs quantum computation for a client while keeping all their inputs, outputs, and intermediate steps completely private. 
A number of error-correction methods can also be adapted to MBQC.
From a more foundational perspective, the role that non-classical resources such as contextuality play for providing quantum advantage may be clearer in the language of MBQC~\cite{raussendorf2013contextuality}. 
On the other hand, MBQC generally requires significantly more qubits for the same algorithms. Additionally, the necessity of measurements in the middle of the computation, as well as the need for adaptive measurements, put stringent limitations on using MBQC for certain hardware approaches.

 Most digital devices can perform MBQC, so that most of the metrics described in this document are relevant, but the order of importance and specifics vary when compared to the circuit model. Of relevance are the following qualities: the ability to prepare high fidelity entangled states is important, the state preparation and measurement (SPAM) errors need to be small, and the number of qubits needs to be large. Beyond these, measures that quantify the performance of a hardware in running applications/algorithms are also relevant. Finally, a crucial metric is the ability and quality of performing mid-circuit measurements, as well as the ability to condition mid-circuit measurements on the outcomes of earlier measurements.
 
\ifbool{showpart1references}{\printbibliography[heading=subbibliography,segment=\therefsegment]}{}
\end{refsegment}

\section{Quantum annealing}

\begin{refsegment}
Quantum annealers constitute a class of devices which have been originally envisioned to solve combinatorial optimization problems~\cite{Kadowaki1998annealing,finilla1994quantumannealing} and related problems in machine learning. These devices rely on directly mapping a problem to a Hamiltonian Function and utilize a variety of physical mechanisms to solve these problems. Currently the most technologically mature and largest scale quantum annealers are those produced by D-Wave Systems Inc.~with over 5,000 superconducting flux qubits at the time of writing~\cite{king2023quantum}. However, it is worth distinguishing the computational paradigm of quantum annealing from the physical devices produced by a single company. In fact, Ryderg-based quantum annealing has become commercially available through the platforms of companies such as QuEra and PASQAL. Recent work has shown that, while not their original intended purpose, the flux-qubit quantum annealers can also act as highly effective simulators of condensed matter systems. In fact, while there is currently no evidence of a scaling advantage for exact optimisation, there has been for approximate optimisation \cite{bauza2024annealingOptAdv}, and compelling evidence for an advantage over path-integral quantum Monte Carlo, which is a classical algorithm (despite the potentially misleading name) used in the simulation of certain condensed matter systems~\cite{King2021advantageQA}.

There are a number of reviews and perspectives on quantum annealing and related topics. A list of some of them along with a short summary can be found below:
\begin{itemize}
\item Mathematically focused review article on topics related to quantum annealing, particularly in the adiabatic limit~\cite{Albash2018AQCreview}
\item Forward-looking perspective on the potential of new methods in quantum annealing, focusing strongly on non-adiabatic methods~\cite{Crosson2021diabatic}
\item Perspective on hybrid quantum-classical algorithms in general, with considerable discussion on hybrid methods for quantum annealing~\cite{Callison2022hybrid}
\item Review of potential industrial applications of quantum annealing~\cite{Yarkoni2021anealingindustry}
\item Review of the mathematics related to quantum annealing~\cite{Berwald2019mathematics}
\end{itemize}

In the simplest version of a typical implementation of quantum annealing, which is the one usually adopted in commercial systems, the objective Hamiltonian is expressed as a classical Ising Hamiltonian,
\begin{equation}
H_{\mathrm{Ising}}=\sum^n_{i,j=1} J_{ij}Z_iZ_j+h_iZ_i ,  \label{eq:H_ising}
\end{equation}
where the problem is encoded into the values of $h$ and $J$, and $Z_i$ is a Pauli $z$ matrix acting on site $i$.
Since the problem of finding the ground state of this Hamiltonian is NP-hard~\cite{barahona1982computational}, any combinatorial optimization problem can be stated in this form\cite{cook1971NPcomplete,Karp1972NPcomplete}. Quantum computation can be achieved by adding fluctuations, the simplest version of which being single-qubit flipping operations, so that one obtains the annealer Hamiltonian
\begin{align}
H_{\mathrm{anneal}}&=A(t)H_{\mathrm{drive}}+B(t)H_{\mathrm{Ising}}, \\
H_{\mathrm{drive}}&=-\sum_{i=1}^n X_i \label{eq:tf_driver},
\end{align}
where $X_i$ is a Pauli $x$ matrix acting on site $i$.
Initializing in an easy-to-prepare state, and using appropriate controls on the time dependent parameters $A(t)$ and $B(t)$, protocols can be constructed to find highly optimal states of $H_{\mathrm{Ising}}$. The conceptually simplest version of this protocol is to initialize in the ground state of $H_{\mathrm{drive}}$, and then slowly reduce $A$ and increase $B$, relying on the adiabatic theorem of quantum mechanics to reach the ground state of $H_{\mathrm{Ising}}$~\cite{Farhi2000AQC}. There are however numerous other mechanisms which could potentially aid a quantum annealer in solving problems. For faster protocols, either a mechanism related to energy conservation~\cite{Callison2021diabatic,banks2024continuous,schulz2024guidedquantumwalk} or a generalization of the adiabatic theorem may play a role~\cite{Crosson2021diabatic}. For some devices, thermal dissipation may also aid computation, and unlike the previous two mechanisms, can do so regardless of the state in which the system was initialized.

Especially in older parts of the literature, adiabatic quantum computation is sometimes used as a synonym for quantum annealing. However adiabatic quantum computing is more properly considered as a subset of annealing protocols, namely those where the adiabatic theorem approximately holds. In practice, assuming NP-hard optimisation problems are not solvable in polynomial time on an annealer (as is widely suspected~\cite{Callison2021diabatic}), then adiabatic quantum computing would require exponentially growing coherence times, which are unlikely to be practically achievable. However, more generally quantum annealing could be performed with a mildly growing runtime (and therefore required coherence time), but sampled many times~\cite{Callison2019walks,Callison2021diabatic}. Since the theory of adiabatic quantum computation is much more mature than general annealing,  more can be proven in that setting~\cite{Albash2018AQCreview}. 

It is important to note that a quantum annealer operating as previously described is not capable of universal quantum computing, because it can only implement Hamiltonians with negative off-diagonal elements, known as stoquastic Hamiltonians. These Hamiltonians cannot capture important quantum phenomena in quantum physics, such as the exchange statistics of electrons. However, quantum annealing can be made universal with a slight generalization of Eq.~\eqref{eq:tf_driver} and introducing time dependent control of $h$ and $J$ in Eq.~\eqref{eq:H_ising}. If in addition to local control, two-body terms are added to the driver, in other words 
\begin{equation}
H_{\mathrm{drive}}=-\sum_i\Delta_i X_i+\sum_{ij}K_{ij}X_iX_j,
\end{equation}
with parameters $\Delta_i$ and $K_{ij}$, then quantum annealing becomes a fully universal model of quantum computation~\cite{Kempe_AQC_universal_2006,Oliveria_AQC_universal_2008,Biamonte_AQC_universal_2008}, able to in principle perform any computation which can be performed on a gate based machine. It is worth noting that the one aspect where quantum annealing is lacking in comparison to gate based quantum computing is that fault tolerance has not been proven to be possible within this model.

The method of mapping optimization or related problems in an analog way and solving them using natural dynamics of a quantum system potentially extends beyond quantum annealing, but it is likely to be worth classifying such related models similarly to quantum annealing. One example is a class of optical devices known as coherent Ising machines, which use non-linear optics and feedback to achieve this type of computation~\cite{Inagaki2016coherentIsing,McMahon2016coherentIsing} (and the related but distinct paradigm of entropy computing \cite{nguyen2024entropycomputing}). Due to experimental limitations, currently all large-scale coherent Ising machines involve feedback based on classical measurements taken directly in the basis which is used to encode the solutions. Such measurements by definition destroy any multi-qubit quantum coherence. In principle this feedback could be performed with delay lines which would maintain coherence, but this is a substantial engineering challenge. Furthermore, this shows the possibility for other techniques, which follow similar operating principles to quantum annealing, to emerge. For this reason it is appropriate to include the possibility of such devices when discussing the future of quantum annealing.
\ifbool{showpart1references}{\printbibliography[heading=subbibliography,segment=\therefsegment]}{}
\end{refsegment}

\section{Boson sampling}
\begin{refsegment}

Boson sampling is an umbrella term for proposals of non-universal quantum computing photonic devices which take advantage of the bosonic statistics of photons. The name originates from the first proposal to demonstrate quantum advantage by Aaronson and Arkhipov in 2010~\cite{Aaronson2010}. Since the main motivation of boson sampling is based on achieving quantum advantage for a specific task rather than on solving practical applications, in what follows we provide a more extended overview of this method and eventually connect it to potential practical applications. The difficulty to link boson sampling to practically useful applications is at difference to the annealing approach presented in the previous section, where the potential for practical applications is the core motivation, while provable quantum advantage is difficult to demonstrate.

Current near-term quantum computing implementations based on quantum optics, similarly as for other platforms, can be decomposed in three stages: (i) initialization of quantum state; (ii) quantum evolution; (iii) measurement at the end of the circuit.
It typically consists of three layers of experimental hardware: a set of parallel sources of quantum states of light, followed by a linear-optic interferometer, and ending with a set of measurements of each output mode. The different variants usually correspond to changing either the sources at the input or the measurements at the output, all sharing the intermediate stage of linear-interferometry. The technology and design of the linear-interferometer can vary, the most commonly used types are integrated circuits, free-space optical experiments and fiber-optics. In the case of quantum optics, the quantum information is encoded into quantum states spanning over multiple modes, but there are other physical systems that can be modeled as harmonic oscillators, and that can therefore be used to implement the same boson sampling experiments. Examples include exploiting the vibration of ions in an ion trap~\cite{ions}, and the microwave resonators in superconducting circuits~\cite{sccircuit}. 
Different near-term quantum optics differentiate themselves via the choice of the quantum states or hardware technology used on each of the three stages of the implementation. As we discuss below the most common initialization is either single-photon sources or squeezed vacuum states, where the measurement are divided into photon number resolving detectors and photo detectors that can only detect the presence of photons but not count them. The quantum evolution usually consists in a linear-interferometer that can be implemented either on free-space, fiber-optics or integrated chips. 

\subsection{Main proposals}
The first theoretical proposal of a boson sampling device was by Aaronson and Arkhipov in 2010, which was followed by a series of alternative proposals~\cite{Gard_2015}. 

\textit{\underline{Boson Sampling}} The original proposal for boson sampling consists in sending exactly $N$ indistinguishable photons over a linear-optic interferometer of $M$ modes. A mode is a degree of freedom of the electromagnetic field where photons can be located, this can be eigenmodes of its spatial distribution, but also polarization, angular momentum, and any other alternative degree of freedom that is experimentally accessible and distinguishable from the other modes. Ideal linear-interferometers are characterized by a unitary that transforms input to output modes, and for boson sampling the unitary is chosen randomly from the uniform distribution of unitaries, also denoted as Haar-random distributed unitaries. The output photons are detected at the output of the linear-interferometer using photodetectors. The proposal requires a regime of few photons per mode $M=O(N^2)$ in order to guarantee that the probability of having two photons leaving at the same output port is negligible. Then, the required detectors do not need to be able to count the number of photons, but only to discriminate zero from one or more (photodetectors), which simplifies the experimental requirement. 

\textit{\underline{Scattershot Boson Sampling}} Until very recently, most photon sources were probabilistic, making the generation of $N$ indistinguishable photons in parallel impractical. To alleviate this experimental problem, scattershot boson sampling was proposed in 2014~\cite{ScattershotBS}. It uses $M$ sources of an entangled two-mode states, one per input of the linear-interferometer. Each two-mode entangled state has a pair of modes, the first (signal) is sent through the interferometer, and the second (idler) is used to herald the presence of a single-photon in the signal mode. Their photon number distribution being governed by a geometric decay, and by properly tuning the average photon number of the two-mode entangled state one can ensure that most heralding detections correspond to single photons in the signal modes and higher number of photons have negligible probability, while at the same time being in a few photon regime of $M=O(N^2)$ modes per heralded single-photon. In practice, scattershot boson sampling can be seen as a version of the original boson sampling, where the input of the $N$ photons is randomized over the $M$ possible input modes, instead of the first $N$ as in the original proposal.

\textit{\underline{Gaussian Boson Sampling}} This proposal consists of measuring the photon number in each of $M$ modes containing a Gaussian state of light~\cite{GBS2017}. The same interference procedure is performed, but the initial states are Gaussian rather than number states, that is they contain an uncertain number of photons but are squeezed in a different way. A Gaussian state corresponds to the ground state of a Hamiltonian composed of quadratic terms in the creation and annihilation operators of the $M$ modes. The name comes from the fact that their phase-space representation is a Gaussian distribution, therefore fully characterized by its first and second order moments. Gaussian operation corresponds to time-evolution with a Hamiltonian also composed of quadratic terms, which evolves a Gaussian state into another Gaussian state. The transformation corresponds to a simple transformation of the phase-space, which can be tracked with simple matrix-matrix and matrix-vector multiplications. It is known that any pure Gaussian state can be generated by a sequence of sources of squeezed states, states given by a superposition of even number of photon of a specific form and obtained through parametric down conversion processes~\cite{RMP2012}, followed by a linear-interferometer and local displacements. The local displacement being a local operation does not add any non-locality, and therefore is usually not used in proposals of quantum advantage, but may play a role in some application of Gaussian boson sampling. See~\cite{RMP2012} for details on Gaussian states.

\subsection{Hardness of Boson sampling}
Boson sampling and its variants are used in a series of proposals for quantum advantage demonstration. The idea behind those proposals is that classically simulating boson sampling devices for randomly selected interferometers has a computational complexity beyond the capability of classical devices. The intuition is that for boson sampling a given output probability is related to a permanent of a submatrix of the unitary representing the linear-interferometer. Because computing permanents of complex matrices is hard, with best known algorithms running in exponential time on the number of photons~\cite{clifford2017classical}, the scaling of its simulation is impractical beyond a given threshold size.
Similarly, in the case of Gaussian boson sampling, the output probability is given by the a Haffnian, another combinatorial object similar to the permanent. In some special cases the Haffnian resulting from a Gaussian boson sampling output reduces to a permanent, as in the case for scattershot boson sampling.

Compared to the intuitive reasons for quantum advantage outlined above, the actual hardness proofs are intricate, and involve technical concepts of computer science. The proof works by contradiction. It is shown that if a classical computer could efficiently solve these sampling problems, it would imply the collapse of the polynomial hierarchy of complexity classes to its 3rd level, which is widely believed not to be the case in a similar way as computer scientists consider the conjecture $N=NP$ to be extremely unlikely. The result is a probabilistic statement, which means that it is proven to be hard on average over a distribution of randomly selected circuits. For more information on quantum computational advantage see~\cite{harrowQuantumComputationalSupremacy2017}.

\textit{\underline{Hardness in practice}} The hardness proofs of boson sampling, but also for similar qubit based proposals, have only limited robustness to small deviation from the output of an ideal circuit. This places current devices with significant levels of errors and imperfections outside the regime of the hardness proofs. It is therefore not excluded that classical algorithms that exploit the presence of noise in real implementations could efficiently simulate current boson sampling devices. 

\textit{\underline{Validation challenge}} Another challenge for quantum advantage demonstration based on the sampling of random circuits is the fact that verifying that we are in the hardness regime is itself not computationally tractable. For example, the cross-linear entropy benchmark (XEB) 
used for qubit random circuits~\cite{Arute2019} needs the computation of the output probabilities of the ideal circuit, which is supposed to be hard in the quantum advantage regime. There is growing evidence that no efficient verification of quantum advantage using solely classical computation is possible, or if possible then it would imply the existence of classical algorithms that can simulate the device~\cite{StilckFranca2022gameofquantum}. It is becoming more clear that to prove quantum advantage, a departure from sampling problems is necessary. Focusing on problems where the solutions can be classically verified or provide a quantifiable advantage for an end-user application seems a more promising and useful route. Despite this, random circuit experiments remain an interesting area for testing the capabilities of quantum computing platforms, including boson sampling devices, to compete against the performance of classical devices.

\subsection{Implementations and imperfections}
While small-scale proof-of-principle demonstration of boson sampling had already been done more than 10 years ago~\cite{Gard_2015}, in the last years there were a number of experimental demonstrations of large-size boson sampling devices~\cite{Jiuzhang1,Jiuzhang2,Xanadu22}. These experiments managed to implement highly complex setups beyond 100s of modes of light with different sets of sources, photon measurements and hardware for the linear-interferometer. The quantum advantage claims made by some of these experiments have been contested~\cite{villalonga2022efficient, oh2023classical}, and the debate of whether they provide quantum advantage remains open. Nevertheless, they are remarkable experimental and technological achievements. We also note that the best known classical algorithms for their simulation are not yet scalable with system size. 

\textit{\underline{Free-space demonstration}}
The first large-scale boson sampling device demonstration was done at USTC in China with a device named Jiuzhang 1 implementing a Gaussian boson sampling demonstration~\cite{Jiuzhang1}. 
It was composed of a single source generating 50 multiplexed squeezed states injected into a free-space interferometer with 100 modes and detected at the output with 41 photon detections. The linear-interferometer was not re-configurable, but had a random configuration chosen during its fabrication. This was later improved to Jiuzhang 2, with now 144 modes and up to 66 photodetections~\cite{Jiuzhang2}, followed by the very recent Jiuzhang 3~\cite{Jiuzhang3}. It has been shown that Jiuzhang 1 and 2 and 3 in the regime of a high number of photons detection can be classically simulated, and a recent report on the simulation of Jiuzhang 2 and 3 using a supercomputer and very intensive computational resources has been published~\cite{oh2023classical}.

\textit{\underline{Fiber-optics platform}}
Free-space interferometers, despite being an ideal way of limiting losses, are not a scalable technology to build a quantum computer. The alternative of using re-configurable integrated chips suffers from serious impact of losses as the size of the circuit is scaled up, which makes boson sampling not scalable in such a platform. Therefore, an alternative approach using fiber-optical loops, initially proposed in~\cite{TNloops2018}, has inspired a novel approach to boson sampling that was recently demonstrated by Xanadu, reaching 216 modes and detecting up to to 219 photons (one can have more than one photon per mode)\cite{Xanadu22}. The advantage of this setup is that it is also a building block for universal-quantum computation using continuous-variables measurement-based quantum computation techniques. To date there is no classical algorithm challenging this quantum advantage demonstration. 

\textit{\underline{Imperfections}}
The main imperfections that occur on a photonic platform used for boson sampling can be divided into three types: losses, thermal noise, and distinguishability of photons. Losses result from the coupling with unwanted modes on the propagating medium, such as optical fibers or waveguides of an integrated chip. Also small imperfections in the coupling between elements of the setup, such as the coupling of the sources to the linear-interferometer, generate losses. Detectors have two main types of errors, detection efficiency (false negative detections) and dark counts (false positive detections), which can be modelled by losses and interaction with a thermal bath, respectively. If one assumes that all imperfections are uniformly distributed among the different components, one can show that using properties of the theory of Gaussian states and operations, the circuit will be equivalent to an ideal circuit where all losses and thermal noise happen right before the detection or after the sources, where one is free to choose the option that makes the theoretical analysis simpler. For example, this can be used to show that a circuit with losses $\eta$ changes the scaling of the classical simulation of $N$ single-photon boson sampling device from $2^N$ to $2^{\eta N}$. Losses therefore decrease the effective hardness of boson sampling from an ideal circuit of size $N$ to at least $\eta N$. Ideal photons are indistinguishable particles, but errors in the source can result in the generation of photons that are partially distinguishable, and this can be modeled by appending additional modes to the ideal circuit to emulate those extra degrees of freedom. It can be shown that this has a similar effect on the hardness of simulating the device. See Ref.~\cite{Moylett_2019} for more details.

\subsection{Applications}
One of the main motivations for manufacturing a quantum computing device is to use it for practical applications. In the last years there was significant development and discussion of several applications that are native to boson sampling devices, and that exploit mathematical connections between the problems of interest and the theoretical description and statistics of the device, such as  with applications to graph optimization, molecular docking, graph similarity, point processes, and quantum chemistry~\cite{Bromley_2020}.

\textit{\underline{Molecular vibronic spectra}}
The vibration of the nuclei in a molecule are modeled by quantum oscillators of bosonic nature. This provides a direct mathematical link between the theory of quantum optics and the vibrations of molecules that can be exploited to design quantum algorithms. 
It was shown in Ref.~\cite{Hu2015} that a relevant problem in spectroscopy of molecular vibronic spectra can be exactly mapped into the statistics of Gaussian boson sampling, with the addition of local displacement. The sampling output is transformed into an energy via a classical post-processing that generates the vibronic spectra of molecules under a harmonic approximation~\cite{fabrizio2007}. The quantum advantage in this scenario is not fully convincing as there are classical algorithms that can efficiently solve problems of similar size as those accessible to current experimental Gaussian boson samplers~\cite{fabrizio2007}.

\textit{\underline{Graph problems}}
Many problems of practical relevance can be stated in terms of a graph. Dense subgraph problems are an example of such graphs, which search for subgraphs that contain many connections between their nodes. These subgraphs may correspond to communities in social networks, correlated assets in a market, or mutually influential proteins in a biological network. Interestingly, Gaussian boson sampling can be used to solve that problem. By encoding a graph into a Gaussian boson sampling device, it samples dense subgraphs with high probability~\cite{Arrazola2018}. Maximum clique, another well-known problem can be seen as a specific instance of dense subgraph problem and can therefore be also addressed using a Gaussian boson sampler~\cite{Banchi2020}. Up to today, there is no study of the potential quantum advantage that these applications could bring, neither a benchmark against classical solvers.

\ifbool{showpart1references}{\printbibliography[heading=subbibliography,segment=\therefsegment]}{}
\end{refsegment}

\chapter{Hardware platform specific performance overview}
\label{chapter:hardwarePlatforms}

Values for individual performance metrics of quantum computers often vary significantly for different hardware platforms. Furthermore, there are metrics that may be relevant only for specific hardware platforms. Therefore, in this section we give a brief overview of some of the widely used hardware platforms, outlining the type of operation and the potential hardware specific aspects of performance metrics.

\section{Superconducting qubits}
\label{sec:superconductingQubits}
\begin{refsegment}
Superconducting quantum circuits are one of the widely used platforms for the development of scalable quantum computers. The main advantage of this technology is that it is based on electrical circuits that can be fabricated with well-established methods used in the semiconductor industry. Since the properties of the qubit are mostly determined by the design, there are many ways to implement a quantum processor, thus giving the designer a great amount of flexibility~\cite{wendin2017quantum}. A large variety of different architectures, qubits and auxiliary building blocks have therefore been proposed and evaluated over the years. Currently the transmon qubit is the most used, but other qubit design such as fluxonium have also shown promise~\cite{krantz_quantum_2019}. The flexibility also means that whilst superconducting qubits are typically used for digital gate-based quantum computing, they can also be used for quantum annealing or to implement various analogue quantum simulators. Another advantage of superconducting quantum processors is that they are controlled and read-out using microwave signals in the gigahertz frequency range. This means that the platform has greatly benefited from readily available high-performance systems, instruments and components originally developed for the telecommunications sector. Several manufacturers also offer products specifically developed to control superconducting quantum processors.  

Since superconducting processors are microfabricated, the technology is, in principle, easy to scale, and systems with hundreds of qubits have been demonstrated~\cite{IBMUnveils400}. However, this freedom does come at a cost: for example, qubits based on trapped ions or other atomic platforms are inherently identical, but this is not the case for superconducting qubits. Small differences in the electrical parameters due to variability in the fabrication process of devices translates into differences between qubits, for example, the qubit frequency. The lack of precise reproducibility therefore is a major issue for superconducting qubits. Another problem is that the coherence properties of superconducting qubits are very sensitive to defects and impurities in the material~\cite{de2022chemical}. Over the last twenty years, much progress has been made, with qubit relaxation times improving by over three orders of magnitude, going from hundreds of nanoseconds to almost a millisecond~\cite{somoroff2023millisecond}. Despite the vast improvement, the coherence times still limit the number of operations that can be performed within the qubit lifetime, since a qubit operation typically takes on the order of 100 \si{\nano\second}. This limits the use of superconducting qubits to rather shallow quantum circuits in the current noisy intermediate-scale quantum (NISQ) era. The defects and impurities are usually collectively referred to as two-level systems (TLSs). As well as causing limitations to coherence times, they also cause the qubit parameters to change over timescales ranging from minutes to hours or even days, which means that frequent system re-calibration is often needed.  

A further challenge is that superconducting qubits must be operated at millikelvin temperatures. This means that the quantum processing unit must be situated at the bottom of a dilution refrigerator. Modern cryogenics can reliable cool down even very large systems, and it is believed that systems with a thousand qubits should be feasible using current methods~\cite{krinner2019engineering}. However, transferring data to and from the processor is still a challenge, since it can require hundreds of microwave lines that all need to be cooled. Future systems will require methods for solving this bottleneck, possibly in combination with placing specially developed cryogenic control electronics inside the cryostat near the processor itself.  

There are many ways to connect superconducting qubits together into a processor. The most straightforward method is to connect each qubit to its nearest neighbour via a fixed or tunable coupler~\cite{krantz_quantum_2019,kjaergaard2020superconducting}. This method restricts connectivity and can slow down certain algorithms~\cite{linkeExperimentalComparisonTwo2017}; recently major efforts have therefore gone into enabling three-dimensional connectivity to facilitate communication between distant qubits~\cite{yost2020solid,rosenberg20193d,foxen2017qubit,chen2014fabrication,rosenberg20173d}.  

Superconducting quantum computing has been very successful in the current NISQ era, with a demonstrated one-qubit Clifford randomized benchmarking gate error of 0.0008 and a two-qubit interleaved Clifford randomized benchmarking gate error of 0.024~\cite{sung2021realization}.  Further improvements will be needed to reliably meet the performance threshold for error corrected logical qubits using QEC codes such as the surface code. Recently, much effort has therefore gone into developing and demonstrating new types of architectures and methods, which allow for error mitigation or error detection or both~\cite{levine2307demonstrating}. This can for example be done by encoding the information not in individual qubits, but in larger system comprising multiple qubits or microwave resonators. These methods aim to enable the creation of logical qubits with much less overhead than would be possible by directly using individual physical qubits.   

\ifbool{showpart1references}{\printbibliography[heading=subbibliography,segment=\therefsegment]}{}
\end{refsegment}

\section{Trapped ions}
\label{sec:trappedions}
\begin{refsegment}

Systems of trapped ions have long been seen as a leading contender for the building of viable quantum computers due to the high degree of precise control over both their internal and external states that has been possible for decades at this point~\cite{ion:wineland}. 
The electronic structure of ions allows for qubits that are well isolated from the environment, as well as a simple readout mechanism based on laser induced fluorescence. Since ions are charged, when trapped in the same potential well they naturally form crystals allowing registers of spatially distinguishable but otherwise identical ions to be easily formed. The different vibrational normal modes of these crystals then provides a shared property that allows controlled interaction between normally isolated qubits, allowing for the implementation of multi-qubit gates. 

Trapped ion quantum computers are almost always based on the radio-frequency (RF) Paul trap~\cite{ion:wineland}, although Penning traps have been used to very successfully demonstrate quantum simulations in large 2D ion crystals~\cite{ion:penning-sim}, and there is ongoing work in using arrays of Penning traps for quantum computation and simulation~\cite{ion:penning-array}. The simplest RF traps for quantum computation consist of macroscopic electrodes, which allow a single linear crystal of ions to be formed. However, there are limits to how many ions can be held in a single potential while still having the precise control required for quantum computation.

To be able to scale to large numbers of ions, ions need to be distributed among a number of trapping potentials, in a so-called quantum charged-coupled device (QCCD). Microfabrication techniques allow traps to be built with the large numbers of electrodes required for creating and controlling multiple potential wells, as well as features such as junctions which allows ions to be re-ordered and re-positioned arbitrarily. Such microfabricated trapping devices can be divided into two categories: two- or three-dimensional devices. Three-dimensional devices~\cite{ion:3d-nist, ion:3d-npl} are similar to the macroscopic trap described in terms of their configuration of RF electrodes, and allow strong confinement of the ions. Two-dimensional surface-electrode traps move the trapping electrodes onto a two dimensional plane to produce trapping potentials above the surface of the device~\cite{ion:surface}. Such an electrode configuration produces weaker confinement compared to a three-dimensional device of similar size. However, fabrication is more straightforward, and it offers potentially easier integration of optics, detection elements, and control electronics into the device~\cite{ion:progress-review}. Depending on the height of the ions above the ion-trap device's surface, cryogenic cooling may be required to reduce motional heating rates to an acceptable rate.

Ions used in trapped ion quantum computers tend to be alkaline earth metals, or other elements that have a single outer-shell electron when singly-ionized. Such ions have a strong dipole-allowed transition permitting easy laser cooling, and often possess long-lived low-lying $D$ states. In addition, isotopes with an odd number of nucleons possess hyperfine structure.  There are many possible ways to embody a qubit in an ion, but the two most common are hyperfine qubits, where a pair of states in the ground-state hyperfine manifold are picked, and optical qubits, where one ground-state level is combined with a level in the $D$ state. Hyperfine qubits formed from states both with $m_F=0$ are extremely insensitive to magnetic fields, leading to extremely stable qubits with coherence times of seconds or longer. 

Optical qubits require extremely narrow linewidth lasers to drive quantum gates, with the coherence time of the light providing an upper limit to the coherence time of the qubit. For hyperfine qubits, stimulated Raman transitions can be used to drive two-qubit gates, removing the requirement for highly coherent laser light. Single qubit rotations can be directly driven between hyperfine states by microwave pulses, allowing gates to be performed without using lasers. However, free-space microwave photons have insufficient momentum to drive two-qubit gates in isolation. By using near-field, oscillating magnetic fields generated in the structure of a surface trap, a sufficiently strong mechanical effect can be obtained, allowing laser-free two-qubit gates to be performed~\cite{ion:osc-grad}. Microwave-field driven two-qubit gates can also be obtained if a static magnetic-field gradient is applied at the ions' positions~\cite{ion:static-grad}.

In addition to the ions used for qubits, a second species of ion may be co-trapped to form mixed-species crystals. These allow processes such as cooling and readout to be performed with sufficient spectral separation as to leave the qubits unaffected by the laser light used to excite this auxiliary species. It is possible to mimic this effect with a single species in the so-called `omg' scheme, where optical and ground-state hyperfine qubits are supplemented by qubits embodied entirely in the metastate $D$ states, which can be used to again spectrally hide the qubit state from laser light during cooling operations~\cite{ion:omg}.

While the original Cirac-Zoller two-qubit gate proposal~\cite{ion:cirac-zoller} kick-started the development of ion trap quantum computing, current trapped ion quantum computers generally make use of variants of the M{\o}lmer-S{\o}rensen gate~\cite{ion:m-s}, or of the closely related dynamic phase gate~\cite{ion:dynamic-phase-gate}. These two gates differ from the Cirac-Zoller gate in that they do not require the motional mode to be cooled to the ground-state, instead working by applying state-dependent classical forces to the ions to produce a state-dependent phase, and thus the two-qubit gate. Such classical forces do not require the ions to be in the ground motional state, relaxing the cooling requirements.

All the required quantum computing primitives have been demonstrated with high-fidelities. Single and two-qubit gate fidelities of 99.9999\%~\cite{PhysRevLett.113.220501} and 99.9\%~\cite{ion:two-qubit1, ion:two-qubit2} respectively have been demonstrated, as well as single-qubit detection fidelities of 99.99\%~\cite{ion:readout}. Note that the two-qubit gate fidelities are obtained using the Bell-state preparation error, which acts as an estimate of the process fidelity~\cite{ion:two-qubit1, ion:two-qubit2}.  
Performance in existing trapped ion quantum processors is still impressive, with Quantinuum for instance reporting typical two-qubit fidelities of approximately 99.8\% in a system with 12 physical qubits~\cite{ion:honeywell}.

The values of metrics directly related to quantum computing primitives such as gate fidelities depend on lower level properties of the device. Hardware specific metrics may include: heating rates, including heating due to shuttling or the splitting and combining of ion crystals; fidelities of conversion between different internal qubit representations; and fidelities of photon-mediated remote ion-ion entanglement. In addition, the degree of crosstalk experienced by spectator ions during the application of native gates is also important to be measured and minimized.

The key challenge for trapped ion quantum computing is scaling to the large numbers of physical qubits that are required to obtain even relatively small numbers of logical qubits, once the overhead of quantum error correction is taken into account. 
While a QCCD approach can be used to go beyond the limitations of a single chain of ions, as the number of ions increases, it is increasingly difficult to individually address ions with light delivered externally from the chip. This is the motivator behind increased integration of optical delivery into the chip, as well as a driver towards laser-free gates, again driven by electrodes integrated into the QCCD.
To go beyond the number of ions that can be accommodated on a single QCCD device a number of different approaches have been proposed. One is a modular approach of individual QCCD devices, with ion-photon interconnects used to distributed entangled states between modules\cite{ion:modular}. Another approach suggests tiling individual QCCD devices and shuttling ions across the gaps between individual devices to produce a quasi-monolithic system~\cite{ion:blueprint}. These approaches are not mutually exclusive - it is possible to imagine increasingly large and powerful individual systems linked by photons, or the use of photonic interconnects to provide connectivity between physically distant ions in a single monolithic device.

\ifbool{showpart1references}{\printbibliography[heading=subbibliography,segment=\therefsegment]}{}
\end{refsegment}

\section{Neutral atoms}
\label{sec:neutralAtoms}
\begin{refsegment}
Neutral atom quantum computers have emerged as an exciting candidate for the development of scalable quantum hardware by offering the advantages of large numbers of identical qubits, implementing parallel two-qubit and multi-qubit gate operations, and the option to perform both digital and analogue quantum algorithms~\cite{saffman10,morgado21}.

Using individual atoms trapped in tightly focused optical tweezers, it is possible to create large qubit registers in one~\cite{endres16}, two~\cite{barredo16} or three~\cite{barredo18} dimensions, with arrays of over 1000 sites demonstrated~\cite{huft22}. Similar to trapped ions, qubit properties are defined by atomic physics, with qubits encoded in the hyperfine-ground states for the alkali atoms, or on optical clock transitions with alkaline-earth species, offering long coherence times of up to 40\si{\second}~\cite{barnes22}. Single qubit gates can be implemented using microwave or optical fields, with Clifford randomized benchmarking average gate errors below $10^{-4}$ demonstrated for large-scale systems~\cite{nikolov23}. 

To couple qubits together, atoms are excited to high principal quantum number Rydberg states, which experience strong, long-range dipole-dipole interactions. For atoms sufficiently close together, typically less than about $5-10~\mu$m, the interactions prevent more than a single Rydberg excitation being created, leading to an effect known as Rydberg blockade. This blockade can be exploited to implement high-fidelity two or three qubit gate operations~\cite{jaksch00,lukin01,shi22}, with the state-of-the-art Bell-state fidelities exceeding 99.5\% for two-qubit controlled-$Z$ gates~\cite{evered23}. 

Experimental progress advancing qubit number and gate fidelity has enabled first demonstrations of small-scale algorithms~\cite{graham2022multi}, however combining these interactions with the ability to dynamically reconfigure qubits using mobile tweezers, it is possible to implement arbitrary connectivity's for creating cluster states or topological couplings~\cite{bluvstein22}. This provides a route to future implementation of constant-overhead fault-tolerance~\cite{xu23}, further enabled by recent demonstrations of mid-circuit state readout~\cite{graham23,lis23}. 

Two limitations of the neutral atom hardware come from the susceptibility to loss during the qubit operations, and leakage from the computational basis. Optical traps are typically on the order of millikelvin deep (compared to Kelvin for ion traps) and the major source of atom loss is from collisions with residual background gases. Typical vacuum lifetimes at room temperature are around 10~s, however this loss can be suppressed by moving to cryogenic platforms offering lifetimes exceeding 10~minutes~\cite{schymik2021single}. Leakage from the qubit basis is caused by light-scattering during qubit operations or errors from finite blockade strength. For the alkali atoms, this can be reduced by performing bias-preserving gate operations with the ability to repump atoms into the qubit basis~\cite{cong22}. However, for the alkaline-earth atoms using clock-state qubit encodings, it is possible to implement erasure conversion to eject atoms outside of the qubit basis and replace with new qubits from a reservoir register~\cite{wu22}.

A further challenge for neutral atom systems is achieving fast, high-fidelity readout. Typically this is performed using fluorescence imaging over 1-10~ms timescales limited by the finite scattering rate, with state-selection achieved by first ejecting atoms in $\vert 1\rangle$ from the trap. An alternative approach is to perform non-destructive state-selection, demonstrated on 10~$\mu$s timescales for few atom systems coupled to optical cavities or detected using single photon counters, recently implemented for 49 atoms using an sCMOS camera showing routes to scalable readout~\cite{nikolov23}. Recent proposals for parallel readout using ancilla qubits provide a route to reduce this to microsecond timescales~\cite{petrosyan24,corlett24}.

Beyond the use for gate-base quantum computing, the same neutral atom hardware provides a powerful tool for quantum simulation~\cite{browaeys20}, enabling studies of quantum magnetism and topological spin liquids~\cite{semeghini21}, as well as analogue quantum computing. The native Rydberg interactions can be used to encode classical optimization problems such as the maximum independent set~\cite{pichler18}, as recently demonstrated on hundreds of atoms~\cite{ebadi22}, with further theoretical work extending the application to a broader range of problems including such as weighted graph optimisation~\cite{nguyen23,lanthaler23,deoliveira24}, QUBO problems~\cite{nguyen23}, and integer factorisation~\cite{nguyen23,park24}.

\ifbool{showpart1references}{\printbibliography[heading=subbibliography,segment=\therefsegment]}{}
\end{refsegment}

\section{Photonic devices}
\label{sec:photonic_devices}
\begin{refsegment}
The development of a photonic quantum computer is an attractive alternative to other approaches as it offers the potential of room temperature operation and the promise of scalability. The advent of miniaturized linear optic circuits on photonic integrated circuits (PICs) has increased the interest in developing fault-tolerant photonic quantum computing based on the approach developed in the 2001 article of Knill, Laflamme, and Milburn (KLM)~\cite{KLM2001}, and also the Koashi, Yamamoto, and Imoto (KYI) 2001 article~\cite{Koashi2001}, which demonstrated polarization entangled qubits.
PICs have the advantage that they are easier to couple directly to fiber communication networks and could form an integral part of quantum networks under development, where photonic qubits can be exchanged between remote locations. This makes the existence of a distributed quantum computing resource a viable possibility. Instead of PICs one may also use optical fibers~\cite{Walmsley2023}, which build on commercially available telecommunication components. This does not result in as compact a form factor as a PIC but the architecture can be simpler, as the detectors do not have to share the same space as the linear optics and most of the systems can be operated at room temperature. The utility of this approach is based on a high speed (GHz) noise free optical memory~\cite{Kaczmarek2018}, where two-photon off-resonant cascaded absorption (ORCA) can be used as optical memory inside a hollow core fiber to allow compatibility with telecom components.

An alternative to the KLM universal quantum computation with single-photons and linear optics proposal or the KYI polarization entangled qubit approach is based on measurement-based quantum computation using a Gaussian cluster states and photon-counting detectors~\cite{RMP2012}. This is a non-gate-based quantum computing approach. 
A further alternative to deterministic unitary entangling gates as the base unit of quantum computing, which are difficult to implement in photonic systems, is fusion based photonic quantum computing~\cite{Bartolucci2023}. This model uses entangling measurements referred to as ``fusions”, which are enacted on the qubits produced by entangled resource states, and which are comprized of low numbers of photons. It employs destructive measurements of entanglement to enable fault-tolerant computation.
This differentiates fusion based photonic quantum computing from one-way photonic quantum computing and measurement-based quantum computing, as fusion based can be considered to occupy the space between these other forms. This structure is scalable and requires less classical processing support than other photonic quantum computing implementations.

There are several companies attempting to produce a universal quantum computing platform using photons, each one championing their own variants of the methodologies. Some approaches to this are to incrementally develop linear optics platforms realized on PICs.
Recent articles have claimed fidelities as high as 99.69\%~\cite{Xanadu22, Shi2022, HibatAllah2024}.
A commercial system has claimed quantum computational advantage~\cite{Xanadu22} using both Gaussian boson Sampling (GBS) and squeezed states of light for differing modes to produce Gottesman, Kitaev and Preskill (GKP) qubits. This hybrid methodology aims to produce a fault-tolerant, scalable architecture which seeks to solve the probabilistic problems of successfully making a GKP qubit available on demand rather than having to use a multiplexing approach with many GBS devices to produce a GKP qubit probabilistically~\cite{Bourassa2021}.

As photons are not affected by the heat potential of room temperature, scalable quantum computing is offered by this platform if the current cryogenic requirement for efficient single photon detection can be overcome. The lack of interaction between photons removes crosstalk issues, but conversely means that there is no conditional interaction as used by other platforms that can be used for a gate model. Instead, photonic approaches to quantum computing must use measurement and quantum interference. This approach is by definition probabilistic, hence requiring indistinguishability of the photons used for measurement. Currently, the single photon detectors required for these systems must be operated at cryogenic temperatures, but these temperatures are much higher than those needed for other hardware platforms.

The scalability of quantum computers is an essential attribute in the argument for their eventual utility. A network to integrate remotely located quantum computers which can have different physical manifestations of qubits, for example trapped ion qubits or superconducting qubits, allows the separate computing units to operate as a single computing entity with a commensurate increase in performance. Photonic interconnects are the main platform considered to achieve such networked quantum computers. The quantum networks currently emerging use entangled photons, superposition and quantum measurement to allow qubits from one system to be transmitted to another, either in free space~\cite{Floyd2023} or through optical fiber~\cite{Claude2022, Kozlowski2023}. These systems can be very simple and use a single prepare-and-measure state to transmit a qubit, or can be complex systems capable of exchanging multiple qubits at high speed. The current issues with the low efficiency of transduction from matter qubits, for quantum computing other than photonic based systems, to flying qubits to be transmitted though the quantum network are under improvement and theoretical studies show that high efficiency is possible~\cite{dAvossa2023}. A quantum network can additionally be used for quantum key distribution (QKD) to transmit encryption keys in a demonstrably secure manner. These systems also have application for long baseline telescopes~\cite{Gottesman2012}, secure cloud based computation~\cite{Fitzsimons2017}, quantum enhanced measurement networks~\cite{Giovannetti2004}. Currently the distance between nodes of a quantum network is limited by the decoherence of the qubits unless reliable quantum repeaters can be produced that support the end-to-end generation of quantum entanglement and thus allow a maximally entangled Bell state to be used to continue transmitting the qubit. A reliable quantum network will require well-defined single photon detection, entangled photon production, repeater technology and Bell state measurement to ensure that maximally entangled states are maintained through the network via fiber connections and other currently used telecom components such as Dense Wavelength Division Multiplexing (DWDM). Existing fiber networks could be used, but technologies such as hollow core fiber show great promise for low loss, transmission over a wide spectral range and low latency~\cite{Fokoua2023}. The latest iteration of this type of fiber is Double Nested Antiresonant Nodeless Fiber (DNANF) with less than 0.11 dB/km loss, which is the lowest attenuation reported in an optical fiber~\cite{Chen2024}. 

\ifbool{showpart1references}{\printbibliography[heading=subbibliography,segment=\therefsegment]}{}
\end{refsegment}

\section{Spin qubits}
\label{sec:spin_qubits}
\begin{refsegment}

Spin-based qubits are gaining ground as an appealing platform for the realization of quantum computers. They are primarily implemented in semiconductor nanoscale systems, even though the host material (e.g. silicon, diamond, silicon carbide, germanium), the confinement architecture (e.g. crystal defect, dopant, quantum dot) and the logical encoding (e.g. electronic spin, hole spin, singlet-triplet, exchange only) can vary significantly~\cite{RevModPhys.95.025003, chatterjee2021semiconductor}. A review of performance metrics for spin qubits in semiconductor nanostructures is given in Ref. \cite{2022_SpinQubitsReview_Stano}.
Among the different types, those which have attracted most recent attention for their promise of scalability through compatibility with commercial manufacturing methods, such as the Complementary Metal-Oxide-Semiconductor (CMOS) technology, are based on electrostatically defined quantum dots (QDs) in silicon~\cite{De_Michielis_2023}. Lately, such qubits have been produced in increasingly large volumes in industrial foundries (e.g. at Intel, IBM, LETI, IMEC) by adapting the well-established process used for conventional transistor technology nodes~\cite{zwerver2022qubits}. In fact, it has been demonstrated that mass-produced transistors realized in silicon-on-insulator substrates can host spin-qubits when operated at cryogenic temperature~\cite{maurand2016cmos,camenzind2022hole}. However, despite compatibility with standard manufacturing methods, spin qubits are subject to large variability and lack of reproducibility in terms of operation points, noise susceptibility and performances. This mostly stems from unavoidable atomic-scale differences among nominally identical qubits due to the presence of random impurities in the material stack or unintentional variations in the lithographic patterns. In order to improve the understanding and impact of these reproducibility issues, extensive statistical studies on large numbers of samples are needed. Techniques to carry this out efficiently at scale are actively being developed for the cryogenic environment~\cite{10.1063/5.0139825,neyens2024probing, eastoe2024efficient, thomas2023rapid}.

Although spin qubits are typically operated at millikelvin temperature in dilution refrigerators, the temperature constraints are much less stringent than for superconductor qubits. In fact, recent results have shown that it is possible to operate transistor qubits at temperatures above 1\si{\kelvin}~\cite{camenzind2022hole,yang2020operation}. The relaxed temperature requirement sets promises for developing chips where the power-hungry control electronics are monolithically integrated or heterogeneously co-located near the qubit layer thanks to the large cooling power available at the few-Kelvin stage of pulse tube refrigerators~\cite{xue2021cmos, bartee2024spin}. Having the control and readout electronics physically close to the quantum processor could prove a step change towards solving the wiring bottleneck issues that have so far limited scalability~\cite{vandersypen2017interfacing}.

A distinctive advantage of spin qubits is that the host crystal can be designed to minimize sources of magnetic noise. In fact, hyperfine interactions with nuclear spins in the host crystal do limit the coherence time. However, for silicon systems the most abundant isotope (28Si: 92.23\%) carries no nuclear spin, whilst the second most abundant isotope (29Si: 4.67\%) does. Through epitaxial growth of isotopically purified crystals, the relative 28Si content can be enhanced (>99.99\%). This has led to orders of magnitude improvements in the coherence times of qubit gates, achieving several milliseconds~\cite{veldhorst2014addressable}. 

The highest average gate fidelities obtained by Clifford randomized benchmarking are up to 99.96\% for single qubit~\cite{yang2019silicon} and up to 99.81\% for two qubits~\cite{doi:10.1126/sciadv.abn5130}.  Gate times as fast as 2.5\si{\nano\second} have been obtained for the singlet-triplet encoding~\cite{jock2022silicon}. The fidelities for initialization and measurement can also be as high as 99\%~\cite{doi:10.1126/sciadv.abn5130,watson2018programmable}.

Multi-qubit connectivity has proved challenging. Spin qubits in silicon have been primarily coupled via nearest neighbour techniques. To this end, the exchange interaction can be controlled via dedicated electrodes and is used to execute two-qubit gates between spins in adjacent QDs. The current state of the art is a linear array of six QDs~\cite{philips2022universal}. While it is important to go beyond nearest-neighbour connectivity, this is difficult to achieve since the exponential decay of the exchange interaction is determined by that of the wavefunction. One possibility is physical shuttling of the electron to a remote location~\cite{PRXQuantum.4.030303}. Another approach is to couple distant spins via an on-chip microwave cavity, employing an intrinsic or induced spin-orbit interaction ~\cite{mi2018coherent}. However, these demonstrations have limited scaling prospects because they require control lines for every qubit, eventually leading to impractically large footprints or interconnect bottlenecks. To solve these issues, one could leverage the commonalities between silicon quantum devices and conventional integrated circuits. For example, proposals currently under scrutiny suggest using a crossbar architecture~\cite{veldhorst2017silicon,doi:10.1126/sciadv.aar3960} based on combination of row and column lines used to address qubit nodes on a grid, a solution similar to that adopted in classical random-access memory technology.  

\ifbool{showpart1references}{\printbibliography[heading=subbibliography,segment=\therefsegment]}{}
\end{refsegment}

\newpage
\vspace*{-4\baselineskip}
\chapter{Concepts and methodologies used for multiple metrics and benchmarks}
\label{chapter:methodologies}
This section presents all the mathematical descriptions of the concepts introduced in the earlier sections that are used for the definitions and methodologies of the metrics and benchmarks presented in Sec. \ref{part:metrics}. It first presents the used mathematical definitions of qubit operations and noise descriptions. It then outlines volumetric benchmarking, a general approach that is common across a number of benchmarks. This is followed by a technical description of gate set tomography, which is used for a number of metrics in Sec. \ref{part:metrics}. It is noted that the metrics in Sec. \ref{part:metrics} are presented in a self-contained approach, so that it is not necessary to read this technical section before reading the metrics in Sec. \ref{part:metrics}. It rather provides all the technical and mathematical background as reference. At the end of this section a list of symbols used throughout the document is given as additional lookup reference.

\section{Qubit states and operations in presence of noise}
\label{sec:methodsmaths}

\subsection{Noiseless qubit states and operations}
\begin{refsegment}

\begin{figure}[!htpb]
    \centering
 \includegraphics[width=\linewidth]{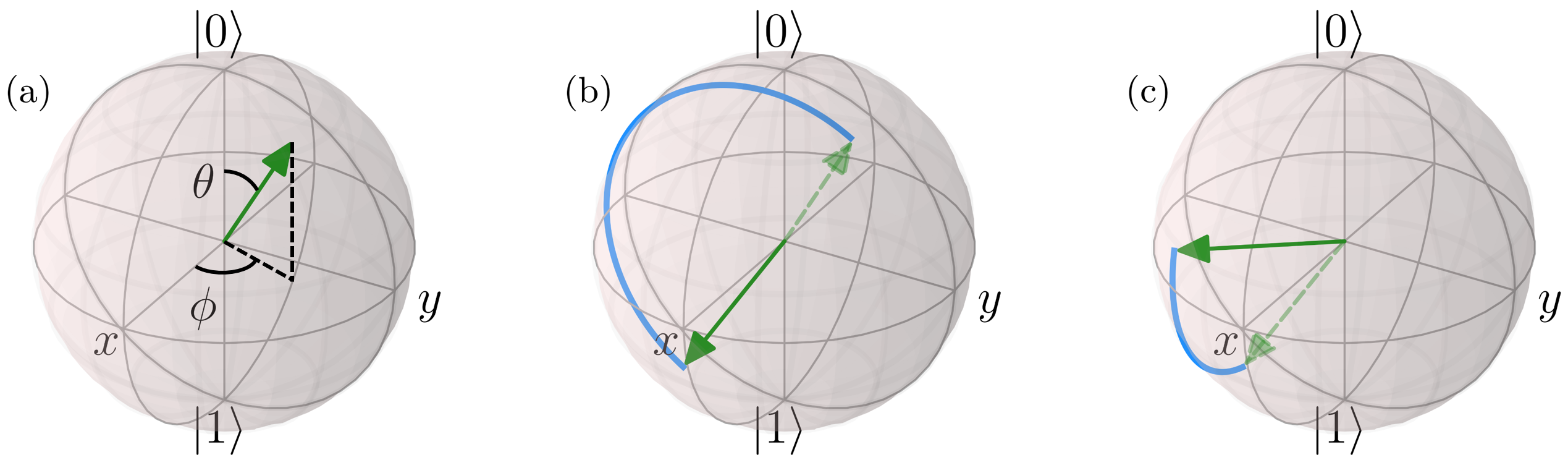}
\caption{Bloch sphere visualizations of (a) a general qubit state $\ket{\psi} = \cos{(\theta)}\ket{0} + e^{i\phi}\sin{(\theta)}\ket{1}$ as green arrow, (b) an $X$ gate applied to this $\ket{\psi}$, leading to $\ket{\psi_1} = X\ket{\psi}$, and (c) an $R_y(-\pi/2)$ gate applied to $\ket{\psi_1}$, leading to $\ket{\psi_2} = R_y(-\pi/2)\ket{\psi_1}$.}
\label{fig:bloch}
\end{figure}

In Sec. \ref{sec:gate_based_quantum_computing} an overview of gate-based quantum computing is given. Here the mathematical descriptions of the gate operations is provided. Following on from the relations introduced in Sec.  \ref{sec:gate_based_quantum_computing}, ignoring a physically irrelevant global phase one may write a general single qubit pure state as $\ket{\psi} = \cos{(\theta/2)}\ket{0} + e^{i\phi}\sin{(\theta/2)}\ket{1}$, where $\theta$ and $\phi$ are real valued rotation angles. One may represent such a general qubit state as a vector on a unit sphere, named the Bloch sphere, as depicted in Fig.~\ref{fig:bloch}(a). 
Some commonly used superposition states are denoted as:
\begin{itemize}
    \item $\ket{+} = \frac{1}{\sqrt{2}} (\ket{0} + \ket{1})$,
    \item $\ket{-} = \frac{1}{\sqrt{2}} (\ket{0} - \ket{1})$,
    \item $\ket{R} = \frac{1}{\sqrt{2}} (\ket{0} + i\ket{1})$,
    \item $\ket{L} = \frac{1}{\sqrt{2}} (\ket{0} - i\ket{1})$.
\end{itemize}
Of these four example states two are real state vectors ($\ket{+}$ and $\ket{-}$, as they only have real-valued coefficients) and two are complex state vectors ($\ket{R}$ and $\ket{L}$, as they have imaginary-valued coefficients $i/\sqrt{2}$). In general, all qubit states live in a complex vector space that is equipped with an inner product and is known as the Hilbert space. 

Quantum gate operations on qubits are described by unitary operators. Denoting a unitary operator by $U$, it evolves a state vector $\psi$ into $U\ket{\psi}$. Unitary operators preserve the length of a complex vector, so that single qubit gates can be interpreted as rotations in the Bloch sphere representation. A widely used set of gates are the so called Pauli gates $X$, $Y$ and $Z$. These correspond to Pauli matrices $\sigma_x$, $\sigma_y$ and $\sigma_z$, respectively, acting on a qubit:
\begin{align}
    X = \sigma_x = \begin{bmatrix} 0 & 1 \\ 1 & 0 \end{bmatrix},\;
    Y = \sigma_y = \begin{bmatrix} 0 & -i \\ i & 0 \end{bmatrix},\;
    Z = \sigma_z = \begin{bmatrix} 1 & 0 \\ 0 & -1  \end{bmatrix}.
\end{align}
In the Bloch sphere representation, these gates correspond to a rotation of $\pi$ around the $x$, $y$ or $z$ axes, respectively. In particular, the $X$ gate is also a quantum generalization of the classical logical NOT gate, since it transforms as $\ket{0}$ state to a $\ket{1}$ state and vice versa. Arbitrary rotations by an angle $\theta$ around the three axes are generated by matrix exponentiation of the corresponding Pauli matrices. For example, an arbitrary rotation around the $x$-axis is given by $R_x(\theta) = \exp[-iX\theta/2] = \cos(\theta/2)I-i\sin(\theta/2)X$, where $I$ is the identity matrix. Example Bloch sphere visualizations of the $X$ and $R_y(\pi/4)$ gates are shown in Fig.~\ref{fig:bloch}(b, c), respectively.
Other commonly used single qubit gates include the Hadamard gate, $H$, and the phase gate, $S$:
\begin{align}
\label{eq:hadamard_phase_gate_defs}
    H = \frac{1}{\sqrt{2}} \begin{bmatrix} 1 & 1 \\ 1 & -1 \end{bmatrix},\;\;
    S = \sqrt{Z} = \begin{bmatrix} 1 & 0 \\ 0 & i \end{bmatrix}.
\end{align}

As described in Sec. \ref{sec:gate_based_quantum_computing}, for a general qubit state $\ket{\psi} = \alpha \ket{0} + \beta \ket{1}$, with $\alpha$ and $\beta$ complex numbers and $|\alpha|^2 + |\beta|^2 = 1$, a measurement of the qubit state in the $\ket{0}$ and $\ket{1}$ basis causes the qubit's wavefunction to collapse, outputting $\ket{0}$ with probability $|\alpha|^2$, or outputting $\ket{1}$ with probability $|\beta|^2$. While measurements in different bases of the qubit state are possible, the $\ket{0}$ and $\ket{1}$ basis is the most commonly used, and is referred to as the computational basis. 

Mathematically, a measurement involves projecting the state of the qubit to a particular basis by applying a complete set of measurement projectors $\{M_j=\ket{j}\bra{j}\}$ to the qubit state, so that the projectors sum to the identity matrix, $\sum_j M_j=I$. An example is the measurement in the computational basis, where the set of projectors is given by $\{M_0=\ket{0}\bra{0}, M_1=\ket{1}\bra{1}\}$. Because $\ket{0}$ and $\ket{1}$ are eigenstates of the Pauli-$Z$ operator, this measurement is also called a $Z$-basis measurement.
It is sometimes useful to measure the qubit in different bases. This can be done by adding a basis change operation before the computational basis measurement. For example, to perform a measurement in the $X$ basis, a Hadamard gate $H$ is used before the measurement. To measure in the $Y$ basis, an $S^\dagger$ gate followed by a $H$ gate is applied before measurement~\cite{adedoyin2018quantum}. These basis change operations can also be used to prepare the qubit in the different bases. For example, the $H$ gate can be applied to the $\ket{0}$ state to obtain the $\ket{+}$ state.

The measurements of the qubit can be used to calculate expectation values of the different Pauli operators. For example, the expectation value of the Pauli-$Z$ operator, $\expval{\sigma_z}$, is given by $\expval{\sigma_z} = 1 - 2p_1$, where $p_1$ is the probability of the measurement outcome being $1$, corresponding to the eigenstate of the Pauli-$Z$ operator with eigenvalue $-1$. The values of $\expval{\sigma_x}$ or $\expval{\sigma_y}$ can be obtained using the appropriate basis change operations before measurement.

For systems consisting of more than one qubit, states can be represented by tensor products. For example, in a two qubit system, the two qubit state can be obtained from the tensor product of single qubit states: 
\begin{align}
    \ket{\psi} = \ket{\psi_1} \otimes \ket{\psi_2} = \begin{bmatrix} \alpha_1 \\ \beta_1 \end{bmatrix} \otimes \begin{bmatrix} \alpha_2 \\ \beta_2 \end{bmatrix} = \begin{bmatrix} \alpha_1 \alpha_2 \\ \alpha_1 \beta_2 \\ \beta_1 \alpha2 \\ \beta_1 \beta_2 \end{bmatrix}.
\end{align}
For $N_\mathrm{q}$ qubits the dimension of the state vector becomes $2^{N_\mathrm{q}}$, and hence grows rapidly with number of qubits.

Commonly used two-qubit gates include the swap gate, SWAP, the controlled-NOT gate, $\mathrm{C}_X$, also known as the CNOT gate, and the general controlled-unitary gate $\mathrm{C}_U$:
\begin{align}
    \text{SWAP}=\begin{bmatrix} 1 & 0 & 0 & 0 \\ 0 & 0 & 1 & 0 \\ 0 & 1 & 0 & 0 \\ 0 & 0 & 0 & 1 \end{bmatrix},\;
    \mathrm{C}_X=\begin{bmatrix} 1 & 0 & 0 & 0 \\ 0 & 1 & 0 & 0 \\ 0 & 0 & 0 & 1 \\ 0 & 0 & 1 & 0 \end{bmatrix},\;
    \mathrm{C}_U =\begin{bmatrix} 1 & 0 & 0 & 0 \\ 0 & 1 & 0 & 0 \\ 0 & 0 & U_{00} & U_{01} \\ 0 & 0 & U_{10} & U_{11} \end{bmatrix},
\end{align}
where the block matrix $U=\begin{bsmallmatrix} U_{00} & U_{01} \\ U_{10} & U_{11}\end{bsmallmatrix}$
is a single-qubit unitary that acts on the target qubit only if the control qubit is in the $\ket{1}$ state. The SWAP gate exchanges the states of two qubits, and the C-$U$ or CNOT gates are often used to entangle two qubits. When qubits are entangled, their compound state can no longer be represented by tensor products, and measurements performed on one qubit restricts possible measurement outcomes on other entangled qubits. For example, applying a $\mathrm{C}_X$ gate, with the second qubit as target qubit, to a two-qubit non-entangled state $\frac{1}{\sqrt{2}}\qty(\ket{0}+\ket{1})\otimes\ket{0}$ results in the entangled state $(\ket{00}+\ket{11})/\sqrt{2}$. This state cannot be written as a tensor product of single-qubit states, and measuring one of the qubit guarantees that the other qubit must be measured in the same state, i.e., both are $\ket{0}$ or both are $\ket{1}$. The capability of quantum computers to generate entangled states is one of the requirements for quantum advantage.

Typically, quantum computing hardware can only directly implement one- and two-qubit operations. However, for $N_\mathrm{q}$ qubits, any $N_\mathrm{q}$-qubit operation can be decomposed into several one- and two-qubit gates by mathematically decomposing the unitary matrix into products of smaller ones, enabling the execution of arbitrary $N_\mathrm{q}$-qubit gates. Operations that apply to a subset of qubits are written as tensor products with identities on remaining qubits. For example, in a four qubit system, an operation that applies an $X$ gate on the first qubit and a $Y$ gate on the third qubit is written as $X \otimes I \otimes Y \otimes I$, following a convention of qubit indices ordered from left to right. 

Often, random unitary gates are used as part of a quantum circuit~\cite{dimatteoUnderstandingHaarMeasure2021, crossValidatingQuantumComputers2019, magesanScalableRobustRandomized2011}. When selecting unitary gates at random, it is important to ensure that all possible unitary gates have an equal probability of being selected. In order to ensure that this is the case, random unitary gates are selected from a probability distribution over unitary operations, known as the Haar measure~\cite{Watrous_2018, Elizabeth2019}.

\end{refsegment}

\subsection{Effects of noise}
\label{sec:noise_effects}
\begin{refsegment}

When there is some classical randomness in the state of a qubit, such as when the qubit can be in different quantum states $\ket{\psi_i}$ with classical probabilities $p_i$, then its state can be described using the so called density matrix formalism. In this formalism the state is represented as $\rho=\sum_i p_i \ket{\psi_i}\bra{\psi_i}$, where $\rho$ is called the density matrix or the density operator of the qubit, and $p_i$ is the probability of state $\ket{\psi_i}$ to occur.
If no classical randomness is involved, a state is said to be a pure state with its density matrix $\rho=\ket{\psi}\bra{\psi}$, and it can be fully described using its state vector $\ket{\psi}$ alone. On the other hand, if classical randomness is present, then the state is said to be a mixed state. In practice, mixed states often arise as a result of noise in the quantum system, where the classical probabilities in $\rho$ reflect loss of quantum information regarding the exact quantum state the qubit is in. The density matrix formalism is therefore often used in quantum computer benchmarking methods, since a number of these characterize the effects of noise on the quantum computation. 

Given an arbitrary quantum state described by a density matrix $\rho \in \mathcal{H}$, where $\mathcal{H}$ is an arbitrary Hilbert space, $\rho$ transforms as follows~\cite{nielsen2002quantum}:
\begin{equation}
    \rho' = \Phi (\rho),
\end{equation} where $\Phi$ is a quantum operation, also known as quantum process, that maps $\rho$ to a new density matrix $\rho'$. Quantum operations can be used to describe both unitary gates and non-unitary noise processes in quantum computers. A unitary gate operation is a special example of a possible quantum operation, and evolves a density matrix $\rho$ into $U \rho U^\dag$.

The quantum operation $\Phi$ must meet the following requirements in order to represent a physical process. For any initial state of the system and environment, the transformed density matrix after applying the quantum operation must be positive semidefinite. This means that the final density matrix must have non-negative eigenvalues. This requirement is known as complete positivity (CP). Another common requirement of quantum operations is that they must also be trace preserving, such that $\Tr[\Phi(\rho)] =  \Tr[\rho]$. This can be understood as the conservation of probability~\cite{nielsen2002quantum}. When considering quantum operations on qubits, this requirement may be relaxed in case of transitions outside the qubit subspace; in this scenario, the total probability cannot increase~\cite{bhandari2016general}.

By considering only the system and not the environment, one can write a CP quantum operation as
\begin{equation}
    \Phi(\rho) = \sum_i^{M} K_i \rho K_i ^{\dagger},
\end{equation}where $M \leq d^2$, where $d=2^{N_\mathrm{q}}$ is the Hilbert space dimension and $N_\mathrm{q}$ is the number of qubits. This is called the operator-sum or Kraus representation. The operators $K_i$ are called operator elements or Kraus operators, and they have the same dimension as $\rho$. These Kraus operators do not need to be unitary, Hermitian or invertible. Additionally, the Kraus operators for a quantum operation are not necessarily unique~\cite{nielsen2002quantum}. If the quantum operation is trace preserving, then $\sum_i K^{\dagger}_iK_i = I$, where $I$ is the identity.

Quantum operations can also be represented in other ways. One widely used representation is the Pauli transfer matrix form~\cite{nielsenGateSetTomography2021, greenbaum_introduction_2015}. It is calculated by expanding $\Phi$ in the $N$-qubit Pauli basis $P_i \in \{I,X,Y,Z\}^{\otimes N}$ such that~\cite{greenbaum_introduction_2015}
\begin{equation}
    \qty(S_{\Phi})_{ij} = \frac{\Tr[P_i \Phi(P_j)]}{2^{N_\mathrm{q}}}.
\end{equation}Here, $S_{\Phi}$ is the Pauli transfer matrix (PTM) of $\Phi$. 

Noise in quantum computers can be described as coherent or incoherent, or as a mixture of both~\cite{Iverson_2020}. Noise can be described as coherent when only a single Kraus operator is required to describe the noise. Incoherent noise is stochastic and requires more than one Kraus operator to describe the noise process~\cite{nielsen2002quantum}.
Coherent noise is noise that is defined by a quantum operation that is described by a single Kraus operator\cite{Iverson_2020}. For example, an over-rotation error around the $x$-axis of angle $\theta$ has a Kraus operator of $e^{i \frac{\theta}{2}X}$ where $X$ is a Pauli operator.
In the case of a noiseless ideal quantum gate, $\Phi_{\text{Ideal}}$, there is only a single Kraus operator, which is given by the unitary matrix of the target gate, $U$, so that
\begin{equation}
    \Phi_{\text{Ideal}}(\rho) = U \rho U^{\dagger}.
\end{equation}
The Kraus operator representation of the quantum operation assumes that the noise is Markovian, which corresponds to a memoryless environment. This means that the noise is assumed to be independent of time during the application of the quantum operation. In physical qubits a number of sources of noise cause memory effects. When such effects are present, one needs to consider non-Markovian noise descriptions, which can account for memory effects arising due to the environment~\cite{agarwal2023modelling}. Coupling to coherent two level systems and slow fluctuations of qubit frequency are examples of mechanisms that can lead to such memory effects. It is therefore important to also consider benchmarks that characterize the amount of non-Markovianity of the noise.

In what follows we present a number of commonly occurring noise contributions with their associated Kraus operators.

\textit{\underline{Amplitude and Phase Damping}}
Amplitude damping describes energy dissipation from the qubit. For example, in trapped ion qubits, amplitude damping occurs due to spontaneous emission of photons from the ion~\cite{PhysRevA.107.052409}. The Kraus operators for the amplitude damping channel are given by
\begin{equation}
    K_0^{\mathrm{AD}} = \begin{pmatrix}
1 & 0\\
0 & \sqrt{1-\gamma^{\mathrm{AD}}}
\end{pmatrix}, \; K_1^{\mathrm{AD}} = \begin{pmatrix}
0 & \sqrt{\gamma^{\mathrm{AD}}}\\
0 & 0
\end{pmatrix}.
\end{equation} where $\gamma^{\mathrm{AD}}$ is the probability of decay from excited state to ground state. Here $K_1^{\mathrm{AD}}$ acts on the qubit and induces a transition from $\ket{1}$ to $\ket{0}$. The operator $K_0^{\mathrm{AD}}$ describes how the state changes if there is no transition. 

Phase damping noise, also called dephasing noise, describes the decay of off-diagonal elements in the qubit density matrix. For example, in superconducting qubits, dephasing noise arises due to interaction with charge  ~\cite{krantz_quantum_2019,agarwal2023modelling}. The Kraus operators for dephasing are represented by a phase-flip channel~\cite{nielsen2002quantum} and are given by
\begin{equation}
    K_0^{\mathrm{PD}} = \sqrt{1-\frac{\gamma^{\mathrm{PD}}}{2}} I, \; K_1^{\mathrm{PD}} = \sqrt{\frac{\gamma^{\mathrm{PD}}}{2}} Z,
\end{equation} where $I$ and $Z$ are the identity and Pauli Z operators. Here $K_1^{\mathrm{PD}}$ projects the qubit state onto the $z$-axis in the Bloch sphere with probability $\gamma^{\mathrm{PD}}$, and $K_0^{\mathrm{PD}}$ does nothing with probability $1-\gamma^{\mathrm{PD}}$.

The Kraus operators for the combined amplitude and phase damping channel are given by~\cite{PhysRevA.86.062318},
\begin{align}
\nonumber
    K_0^{\mathrm{AP}} &= \begin{pmatrix}
1 & 0\\
0 & \sqrt{1-\gamma^{\mathrm{AD}} - (1 - \gamma^{\mathrm{AD}})\gamma^{\mathrm{PD}}}
\end{pmatrix}, \\
\nonumber
K_1^{\mathrm{AP}} &= \begin{pmatrix}
0 & \sqrt{\gamma^{\mathrm{AD}}}\\
0 & 0
\end{pmatrix}, \\ K_2^{\mathrm{AP}} &= \begin{pmatrix}
0 & 0\\
0 & \sqrt{(1 - \gamma^{\mathrm{AD}})\gamma^{\mathrm{PD}}}
\end{pmatrix}
\label{eq:ap_noise}
\end{align}

One can also represent the amplitude and phase damping parameters in terms of the qubit relaxation time, $T_1$ (metric~\ref{sec:t1}), and of the qubit dephasing time, $T_2$ (metric~\ref{sec:t2}), with the following relation~\cite{PhysRevA.86.062318}:
\begin{align}
\nonumber
    1 - \gamma^{\mathrm{AD}} = e^{-t/T_1}\\
\label{eq:damping_params_to_t1}
    \sqrt{(1 - \gamma^{\mathrm{AD}})(1 - \gamma^{\mathrm{PD}})} = e^{-t/T_2}
\end{align}
These qubit relaxation times are commonly used metrics to characterize the quality of a single qubit.

\textit{\underline{Depolarizing noise}}
Depolarizing noise describes the noise process, where the qubit transforms to the maximally mixed state with probability $\gamma^{\mathrm{D}}$, and does nothing with probability $1-p$~\cite{nielsen2002quantum}. This describes a noise process of information in the qubit being lost. The Kraus operators for depolarizing noise are given by
\begin{equation}
     K_0^{\mathrm{D}} = \sqrt{1-\frac{3\gamma^{\mathrm{D}}}{4}} I, \; K_1^{\mathrm{D}} = \frac{\sqrt{\gamma^{\mathrm{D}}}X}{2}, \; K_2^{\mathrm{D}} = \frac{\sqrt{\gamma^{\mathrm{D}}}Y}{2}, \; K_3^{\mathrm{D}} = \frac{\sqrt{\gamma^{\mathrm{D}}}Z}{2},
     \label{eq:depolnoise}
\end{equation}where $I$ is the identity, and $X,Y,Z$ are the the Pauli operators.

\end{refsegment}

\subsection{Noise model used in quantum computing emulator runs of metrics}
\label{sec:noise_model}
\begin{refsegment}

In a number of  of the metrics in Sec. \ref{part:metrics} results are given using an emulator of a quantum computer including the effects of noise. In this section the details of the emulator runs and of the used parameters and noise model are provided.

The quantum computer is emulated with all-to-all connectivity (see metric ~\ref{sec:connectivity}). The gates used in the emulator gate set are $\{I, R_x(\pi/2), R_z(\theta) , C_X\}$, where  $R_z(\theta)$ is a parameterized rotation gate with rotation angle $\theta$.
To this aim the emulator within the Qiskit open-source quantum computing software library is used~\cite{qiskit2024}. The noise model in Qiskit allows the user to specify the exact type of noise applied when executing a quantum circuit. 
The noise model used in this documents includes the noise contributions specified in Sec.~\ref{sec:noise_effects}, where for amplitude damping and depolarizing noise Eqs. \ref{eq:ap_noise} are used.
In order to obtain meaningful amplitude damping and phase damping parameters, Eq.~\ref{eq:damping_params_to_t1} was used after specifying the $T_1$ and $T_2$ times of the qubits. The $T_1$ and $T_2$ times are selected from a random normal distribution with a mean of 50\si{\micro\second} and 70\si{\micro\second} respectively, and a standard deviation of 1\si{\micro\second} for both. To ensure repeatability, a random seed is set such that the T1 times selected for each qubit remain the same. In order to apply amplitude and phase damping noise when executing a circuit, the gate times must also be known. The following gate times are used: time for idle gate ($I$) is 50\si{\nano\second}, time for $R_x(\pi/2)$ gate is 50\si{\nano\second}, time for $C_X$ gate is 300\si{\nano\second}, and time for measurement is 1000\si{\nano\second}.
 These values are motivated by typical gate times found in superconducting qubits (see Sec.~\ref{sec:superconductingQubits}).
There is no noise applied on the $R_z(\theta)$ as it is modelled to be a virtual gate that is applied by adding a phase to the following gates~\cite{PhysRevA.96.022330}.

For all of the $R_x(\pi/2)$ gates applied, after the ideal $R_x(\pi/2)$ gate the following noise contributions are added:
\begin{itemize}
    \item an over-rotation around the $x$-axis of $\pi/100$ to simulate coherent calibration errors,
    \item a rotation about the $z$-axis of $\pi/120$ to simulate the coherent phase error occurring due to the applied pulse being detuned from the qubit frequency,
    \item a depolarizing noise channel to approximate effectively averaged noise in a large quantum circuit~\cite{PhysRevLett.127.270502}. The depolarizing parameter used in Eq. \ref{eq:depolnoise} for this gate is $\gamma^{D}=0.0005$.
\end{itemize}.
For all of the 2-qubi C$_X$ gates applied, after the ideal C$_X$ gate the following noise contributions are added:
\begin{itemize}
    \item an $e^{\frac{-i}{2}\sigma_z \otimes \sigma_x \theta_{zx}}$ operation and an $e^{\frac{-i}{2}\sigma_z \otimes \sigma_z \theta_{zz}}$ operation on the 2-qubit subspace the $C_X$ gate acts on. The parameters $\theta_{zx}$ and $\theta_{zz}$ are both set to $\pi/100$. The $zx$- and $zz$- rotation axes are chosen to reproduce some of the dominant sources of coherent error when applying a cross-resonance gate in superconducting qubits~\cite{PhysRevA.101.052308, PhysRevA.102.042605},
    \item a depolarizing noise channel, with depolarizing parameter in Eq. \ref{eq:depolnoise} set to $\gamma^{D}=0.005$. It is larger than the value used for single qubit gates, as two-qubit gates typically have larger average errors.
\end{itemize} 

\ifbool{showpart1references}{\printbibliography[heading=subbibliography,segment=\therefsegment]}{}
\end{refsegment}

\section{Volumetric benchmarking}
\label{sec:volumetric_benchmarking}
\begin{refsegment}

Volumetric benchmarking is a method to probe the overall performance of a quantum computer. The method is based on the execution of quantum circuits with varying number of qubits, $N_\mathrm{q}$, and with varying depth, $d_c$~\cite{blume-kohoutVolumetricFrameworkQuantum2020, crossValidatingQuantumComputers2019, lubinskiApplicationOrientedPerformanceBenchmarks2021}. In the context of volumetric benchmarking, the number of qubits is often also denoted as circuit width, $w$. It allows one to evaluate how noise affects the results of quantum circuits when progressively increasing $w$ and $d_c$. For each pair $(w,d_c)$ a number of test circuits are specified, along with a criterion for when the execution of the quantum circuit is considered to be successful. By executing the sets of circuits on quantum hardware, and comparing results with those expected for noiseless quantum computers, one can quantify how successful a given device is in executing the specified tasks. The results are usually depicted via a graphical summary that illustrates the performance of the device for different values of $w$ and $d$~\cite{blume-kohoutVolumetricFrameworkQuantum2020, lubinskiApplicationOrientedPerformanceBenchmarks2021}. 

There are multiple metrics that use the volumetric benchmarking framework, which include:
\begin{itemize}
    \item \hyperref[sec:quantum_volume]{quantum volume (metric}~\ref{sec:quantum_volume}),
    \item \hyperref[sec:mirror_circuits]{mirrored circuits average polarization (metric}~\ref{sec:mirror_circuits}),
    \item \hyperref[sec:algorithmic_qubits]{algorithmic qubits (metric}~\ref{sec:algorithmic_qubits}).
\end{itemize}

\noindent A volumetric benchmarking process has the following components:
\begin{enumerate}
    \item A selection of a family of quantum circuits $C(w,d_c)$ that are to be tested for each $w$ and $d_c$, along with a rule of how to sample from this family. 
    \item An experimental design of how the circuits are to be run, including for example setting the number of shots, $n_{\mathrm{s}}$.
    \item A rule restricting how the specified circuits are to be compiled to the native gates of the hardware tested.
    \item A rule to measure how successful a run of the circuit is. It takes the measurement outcomes of a $w$-qubit circuit with $n_{\mathrm{s}}$ shots as input, and returns either a single bit $\{0,1\}$ that represents fail or pass, respectively, or a real number $\in [0,1]$ that quantifies how well the hardware scored in the test.

    There are many different ways of measuring the success of running a circuit. For example, one measure can be that a quantity needs to be greater than a given threshold, which is an approach used within the quantum volume metric presented in Sec.~\ref{sec:quantum_volume}.
    Alternatively, one may compare the measurement outcome probability distribution with the ideal distribution, which is an approach used for the algorithmic qubits metric in Sec.~\ref{sec:algorithmic_qubits}.
    \item A measure of overall success over the ensemble of circuits that are tested. Similar to measuring the success of a circuit run, there are different ways to measure if an ensemble of circuits, $C(w,d_c)$, is executed successfully.
    
    One measure is to require that a specified fraction of the total number of circuits in $C(w,d_c)$ runs successfully in order for $C(w,d_c)$ to pass. Alternatively, a stricter requirement can be that all circuits in $C(w,d_c)$ need to pass in order for $C(w,d_c)$ to pass.
    \item A specified approach for the analysis of the results, which can be the extraction of a single number for a metric, or a plot of all the results for a more general analysis. 
\end{enumerate}

\subsection{Measurement procedure}

Since this is a framework, the specific measurement procedure depends on the chosen algorithms to be used as test circuits. Here a general procedure is presented:
\begin{enumerate}
    \item Select the dimensions $(w,d_c)$ and the family of circuits $C(w,d_c)$ to be tested. Define a method to sample from $C(w,d_c)$, for example by specifying a distribution. Additionally, define the success criteria for each circuit, and define the success criteria for the circuit ensembles $C(w,d_c)$.
    \item Select one circuit randomly according to the defined sampling method, and compile it according to the native gates of the hardware.
    \item Implement the circuit on the hardware and store the measurement outcome. 
    \item Choose a number of shots to be used, $n_{\mathrm{s}}$. Using the $n_{\mathrm{s}}$ measurement outcomes, compute the success value of the hardware for the circuit performed using the success criteria defined in step 1. 
    \item Repeat steps 2-4 and collect the success values for each circuit in $C(w,d_c)$. A success value for the overall performance of the hardware at the dimension tested can be computed using the success criteria for the circuit ensemble defined in step 1.
    \item Collect the results for different dimensions; these are usually plotted in a 2D figure expressing how the hardware performs for different widths and depths.
    \item Optionally, based on the collected results, an overall single metric number is extracted based on a specified criterion. 
\end{enumerate}

\subsection{Assumptions and limitations}

\begin{itemize}
    \item One common assumption is that the performance with respect to the property tested does not vary significantly over time, especially during the testing of different circuits and dimensions.
    \item This framework relies on a meaningful choice of the success criteria for circuits, as well as of the circuit ensembles. This choice is largely arbitrary, and hence leads to the potential problem that the chosen success criteria may be biased, and may favour one hardware platform over another. Since there is no specific set of rules that define whether a success criterion is a relevant one, it has the potential to be biased. Any results using volumetric benchmarking should precisely explain and motivate the success criteria used.
    \item Whether or not this framework is scalable depends on the success criterion used. Typically it requires comparison to the ideal measurement outcomes, in which case it is only scalable if these outcomes can be classically computed efficiently also for larger qubit numbers and circuit depths. For systems where the circuits selected for the volumetric benchmarking are generated randomly, the number of possible quantum circuits to sample from may increase steeply with the number of qubits. For these cases the results can depend on the used sampling strategy. 
\end{itemize}

\ifbool{showpart1references}{\printbibliography[heading=subbibliography,segment=\therefsegment]}{}
\end{refsegment}

\section{Gate set tomography}
\label{sec:gst}
\begin{refsegment}

Gate set tomography (GST) gives a matrix representation for each gate in a user defined set of gates, which is denoted as the process matrix, and also provides estimates for the initial state preparation and for measurement accuracy in a quantum computer~\cite{greenbaum_introduction_2015, nielsenGateSetTomography2021}. Typically GST is used only for very few qubits, often just one or two, since the parameter space needed to define larger systems grows exponentially~\cite{nielsenGateSetTomography2021}.

GST is a methodology to characterize operations on qubits, and quality metrics obtained using GST are various measures for the closenesses of the process matrices obtained on hardware to the ideal targets on a perfect noiseless quantum computer. Rather than comparing the process matrices themselves, their differences are typically quantified with more intuitive single-valued metrics, which include the following:
\begin{itemize}
    \item \hyperref[sec:process_fidelity]{process fidelity (metric}~\ref{sec:process_fidelity}),
    \item \hyperref[sec:diamond_norm]{diamond norm of a quantum gate (metric}~\ref{sec:diamond_norm}),
    \item \hyperref[sec:spam_fidelity]{state preparation and measurement fidelity (metric}~\ref{sec:spam_fidelity}).
\end{itemize}

Quantum tomography is a general term used to encompass procedures for completely reconstructing a quantum state or process. This is commonly done by running many different circuits to get complete information of the state or process. 
Quantum state tomography is used to reconstruct a quantum state~\cite{PhysRevA.40.2847}. It works by preparing many identical states, and then measuring them in different bases. For example, for a single qubit state, one measures in the $X$, $Y$, and $Z$ bases to obtain all elements of the density matrix of the state being characterized, where the bases are constituted by the eigenstates of the corresponding Pauli gate.

In order to characterize the gates in a quantum computer, quantum process tomography is used~\cite{chuang1997prescription}. In an ideal noiseless quantum computer, a quantum gate is a unitary operation. However, when physically realized, these gates are noisy, and may no longer be unitary operations. Instead, they are described by quantum processes. These are completely positive linear maps, which map a density matrix to another density matrix, whilst also ensuring that the defining properties of the density matrix are preserved, such as the fact that it is positive semi-definite and that its trace is equal to one. Positivity of a density matrix means that all its eigenvalues are greater than or equal to 0. 
Quantum process tomography amounts to reconstructing the matrix corresponding to the action of the quantum gate being applied to the qubit. In order to do this, the quantum gate is applied to different initial states, and then measured in different measurement bases. This then allows for the quantum gate to be completely characterized if the set of states, and measurement bases, form a basis on the Hilbert space of the system. 

An important drawback of quantum state and process tomography is that they assume that the state preparation and also the measurement processes are error-free, while in practice these errors can be significant~\cite{merkel_self-consistent_2013}. GST mitigates this issue by taking these errors into account, and by also providing an estimate for the magnitude of such errors.

A commonly used implementation of GST is long-sequence GST (LSGST), where estimations of gate set parameters are possible to a precision of $O(1/d_{c_{\mathrm{max}}})$, where $d_{c_{\mathrm{max}}}$ is the maximum depth of the set of circuits generated by the protocol~\cite{nielsenGateSetTomography2021,}.
LSGST treats the qubit as a black box with operation buttons, one for initialization, one for measurement, and the rest for gate operations. It then self-consistently determines estimates for: the initial state of the qubits, denoted as $|\rho\rangle\rangle$~\cite{nielsenGateSetTomography2021}; the set of native measurement projectors, denoted as $\lbrace{\langle\langle M_i|}\rbrace$, where $i$ is an integer that spans all the native measurement projectors; the set of gates of the quantum computer, denoted as $\{G_k\}$, where $k$ is an integer that spans all gates in the basis. 
The notation $|\rho\rangle\rangle$ for the initial density matrix $\rho$ is used to indicate that it is represented in a vectorized form, where the vector entries are in the Pauli tensor product basis, and represented by a real column vector $|\rho\rangle\rangle$,
called a superket. Similarly, each measurement projector ${ M_i}$ is vectorized in the Pauli basis and represented by a real row vector ${\langle\langle M_i|}$, called a superbra.
Consequently, quantum gates are represented with $4^{N_\mathrm{q}} \times 4^{N_\mathrm{q}}$ matrices,
denoted as superoperators, where $N_\mathrm{q}$ is the number of qubits that act on superkets.
Thus, for a noisy quantum gate, defined by the quantum process $\Phi$, the superket $|\rho\rangle\rangle$ is mapped to
$S_{\Phi}  |\rho\rangle\rangle$, 
where $S_{\Phi}$ is the superoperator representation of $\Phi$. Using this superoperator formalism simplifies the mathematical representation of maps, since they can be written as a matrix-vector multiplication. This simplification is the main reason for the use of the superoperator formalism in in the GST framework. Since the Pauli basis is used to vectorize the states and measurement projectors, the superoperator that characterizes a quantum operation is also defined by the Pauli basis, and is thus referred to as the Pauli transfer matrix (PTM) of $\Phi$~\cite{wood2011tensor}. 

GST requires the choice of a gate set, $\mathcal{G}$. This includes the target gates to be characterized, and also the state preparation and measurement gates, as well as the null gate, which performs the identity operation on an ideal qubit. All the information required for GST is therefore included in the gate set, which can be written as $\mathcal{G} = \{|\rho\rangle\rangle, \{\langle\langle M_i|\}, \{G_k\}\}$.

In order to estimate the experimental process matrices for the gate set, the LSGST protocol creates a set of circuits that amplify errors in the gate set. It does this by generating a list of so-called germ circuits and fiducial circuits. The germ circuits are chosen such that they amplify all the different possible errors from the target gate set. The fiducial circuits are chosen such that they create a set of states that can be used to fully characterize the germ circuits. They are used to prepare the states and to rotate the states before measurement. As a result, they effectively sandwich the germ circuits between fiducial circuits, with fiducial circuits being applied before and after a germ circuit. This then ensures that all possible errors in the gate set can be fully characterized~\cite{nielsenGateSetTomography2021}. 

\begin{figure}[ht]
    \centering
\includegraphics{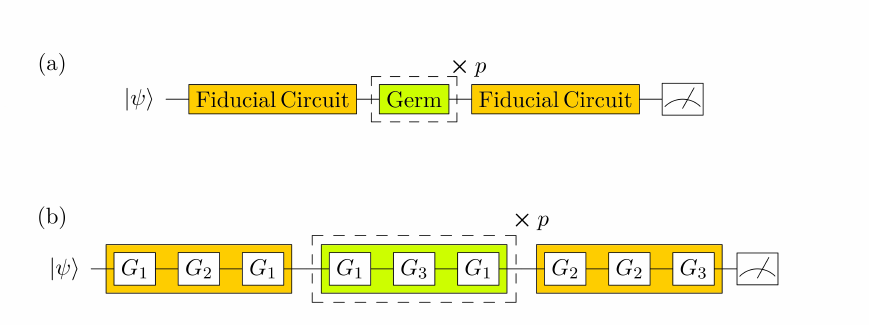}
    \caption{Circuit diagrams showing the circuit structure of all long-sequence gate set tomography (LSGST) circuits for an arbitrary number of qubits. For both circuits, each fiducial circuit is colored orange, and the germ circuit is colored green. (a) First the fiducial circuit is applied to the qubits, then the germ circuit is applied $p$ times, and then another fiducial circuit is applied before measurement. (b) An example circuit, where each fiducial and germ circuit is constructed from the native gates $G_1$, $G_2$, and $G_3$ in the gate set $\mathcal{G}$. Each fiducial and germ circuit in the list of germ and fiducial circuits is constructed from a unique sequence of native gates. The depth of the LSGST circuit is then calculated by counting the number of native gates in the circuit.}
    \label{fig:gst_circuits}
\end{figure}

The term germ refers to the fact that these circuits are like seeds, which is used as a base that is repeated $p$ times, where $p$ is referred to as the germ power. They are usually repeated in exponentially increasing numbers, for example, $p \in \{1,2,4,8,16,32,...\}$~\cite{nielsenGateSetTomography2021, blume2017demonstration}. The larger the power, the more the errors are amplified, and, as a result, more precise estimates are possible. Given the germs, fiducials, and the germ powers, within LSGST a list of experiments is generated to obtain the process matrices. As shown in Fig.~\ref{fig:gst_circuits}a, in each experiment, first a fiducial circuit is performed. Then the germ circuit is applied $p$ times in series. Finally another fiducial circuit is applied, and then a native measurement takes place. Each experiment is then a circuit built from native gates, as shown for a representative example in Fig.~\ref{fig:gst_circuits}b. Using this circuit one can calculate the depth $d_c$ of each circuit by counting the maximum number of native gates applied to a qubit in the circuit. The maximum depth $d_{c_{\mathrm{max}}}$ is the maximum depth of any circuit in the LSGST list of experiments. The value of $d_{c_{\mathrm{max}}}$ determines the maximum precision for the GST estimate. Note that this is different from $p$, since a germ circuit may involve multiple operations on a qubit.

Once the list of experiments is generated, they are run on the quantum computer, and the outcomes for each experiment are saved to a data set. For a given an initial state $\rho$, measurement projector $M_i$ and noisy gate $\Phi$, the experimentally measured probability of outcome $i$ is given by~\cite{nielsenGateSetTomography2021}
\begin{equation}
    \label{eq:ptm_representation}
    p_i = \langle\langle M_i| S_{\Phi}  |\rho\rangle\rangle = \Tr(M_i S_{\Phi} \rho).
\end{equation}

Using the so obtained data set, a gate set estimate is found, denoted as $\Tilde{\mathcal{G}}$, by using maximum likelihood estimation (MLE), where the probability that the experimental data was generated given $\Tilde{\mathcal{G}}$ is maximized by varying over different gate set estimates.
The MLE works by parameterizing the PTM representation for each element in $\mathcal{G}$ to generate a gate set estimate, $\Tilde{\mathcal{G}}$. Using $\Tilde{\mathcal{G}}$, the LSGST circuits are simulated, and outcomes from the simulations are then used to evaluate the likelihood function. An optimizer is used to maximize this likelihood by adjusting the parameters that define $\Tilde{\mathcal{G}}$. This then gives an optimal $\Tilde{\mathcal{G}}$, the estimated gate set with the optimized gates from the MLE. 

The gate sets estimated in this way are determined up to an arbitrary unitary transformation. For example, given a unitary transformation determined by a matrix $B$, the estimate of the gate set can be transformed by $\langle\langle M_i| \rightarrow \langle\langle M_i^{'}| = \langle\langle M_i| 
B$, and  $|\rho\rangle\rangle \rightarrow |\rho^{'}\rangle\rangle = B^{-1}|\rho\rangle\rangle$, and $S_{\Phi} \rightarrow S_{\Phi}^{'}=B^{-1}S_{\Phi}B$. Upon such a transformation the expectation values in Eq. \ref{eq:ptm_representation} remain unchanged, $ \langle\langle M_i^{'}|S_{\Phi}^{'}|\rho^{'}\rangle\rangle = \langle\langle M_i|S_{\Phi}|\rho\rangle\rangle$. The transformation matrix $B$ therefore does not change any observable quantity that can be obtained with the gate set, and thus cannot be measured. The matrix $B$ is often called a gauge transformation, and the fact that the gate set can be transformed by $B$ without altering expectation values is referred to as the gate set having a gauge freedom.

When comparing the PTMs obtained with GST with their ideal targets, changing the gauge can significantly alter the results~\cite{nielsenGateSetTomography2021}. However, changing the gauge does not alter any predicted circuit outcome probabilities. Therefore, the gauge needs be set such that gate set estimates are as similar as possible to their ideal target operations, while still describing the experimental observations. This then allows interpreting the remaining differences between the PTM of the ideal target gate and the PTM obtained with GST as measures for the accuracy of implementation of the gate on the hardware. They can then be used to estimate the process fidelity between the ideal target gate and its practical hardware implementation~\cite{nielsenGateSetTomography2021} (Sec.~\ref{sec:process_fidelity}).  To find this optimal gauge, an optimization is performed, which adjusts the parameters of $B$ to minimize the trace norm between the the estimated gate set and the ideal target gate set. The gauge optimization is performed after a gate set compatible with the observed data is estimated with MLE. Once it is found, the final PTM obtained with GST is the one for this optimized gauge.

\subsection{Measurement procedure}

The general procedure for implementing LSGST is as follows:
\begin{enumerate}
    \item Select the gate-set to characterize, find fiducial and germ circuits that allow the full characterization of all errors possible for the target gate set, define the list of the germ power $p$ to generate circuits for, and then generate the complete experiment list for the LSGST protocol.
    \item Run the generated experiments on the quantum computing hardware.
    \item Run an optimizer to find the gate set that can produce a data set most similar to experimental data set using maximum likelihood optimization.
    \item Run gauge optimization to find gate set estimate that is closest to the target gate set by adjusting the parameters of $B$.
\end{enumerate}   
The output provides the PTMs of the gates in the gate set. The difference between the PTMs obtained by GST and the target ones obtained for the ideal noiseless quantum computer is the metric that allows quantifying the quality of the hardware operations. As mentioned earlier, the PTMs are typically further processed to calculate a set of more intuitive single-valued metrics, which are listed in the first part of this section and described in the following sections.

\metricfig{
Example PTMs of an ideal $\mathrm{C}_X$ gate and a noisy $\mathrm{C}_X$ gate obtained by GST are shown in Fig.~\ref{fig:PTM_CNOT}.
\begin{figure}[htpb]
    \centering
     \includegraphics{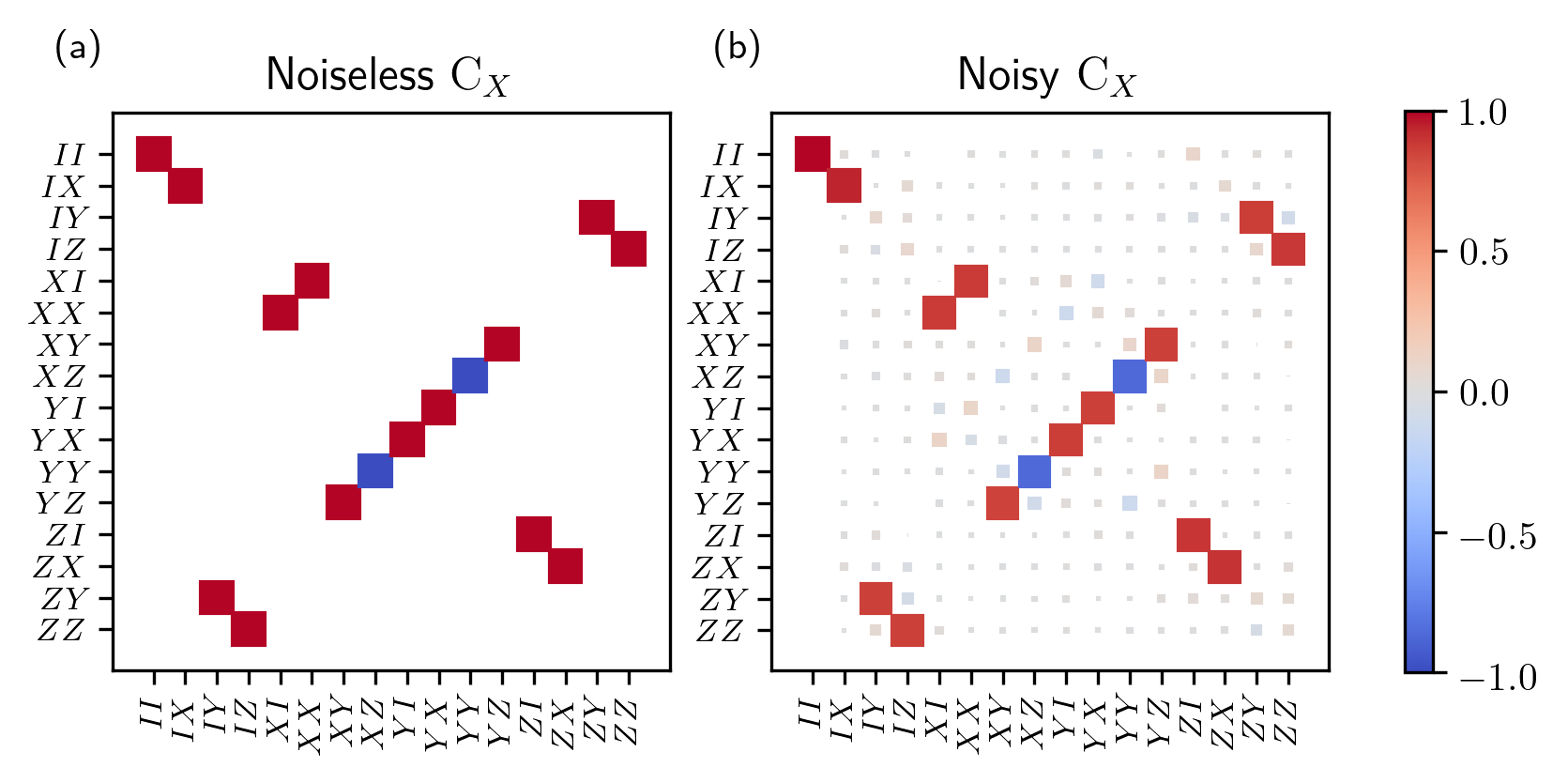}
    \caption{Example PTMs of (a) an ideal $\mathrm{C}_X$ gate, and (b) a noisy $\mathrm{C}_X$ gate, obtained by GST. The noisy $\mathrm{C}_X$ has many small non-zero elements in the PTM that arise due to noise.  The color represents the value of the PTM element. For increased clarity, the size of the squares is set to be proportional to the absolute value of the PTM element.}
    \label{fig:PTM_CNOT}
\end{figure}
}

\subsection{Assumptions and limitations}
\begin{itemize}
    \item The GST model assumes that all noise is Markovian~\cite{nielsenGateSetTomography2021}. Markovian noise corresponds to a memoryless environment. In contrast, non-Markovian noise can account for memory effects arising due to the environment~\cite{agarwal2023modelling}. Coupling to coherent two level systems and slow fluctuations of qubit frequency are examples of mechanisms that can lead to such memory effects. When non-Markovian noise is present in the experimental hardware, it generally leads to a worse agreement between the GST estimates and the experimental data. Hence, if fitting experimental data with GST leads to a fit with a large mean squared error, it is usually attributed to the presence of non-Markovian noise~\cite{nielsenGateSetTomography2021}.
    \item It is a very time consuming process to run the GST protocol, and currently to run 2-qubit LSGST requires on the order of $10^5$ circuits. The development of more efficient ways to perform GST is an active area of research.
\end{itemize}

\subsection{Source code}
A python notebook demonstrating how to run gate set tomography can be found at \sourceurl{gate_execution_quality_metrics/gst_based_gate_execution_quality_metrics}. It is based on the PyGSTi software package~\cite{Nielsen_2020}.

\ifbool{showpart1references}{\printbibliography[heading=subbibliography,segment=\therefsegment]}{}
\end{refsegment}

\section{Symbols and notation used for the metrics and benchmarks}

In this section a list of symbols used throughout the document is given as 
lookup reference.

\subsection{Quantum operations and circuits}
\begin{longtable}{@{}  p{.22\textwidth} p{.80\textwidth} @{}} 
$\mathcal{H}$ & Hilbert space\\
$\rho$ & density matrix or density operator\\
$\{M_i\}$ & set of measurement projectors \\
$I$ & the identity operator \\
$\sigma_x~\equiv~X$ & Pauli X operator \\
$\sigma_y~\equiv~Y$ & Pauli Y operator \\
$\sigma_z~\equiv~Z$ & Pauli Z operator \\
$\Phi (\rho)$ & quantum operation, also known as a quantum process or channel \\
$|\rho\rangle\rangle$ & density matrix vectorized in Pauli basis \\
$\lbrace{\langle\langle M_i|}\rbrace$ &  set of native measurement projectors vectorized in Pauli basis \\
$S_{\Phi}$ & Pauli transfer matrix of $\Phi$\\
$\rho_{\Phi}$ & Choi matrix of quantum operation $\Phi$\\
$N_\mathrm{q}$ & number of qubits \\
$d=2^{N_\mathrm{q}}$ & Hilbert space dimension \\
$K_i$ & Kraus operator with integer index $i$\\
$U$ & unitary operation \\
$\psi$ & state vector\\
$R_u(\theta)$ & single qubit rotation by angle $\theta$ around arbitrary rotation axis $u$ \\

$\rho_{\mathrm{init}} $& initial state of the qubits in quantum computer \\
$U_{\mathrm{1q}}(\theta, \phi, \lambda)$ & general single-qubit unitary with parameters $\theta$, $\phi$, and $\lambda$ \\
$G$ & Clifford gate \\
$C$ & quantum circuit \\
$\mathrm{SWAP}$ & SWAP gate \\
$\mathrm{PERM}$ & qubit permutation gate \\
$\mathrm{C}_U$ & controlled unitary gate \\
$\mathrm{C}_X$ & controlled $X$ gate, also known as CNOT gate \\
$n_s$ & the number of shots for the circuit \\
$w$ & width of circuit, typically equal to $N_q$\\
$d_c$ & depth of circuit\\ 
$C(w,d_c)$ & a family of circuits with width $w$ and depth $d_c$
\end{longtable}

\subsection{Probabilities}

The notation here uses $p$ to denote a single probability, while the notation of $Q$ denotes a probability distribution.

\begin{longtable}{@{}  p{.22\textwidth} p{.80\textwidth} @{}} 
$p_i$ &  probability of measuring the qubit in state $\ket{i}$\\
$p_1(t)$& probability of measuring 1 state in $T_1$ experiment at time $t$\\
$p_1^{\mathrm{R}}(t)$ & probability of measuring 1 state in Ramsey $T_2^*$ experiment at time $t$\\
$p_0^{\mathrm{E}}(t)$ & probability of measuring 0 state in Hahn echo $T_2$ experiment at time $t$\\
$p_{\text{survival}}$ & survival probability in randomized benchmarking \\
$p_0(m)$ & probability of measuring the qubit in the $\ket{0}$ state after application of $m$ pseudo-identities\\
$p_{\mathrm{success}}$ & success probability\\
$p_{x,\text{ideal}}$ & probability of obtaining output bitstring $x$ in measurement of ideal circuit\\
$Q_{\text{ideal}}$ & probability distribution of output bitstrings from measuring ideal circuit \\
$p_{x,\text{output}}$ & probability of obtaining output bitstring $x$ in measurement of experimental circuit\\
$Q_{\text{output}}$ & probability distribution of output bitstrings from measuring experimental circuit \\
$p_{(\text{HYP})}(\bm{s}_k)$ & the probability of sample $\bm{s}_k$ under hypothesis $\text{HYP}$ in boson sampling \\
$p_{(\text{HYP})}(N_c)$ & the grouped probability of obtaining $N_c$ clicks in total under the hypothesis $\text{HYP}$ in boson sampling
\end{longtable}

\subsection{Fidelities}

All fidelities use $F$ as the symbol, and can be distinguished by their subscript.

\begin{longtable}{@{}  p{.22\textwidth} p{.80\textwidth} @{}} 
$F(\rho, \sigma)$ &  state fidelity between two density matrices\\
$F_{\text{pro}}(\Phi_A,\Phi_B)$ & process fidelity between two quantum processes\\
$F_{ \rho_{\mathrm{init}}}$ & the fidelity of the initial state  \\
$F_{M_0}$ & fidelity of the $\ket{0}$ measurement projector for a single qubit \\
$F_{M_1}$ & fidelity of the $\ket{1}$ measurement projector for a single qubit \\
$F_\mathrm{r}$ & readout fidelity \\
$F_{\text{CB}}$ & composite process fidelity obtained using cycle benchmarking\\
$F_c$ & fidelity of output probability distributions\\
$F_{\mathrm{norm}}$ & normalized fidelity between two probability distributions\\
\end{longtable}

\subsection{Times}
All symbols with the units of time are denoted with the symbol $t$. The sub and superscripts are used to clarify their meanings.
\begin{longtable}{@{}  p{.22\textwidth} p{.80\textwidth} @{}} 
$t_{UN_\mathrm{q}}$ & time taken to execute a general $N_q$-qubit unitary\\
$t_{R_x}$ & time taken to execute a single $R_x$ gate\\
$t_O$ & total execution time that are independent of the number of gates, for example state preparation and measurement time \\
$t_{\mathrm{tot}}$ & total computation time\\
$t_o$ & overhead time which occurs independently of the circuit, which is different to $t_O$ \\
$t_C$ & actual time to run a circuit $C$ \\
$t_m$ & maximum of the measurements times of all the qubits \\
$t_{\mathrm{reset}}$ & time to reset device\\
\end{longtable}

\subsection{Metrics}
These are the symbols used for the metrics presented in Sec. \ref{part:metrics}. Note that some performance indicators presented in that section may not have a single value as their output and instead may describe the hardware architecture properties of the quantum computer.

\begin{longtable}{@{}  p{.22\textwidth} p{.80\textwidth} @{}} 
$N_{\text{max}}$ & number of usable qubits \\
$T_{1}$ & qubit relaxation time \\
$T_2$ & qubit dephasing time from Hahn echo experiment\\
$T_2^*$ & qubit dephasing time from Ramsey experiment \\
$\omega_\mathrm{max}$ & idle qubit purity oscillation frequency \\
$F_{\text{pro}}(\Phi_A,\Phi_B)$ & process fidelity between two quantum processes\\\\
$\lVert \Phi_{A} - \Phi_{B} \rVert_{\diamond}$ & diamond norm between two quantum operations \\
$r$ & Clifford randomized benchmarking average gate error \\
$r_{G_{\mathrm{target}}}$ & interleaved Clifford randomized benchmarking  gate error for target gate\\
$\theta_{\mathrm{err}}$ & amount of over- or under- rotation \\
$F_{ \rho_{\mathrm{init}}}$ & the state fidelity of the initial state  \\
$F_{M_i}$ & fidelity of the $\ket{i}$ measurement projector\\
$F_{CB}$ & composite process fidelity obtained using cycle benchmarking\\
$V_Q$ & quantum volume \\
$J_{\text{ave}}$ & mirrored circuits average polarization\\
$\#\mathrm{AQ}$ & the number of algorithmic qubits \\
$b$ & upper bound on the variation distance \\
$E_{\mathrm{diff}}$ & difference between the exact energy of wavefunction and VQE computed energy \\
Q-score & the largest size of graph for which the hardware can solve the MaxCut problem using QAOA sufficiently accurately on average \\
$F_{\mathrm{norm}}(Q_{\text{ideal}}, Q_{\text{output}})$ & normalized fidelity between two probability distributions\\
$t_{UN_\mathrm{q}}$ & time taken to execute a general single- or multi-qubit gate\\
$t_{\mathrm{reset}}$ & time to reset qubits \\
\end{longtable}

\newpage

\part{Metrics and Benchmarks: definitions, methodology and software}
\renewcommand{\thechapter}{\Roman{chapter}}
\setcounter{chapter}{3}
\unchapter{Metrics and Benchmarks: definitions, methodology and software}
\renewcommand{\thechapter}{\arabic{chapter}}
\setcounter{chapter}{6}
\label{part:metrics}
In this part, the metrics are presented following the categorization introduced in Sec. \ref{chapter:overviewmetrics}. As the first section in this part, we also include a description of benchmarking and characterization methodologies used for the computation of a number of the metrics. The listed metrics include both well defined and widely used metrics, as well as emerging performance measures. Most metrics are characterized by a single number, for example the various fidelities. Some metrics, especially architectural metrics such as the connectivity or the available gate-set, are specified by lists or graphs.

For each metric we present a short set of assumptions and limitations. There are also implicit assumptions and limitations that apply more generally to most metrics. These general assumptions and limitations include but are not limited to the following:
\begin{itemize}
    \item For many benchmarking approaches it is implicitly assumed that there is no leakage outside the qubit subspace to higher-order levels in the hardware.
    \item For all qubit quality metrics and gate execution quality metrics, with the exception of the \hyperref[sec:amount_of_1q_nonmarkovian_noise]{idle qubit purity oscillation frequency (metric}~\ref{sec:amount_of_1q_nonmarkovian_noise}), Markovian noise is assumed. Note that non-Markovianity does not negate the usefulness of the metrics, however, more care needs to be taken when interpreting the metrics and using them for comparisons if there is significant non-Markovianity.
    \item Most metrics require running circuits repeatedly to obtain the probability distributions of the outcomes. The number of repetitions, also called the number of shots, should be chosen based on the desired precision of the metric, and is assumed to be reported with the metric value together with other relevant circuit execution and analysis parameters.
    \item It is assumed that the circuits are run on hardware as specified, and that where circuit recompilations or classical emulations of circuits to obtain results are performed, these are reported with the metrics values.
\end{itemize}
We will not explicitly refer to these general assumptions and limitations in each metric description.

For most of the metrics, we develop tutorial code on how to run and calculate the metric, and make it publicly available in an online open-source repository, with links provided. All tutorial code is written using open-source Python packages with permissive licenses. We have developed a generic framework for submitting quantum circuits to a user-specified backend, which can be emulators or real quantum hardware. We have run most metrics for gate-based quantum computers on an emulator with a consistent noise model, which is described in Sec.~\ref{sec:noise_model}, and the results are given as examples in the description of each metric. Users can both adapt the code relevant to the metrics independently of the backend, and change the targeted backend independently of the metrics.To execute circuit on a specific quantum hardware the users need to include the hardware-specific backend interfaces to enable running the circuits generated by the metrics code and also fetching the results. We have included example backend interfaces for quantum computers that were accessible through the the Amazon Braket interface via Amazon Web Services (AWS) at the time of writing. Standardization of such low-level hardware interfaces may progress independently as hardware platforms mature.

Below, we provide the template that each metric in our database follows. Each metric includes a short and then a more extended description, followed by the measurement procedure. This is followed by an outline of the main assumptions and limitations for the measurement and use of the metric. Finally, a link to open source code that implements the measurement procedure is provided. The references used for the metric description are provided at the end of each metric section. The format for the metric description is designed to provide an overview of the metric, and for a more in-depth discussion the provided literature and references therein can be consulted. Note that metric descriptions in Sec.~\ref{chapter:boson_sampling} sway from this template due to the relative nascency of the metrics presented in them.


\tcbset{
    titlebox/.style={
        colframe=gray,
        colback=gray,
        arc=0pt, outer arc=0pt,
        top=1mm,
        bottom=1mm,
        left=\myleftmargin
    }
}

\makeatletter
\ExplSyntaxOn
\cs_set:Npn \myleftmargin {
    \dim_eval:n {1in + \hoffset + \oddsidemargin}
}
\cs_set:Npn \myfullwidth {
    \dim_eval:n {1in + \hoffset + \oddsidemargin + \textwidth}
}

\box_new:N \l_title_tmpa_hbox

\renewcommand{\sectionlinesformat}[4]{
    \Ifstr{#1}{section}{
        \hbox_set:Nn \l_title_tmpa_hbox {
            \@hangfrom{\textcolor{white}{#3}}{\textcolor{white}{#4}}
        }
        \dim_compare:nNnTF {\box_wd:N \l_title_tmpa_hbox} < {\textwidth} {
            \adjustbox{lap=-\myleftmargin}{
                \tcbox[titlebox]{\@hangfrom{\textcolor{white}{#3}}{\textcolor{white}{#4}}}
            }
        } {
            \adjustbox{lap=-\myleftmargin}{
                \begin{tcolorbox}[titlebox, width=\myfullwidth]
                    \@hangfrom{\textcolor{white}{#3}}{\textcolor{white}{#4}}
                \end{tcolorbox}
            }
        }

    }{
        \@hangfrom{\hskip#2#3}{#4}%
    }
}
\ExplSyntaxOff
\makeatother

\section*{Metric title}
\begin{refsection}
Short definition of the metric.

\subsection{Description}
More detailed description of the metric.

\subsection{Measurement procedure}
Description of the methodology to measure the metric.

\subsection{Assumptions and limitations}
Where relevant, a list of assumptions underpinning the metric is provided, as well as practical limitations.

\subsection{Source code}
Where available, this section provides links to free open source libraries suitable for measuring the metric.

\subsection{Metric references}
At the end of each entry, a short list of relevant literature is provided.

\end{refsection}

\let\theoldchapter\thechapter
\newpage
\vspace*{-4\baselineskip}
\def\thechapter{M1}
\chapter{Hardware architecture properties}
\label{chapter:hardware_architecture_properties}

\section{Number of usable qubits}
\label{sec:number_of_usable_qubits}
\begin{refsegment}
This metric gives the total number of qubits available to use in a quantum computer.

\subsection{Description}
The number of usable qubits, $N_{\text{max}}$, in a quantum computer refers to the maximum number of qubits that can be used in a single quantum circuit, so that an $N_\mathrm{max}$-qubit quantum state can be represented on the hardware. A device may have more than $N_\mathrm{max}$ qubits in total, but some of these may not be available for use in a quantum circuit. Qubits may be disabled for various reasons, for example when individual qubits are defective due to problems in the fabrication process. The aim of this metric is not to determine the quality of the qubits, only that they exist, and they can be used together to run quantum circuits.

\subsection{Measurement procedure}
The number of qubits available to use needs to be reported by the hardware manufacturer. The procedure below is a way for the end-user to verify that the number provided by the hardware manufacturer is indeed usable in a single quantum circuit:
\begin{enumerate}
    \item Initialize $N_\mathrm{q}$ qubits and perform a single-qubit gate, such as the $X$ gate, on each of the $N_\mathrm{q}$ qubits. Then measure all qubits at the end.
    \item If the circuit is executed and an $N_\mathrm{q}$-bit output corresponding to the measurement results is received without an error being raised by the hardware or by its software interface, one can infer that the hardware contains at least $N_\mathrm{q}$ usable qubits.
    \item Increase $N_\mathrm{q}$ until an error is raised by the hardware, or generally if the measurement over all qubits returns fewer than $N_\mathrm{q}$ values.
    \item The maximum number of usable qubits $N_\mathrm{max}$ is given by the largest value of $N_\mathrm{q}$, for which the circuit was executed and resulted in an $N_\mathrm{q}$-bit output without raising an error.
\end{enumerate}
In this procedure it is assumed that when a user tries to run a circuit with more qubits than those that are available, the hardware either raises an error that explains to the user that the number of qubits available is lower than the requested one, or that the number of bits in the output is lower than the requested one.
  
\subsection{Assumptions and limitations}
\begin{itemize} 
    \item The connectivity of the qubits is not taken into account. Therefore, a device with $N_\mathrm{max}$ qubits, but which does not allow two-qubit gates, has $N_\mathrm{max}$ as the number of usable qubits, even though entangled states cannot be generated.
    \item The number of usable qubits in a device by itself may not be representative of how many qubits can be used concurrently in a useful manner due to the noise in the device.
    \item If a subset of qubits has significantly higher error rates than the rest of the qubits, then an application may not be able to usefully include these large-error qubits in a quantum circuit, even though they are counted in this metric.
    \item  For some hardware platforms, the number of available qubits may be probabilistic, depending on the state of the device at a given time~\cite{PhysRevA.102.063107, wurtz2023aquila}. For example, this may be the case for neutral atom systems (see Sec.~\ref{sec:neutralAtoms}).
\end{itemize}

\metricbibliography
\end{refsegment}

\section{Pairwise connectivity}
\label{sec:connectivity}
\begin{refsegment}
This is an architectural property that specifies which pairs of qubits can physically perform two-qubit gates between each other. 

\subsection{Description}
The connectivity outlines which pairs of qubits are directly connected, and is typically expressed as a graph or list of pairs. Two qubits are considered to be directly connected if a two-qubit gate between these two qubits can be executed directly on the hardware without the need to resort to SWAP gates. For some hardware platforms, the two qubits may only be connected in a unidirectional manner, where only one of the two qubits can be the target qubit. For example, Fig.~\ref{fig:connectivity}a shows the connectivity of the Oxford Quantum Circuits (OQC) quantum computer ``Lucy'', where the arrows between qubits represent the direction of the native two-qubit gate. The arrow starts at the control qubit and is pointed towards the target qubit. Applying two-qubit gates between qubits that are not directly connected may be achieved by applying a chain of SWAP gates or by physically moving qubits to different locations in architectures such as trapped ions or neutral atoms. Such operations increase both the execution time as well as the induced noise error~\cite{Holmes_2020}.  

A better connectivity can reduce algorithmic complexity~\cite{Holmes_2020, cowtan2019qubit}. Error correction codes typically require specific types of connectivity between physical qubits~\cite{doi:10.1080/00107514.2019.1667078,doi:10.1126/sciadv.abn1717}.

\metricfig{
For example, Fig.~\ref{fig:connectivity} shows the connectivity of the OQC device ``Lucy'' and the IonQ device ``Aria 1''. 

\begin{figure}[htpb]
\centering
  \includegraphics{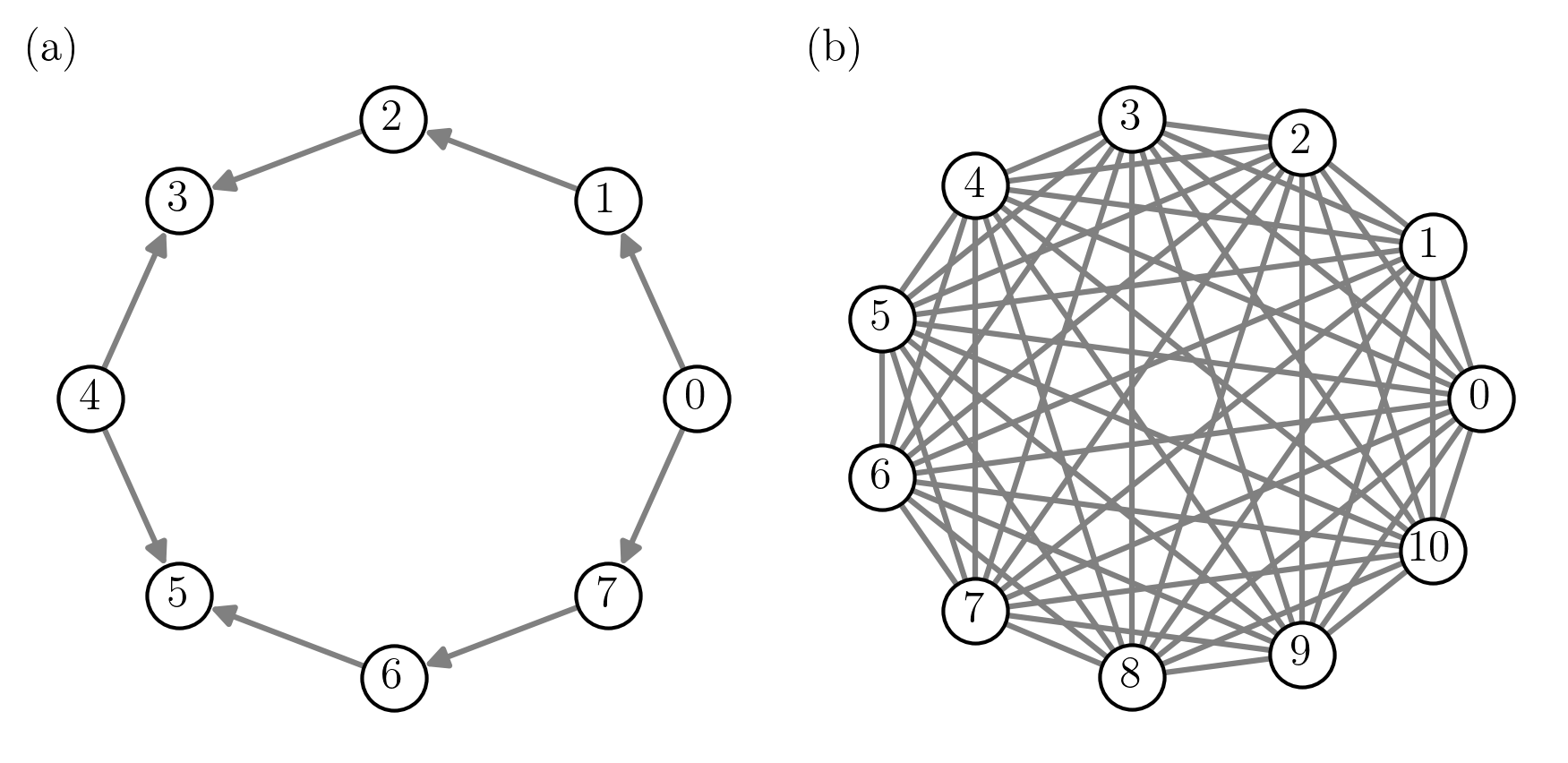}

\caption{The connectivity of (a) the OQC superconducting circuit quantum computer ``Lucy'' and (b) the IonQ trapped-ion quantum computer ``Harmony''. The qubits are indexed with numbers starting from 0. (a) There are 8 qubits in OQC Lucy with circular connectivity, where an arrow between a pair of qubits indicates the control to target direction of applicable two-qubit controlled gates. (b) There are 11 qubits in IonQ Harmony with all-to-all connectivity.}
\label{fig:connectivity}
\end{figure}
}

\subsection{Measurement procedure}

As connectivity is hardware specific, it needs to be reported by the hardware manufacturer. To verify the correctness of the reported connectivity, as user one can execute a set of quantum circuits that apply two-qubit gates between all pairs of reportedly connected qubits. Circuit transpilation needs to be disabled for these runs, since it would implement two-qubit gates also between non-connected qubits by including SWAP gates. 

\subsection{Assumptions and limitations}

\begin{itemize}
    \item The information on the connectivity needs to be provided by the manufacturer. Not all manufacturers may be willing to provide this information.
    \item Whilst connectivity determines which qubits are physically connected to each other, the connectivity map may not be a good representation of the physical layout of the device. The physical layout is important in problems such as analysing crosstalk errors, where qubits that are physically closer may create more mutual crosstalk noise~\cite{10.1145/3373376.3378477}. 
    \item While a device may claim all-to-all connectivity, this does not necessarily mean that the qubits are always connected. For example, in the trapped ion architecture (see section~\ref{sec:trappedions}), a two-qubit gate between arbitrary qubits is implemented by physically moving the two selected qubits together for a two-qubit gate~\cite{crainHighspeedLowcrosstalkDetection2019,PhysRevLett.113.220501}.
    \item Some hardware platforms, such as neutral atoms or trapped ion devices, also have native multi-qubit gates that operate simultaneously on more than two qubits. 
\end{itemize}

\metricbibliography
\end{refsegment}

\section{Native gate set}
\label{sec:native_gate_set}
\begin{refsegment}
This architectural metric corresponds to a list of the native gates that are available on the quantum processor. The native gates are those that are physically executed on the quantum processor. Quantum circuits not specified using native gates need to be compiled to circuits using the native gates in order to be executed, which is also referred to as circuit transpilation or decomposition. An important information provided by the metric is whether the native gate set allows for a universal quantum computation. 

\subsection{Description} 
The native gate set that a quantum processing unit (QPU) uses can vary between different hardware platforms. Furthermore, individual algorithms can benefit from the availability of a specific native gate set, as transpilation into the required gates may typically lead to slower performance. The native gates need to correspond to physical operations on the qubit hardware, and therefore different types of quantum processors have different native gate sets~\cite{10.1145/3307650.3322273}. For example, the single-qubit native gates may only consist of calibrated $R_x(\pi/2)$ pulses, which combined with virtual $R_z$-gates can give arbitrary rotations~\cite{PhysRevA.96.022330}; alternatively, the $Z$-gates can also be implemented in hardware rather than executed virtually~\cite{PhysRevA.82.042339}.
For two-qubit gates, a superconducting QPU may have controlled-$Z$ gates as its native two qubit gate~\cite{krantz_quantum_2019}, whereas an ion trap QPU may use an $XX(\theta)$ gate~\cite{ion:progress-review}, which is defined by $e^{- i\theta  X \otimes X / 2 }$, and referred to as Mølmer-Sørenson gate~\cite{ion:m-s}.

\subsection{Measurement procedure}
The information on the native gate set needs to be provided by the hardware manufacturer. Here we list a number of possibilities, with which it can be extracted from the device:
\begin{enumerate}
    \item The hardware provider documentation can directly provide the details on the native gate set.
    \item Performing only the transpilation step before executing a quantum circuit allows extracting the native gates that the circuit uses to run on hardware.
    \item A list of gates can be sent to the hardware for execution without transpilation: if a circuit runs successfully for a given gate, then that gate is a native gate. If an error is reported by the system, then that gate is not part of the native gate set. 
\end{enumerate}

\subsection{Assumptions and limitations}
\begin{itemize}
    \item Hardware manufacturers may not make the information required to determine the native gate set publicly available.
\end{itemize}

\subsection{Source code}

A tutorial showing how to fetch the set of native gates of a device is provided in \sourceurl{hardware_architecture_properties/native_gates}.

\metricbibliography
\end{refsegment}

\section{Capability to perform mid-circuit measurements}
\label{sec:capability_mid_circuit_measurements}
\begin{refsegment}
This architectural metric states whether or not a quantum computer has the capability to perform mid-circuit measurements, where one or more qubits can be measured multiple times with the results of the measurements recorded, while the circuit is still being executed without measurement on other qubits, so that, in absence of noise in the device, these preserve their quantum coherence during the full circuit execution.

\subsection{Description} 
Mid-circuit measurements are required for many quantum error correction protocols~\cite{google2023suppressing, ion:honeywell}. They can be used to reduce the number of qubits in other algorithms, such as Shor's algorithm~\cite{shor1997polynomial}, where the total number of qubits is one-third of that required in the  protocol without mid-circuit measurements~\cite{martin2012experimental}. They can also be used to reduce the number of physical qubits required to run quantum algorithms such as the quantum approximate optimization algorithm (see metric~\ref{sec:Q-Score}) by measuring qubits as early as possible, and reusing them
elsewhere in the circuit~\cite{PhysRevX.13.041057}. Additionally, mid-circuit measurements can be used to drive conditional operations later in the quantum circuit which can enable measurement-based quantum computation~\cite{Browne_2007}. This binary metric provides the information on whether such mid-circuit measurements can be performed. 

\subsection{Measurement procedure}
The hardware vendor is to provide the information as to whether mid-circuit measurements are possible. If they are possible, also the software commands to execute them are to be provided. The user can then execute such commands to verify the functionality of the mid-circuit measurements. To verify and test the information provided by the hardware provider, a user can apply the following quantum circuit:
\begin{enumerate}
    \item Initialize two qubits to the $\ket{00}$ state.
    \item Apply a Hadamard gate to the first qubit, then apply a $C_X$ gate between the two qubits, with the first qubit being the control and the second qubit being the target.
    \item Perform a mid-circuit measurement on the second qubit, and record the result.
    \item Apply a second $C_X$ gate between the two qubits, again with the first qubit being the control and the second qubit being the target.
    \item Measure both qubits and record the results.
    \item Repeat steps 1-5 for a set number of shots.
\end{enumerate}
If an error is returned by the sequence of operations, then it means the device does not accept mid-circuit measurements. Alternatively, if there is no error, one must check that there are no measurement inconsistencies to ensure that the mid-circuit measurements are working as expected.

We note that this metric only evaluates the capability to perform a mid-circuit measurement, and does not evaluate its fidelity. Nevertheless, the results obtained with the methodology outlined above may also form the basis for the evaluation of the fidelity of the mid-circuit measurement. In an ideal noiseless case, the first measurement in the quantum circuit gives states $\ket{0}$ or $\ket{1}$ with equal probabilities, and the second measurement gives $\ket{00}$ if the first measurement gives $\ket{0}$, or it gives $\ket{10}$ if the first measurement gives $\ket{1}$. The difference between the measured probabilities of the three measurements to these ideal values give an estimate of the fidelity of the whole circuit including the mid-circuit measurements. Deviations of the results obtained on hardware from the ideal result are also induced by noise in state preparation and measurement, as well as in circuit execution. One may separate out the errors induced by the mid-circuit measurements from the other noise-induced errors by running the same circuit, but without performing the mid-circuit measurement. The final output for a noiseless device in this case is also either $\ket{00}$ or $\ket{10}$ with equal probability. By comparing the results with and without mid-circuit measurements, one can then infer the infidelity in the final probabilities induced by the mid-circuit measurement. 

\subsection{Assumptions and limitations}
\begin{itemize}
    \item Beyond the ability to perform mid-circuit measurements, it is also important to quantify the errors that such measurement cause, both on the measured qubit as well as on other qubits, which are a benchmark for the quality of mid-circuit measurements~\cite{goviaRandomizedBenchmarkingSuite2022, PhysRevApplied.17.014014}. Information on the quality of the mid-circuit measurement can be provided through further metrics, such as circuit based metrics that include mid-circuit measurements~\cite{PhysRevApplied.17.014014}.
    \item Further useful information that the hardware manufacturer should provide is if control on qubits can be performed based on the outcome of mid-circuit measurements during circuit execution.
\end{itemize}

\subsection{Source code}

A tutorial for  testing if a device supports mid-circuit measurements is provided in \sourceurl{hardware_architecture_properties/mid-circuit_measurements}.

\metricbibliography
\end{refsegment}

\newpage
\vspace*{-4\baselineskip}
\def\thechapter{M2}
\chapter{Qubit quality metrics}
\label{chapter:qubit_quality_metrics}

\section{Qubit relaxation time (\texorpdfstring{$T_1$}{T1})}
\label{sec:t1}
\begin{refsegment}

The qubit relaxation time, usually referred to as $T_{1}$ time, is a metric for the timescale at which a qubit decays from its excited state, $\ket{1}$, to the ground state, $\ket{0}$~\cite{krantz_quantum_2019}. 

\subsection{Description}

Energy exchange with the environment can lead to the qubit spontaneously going from the excited state, $\ket{1}$, to the ground state, $\ket{0}$, or vice versa. The steady-state population of the states depends on the temperature and can be derived using Boltzmann statistics~\cite{krantz_quantum_2019}. Typically, qubits are operated at low temperatures, such that the rate of excitation is significantly suppressed compared to the rate of decay~\cite{PhysRevLett.97.150502}. In this regime, the steady-state of the qubit is the $\ket{0}$ state. The probability that a qubit prepared in the $\ket{1}$ state at time zero has decayed to the $\ket{0}$ at time $t$, denoted as $p_1(t)$, is given by
\begin{equation}
\label{eq:t1_time}
    p_1(t) = e^{-t/T_1},
\end{equation}
where $T_1$ is the qubit relaxation time. This is also sometimes referred to as the longitudinal relaxation time~\cite{krantz_quantum_2019}.

\subsection{Measurement procedure}
    \begin{enumerate}
        \item Prepare the qubit in the $\ket{1}$ state.
        \item Leave the qubit idle for a set time, $t$.
        \item Measure the qubit in the $\{\ket{0},\ket{1}\}$ computational basis and store the outcome.
        \item Repeat steps 1-3 for a certain number of shots to calculate the probability of the qubit being in the $\ket{1}$ state. The number of shots should be chosen based on the desired benchmarking precision.
        \item Repeat steps 1-4 for different values of $t$ in step 2.
        \item Fit the observed probabilities to Eq. \ref{eq:t1_time} to obtain $T_1$.
    \end{enumerate}

\metricfig{
An example of the results obtained on an emulator is shown in Fig.~\ref{fig:t1}. 
\begin{figure}[htpb]
    \centering
    \includegraphics{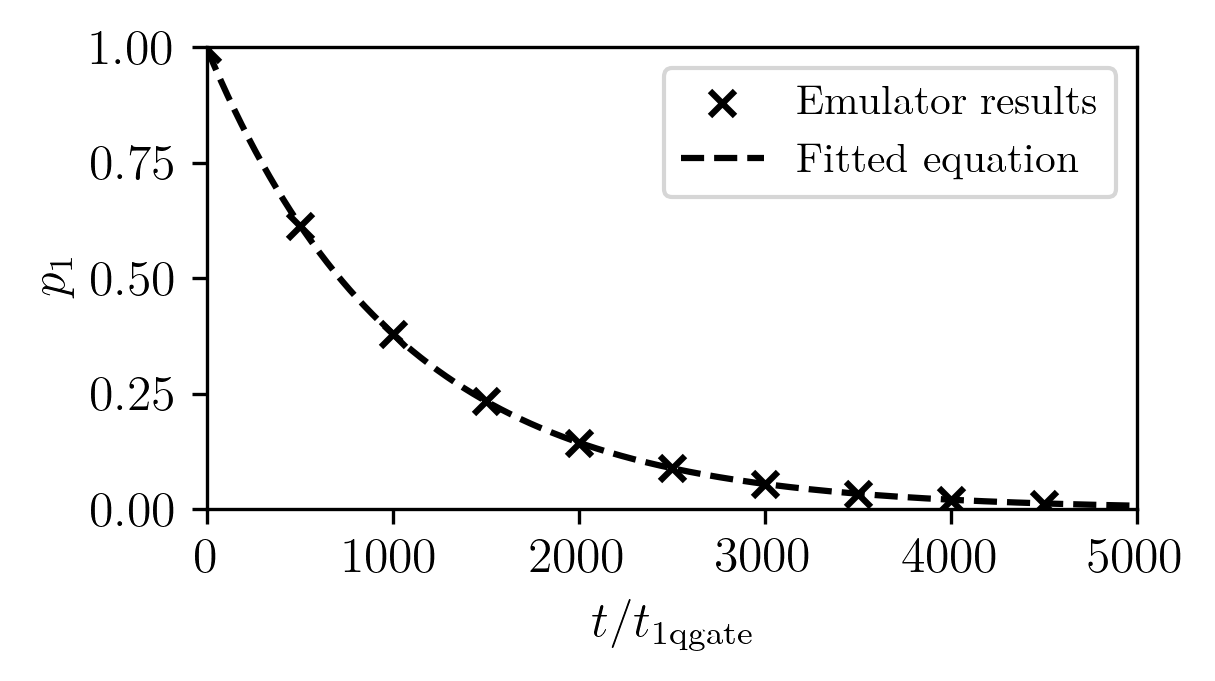}
    \caption{Results of an emulator run with the methodology outlined in this section to obtain the
    $T_1$ time, where the noise model specified in Sec.~\ref{sec:noise_model} is used. The black crosses indicate the measured probabilities of the $\ket{1}$ state, $p_1$, at each qubit idle time, $t$, where the time is in units of the single-qubit gate time, $t_{\mathrm{1qgate}}$. An exponential decay curve following Eq.~\ref{eq:t1_time} is fitted to obtain $T_1=1035\, t_{\mathrm{1qgate}}$. This value is in good agreement with the parameters set in the used noise model, which give a theoretical value of $T_1=1035\, t_{\mathrm{1qgate}}$. If $t_{\mathrm{1qgate}}$ is known, then one can obtain $T_1$ in units of seconds.
    }
    \label{fig:t1}
\end{figure}
}

\subsection{Assumptions and limitations}
\begin{itemize}
    \item It is assumed that the $\ket{1}$ state is the higher-energy state, and that the operating temperature of the qubit is low enough that the steady state of the qubit is mostly the $\ket{0}$ state.
    \item At the beginning of the measurement procedure the qubit is typically in the steady state $\ket{0}$, and the $\ket{1}$ state is then set by applying a gate to rotate the qubit from the $\ket{0}$ state to the $\ket{1}$ state as part of the initialization. The result depends on the accuracy of the calibration of this gate. A badly calibrated gate only approximately rotates the $\ket{0}$ state to the $\ket{1}$ state in the first step completely, which adds to the overall measurement uncertainty of the $T_1$ estimation.
    \item It is assumed that one either has precise knowledge of, or has precise control over, the timings of the gates applied to the qubits. When this information is not provided by the hardware vendor, it may be challenging to estimate the durations and timings of the gates.
    \item The $T_1$ time of single qubits is also meaningful in an annealing setting,  but may be misleading. In particular, this quantity does not take into account that low-temperature dissipation can actually restore coherence in an annealing setting, and as a result quantum annealing can operate successfully even when the annealing time far exceeds the single qubit $T_1$.
\end{itemize}

\subsection{Source code}

A tutorial for measuring $T_1$ time is provided in \sourceurl{qubit_quality_metrics/t1}.

\metricbibliography
\end{refsegment}

\section{Qubit dephasing time (\texorpdfstring{$T_2$}{T2})}
\label{sec:t2}
\begin{refsegment}

The qubit dephasing time, usually referred to as $T_{2}$, is a metric for the timescale at which the phase information of a state in superposition is lost~\cite{krantz_quantum_2019, PhysRevB.72.134519}. 
 
\subsection{Description}
The qubit dephasing time, also sometimes referred to as the transverse relaxation time~\cite{krantz_quantum_2019}, can be reported through both the so-called $T_2$ and the $T_2^*$ values. Both metrics quantify the rate of phase information loss, but are different in that $T_2^*$ is highly sensitive to low-frequency fluctuations in qubit frequency, while the measurement procedure of $T_2$ makes it less sensitive to such fluctuations~\cite{krantz_quantum_2019,PhysRevB.72.134519}. 
Note that the qubit dephasing times include the effects of both amplitude damping and pure dephasing (see Sec.~\ref{sec:noise_effects}).
    
\subsection{Measurement procedure}

First, the procedure for measuring the $T_2^{*}$ time, denoted as the Ramsey experiment~\cite{krantz_quantum_2019}, is described. This experiment is carried out with intentional detuning of the qubit reference frame, which leads to the qubit precessing about the $Z$-axis even when no quantum gates are being applied. The physical implementation of this detuning depends on the hardware platform being used. Slow oscillations of the qubit due to unintentional detuning caused by a lack of accuracy in the frequency calibration can be misinterpreted as a decay, thus a large intentional detuning, which leads to clearly visible fast oscillations, is used to avoid this.
\begin{enumerate}
    \item With the qubit starting in the $\ket{0}$ state, apply an $R_x(\pi/2)$ rotation gate to rotate the qubit into the equator of the Bloch sphere.  
    \item Leave the qubit idle for a certain amount of time, $t$.
    \item Apply a $R_x(\pi/2)$ gate, which rotates the qubit back from the equator of the Bloch sphere into the $Z$-axis. 
    \item Measure the qubit in the computational basis.
    \item Repeat steps 1-4 for a specified number of shots to calculate the probability of the qubit being in the $\ket{1}$ state. The number of shots should be chosen based on the desired benchmarking precision.
    \item Repeat steps 1-5 for different idle time durations $t$.
    \item By estimating the probability for different $t$, the oscillations between the $\ket{0}$ and $\ket{1}$ can be fitted to the equation
    \begin{equation}
    \label{eq:ramsey_fit}
    p_1^{\mathrm{R}}(t) = a\,e^{-t/T_{2}^*}\cos(2\pi f t + \phi) + b,
\end{equation}
where $p_1^{\mathrm{R}}(t)$ is the probability of measuring the $\ket{1}$ state at a given delay time $t$, $a$ is the amplitude of the decaying cosine, $f$ is the frequency of the oscillations, which depends on the detuning of the qubit reference frame, $\phi$ is the offset of the cosine, and $b$ is the baseline offset of the cosine. This allows one to find $T_2^*$.
\end{enumerate}

To measure the $T_2$ time, the same procedure as the Ramsey experiment is followed, with the difference that in step 2, the qubit is left idle for time $t/2$ , then an $R_x(\pi)$ gate is applied, and then the qubit is left idle again for time $t/2$. This is denoted as a Hahn echo experiment~\cite{krantz_quantum_2019, PhysRevB.72.134519}.  
The Hahn echo experiment is fitted to the following equation:
\begin{equation}
    \label{eq:hahnecho_fit}
    p_0^{\mathrm{E}}(t) = A\, e ^{-t/T_{2}} + B,
\end{equation}
 where $p_0^{\mathrm{E}}(t)$ is the probability of measuring the $\ket{0}$ state versus the idle time $t$, $A$ sets the amplitude, and $B$ is the baseline offset. This allows one to find $T_2$.

The naming for these experiments has been adopted from nuclear magnetic resonance (NMR) spectroscopy~\cite{PhysRev.80.580, PhysRevB.72.134519}. 
Since the $T_2$ time is measured using an echo experiment, it is often labeled $T_\mathrm{2E}$. Similarly, the $T_2^*$ time is often labeled $T_\mathrm{2R}$, as it is obtained using a Ramsey experiment.

The choice of the idle durations $t$ can either be made by using prior knowledge of the expected $T_2^*$ time, or by trial-and-error. To do the latter, one chooses the longest $t$, denoted as $t_{\mathrm{max}}$, by repeating the experiment and evaluating the time required for the probability to measure $\ket{1}$ to be approximately $\frac{1}{2}$. Then, the number of idle durations between $0$ and $t_{\mathrm{max}}$ must be chosen to allow for a sufficiently good fit of the resulting decay envelope.

\begin{figure}[htpb]
    \centering
    \includegraphics{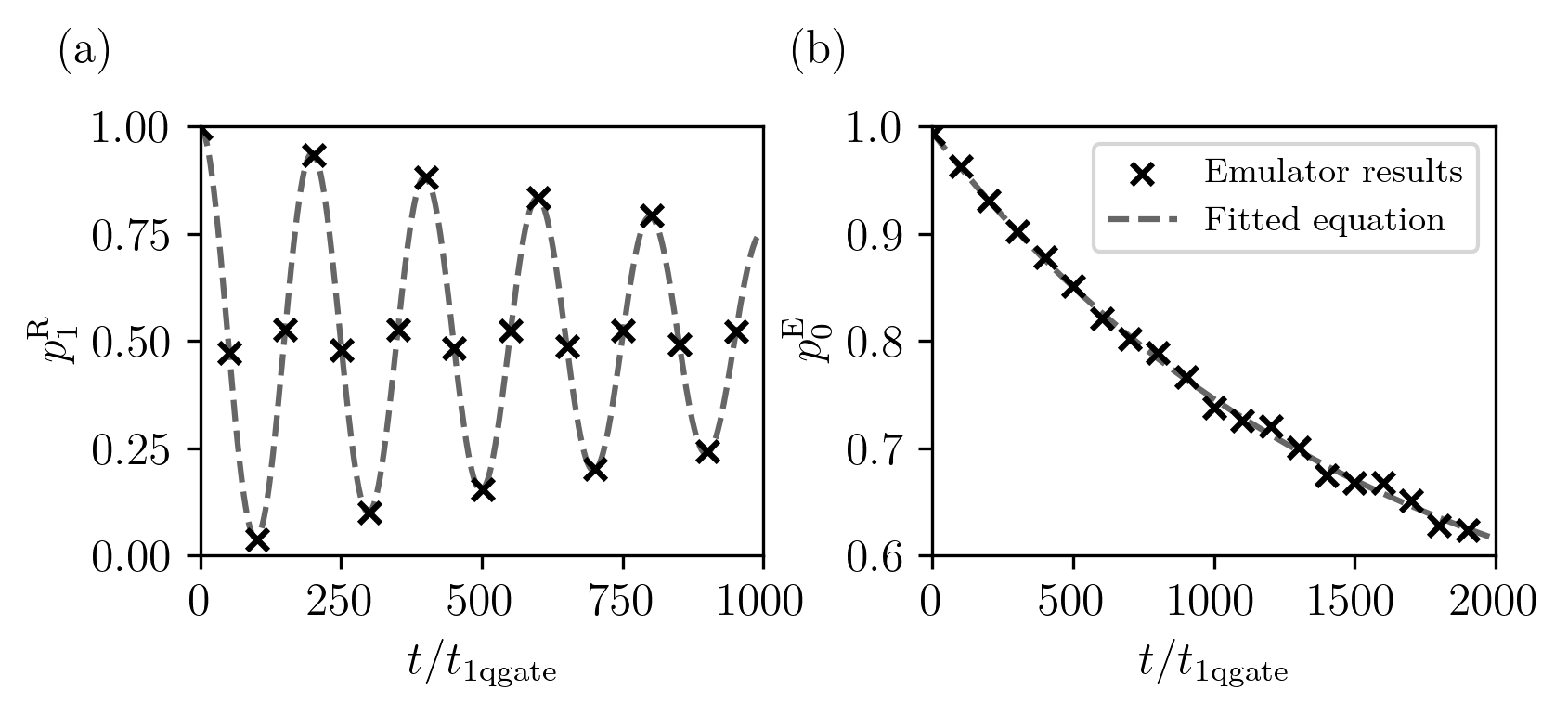}
    \caption{Results of emulator runs with the methodology outlined in this section to obtain the $T_2^*$ and $T_2$ times, where the noise model specified in Sec.~\ref{sec:noise_model} is used. Here the crosses represent the measured results on the emulator, and the dashed curves are fits to the data.
    The time is in units of the single-qubit gate time, $t_{\mathrm{1qgate}}$. 
    (a) Results of the Ramsey experiment: probability of the $\ket{1}$ state, $p_1^\mathrm{R}$, at each qubit idle time, $t$, where the time is in units of $t_{\mathrm{1qgate}}$; Eq.~\ref{eq:ramsey_fit} is fitted to the data to obtain a $T_2^* = 1444.3 \, t_{\mathrm{1qgate}}$. (b) Results of the Hahn echo experiment: probability of the $\ket{0}$ state, $p_0^\mathrm{E}$ as function of time; here Eq.~\ref{eq:hahnecho_fit} is fitted to the data to obtain a $T_2 = 1554.6 \, t_{\mathrm{1qgate}}$. This value is in approximate agreement with the parameters set in the used noise model, which give a theoretical value of $T_2=1437\, t_{\mathrm{1qgate}}$. If $t_{\mathrm{1qgate}}$ is known, then one can obtain the times in units of seconds.}
    \label{fig:t2}
\end{figure}

An example of the results one might obtain when measuring the $T_2^{*}$ and the $T_2$ times is shown in Fig.~\ref{fig:t2}a and Fig.~\ref{fig:t2}b, respectively.

\subsection{Assumptions and limitations}
\begin{itemize}
    \item It is assumed that one either has precise knowledge of, or has precise control over, the timings of the gates applied to the qubits. When this information is not provided by the hardware vendor, it may be challenging to estimate the durations and timings of the gates.
    \item The $T_2$ time of single qubits is also meaningful in an annealing setting, but may be misleading. In particular this quantity does not take into account that low-temperature dissipation can actually restore coherence in an annealing setting, and as a result quantum annealing can operate successfully even when the annealing time far exceeds the single qubit $T_2$.
    \item Inaccuracies in gate calibrations can add to the overall measurement uncertainty of the $T_2$ estimation.
\end{itemize}

\subsection{Source code}
A tutorial for measuring the $T_2^*$ time is provided in \sourceurl{qubit_quality_metrics/t2}.

\metricbibliography
\end{refsegment}

\section{Idle qubit purity oscillation frequency}
\label{sec:amount_of_1q_nonmarkovian_noise}
\begin{refsegment}
An idle qubit periodically losing and regaining coherence can be a signature of non-Markovian noise~\cite{agarwal2023modelling, gulacsi2023smoking}. The oscillation frequency of the qubit purity can quantify the effect of non-Markovian noise induced coherence revivals, where a higher frequency indicates larger non-Markovianity of the noise.

\subsection{Description} 
Non-Markovian noise can lead to the qubit purity oscillating as a function of time, even for an idle qubit\cite{agarwal2023modelling, gulacsi2023smoking}. This noise is challenging to correct with quantum error correction because it leads to time-correlated error~\cite{osti_1671379}. The qubit purity, $\zeta$, is a measure of the coherence of a qubit, and is defined as $\zeta = \Tr\left[\rho^2\right]$, where $\rho$ is the qubit density matrix. Measuring the qubit observables corresponding to the expectation values of $\sigma_x, \sigma_y,$ and $\sigma_z$, allows for the calculation of the qubit purity using the equation
\begin{equation}
\zeta = \frac{1}{2} \bigl( 1 + \expval{\sigma_x}^2 + \expval{\sigma_y}^2 + \expval{\sigma_z}^2\bigr),
\label{eq:nonMarkovianPurity}
\end{equation}
where $\expval{\sigma_i} = \Tr{\sigma_i \rho}$. The values of $\expval{\sigma_i}$ can be measured by applying a change of basis operation before the qubit measurement.

In this metric the single-qubit non-Markovian noise induced purity oscillation frequency is characterized for the idle qubit. Since purity oscillations in general depend on the initial state of the qubit, it is necessary to prepare the qubit in the $x$, $y$, and $z$ bases by initializing the qubit in the $\ket{0}$, $\ket{+} = (\ket{0} + \ket{1})/\sqrt{2}$, and $\ket{R} = (\ket{0} + i\ket{1})/\sqrt{2}$ states, respectively, and then perform the calculations in each basis. The native measurement occurs in the $z$ basis, whilst measuring in the $x$ basis requires a Hadamard gate before the native measurement operation, and measuring in the $y$ basis requires applying the conjugate transpose of the $S$ gate followed by a Hadamard gate before native measurement. 

When measuring the qubit purity for different initial states, oscillations in purity as a function of idle time appear. These can be fitted with the following decaying oscillation function
\begin{equation}
\label{eq:non_Markivian_1q_purity_oscillations}
    \zeta(t) = a + b\; e^{-\lambda\,t}\cos(\omega\, t),
\end{equation}
where fitting parameters $a,b$ include the effects of readout and state preparation error, respectively. The fitting parameter $\lambda$ includes the effect of decoherent errors, and the fitting parameter $\omega$ corresponds to purity oscillation frequency. The result of this procedure gives a value for the qubit purity oscillation frequency for each of the different initial states of the qubit. To estimate the worst-case effect on the qubit performance, the maximum of the three fitted frequencies, $\omega_\mathrm{max}$, is taken as a metric for the amount of single-qubit non-Markovian noise induced purity oscillations. Markovian qubit dynamics correspond to $\omega_\mathrm{max} = 0$.

\subsection{Measurement procedure}
\begin{enumerate}
    \item Select the time step $\Delta t$ and maximum time $t_\mathrm{max}$ to leave the qubit idle to get the set of idle times $T =\{ 0, \Delta t, 2\Delta t, ..., t_\mathrm{max}\}$ for which to run the circuits.
    \item Then, for each initial state $\ket{i} \in \{ \ket{0}, \ket{+}, \ket{R}\}$, perform the following operations:
    \begin{enumerate}
        \item For each idle time $t \in T$:
        \begin{enumerate}
            \item For each measurement basis $b \in \{x,y,z\}$:
                \begin{enumerate}
                    \item Prepare the qubit in the state $\ket{i}$.
                    \item Leave the qubit idle for time $t$.
                    \item Measure the qubit in basis $b$ by applying the corresponding basis change gates and then measuring in the computational basis.
                \end{enumerate}
            \item Calculate the purity $\zeta$ using Eq.~\ref{eq:nonMarkovianPurity}.
        \end{enumerate}
        \item For each initial state $\ket{i}$, fit the function in Eq.~\ref{eq:non_Markivian_1q_purity_oscillations} to a decaying oscillation and store the estimated oscillation frequency $\omega_i$.
    \end{enumerate}
    \item The largest oscillation frequency from all three initial states gives the single-qubit non-Markovian noise induced purity oscillation frequency, $\omega_\mathrm{max}$.
\end{enumerate}
The value of $\Delta t$ must be chosen to be small enough to avoid aliasing effects in oscillations, and $t_\mathrm{max}$ must be chosen to be large enough to either see oscillations, or to ensure $\zeta(t)$ has reached its steady state value in the absence of oscillations. 

\metricfig{
An example of the results one might obtain is shown in Fig.~\ref{fig:purity_oscillation}.

\begin{figure}[htp]
\centering
    \includegraphics{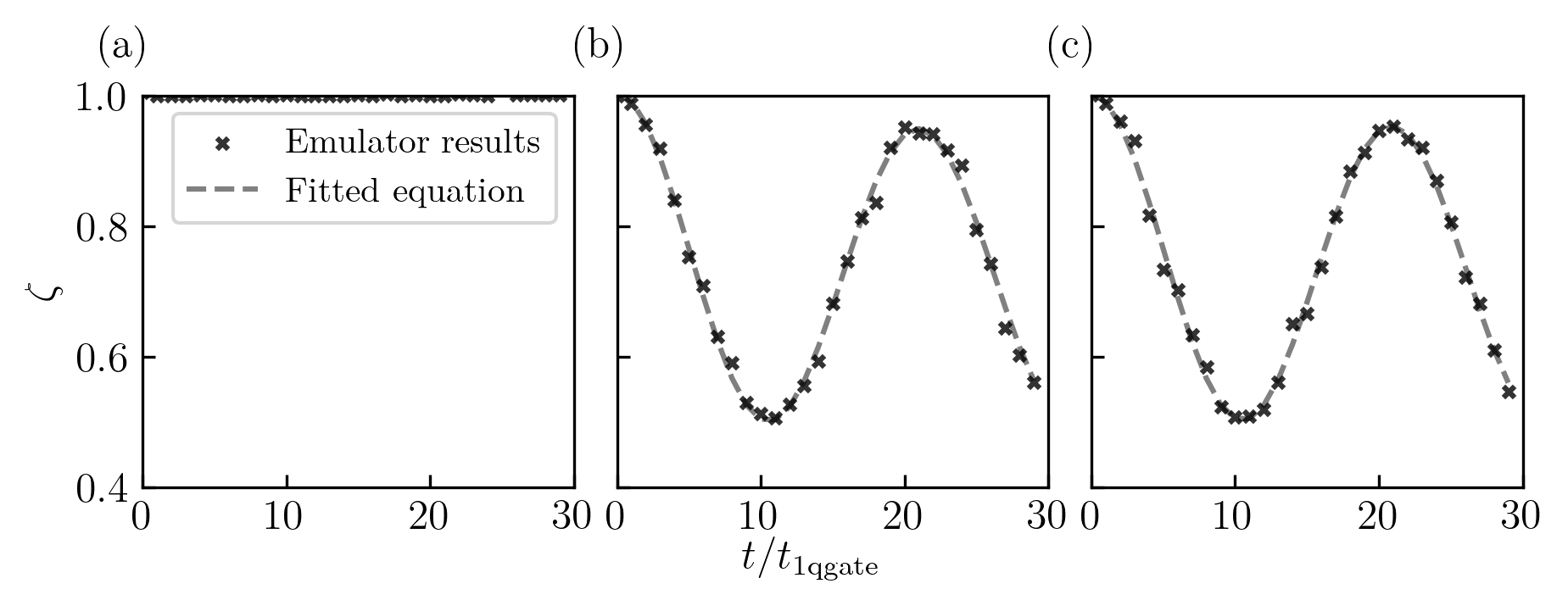}

    \caption{Results of emulator runs using the methodology outlined in this section to obtain the idle qubit purity oscillation frequency, $\omega_\mathrm{max}$, where the noise model used is specified in Sec.~\ref{sec:noise_model}. Since that noise model does not include non-Markovian noise, in order to also model the non-Markovian noise, a second qubit is added to the emulator to act as the environment. Then, a two qubit $R_{ZZ}(\theta)$ gate is applied at each time-step between the two qubits, with $\theta=0.15$ to emulate $ZZ$-crosstalk noise. Tracing out the additional qubit, which is acting as the environment, leads to non-Markovian dynamics of the qubit being measured~\cite{agarwal2023modelling}. The theoretical value is then $\omega_\mathrm{max} =0.15 / t_{\mathrm{1qgate}}$.  The three plots show the qubit purity $\zeta(t)$ calculated using Eq.~\ref{eq:nonMarkovianPurity} for each initial state, (a) $\ket{0}$, (b) $\ket{+}$ and (c) $\ket{R}$, respectively. The black crosses are the calculated $\zeta$. The gray dashed curves are the fits to Eq.~\ref{eq:non_Markivian_1q_purity_oscillations}. The estimated oscillation frequencies are (a) 0, (b) 0.1490 and (c) 0.1496, respectively, hence the largest oscillation frequency gives the single-qubit non-Markovian noise induced purity oscillation frequency, $\omega_{max}=0.1496/t_{\mathrm{1qgate}}$. This is in good agreement with the theoretical value obtained directly from the noise model parameters.}
    \label{fig:purity_oscillation}
\end{figure}
}

\subsection{Assumptions and limitations}
\begin{itemize}
\item This method only measures the effect of non-Markovian noise which leads to purity oscillations. There may be other kinds of non-Markovian noise effects not accounted for by this procedure~\cite{Rivas_2014,hall2014canonicalformofmasterequations}.
\item This metric can indicate the presence of effects that cannot be describe by a purely Markovian single qubit model, however, it cannot distinguish between environment induced sources of non-Markovian noise and effects such as crosstalk without further measurements.
\item In the fitting procedure it is assumed that the purity has at most one oscillation frequency. If multiple sources of non-Markovian noise are present, the purity may exhibit multi-frequency oscillations as well.
\item State preparation and measurement (SPAM) errors can lead to significant errors in the estimation of the qubit purity, $p$, but they will not affect the purity oscillation frequency. However, if the SPAM errors are large, they can make the fitting procedure to calculate oscillation frequency challenging. In this case, precise experiments with a large number of shots might be needed.
\end{itemize}

\subsection{Source code}
A tutorial showing how to obtain the amount of single-qubit non-Markovian noise induced purity oscillations is provided in \sourceurl{qubit_quality_metrics/idle_qubit_purity_oscillation}.

\metricbibliography
\end{refsegment}

\def\thechapter{M3}
\chapter{Gate execution quality metrics}
\label{chapter:gate_quality_metrics}

\section{Gate set tomography based process fidelity}
\label{sec:process_fidelity}

\begin{refsegment}

The process fidelity is a measure of how close one quantum process is to another quantum process. When referring to the process fidelity of a quantum process, this means that it compares the closeness of the quantum process implemented on a noisy quantum computer to its ideal noiseless and perfectly calibrated target process~\cite{PhysRevA.60.1888, schumacher1996sending, gilchrist2005distance}. Often, the resulting process infidelity is given rather than the fidelity, which is given by one minus the fidelity. For this specific metric, gate set tomography (GST) is used to compute the fidelity (section \ref{sec:gst}), but also different process tomography approaches may be used.

\subsection{Description} 

The evaluation of the fidelity between two processes is based on the methodology to obtain the fidelity of two states~\cite{gilchrist2005distance}. The fidelity between two states, represented by their density matrices $\rho$ and $\sigma$, is given by~\cite{jozsa1994fidelity}
\begin{equation}
    \label{eq:state_fidelity}
    F(\rho, \sigma) = \qty(\Tr[\sqrt{\sqrt{\rho}\,\sigma\sqrt{\rho}}])^2.
\end{equation}
This is a measure for the closeness between the two generally mixed states $\rho$ and $\sigma$. If one of the two states is a pure state, then the equation for the fidelity is simplified to
\begin{equation}
    \label{eq:state_fidelity_1pure}
    F(\rho, \sigma) = \Tr\left[\rho\sigma\right].
\end{equation}

To determine the fidelity of a quantum process, denoted as $\Phi$, one can use these equations as well, where the process is represented as a valid density matrix, $\rho_{\Phi}$. For a given $S_{\Phi}$, the Pauli transfer matrix (PTM) of $\Phi$ as described in section~\ref{sec:gst}, one can find $\rho_{\Phi}$ using the following equation~\cite{merkel_self-consistent_2013}
\begin{equation}
    \label{eq:choi_matrix}
    \rho_{\Phi} = \frac{1}{d^2} \sum_{i,j=1}^{d^2} S_{\Phi_{ij}} P^{\dagger}_j \otimes P_i,
\end{equation}
where $P_i$ and $P_j$ are tensor products of Pauli matrices, and $i,j$ iterate through all possible $P \in \{I,X,Y,Z\}^{\otimes N_{\mathrm{q}}}$, where $N_{\mathrm{q}}$ is the number of qubits and $d=2^{N_{\mathrm{q}}}$.
The matrix $\rho_{\Phi}$ is denoted as the Choi matrix of a quantum channel.
Using this representation of a quantum process, one can calculate the process fidelity between two quantum processes, $\Phi_A$ and $\Phi_B$, denoted as $F_{\text{pro}}(\Phi_A,\Phi_B)$, using Eq.~\ref{eq:choi_matrix} and Eq.~\ref{eq:state_fidelity} to get~\cite{gilchrist2005distance}
\begin{equation}
\label{eq:process_fidelity_generalised}
     F_{\text{pro}}(\Phi_A,\Phi_B) =  F(\rho_{\Phi_A}, \rho_{\Phi_B}).
\end{equation}
This equation is valid for all quantum processes. In the case where one wants to calculate the process fidelity where the target process $\Phi_B$ is unitary, then one can use Eq. \ref{eq:state_fidelity_1pure} to obtain a simpler equation~\cite{NIELSEN2002249}:
\begin{equation}
    \label{eq:process_fidelity}
    F_{\text{pro}}(\Phi_A,\Phi_B) = \frac{1}{d^2}\Tr\left[S_{\Phi_A} S_{\Phi_B}^{\dagger}\right].
\end{equation}

To obtain the process fidelity, the PTMs for the processes are needed, so that one has to run process tomography to obtain the PTMs. In order to mitigate for state preparation and measurement (SPAM) errors, typically gate set tomography (GST) is run, which is a special type of process tomography that can correct for SPAM errors at an approximate level (see section~\ref{sec:gst}).

\subsection{Measurement procedure}

The procedure to calculate the process fidelity of a single experimental gate is as follows:
\begin{enumerate}
    \item Select a gate to characterize, $\Phi_{\text{exp}}$, with the ideal quantum process matrix given by $\Phi_{\text{ideal}}$, which typically is a unitary operation for a noiseless quantum computer.
    \item Find $S_{\Phi_{\text{exp}}}$, the PTM of the gate $\Phi_{\text{exp}}$, using GST, as described in section~\ref{sec:gst}. Also determine $S_{\Phi_{\text{ideal}}}$, the PTM of the ideal gate.
    \item Using Eq.~\ref{eq:process_fidelity} find the process fidelity $F_{\text{pro}}(\Phi_{\text{exp}}, \Phi_{\text{ideal}})$, where it is assumed that the ideal target process is unitary. If it is not, then the more general Eq. \ref{eq:process_fidelity_generalised} needs to be used.
\end{enumerate}

\subsection{Assumptions and limitations}
    \begin{itemize}
        \item Since the measurement procedure is based on GST, the assumptions and limitations for GST in section~\ref{sec:gst} also apply here.
    \end{itemize}

\subsection{Source code}
A tutorial for calculating the single qubit process fidelity for a given gate, also shortened as gate fidelity, is provided in:
\sourceurl{gate_execution_quality_metrics/gst_based_gate_execution_quality_metrics}.

\metricbibliography
\end{refsegment}

\section{Diamond norm of a quantum gate}
\label{sec:diamond_norm}
\begin{refsegment}
The diamond norm of a quantum gate gives the probability that, given only one application of the gate, the implemented quantum gate can be distinguished from its ideal target gate~\cite{aharonov1998quantum}. The diamond norm, also known as completely bounded trace norm, is a norm on the space of quantum operations on density matrices, and is the metric in which fault-tolerant thresholds are typically presented~\cite{PhysRevLett.117.170502}.

\subsection{Description} 

Given two different quantum processes, if a quantum state is randomly passed through one of the two processes and then measured, the maximum probability of successfully determining which of the two quantum processes was applied is determined by the diamond norm of the difference between the two quantum processes ~\cite{aharonov1998quantum}. A quantum gate corresponds to a quantum operation, which is also referred to as a quantum process. The diamond norm of a quantum gate is calculated by taking the ideal gate and its physically realized noisy quantum gate as quantum operations, $\Phi_A$ and $\Phi_B$, respectively. A quantum operation is any operation that takes one density matrix to another one, and does not need to be unitary. The resulting diamond norm gives the maximum probability of distinguishing between the ideal gate and its physically realized noisy version in a single measurement. This is sometimes referred to as obtaining the worst case error of the noisy gate~\cite{A_Yu_Kitaev_1997, hashim2023benchmarking}. When comparing to metrics such as the average gate fidelity, the relation with the diamond norm is generally only loose~\cite{Sanders_2016}. Hence computing the diamond norm provides additional independent information from the fidelity based metrics. 

The diamond norm between two quantum operations, $\Phi_A,\Phi_B \in \mathcal{H}$, is defined as
\begin{equation}
\lVert \Phi_{A} - \Phi_{B} \rVert_{\diamond} = \max_{\rho \in \mathcal{H} \otimes \mathcal{H}} \lVert \qty(\Phi_{A} \otimes I)\qty[\rho] -  \qty(\Phi_{B} \otimes I)  \qty[\rho] \rVert_1 ,
\label{eq:diamondnorm}
\end{equation}
where the norm $||X||_1 = \Tr[\sqrt{X^* X}]$ for some matrix $X$, $I \in \mathcal{H}$ is the identity operator, and $\rho \in \mathcal{H} \otimes \mathcal{H}$ is a density matrix. The quantum gate operates on the entangled state $\rho$, which can make it easier to distinguish between $\Phi_{A}$ and $\Phi_{B}$~\cite{wilde2011classical}.
The density matrix $\rho$ is parameterized, and each element is optimized to maximize the distance $\lVert \Phi_A - \Phi_B \rVert_{\diamond}$, whilst also ensuring $\rho$ remains a valid density matrix.

It is possible to compute the diamond norm by first obtaining a full characterization of a quantum hardware gate using gate set tomography (GST), and then using a classical computer to perform the maximization task in Eq. (\ref{eq:diamondnorm}) via a semi-definite program (SDP)~\cite{watrous2012simpler}. In general, a SDP solves an optimization problem that is defined by a linear objective function, while also satisfying a set of constraints defined by the problem. For example, the density matrices need to satisfy the constraint that they are positive semi-definite Hermitian matrices with trace one. SDP also requires that the solution matrix is positive semi-definite, meaning that its eigenvalues are greater than or equal to zero.

\subsection{Measurement procedure}

Calculating the diamond norm of a quantum gate requires a complete characterization of the gate. Typically, this means that the gate estimate is given using gate set tomography (GST), where the output is the Pauli transfer matrix (PTM). Details on GST and the PTM are given in section~\ref{sec:gst}. 

To calculate the diamond norm of a quantum gate, the procedure is as follows:
\begin{enumerate}
    \item Characterize the gate of interest $\Phi_{\text{exp}}$ using GST to obtain the PTM, $S_{\Phi_{\text{exp}}}$, for the noisy gate. See section~\ref{sec:gst} for information on how to run GST. Additionally, obtain the PTM for the ideal gate using a classical computer,  $S_{\Phi_{\text{ideal}}}$.
    
    \item Find the Choi matrix, $\rho_{\text{diff}}$, of $S_{\text{diff}} = S_{\Phi_{\text{exp}}} - S_{\Phi_{\text{ideal}}}$ using Eq.~\ref{eq:choi_matrix} in metric~\ref{sec:process_fidelity}; $\rho_{\text{diff}}$ is a density matrix in the space $\mathcal{H} \otimes \mathcal{H}$, 

    \item Run the following SDP, as outlined in section 3.2 in Ref.~\cite{watrous2012simpler}:
    \begin{align}
    \mathrm{Maximize} \:  \frac{1}{2} \qty( \Tr[\rho_{\text{diff}}^{\dagger} X ] + \Tr[\rho_{\text{diff}} X^{\dagger}] ), \;
    \mathrm{\text{subject to}} \: \begin{pmatrix}
            I \otimes \rho_0 & X \\
            X^{\dagger} &  I \otimes \rho_1
            \end{pmatrix} \geq 0,
    \end{align}
     where $\rho_0$, $\rho_1$ are density matrices acting on the space $\mathcal{H}$, and $X$ is a linear operator acting on the space $\mathcal{H} \otimes \mathcal{H}$. Here $X$, $\rho_0$, $\rho_1$ are variables that are optimized in the SDP. The SDP also ensures that  $\rho_0$ and $\rho_1$ are valid density matrices, meaning that they must have $\Tr[\rho]=1$ and that they must be positive semi-definite Hermitian. 
    \item The final output of the SDP optimization gives the diamond norm of the quantum gate.
\end{enumerate}

\subsection{Assumptions and limitations}
\begin{itemize}
    \item For high quality gates the diamond norm of a single gate can be very small, and hence difficult to measure accurately. In this case it is often more useful to compute $\lVert \Phi_{A}^{n} - \Phi_{B}^{n} \rVert_{\diamond}$, which corresponds to the application of the gate $n$ times, which amplifies the noise induced errors.
    \item For larger number of qubits the resources needed for the computation of the diamond norm scale exponentially in the number of qubits. This is due to the scaling of GST which is required to fully characterize the PTM of the operation of interest and not the classical optimization time using semi-definite programming which can handle matrices with $10^{}14$ entries~\cite{doi:10.1137/19M1305045}.
\end{itemize}

\subsection{Source code}
A tutorial showing how to use gate set tomography to calculate the diamond norm is provided in \sourceurl{gate_execution_quality_metrics/gst_based_gate_execution_quality_metrics}. \\It is based on the open-source PyGSTi software package~\cite{Nielsen_2020}.

\metricbibliography
\end{refsegment}

\section{Clifford randomized benchmarking average gate error}
\label{sec:RB}
\begin{refsegment}
The Clifford randomized benchmarking (RB) average gate error metric provides an estimate of the average gate error of a set of single- and multi-qubit Clifford gates in a quantum computer~\cite{emerson2005scalable,knillRandomizedBenchmarkingQuantum2008}.

\subsection{Description} 
RB estimates the average gate error of a set of gates in a way that is robust to state preparation and measurement (SPAM) errors, giving an estimate on their average performance in a quantum computer.
RB runs random Clifford circuits at increasing depths, 
where the final Clifford gate is the inverse of all previous operations. 
A Clifford gate, denoted as $G$, is a gate that transforms any Pauli matrix, $P$, into another Pauli matrix, $P'$, by $GPG^{\dagger} = P'$. Note that $P$ and $P'$ may not necessarily be different. The set of Clifford gates is called the Clifford group.

The random Clifford gate sequence used within RB is expressed as
\begin{equation*}
    G_1 G_2 G_3 \dots G_{m-1}G_{m} G_{\text{inverse}} ,
\end{equation*} 
where each $G_i$ is a Clifford gate, and $G_{\text{inverse}}$ inverts all the previous gates, so that $G_{\text{inverse}}=G_m^{-1} G_{m-1}^{-1} \dots G_{3}^{-1}G_{2}^{-1}G_1^{-1}$; $m$ corresponds to the number of Clifford gates applied, also called sequence length or depth. Since all gates are Clifford gates, $G_{\text{inverse}}$ can be computed efficiently on a classical computer. In the ideal noiseless case, the RB operations reduce to the identity, and the measurement of the circuit gives the initial state that is set at the start of the circuit. 
In a realistic noisy case, the probability of the state to remain in its initial state decreases with increasing number of operations due to noise in the hardware. This probability is called the survival probability, $p_{\text{survival}}$. For example, if after running the RB circuit with an initial state of $\ket{0}$, $80\%$ of the outcomes are measured to be in the $\ket{0}$ state, then $p_{\text{survival}}=0.8$.

Long sequences of random Clifford gates for $N_{\mathrm{q}}$ qubits
uniformly sampled from the Clifford group result in an exponential decay of the survival probability~\cite{emerson2005scalable}. The decay can then be written in the form of 
\begin{equation}
    \label{eq:rb_fit}
    p_{\text{survival}} = A_0\; \alpha ^m + B_0,
\end{equation} 
where $A_0$, $\alpha$, and $B_0$ are the fitting parameters corresponding to amplitude, decay, and baseline, respectively. Here $\alpha$ represents the error per Clifford gate. Using $\alpha$, the estimate for the Clifford RB average gate error, $r$, can be computed as~\cite{magesanScalableRobustRandomized2011}
\begin{equation}
\label{eq:rb_error}
    r = 1- \alpha - \frac{(1- \alpha)}{d},
\end{equation}
where $d=2^{N_{\mathrm{q}}}$ is the dimension of the Clifford gates, and $N_{\mathrm{q}}$ is the number of qubits. For the single-qubit case $N_{\mathrm{q}}=1$ and thus $d=2$. For the multi-qubit case the random Clifford gates are always of dimension $2^{N_{\mathrm{q}}}$, and are then transpiled into the native gate set to obtain a circuit that can be executed on the hardware. 

For multi-qubit RB the set of possible Clifford gates that one needs to sample grows exponentially~\cite{PhysRevLett.122.200502} with the number of qubits in the multi-qubit operation, which limits scalability.
Furthermore, to obtain accurate results one needs to run a large number of circuits, making it difficult to implement RB on more than a few qubits, with the largest published result being for three qubits~\cite{PhysRevLett.122.200502}. 
For a given number of qubits, $N_{\mathrm{q}}$, the $N_{\mathrm{q}}$-qubit Clifford gates need to be decomposed into the native gate-set available on the hardware platform, which typically consists of single- and two-qubit gates. The resulting circuit depth of an individual $N_{\mathrm{q}}$-qubit Clifford gate become very large as $N_{\mathrm{q}}$ increases. Hence, the survival probability can eventually become negligible even for a single $N_{\mathrm{q}}$-qubit Clifford gate, so that the decay of the survival probability with increasing number of Clifford gates cannot be obtained ~\cite{PhysRevResearch.2.013317}.

As a way to run RB on large devices with many qubits, one can run RB on subsets of qubits on the device in parallel. Note that in this case the number of circuits needed still grows exponentially with the size of the multi-qubit operations, but does not scale with the total number of qubits in the device.
Running RB on multiple sets of qubits in parallel may introduce additional crosstalk errors due to the qubits being operated in parallel. This can be used to evaluate the amount of crosstalk errors~\cite{PhysRevLett.109.240504} in the device.

\subsection{Measurement procedure}
\begin{enumerate}
    \item Select a set of sequence lengths over which the expected survival probability decays significantly.
    \item Choose a number $n_{\mathrm{circ}}$. For each sequence length, $m$, construct $n_{\mathrm{circ}}$ circuits in the following way. First generate a sequence of Clifford gates, uniformly randomly selected from the Clifford group. Then add a last gate to the circuit, which should be the inverse of all previous gates before it, so that the circuit applied performs an identity operation for an ideal noiseless quantum computer. Note that when running RB on $N_{\mathrm{q}}$ qubits, then $N_{\mathrm{q}}$-qubit Clifford gates are required. The Clifford gates are decomposed into the native gates of the quantum computer at time of execution, which are typically single- and two-qubit gates.
    \item Run each of these RB circuits on hardware with a specified number of shots. The number of shots should be chosen based on the desired benchmarking precision.
    \item Calculate $p_{\text{survival}}$ for each sequence length $m$.
    \item Fit the exponential decay with Eq.~\ref{eq:rb_fit} and obtain the error per Clifford gate, $\alpha$.
    \item Determine the Clifford RB average gate error, $r$, using Eq.~\ref{eq:rb_error}.
\end{enumerate}

\metricfig{
An example of the results one might obtain is shown in Fig.~\ref{fig:rb}.
\begin{figure}[htpb]
    \centering
    \includegraphics{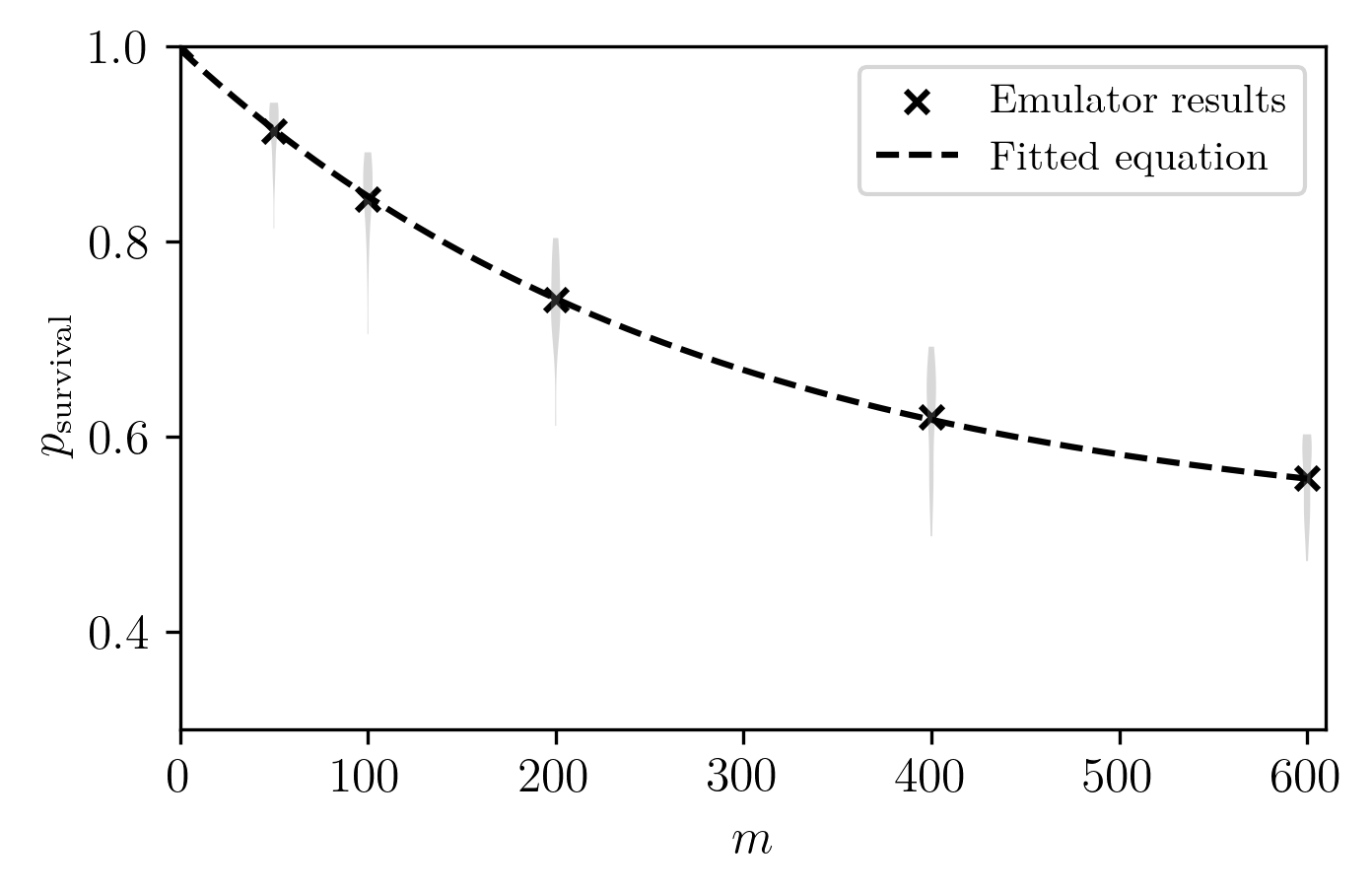}
    \caption{Results of emulator runs of 1-qubit Clifford randomized benchmarking (RB) using the methodology outlined in this section and the noise model specified in Sec.~\ref{sec:noise_model}. The results are shown as a violin plot. The gray regions use kernel density estimation to show the distribution of survival probabilities obtained from the $n_{\mathrm{circ}}$ circuits run for each sequence length of Clifford gates, $m$, where the average survival probability for each sequence length is marked with a cross. Here, $n_{\mathrm{circ}} = 10$. The black dashed curve shows the fitted Eq.~\ref{eq:rb_fit}, from which the Clifford RB average gate error, $r$, is calculated using Eq.~\ref{eq:rb_error} to obtain $r=0.0018$.}
    \label{fig:rb}
\end{figure}
}

\subsection{Assumptions and limitations}
    \begin{itemize}
        \item RB for Clifford gates only spans a subset of all possible quantum gates. Clifford gates can be simulated efficiently using a classical computer. This may not be fully representative of the quantum computer performance, as gates from the Clifford group do not form a universal gate set.
        \item When implementing RB experimentally, the Clifford gates need to be decomposed into the native gates of the device. There can be differences in reported errors depending on the decomposition of the Clifford gates. To estimate the average gate error of native gates after obtaining the RB average gate error $r$, one needs to calculate the average number of native gates required to decompose Clifford gates. 
        \item Clifford RB itself has limited scalability. As discussed above, modifications with improved scalability are required for large numbers of qubits.
        \item When benchmarking larger systems, decomposing the $N_{\mathrm{q}}$-qubit Clifford requires increasingly more two-qubit gates. The error per Clifford increases, so that the survival probability decays more rapidly and is only non-zero for a small number of layers, leading to larger uncertainties in fitting the decay parameters. 
        \item This protocol assumes that all noise is  Markovian and gate independent~\cite{knillRandomizedBenchmarkingQuantum2008, PRXQuantum.2.040351}. Under this assumption, the Clifford RB gate error is linearly related to the process fidelity averaged over the $N_{\mathrm{q}}$-qubit Clifford group~\cite{helsen2022general}. In practice, the noise from a gate application can depend on gates applied before it or have some time-dependent noise and the relationship between the Clifford RB gate error and the process fidelity averaged over the $N_{\mathrm{q}}$-qubit Clifford group becomes less clear~\cite{helsen2022general}. Additionally, if the survival probability deviates from the exponentially decay described in the description, this may indicate the presence of non-Markovian noise~\cite{helsen2022general,Wallman2018randomized}.
        \item The RB protocol assumes that the average error converges to the average over all possible Clifford sequence for small sample sizes because only a small fraction of possible sequence lengths are able to be implemented~\cite{PhysRevA.89.062321}.
        \item The protocol assumes there is not much variation in errors so that most errors are close to the average~\cite{PhysRevA.89.062321}.
    \end{itemize}

\subsection{Source code}
A tutorial demonstrating how to calculate the Clifford RB average gate error for a single qubit is provided in
\sourceurl{gate_execution_quality_metrics/randomized_benchmarking/clifford_randomized_benchmarking}.
\metricbibliography
\end{refsegment}

\section{Interleaved Clifford randomized benchmarking gate error}
\label{sec:1qinterleaved_rb}
\begin{refsegment}
The interleaved Clifford randomized benchmarking (RB) gate error provides an estimate for the average error of a target Clifford gate in a gate set~\cite{magesanEfficientMeasurementQuantum2012, PhysRevLett.108.260503}.

\subsection{Description} 

As described in metric~\ref{sec:RB}, Clifford RB provides the average gate error over all of the gates in the gate set used to decompose the Clifford gates. In contrast, interleaved Clifford RB characterizes the average error of a specific Clifford gate, which is interleaved through the random Clifford sequences used in RB. If the RB random Clifford gate sequence is expressed as (see section ~\ref{sec:RB})
\begin{equation*}
    G_1 G_2 G_3 \dots G_{m-1}G_{m} G_{\text{inverse}} ,
\end{equation*} 
where each $G$ is a Clifford gate, then the interleaved RB sequence can be written as 
\begin{equation*}
    G C'_{\text{inverse}},
    G_1 G_{\mathrm{target}} G_2 G_{\mathrm{target}} G_3 G_{\mathrm{target}}\dots G_{m-1}G_{\mathrm{target}}G_{m}G_{\mathrm{target}} G'_{\text{inverse}} ,
\end{equation*}
where $ G_{\mathrm{target}}$ is the gate to characterize and must also be part of the Clifford group. Additionally, the final inverting gate must be updated to invert the full gate sequence including $G_{\mathrm{target}} $. The specific gate error is obtained by comparing the decay parameters in Eq.~\ref{eq:rb_fit} for both the non-interleaved and interleaved RB results.

The fitted decay parameters from the non-interleaved RB and the interleaved RB are denoted by $\alpha$ and $\alpha_{G}$, respectively. Then, the interleaved RB gate error for $ G_{\mathrm{target}}$ is found using~\cite{magesanEfficientMeasurementQuantum2012, PhysRevLett.108.260503}
\begin{equation}
    \label{eq:irb_gatefidelity}
    r_{G_{\mathrm{target}}} = \frac{\qty(d-1)\qty(1 -\frac{\alpha_{G}}{\alpha})}{d},
\end{equation}
where $d = 2^{N_\mathrm{q}}$, and $N_\mathrm{q}$ is the number of qubits on which the randomized benchmarking is performed. For the single-qubit case $N_\mathrm{q}=1$ and thus $d=2$. 

\subsection{Measurement procedure}
\begin{enumerate}
    \item Select the target gate, $ G_{\mathrm{target}}$, for which the average gate error needs to be estimated. Then, select a set of sequence lengths over which the expected survival probability decays significantly.  
    \item Generate the circuits for the varying depths used to obtain the non-interleaved Clifford RB average gate error, as outlined in metric~\ref{sec:RB}.
    \item For each sequence length generate another sequence of Clifford gates, uniformly randomly selected from the Clifford group, but this time with the target gate interleaved in each sequence. The last gate should be the inverse of all previous gates before it, so that the circuit applied acts as an identity operation for a noiseless quantum computer.
    \item Execute each of these non-interleaved and interleaved RB sequences on hardware.
    \item Calculate the survival probability for each sequence length for both.
    \item Fit the exponential decay to model the data for the non-interleaved and for the interleaved RB gate sequences using Eq.~\ref{eq:rb_fit}.
    \item Convert decay parameters extracted from the fits into Clifford RB average gate error, $ r_{G_{\mathrm{target}}}$, using Eq~\ref{eq:irb_gatefidelity}.
\end{enumerate}

\subsection{Assumptions and limitations}
    \begin{itemize}
        \item Interleaved Clifford RB has the same assumptions and limitations as for the Clifford RB average gate error estimation (see metric~\ref{sec:RB}).
        \item The protocol may underestimate the interleaved gate error if the error on the interleaved gate partially inverts the error on the previous Clifford gate~\cite{PhysRevA.89.062321}.
    \end{itemize}

\subsection{Source code}
A tutorial demonstrating how to calculate the interleaved Clifford randomized benchmarking gate error is provided in
\sourceurl{gate_execution_quality_metrics/randomized_benchmarking/interleaved_clifford_randomised_benchmarking}.

\metricbibliography
\end{refsegment}

\section{Cycle-benchmarking composite process fidelity}
\label{sec:cycle_benchmarking}
\begin{refsegment}

The cycle-benchmarking (CB) composite process fidelity quantifies the total effect of errors when a full clock cycle of specified gates is applied on a quantum computer~\cite{erhardCharacterizingLargescaleQuantum2019a}. A clock cycle refers to the application of a specified sequence of gates acting on a given number of qubits, and may consist of single and multi-qubit gates.

\subsection{Description}
CB evaluates the performance of a quantum hardware in performing full cycles of operations. The specific sequence of the gates in such a cycle is specified by the user, and must be reported by the user together with the obtained composite process fidelity. A given quantum process performed with a cycle in CB, denoted as $\mathcal{G}$, consists of Clifford operations, and has the property that when repeated sufficient times it returns the identity for an ideal noiseless quantum computer. Therefore, if the cycle for an ideal operation is denoted as $\bar{\mathcal{G}}$, one has $\bar{\mathcal{G}}^{m}=I$, where $m$ is an integer number specifying the number of repetitions of the cycle. The CB composite process fidelity can then be written as $F_{\text{pro}}(\mathcal{G}, \bar{\mathcal{G}})$. The process fidelity is presented in detail in Sec. \ref{sec:process_fidelity}, and given in mathematical form in Eq. \ref{eq:process_fidelity}.
In order to obtain $F_{\text{pro}}(\mathcal{G}, \Bar{\mathcal{G}})$ the Pauli transfer matrix (PTM) of $\mathcal{G}$ needs to be estimated. The procedure to obtain the PTM using gate set tomography (GST) is presented in Sec. \ref{sec:gst}.
Since the evaluation of the PTM using GST is very time-consuming, especially for larger number of qubits, within CB a number of approximations are performed, which allow for improved efficiency and increased scalability. To emphasize that the CB composite process fidelity is based on a defined set of approximations, it is denoted as $F_{\text{CB}}(\mathcal{G}, \bar{\mathcal{G}})$ to distinguish it from the exact fidelity. 

The main component that CB uses to increase the efficiency and scalability is the inclusion of randomized compiling (RC). Within RC one applies random single qubit Pauli gates to all qubits before each application of a target gate sequence, in this case each individual $\mathcal{G}$, and then applies single qubit Pauli gates afterwards, such that the overall unitary remains unchanged for the noiseless case~\cite{Wallman-EmersonPRA2016}. When using RC one generates a set of $L$ randomized circuits. It can be shown that when evaluating expectation values averaging over all the $L$ randomized gate sequences, RC converts noise with arbitrary coherence and spatial correlations into stochastic Pauli noise~\cite{hashim2023benchmarking}. RC therefore ensures that off-diagonal noise in the PTM is transformed to diagonal noise. This is referred to as noise tailoring. Since the noiseless cycle satisfies $\bar{\mathcal{G}}^{m}=I$, its PTM is purely diagonal. Importantly, the addition of stochastic Pauli noise to such a diagonal PTM keeps the PTM diagonal. The larger the value of $L$, the better the noise tailoring performs, and hence the closer the PTM is to a diagonal form. The key advantage of using RC within CB therefore is that it makes the PTM of the noisy quantum process of $\mathcal{G}^{m}$ diagonal, so that only the diagonal elements need to be computed. Using Eq.~\ref{eq:process_fidelity}, and the fact that $\bar{\mathcal{G}}^{m}=I$, one obtains the fidelity of $\mathcal{G}^m$ as
\begin{equation}
    \label{eq:composite_process_fidelity_m}
    F_\mathrm{CB}(\mathcal{G}^m,\bar{\mathcal{G}}^m=I)= \mathrm{Tr}\left[S_{\mathcal{G}^m}\right].
\end{equation}
where $S_{\mathcal{G}^m}$ is the PTM of $\mathcal{G}^m$ when using RC.

Within CB one therefore only needs to compute the diagonal elements of the PTM of $\mathcal{G}^m$. To this aim, a loop is performed over all initial states that span to eigenbasis of the $N_{\mathrm{q}}$-qubit tensor products of Pauli matrices. These $N_{\mathrm{q}}$-qubit Pauli strings are composed of a single arbitrary Pauli gate on each qubit, including the identity, so that the set of all Pauli strings can be written as $\mathbf{P}=\{X, Y, Z, I\}^{\otimes N_{\mathrm{q}} }$. For a given Pauli string, $P_i$, the initial state is set so that on each qubit it corresponds to the eigenstate with $+1$ eigenvalue of the Pauli matrix for that specific qubit. This is prepared by applying an appropriate Clifford gate to the initial zero state on that qubit.
Once the appropriate initial state is prepared on the quantum computer for a given $P_i$, a RC based implementation of $\mathcal{G}^{m}$ is applied in the circuit, which requires running $L$ different randomized sequences of gates as described above. For each of these circuits, the inverse of the state preparation is applied before measurement in the $Z$ basis on all qubits. For each randomized RC circuit with index $l\in[1,L]$, a number of shots is performed to evaluate the expectation value of these measurements. The diagonal element of the PTM of $\mathcal{G}^m$ for the given $P_i$, denoted as $f_{P_i,m,l}$, is then equal to the parity of each of the measurement outcomes averaged over the number of shots.
The parity is calculated by first assigning a value of $+1$ ($-1$) to each qubit measurement outcome of zero (one), with the exception for those qubits that have an identity operation as part of the $P_i$ Pauli string, for which the assigned value is always $+1$. These assigned values are then multiplied over all qubits to obtain the overall parity. 
By averaging the obtained parities over all the $L$ randomized circuits one obtains an estimate of $f_{P_i, m}$ as
\begin{equation}
    \label{eq:fpm}
    f_{P_i, m}=\frac{1}{L}\sum_{l=1}^L f_{P_i,m,l},
\end{equation}
so one obtains the composite process fidelity of $m$ applications of $\mathcal{G}$ using
\begin{equation}
    \label{eq:composite_process_fidelity_m2}
    F_\mathrm{CB}(\mathcal{G}^m,\bar{\mathcal{G}}^m=I)=\frac{1}{N_{\mathbf{P}}} \sum_{P_i\in\mathbf{P}} f_{P_i,m}.
\end{equation}

The $f_{P_i, m}$ computed in this way include the effects of state preparation and measurement (SPAM) errors, which is undesirable. To filter out such effects at an approximate level, one can compute these elements for two different values of $m$, and use the fact that the fidelity for such cycles is expected to decay exponentially with increasing $m$.
The robustness to SPAM errors is therefore achieved by running the cycle for different number of repetitions, and using the fact that the reduction in fidelity increases for increasing $m$, while the SPAM induced infidelity remains approximately constant. One can therefore approximate the diagonal element of the PTM for a single application of $\mathcal{G}$ as
\begin{equation}
    \label{eq:fp}
    f_{P_i}=\left(\frac{f_{P_i, m_2}}{f_{P_i, m_1}}\right)^{\frac{1}{m_2-m_1}},
\end{equation}
where $m_2>m_1$, and the values are chosen in a way that $\bar{\mathcal{G}}^{m_1}=\bar{\mathcal{G}}^{m_2}=I$. Note that one can also use multiple values of $m$ and then fit the results as function of $m$ to an exponential decay:
\begin{equation}
    \label{eq:fp_exp_fit}
    f_{P_i}(m) = a e^{-b f_{P_i, m}} + c,
\end{equation}
where $a, b$ and $c$ are fitting parameters, and they must be chosen such that when $m=0$, $f_{P_i}(m=0)=1$. Then the diagonal element of the PTM for a single application of $\mathcal{G}$ is given by:
\begin{equation}
    \label{eq:fp_1}
    f_{P_i} = f_{P_i}(m=1).
\end{equation}

The overall CB composite process fidelity of a single application of $\mathcal{G}$ is obtained by averaging over all diagonal elements of the PTM: 
\begin{equation}
    \label{eq:composite_process_fidelity}
    F_{CB}(\mathcal{G},\bar{\mathcal{G}})= \frac{1}{N_{\mathbf{P}}} \sum_{P_i\in\mathbf{P}} f_{P_i}.
\end{equation}
Since the number of Pauli strings, $N_{\mathbf{P}}$, scales as $N_{\mathbf{P}} \leq 4^{N_{\mathrm{q}}}$, it becomes infeasible to include all Pauli strings in the sum as the number of qubits becomes large. As an approximation one therefore uses a uniform sampling of Pauli strings at random to produce a set with a user defined number of elements. This generally gives a good approximation of a sum over all Pauli strings, provided a sufficiently large number of elements is sampled at random~\cite{erhardCharacterizingLargescaleQuantum2019a}.

\subsection{Measurement procedure}

\begin{enumerate}
    \item Select a set of $N_{\mathrm{q}}$-qubit Pauli strings, $\mathbf{P}$, with the number of elements denoted by $N_{\mathbf{P}}$. The elements of $\mathbf{P}$ determine which diagonal elements of the PTM are included in the sum to obtain the total CB combined process fidelity.  
    \item Select the cycle of Clifford gates $\mathcal{G}$ to be tested. Either select two lengths, $m_1$ and $m_2$, such that $m_2 > m_1$, or select a number of lengths $\{m_1, m_2, \cdots, m_k\}$, where $m_{j+1} > m_j \: \forall j \in \{1, 2, \cdots, k-1\}$. For all selected lengths $m$, $\bar{\mathcal{G}}^{m}=I$ must be satisfied. 
    \item Select the number of randomizations, $L$, to apply for each Pauli string.
    \item For each $P_i\in\mathbf{P}$, and each length $m\in\{m_1,m_2\}$ or $\{m_1, m_2, \cdots, m_k\}$, repeat the following steps for $l\in\{1,\cdots,L\}$:
    \begin{enumerate}
        \item Select $m+1$ random $N_{\mathrm{q}}$-qubit Pauli strings $R_0,R_1,\cdots,R_m$, to define the randomized circuit, so that $R_i' \bar{\mathcal{G}} R_i = \bar{\mathcal{G}}$, where $R_i'$ is defined through $R_i$ and $\bar{\mathcal{G}}$. This then defines the randomized circuit $C_{P,m,l}(\mathcal{G})$ as
        \begin{equation}
            C_{P,m,l}(\mathcal{G})= B(P_i)^{\dagger}R_m'\mathcal{G}R_mR_{m-1}'\mathcal{G}R_{m-1}\cdots R_1R_0'\mathcal{G}R_0 B(P_i),
        \end{equation} 
        where $B(P_i)$ is the operation to transform the initial zero state into the $+1$ eigenstates on each qubit of the Pauli string $P_i$.  
        \item Run the circuit on the quantum computer for a specified number of shots, and use the circuit outcomes to calculate the Pauli expectation $f_{P,m,l}$. For each shot, the parity of the measurement outcomes bitstrings over all the qubits of $C_{P,m,l}(\mathcal{G})$ is computed, and the results averaged over all shots.  
    \end{enumerate}
    \item Calculate $f_{P_i, m}$ using Eq.~\ref{eq:fpm} for all $P_i$ and $m$.
    \item If two lengths, $\{m_1, m_2\}$, were used, then use Eq.~\ref{eq:fp} to calculate $f_{P_i}$. If more than two lengths were used, then use Eqs.~\ref{eq:fp_exp_fit} and~\ref{eq:fp_1} to calculate $f_{P_i}$.
    \item Calculate the composite process fidelity using Eq.~\ref{eq:composite_process_fidelity}.
  
\end{enumerate}

\metricfig{
An example of the results one might obtain is shown in Fig.~\ref{fig:cb}.
\begin{figure}[htpb]
    \centering
    \includegraphics{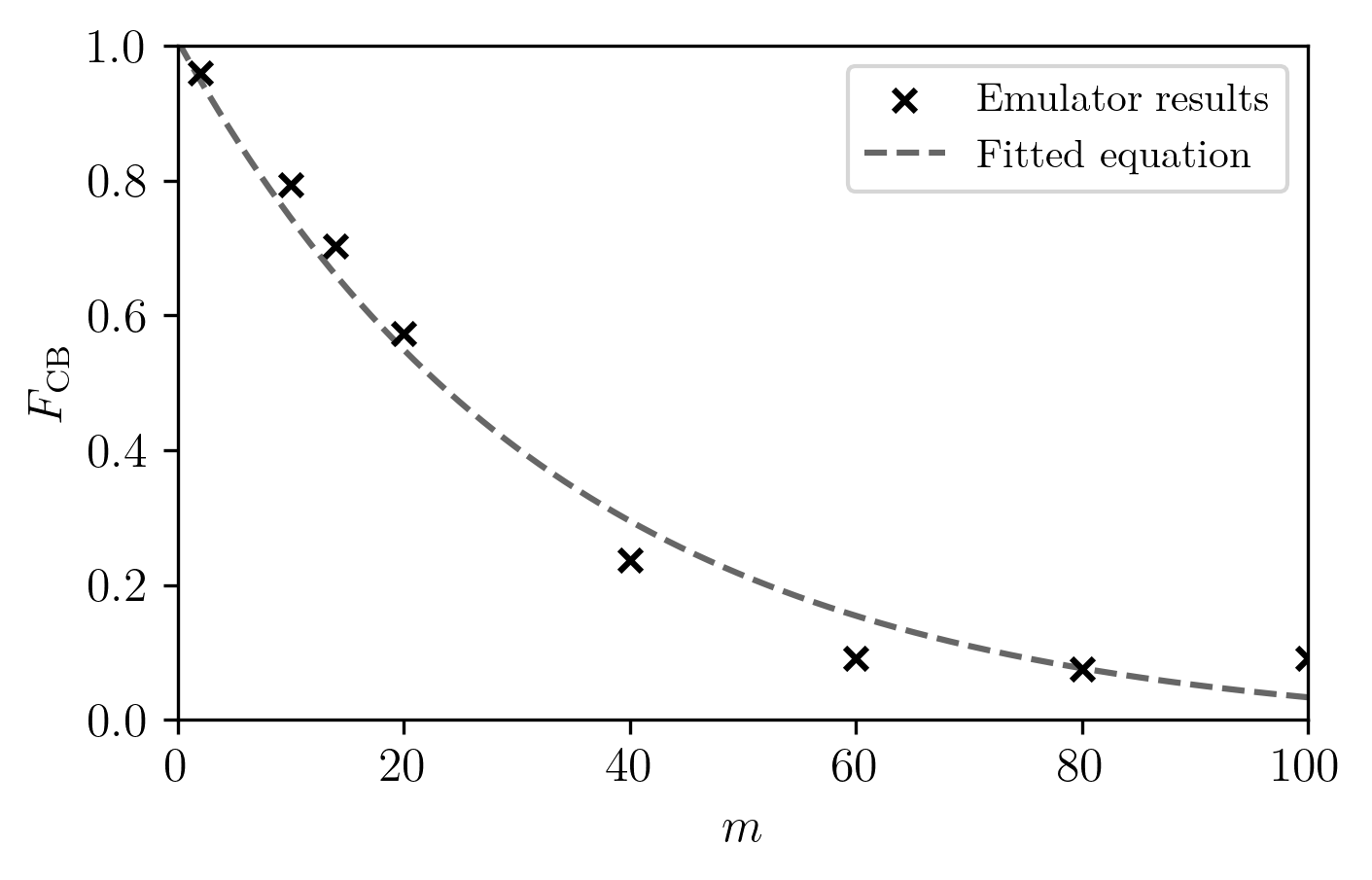}
    \caption{Results of an emulator run with the methodology outlined in this section to obtain the
  cycle-benchmarking composite process fidelity, where the noise model specified in Sec.~\ref{sec:noise_model} is used. Here the cycle consists of a single CNOT gate such that $\mathcal{G}= C_X$.  The cross symbols indicate the calculated diagonal element of the PTM, $f_{P_i,m}$, from the emulator results, for each number of cycles $m$. The gray dashed line is the fit to the data using Eq.~\ref{eq:fp_exp_fit}. Then Eqs.~\ref{eq:fp_1} and~\ref{eq:composite_process_fidelity} are used to obtain the composite process fidelity $F_{\mathrm{CB}} = 0.9781$ for the cycle consisting of a single $C_X$ gate.}
    \label{fig:cb}
\end{figure}
}

\subsection{Assumptions and limitations}

\begin{itemize}
    \item The cycle tested needs to be composed of Clifford gates, and furthermore needs to obey the condition $\mathcal{G}^m=I$ for some $m$. Generalizations that lift these restrictions may be possible by selecting random gates before the cycle and applying the corresponding gates that leave the cycle unitary unchanged using randomized compiling~\cite{Wallman-EmersonPRA2016}. 
    \item The noise is assumed to be Markovian (see metric~\ref{sec:amount_of_1q_nonmarkovian_noise}), meaning that it is not dependent on time; it is also assumed that the noise on Pauli gates is independent of the cycle.
\end{itemize}

\subsection{Source code}
A tutorial showing how to create and run the CB circuits and estimate the CB composite process fidelity is provided in \sourceurl{gate_execution_quality_metrics/cycle_benchmarking_composite_process_fidelity}.

\metricbibliography
\end{refsegment}

\section{Over- or under-rotation angle}
\label{sec:amount_of_over_under_rotation}
\begin{refsegment}
For gates corresponding to single qubit rotations, this metric quantifies the amount of over- or under-rotation compared to the target rotation angle.

\subsection{Description}
Single qubit unitary gates correspond to rotations on the Bloch sphere, $R_u(\theta)$, where $u$ is the rotation axis, and $\theta$ is the rotation angle. Due to gate mis-calibration, drifts present in the device, and limitations in the precision of gate calibration, the actual rotation angle, $\theta'$, of the gate is usually different from the target rotation angle, $\theta$, such that $\theta' = \theta + \theta_\mathrm{err}$, where $\theta_\mathrm{err}$ is the amount of over or under rotation. The value of $\theta_\mathrm{err}$ may be small, so that repeated gates are necessary to amplify the rotation errors in order to reliably quantify them~\cite{SheldonIterativeRandomizedBenchmarking}.
To quantify the amount of over- or under-rotation one can consider angles $\theta = \frac{2\pi}{n}$, so that $n$ repeated applications of $R_u$, corresponding to $R_u^n$, lead to a pseudo-identity operation for the ideal noiseless operation, since it corresponds to a total ideal rotation amount of $2\pi$.

\subsection{Measurement procedure}
In order to extract the rotation error, a number of different circuits are run, where each applies a different number of pseudo-identities. The number of pseudo-identities applied, $m$, starts from 0 and increases with step size $\Delta m$, up to the maximum number of pseudo-identities to apply, $m_\mathrm{max}$. It must also be ensured to prepare the qubit in a state that is orthogonal to the rotation axis of $R_u$. For example, one can prepare the qubit in the $\ket{0}$ state for gates corresponding to rotations about the $x$ or $y$ axes, and in the $\ket{+}= \qty(\ket{0}+ \ket{1})/\sqrt{2}$ state for rotations about the $z$ axis. This maximises the amplitude of the oscillations that will occur due to the over or under rotations, allowing for more precise estimation of the error.

Each circuit is executed for a chosen number of shots in order to find the probability of measuring the qubit in the $\ket{0}$ state after application of $m$ pseudo-identities, which is denoted as $p_0(m)$. An equation for $p_0(m)$ in the presence of over/under-rotation can be derived~\cite{SheldonIterativeRandomizedBenchmarking}, and then generalized to include the effects of state preparation and measurement errors, as well as decoherent errors. This equation is given by 
\begin{equation}
    \label{eq:over/underrotation}
    p_0(m) = \frac{1}{2} + a + \frac{1}{2}(1-b)e^{-\lambda\,m} \cos(m\, \theta_\mathrm{err}).
\end{equation}
Here, the fitting parameters $a,b$ include the effects of state preparation and measurement errors, since these errors affect the vertical offset and the maximum amplitude of the oscillation decay envelope. The fitting parameter $\lambda$ includes the effect of decoherent errors, and the fitting parameter $\theta_\mathrm{err}$ corresponds to the over- or under-rotation error amount. All parameters $a,b,\lambda$, and $\theta_\mathrm{err}$ are $0$ in the absence of noise. Since $\theta_\mathrm{err}$ effectively is determined by an oscillation frequency, its characterization from experimental data is affected little by the exact values of the other parameters. State preparation and measurement errors, as well as decoherent errors, can in principle also be extracted using this equation, although one typically uses dedicated metrics and measurement procedures that are aimed to quantify them (see metrics \ref{sec:t1}, \ref{sec:t2}, \ref{sec:spam_fidelity}). 

The measurement procedure then is as follows:
\begin{enumerate}
    \item To characterize the rotation error for a single-qubit rotation gate, $R_u$, select $\Delta m$, $m_\mathrm{max}$, and $n$. This defines the set of circuits to run, where each pseudo-identity is given by $R_u^n$, and the set of circuits to run with differing pseudo-identities is given by $M = \{0,\Delta m, 2\Delta m,...,m_\mathrm{max}\}$. Additionally, select a state $\ket{\psi}$ that is orthogonal to the eigenstates of $R_u$, and execute the operation $U_\psi$ that prepares this state such that $\ket{\psi} = U_\psi \ket{0}$.
    \item For each $m \in M$: 
    \begin{enumerate}
        \item Prepare the qubit in the state $\ket{\psi}$ by applying $U_\psi$ to a qubit initially in state $\ket{0}$.
        \item Apply gate $R_u$ a total of $n \times m$ times.
        \item Invert the state preparation operation by applying $U_\psi^\dagger$ and measure the qubit in the computational basis.
        \item Repeat steps 2a-2c for a specific number of shots in order to estimate the probability of measuring the qubit in the $0$ state, $p_0(m)$.
    \end{enumerate}   
    \item Fit the results to the function in Eq.~\ref{eq:over/underrotation} to extract the over or under rotation error amount $\theta_\mathrm{err}$. 
    \item To determine whether the rotation error is an under- or over-rotation, proceed as follows:
    \label{step:determing_sign_of_rotation}
    \begin{enumerate}
        \item Using the fit in the previous step, select $m_\mathrm{eq}$, the number of pseudoidentities at which $p_0$ first reaches the mean or equilibrium point of the first full oscillation.
        \item  Run the circuit as before, preparing the qubit in the state $\ket{\psi}$, and then applying $R_u$ a total of $n\times m_\mathrm{eq}$ times. 
        \item Let $\vec{u}$ be the vector in the Bloch sphere corresponding to the positive eigenstate of $R_u$, and let $\vec{\psi}$ be the vector corresponding to $\ket{\psi}$. Evaluate the state $\ket{\phi}$ which is defined as the state corresponding to the Bloch vector $\vec{\phi} =  \vec{u} \cross \vec{\psi}$. Apply the operation $U_\phi^\dagger$, where $U_\phi$ is defined such that $U_\phi \ket{0} = \ket{\phi}$. Finally, measure the qubit in the computational basis.
        \item Repeat the above and define $p_\phi$ as the probability of obtaining the $\ket{0}$ state upon measurement. A value of $p_\phi > 0.5$ corresponds to an over-rotation error while $p_\phi < 0.5$ corresponds to an under-rotation error. 
    \end{enumerate} 
\end{enumerate}
In practice, $\Delta m$ must be chosen to be small enough to avoid aliasing effects in oscillations, and $m_\mathrm{max}$ must be large enough to either see at least half an oscillation, or to ensure $P_0(m)$ has reached its steady state value. 

\metricfig{
An example of the results one might obtain is shown in Fig.~\ref{fig:over_under_rotation}.
\begin{figure}[htpb]
    \centering
    \includegraphics{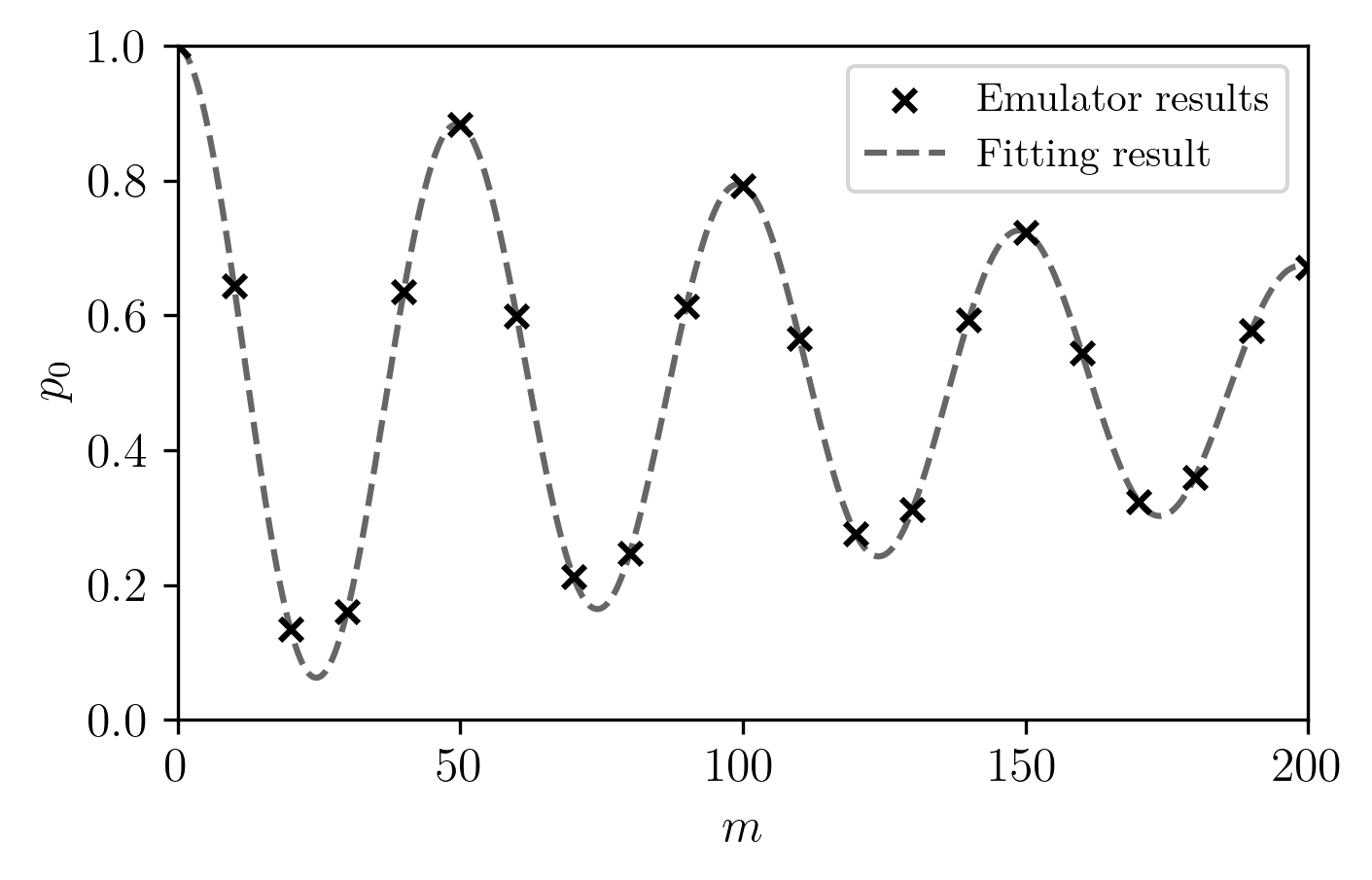}
    \caption{Results of emulator runs with the methodology outlined in this section to obtain the amount of over- or under-rotation, using the noise model specified in Sec.~\ref{sec:noise_model}. Here the pseudo-identity gates are implemented by applying an $R_x(\pi/4)$ four times. The crosses indicate the measured probabilities of the $\ket{0}$ state, $p_0$, at each number of pseudo-identities applied, $m$. The oscillation frequency is found by fitting Eq.~\ref{eq:over/underrotation} to the data. Then using step~\ref{step:determing_sign_of_rotation} in the measurement procedure, the rotation error is determined to be an over-rotation. The calculated over-rotation is found to be $\theta_{err}=0.01\pi$. This is in agreement with the theoretical value as specified in the noise model.}
    \label{fig:over_under_rotation}
\end{figure}
}

\subsection{Assumptions and limitations}
\begin{itemize}
    \item  In the fitting equation it is assumed that $p_0(m)$ has at most a single frequency oscillation. In the presence of large amounts of non-Markovian noise (see metric~\ref{sec:amount_of_1q_nonmarkovian_noise}), this assumption is often violated~\cite{agarwal2023modelling}.
    \item The coherence time of the qubit sets a bound on the smallest rotation error that can be detected. If no oscillations are detected, and only a decay of $p_0(m)$ to its steady state is observed, it suggests that rotation error is smaller than this bound.
    \item The rotation axis $u$ is typically chosen such that the rotation is around the $x$, $y$, or $z$ axis. This means that the single-qubit rotation gates typically characterized are the $R_x$, $R_y$, and the $R_z$ gates.
\end{itemize}

\subsection{Source code}
A tutorial showing how to calculate the over- and under-rotation errors for an $R_x$ rotation gate is provided in
\sourceurl{gate_execution_quality_metrics/over_under_rotations}.
\metricbibliography
\end{refsegment}

\section{State preparation and measurement fidelity}
\label{sec:spam_fidelity}
\begin{refsegment}
 The state preparation and measurement (SPAM) fidelity measures how well qubits can be initialized and measured~\cite{nielsenGateSetTomography2021}. The SPAM operations are represented by the density matrix  of the initial state, $\rho_{\mathrm{init}}$, and by the measurement projectors, $\{M_i\}$. The SPAM fidelity measures how close the SPAM operations, $\{\rho_{\mathrm{init}}, \{M_i\}\}$, are to the ideal initial states and measurement projectors. 

\subsection{Description} 
SPAM, which makes up two of the five criteria for the physical realization of a quantum computer~\cite{divincenzo2000physical}, directly impacts the performance of quantum computers. The method to determine the SPAM fidelity is based on the methodology to obtain general process and state fidelities, outlined in section ~\ref{sec:process_fidelity}.
 
The state preparation errors occur due to the non-ideality of the physical realization of quantum computers. For example, the state of a qubit after initial preparation, usually its equilibrium state, may not fully be the generally targeted $\ket{0}$ state, but instead may have a small chance of being in the $\ket{1}$ state. Similarly, when the measurement errors occur, it means that the wrong outcome is reported by the measurement. For example, the measurement may report a $\ket{0}$ even when the qubit is in the $\ket{1}$ state. The SPAM fidelity is a measure of how close the SPAM operations are to their ideal operations.

For a single qubit, the SPAM components are $\rho_{\mathrm{init}}$, $M_0$, and $M_1$, where $\rho_{\mathrm{init}}$ is the initial density matrix, and $M_0$ is the measurement projector to measure the qubit in the $\ket{0}$ state, and $M_1$ is the projector to measure the qubit in the $\ket{1}$ state. The ideal SPAM components for the single qubit case are denoted by $\bar{\rho}_{\mathrm{init}}$, $\bar{M_0}$, $\bar{M_1}$. Typically the initial state of the qubit is set to the $\ket{0}$ state, and, in this case, the ideal noiseless SPAM operations are given by the following~\cite{nielsen2002quantum}:
\begin{align}
    \bar{\rho}_{\mathrm{init}}=\begin{bmatrix} 1 & 0 \\ 0 & 0 \end{bmatrix},\;
   \bar{M_0} =\begin{bmatrix} 1 & 0 \\ 0 & 0\end{bmatrix},\;
    \bar{M_1}=\begin{bmatrix} 0 & 0\\ 0 & 1 \end{bmatrix},
\end{align}
The number of measurement projectors is dependent on the number of possible outcomes from measurement. For a single qubit, there are two outcomes, and hence there are two measurement projectors. As the number of qubits increases, the number of measurement projectors increase as $2^{N_{\mathrm{q}}}$, where $N_{\mathrm{q}}$ is the number of qubits.
Given all of the SPAM components, there are two fidelities that constitute SPAM fidelity: the fidelity of the initial density matrix, $F_{\rho_{\mathrm{init}}}$, and the measurement fidelity $F_{M}$, which is calculated using estimates of $\{M_i\}$~\cite{Nielsen_2020}.

The SPAM fidelity is different to what is often referred to as readout fidelity, $F_\mathrm{r}$, which gives a fidelity estimate solely based on mis-classification of readout~\cite{chen2023transmon}.

\subsection{Measurement procedure}
\begin{enumerate}
    \item Using gate set tomography (GST) on the device of interest, obtain an estimate of the Pauli transfer matrices of the device gate set (see Sec. \ref{sec:gst}). This GST gate set estimate includes estimates of the state preparation, characterized by the initial density matrices $\rho_0$, and the measurement projectors, characterized $\{M_i\}$ . For a singe qubit one has only two measurement projectors, $\{M_0 , M_1\}$.
    \item Using these estimates, the fidelity of the initial density matrix can be calculated using Eq.~\ref{eq:state_fidelity_1pure}. The measurement fidelity is computed by transforming the measurement projectors into a gate-like quantity~\cite{Nielsen_2020}. Then one can compute the measurement fidelity using Eq.~\ref{eq:process_fidelity}.
\end{enumerate}

$F_\mathrm{r}$, on the other hand, is obtained by preparing all possible non-superposition states in the computational basis, and then measuring and recording the probability of each outcome. This then gives a matrix, where each row corresponds to the prepared state, and each column corresponds to the measured state. The readout fidelity is the trace of this matrix. For example, for a single qubit system, one prepares the $\ket{0}$ and the $\ket{1}$ state, and then immediately measures it to find the outcome probabilities. This gives a $2 \times 2$ matrix, which has the elements $P(0|0), P(1|0), P(0|1)$, and $ P(1|1)$. The readout fidelity is then given by $F_\mathrm{r} = P(0|0) + P(1|1)$. There are limitations to this method, as it assumes the state preparation to be perfect, and as a result, $F_\mathrm{r}$ also includes the state preparation error in the estimate. The SPAM estimate obtained using GST overcomes this limitation.

\subsection{Assumptions and limitations}
\begin{itemize}
    \item Since the measurement procedure is based on GST, the assumptions and limitations for GST (section~\ref{sec:gst}) also apply here.
\end{itemize}

\subsection{Source code}
A tutorial for calculating the SPAM fidelity using gate set tomography is provided in:
\sourceurl{gate_execution_quality_metrics/gst_based_gate_execution_quality_metrics}.
\metricbibliography
\end{refsegment}

\def\thechapter{M4}
\chapter{Circuit execution quality metrics}
\label{chapter:circuit_quality_metrics}

\section{Quantum volume}
\label{sec:quantum_volume}
\begin{refsegment}  
 
 The quantum volume is a metric for the overall capabilities of a noisy quantum computer. It follows the volumetric benchmarking framework (see section~\ref{sec:volumetric_benchmarking}), where the number of layers in the quantum circuit is set equal to the number of qubits. Quantum volume is defined as $2^{n_\mathrm{pass}}$, where $n_\mathrm{pass}$ is the largest number of qubits for which a type of randomized circuits can be successfully run. The success criterion is that the heavy output probability, $p_h$, is greater than $2/3$~\cite{crossValidatingQuantumComputers2019}. 
  The heavy output probability for a circuit is obtained by first performing noiseless emulator runs for the purpose of finding all the measurement outcomes that have probabilities greater than the median of the measurement outcomes. These are collected into a set of selected measurement outcomes. The heavy output probability for a given quantum hardware then is the sum of the measurement probabilities of this set of outcomes when the circuit is executed on the experimental device.
  Evaluation of this quantity requires to run both the experiments on the quantum hardware and the analogous emulator runs for noiseless quantum circuits on classical computers, and hence scalability is limited by the maximum system sizes accessible in the emulator runs. 

\subsection{Description} 

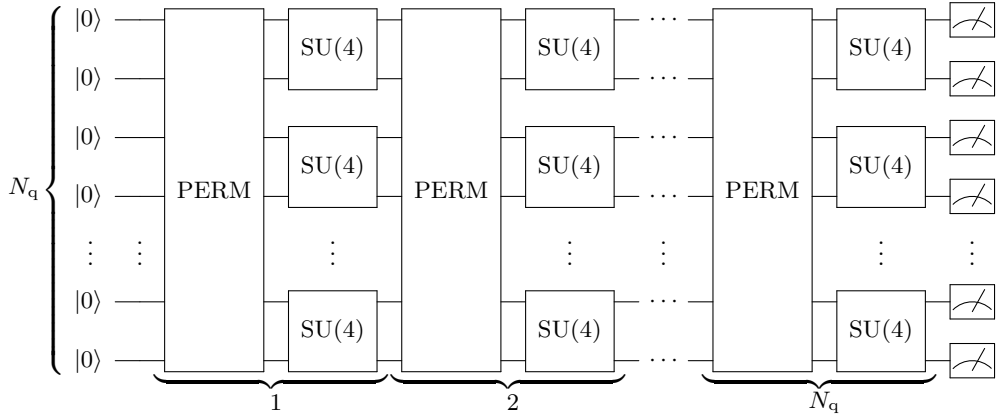
\begin{figure}[ht]
\centering
\small{
\[
\begin{array}{c}
\Qcircuit @C=1em @R=1em {
  && \lstick{\ket{0}} & \qw & \multigate{6}{\text{PERM}}  & \multigate{1}{\mathrm{SU}(4)} & \multigate{6}{\text{PERM}}  & \multigate{1}{\mathrm{SU}(4)} & \qw & \cdots & & \multigate{6}{\text{PERM}}  & \multigate{1}{\mathrm{SU}(4)} & \meter   \\
    && \lstick{\ket{0}}  & \qw & \ghost{\text{PERM}}  & \ghost{\mathrm{SU}(4)} & \ghost{\text{PERM}} & \ghost{\mathrm{SU}(4)} & \qw & \cdots& & \ghost{\text{PERM}}  & \ghost{\mathrm{SU}(4)} & \meter \\
  && \lstick{\ket{0}} & \qw & \ghost{\text{PERM}}  & \multigate{1}{\mathrm{SU}(4)} & \ghost{\text{PERM}}  & \multigate{1}{\mathrm{SU}(4)} & \qw & \cdots & & \ghost{\text{PERM}}  & \multigate{1}{\mathrm{SU}(4)} & \meter   \\
   & & \lstick{\ket{0}}  & \qw & \ghost{\text{PERM}}  & \ghost{\mathrm{SU}(4)} & \ghost{\text{PERM}} & \ghost{\mathrm{SU}(4)} & \qw & \cdots& & \ghost{\text{PERM}}  & \ghost{\mathrm{SU}(4)} & \meter \\
   &\qvdots &  & \qvdots & \nghost{\text{PERM}}     &  \qvdots  & \nghost{\text{PERM}}  &  \qvdots  &  & \qvdots & & \nghost{\text{PERM}}     &  \qvdots & \qvdots \\
  && \lstick{\ket{0}}  & \qw & \ghost{\text{PERM}}  & \multigate{1}{\mathrm{SU}(4)} & \ghost{\text{PERM}}  & \multigate{1}{\mathrm{SU}(4)} & \qw  & \cdots& & \ghost{\text{PERM}}  & \multigate{1}{\mathrm{SU}(4)} & \meter \\
  &  & \lstick{\ket{0}} & \qw & \ghost{\text{PERM}}  & \ghost{\mathrm{SU}(4)} & \ghost{\text{PERM}}  & \ghost{\mathrm{SU}(4)} & \qw & \cdots & & \ghost{\text{PERM}}  & \ghost{\mathrm{SU}(4)}  & \meter \inputgroupv{1}{7}{1em}{6.9em}{N_{\mathrm{q}}}
  \gategroup{1}{5}{7}{6}{.8em}{_\}}
  \gategroup{1}{7}{7}{8}{.8em}{_\}}
  \gategroup{1}{12}{7}{13}{.8em}{_\}} \\
 & & & & \shifttext{5em}{1}& & \shifttext{5em}{2} & & & & & \shifttext{5em}{$N_{\mathrm{q}}$} &
    }
\end{array}
\]
 \caption{Circuit diagram indicating the structure of a quantum volume circuit. There are $N_{\mathrm{q}}$ qubits, and $N_{\mathrm{q}}$ layers of gates, creating a square circuit. Each layer consists of a $N_{\mathrm{q}}$-qubit gate that randomly changes the order of qubits, denoted as $\text{PERM}$ gate, and a column of Haar-random two-qubit gates, denoted as $\mathrm{SU}(4)$.}
\label{fig:quantum_volume_circuit}
}
\end{figure}

Volumetric benchmarking metrics are defined by a task to solve, which determines the type of quantum circuits to run, and by a success criterion. For the quantum volume, $V_Q$, the task to solve is motivated by the work in Ref.~\cite{aaronson2016complexity}, which proposes a method to demonstrate quantum advantage through sampling randomized quantum circuits. The circuit structure is illustrated in Fig.~\ref{fig:quantum_volume_circuit}. 
There are $N_{\mathrm{q}}$ qubits, and $N_{\mathrm{q}}$ layers of gates that create a square circuit. Each layer consists of an $N_{\mathrm{q}}$-qubit gate that randomly changes the order of qubits, followed by a layer of two-qubit gates that apply general two-qubit operations between nearest neighbor qubits. Both the permutations and two-qubit operations are sampled randomly, where the two-qubit gates are sampled from a Haar distribution. 

For each circuit, $p_h$ is obtained as follows. For a given $N_{\mathrm{q}}$-qubit circuit, obtain the ideal probabilities of all the possible $2^{N_{\mathrm{q}}}$ output bitstrings using noiseless simulations. Then sort these probabilities, and collect the half of the bitstrings that have the ideal probability greater than the median ideal probability: these are called the heavy outputs. The sum of the  probabilities of heavy outputs is $p_h$. For the heavy output bitstrings obtained for the noiseless emulators, also calculate the total probability obtained from the experimental measurements on the hardware. The process is repeated for each circuit in the set of randomly selected circuits, and the experimentally measured $p_h$ is averaged over all the circuits to obtain $\langle p_h \rangle$. It is shown in Ref.~\cite{aaronson2016complexity} that for the types of circuits considered here, $\langle p_h \rangle$ for a noiseless quantum computer needs to be larger than $2/3$. The success criterion is therefore set to be that $\langle p_h \rangle > 2/3$. If the qubits on a noisy quantum computer have fully decohered due to noise, then the output probabilities of the quantum circuits converge to an approximately uniform distribution, for which $\langle p_h \rangle = 1/2$. 

The $V_Q$ is then defined as~\cite{crossValidatingQuantumComputers2019}
\begin{align}
    V_Q &= 2^{n_\mathrm{pass}},\\
    n_\mathrm{pass}&={\argmax_{N_{\mathrm{q}}} \, \textrm{min}(N_{\mathrm{q}}, d(N_{\mathrm{q}}))},
\end{align}
where $N_{\mathrm{q}}$ is the number of qubits, and $d(N_{\mathrm{q}})$ is the largest number of circuit layers at which an $N_{\mathrm{q}}$-qubit circuit can produce heavy outputs with probability greater than $2/3$. Due to taking the minimum of $N_{\mathrm{q}}$ and $d(N_{\mathrm{q}})$, it suffices to test only $N_{\mathrm{q}}$-qubit square circuits, which are constructed with $N_{\mathrm{q}}$ layers.

This description follows the definition of quantum volume in Ref.~\cite{crossValidatingQuantumComputers2019}. This is a redefinition of the quantum volume metric introduced in Ref.~\cite{Moll_2018}.

\subsection{Measurement procedure}
\begin{enumerate}
    \item Generate $m\ge100$ circuits for $N_{\mathrm{q}}$ qubits. Each layer consists of a permutation block, and then a block of Haar-random two-qubit gates, as illustrated in Fig.~\ref{fig:quantum_volume_circuit}. The number of layers must be equal to the number of qubits. The manner in which the random permutations are achieved depends on qubit connectivity. For example, for a superconducting qubit device, a block of SWAP gates is required for the qubit order permutations, whereas for a trapped-ion device with all-to-all connectivity, no SWAP gates are necessary. 
    \item Run noiseless simulations for these circuits to find the heavy output bitstrings.
    \item Run the $m$ circuits, and for each circuit calculate $p_h$. 
    \item Compute the $\langle p_h \rangle$ across all $m$ circuits. If it is greater than $2/3$ with a confidence of at least 2 standard deviations, meaning that at least 97.5\% of heavy output probabilities from all circuits fall above 2/3, then the device passes the test, and hence has a quantum volume of at least $2^{N_{\mathrm{q}}}$. If it is smaller, then the quantum volume is less than $2^{N_{\mathrm{q}}}$.
    \item If the test in the previous point fails, then test for a smaller $n$ until it passes. If the test passes, keep using successively larger $N_{\mathrm{q}}$ until the test fails. The quantum volume, $V_Q$, is then determined by the largest $N_{\mathrm{q}}$ that did not fail, $n_\mathrm{pass}$, so that $V_Q = 2^{n_\mathrm{pass}}$.
\end{enumerate}

\metricfig{
An example of the results one might obtain is shown in Fig.~\ref{fig:qv_result}.
\begin{figure}[ht]
    \centering
    \includegraphics{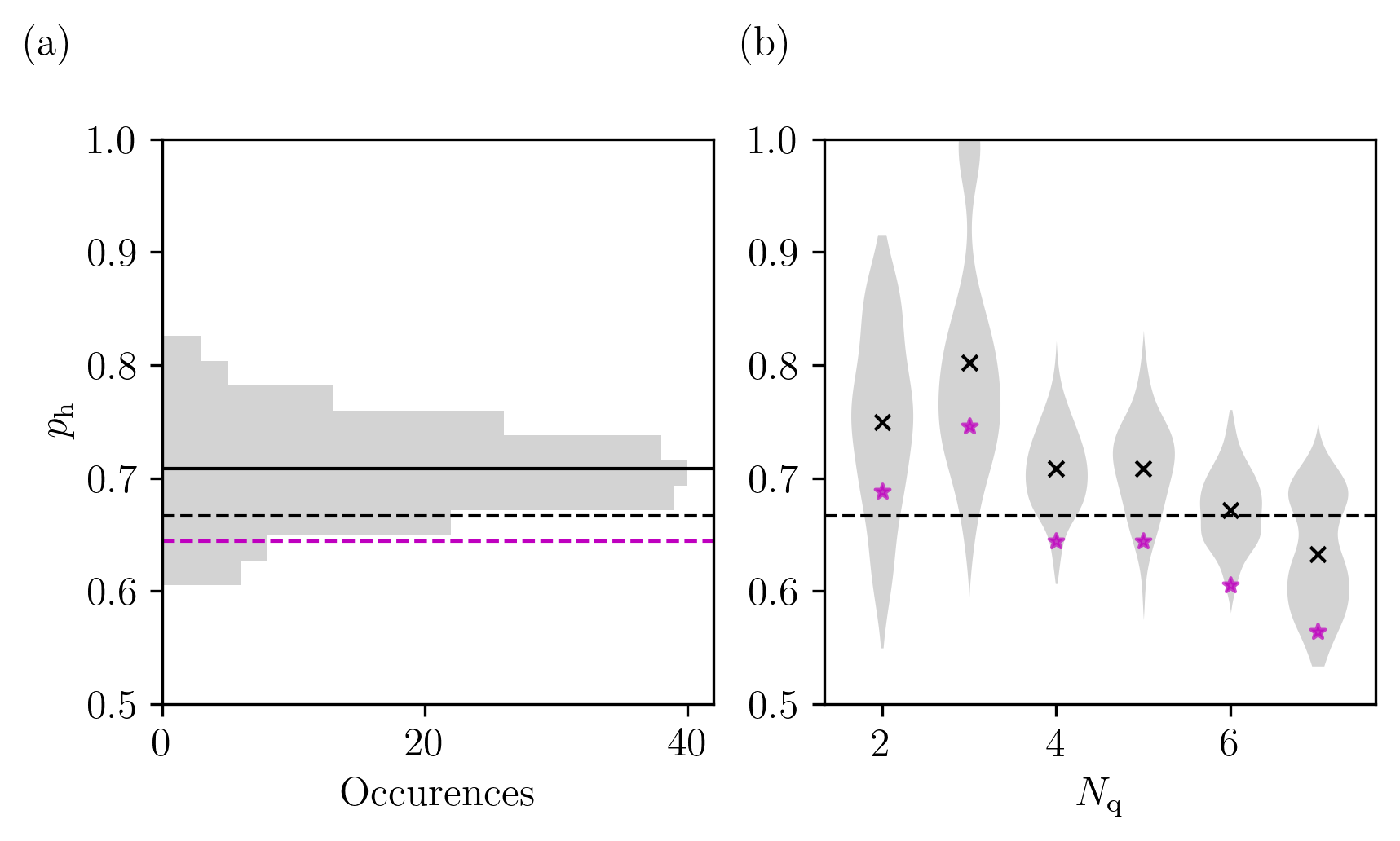}
    \caption{Results of emulator runs using the measurement procedure outlined in this section to obtain the quantum volume using the noise model specified in Sec.~\ref{sec:noise_model}. (a) A histogram is plotted horizontally, showing the distribution of the heavy output probabilities, $p_h$, for $N_{\mathrm{q}}=4$ and $m=200$. The black line represents $\langle p_h \rangle$, the magenta dashed line shows the confidence bound of the mean with 2 standard deviations, and the black dashed line shows the threshold of 2/3. The device fails the test as $\langle p_h \rangle$ with a confidence of 2 standard deviations is less than the required value of $2/3$, that is, the magenta dashed line is below the black dashed line. (b) This violin plot shows the distribution of heavy output probabilities for the $m$ circuits for increasing $N_{\mathrm{q}}$. The distribution of results is plotted using kernel density estimation. The black dashed line indicates the passing criteria of $2/3$. The black cross  represents  $\langle p_h \rangle$ output and the magenta star symbol represents $\langle p_h \rangle$ with a confidence of 2 standard deviations. This shows that for the emulator with the specified noise model in Sec.~\ref{sec:noise_model}, when $N_{\mathrm{q}} \geq 4$ it fails the quantum volume test, therefore its quantum volume is $V_Q = 2^3 = 8$.}
    \label{fig:qv_result}
\end{figure}
}

\subsection{Assumptions and limitations}

\begin{itemize}
    \item Computing the quantum volume requires classical simulation of the circuits, which have a computational cost that scales exponentially with qubit number; this puts a limit on the size of the systems that can be run at around 30 qubits~\cite{baldwin2022re}.
    \item The randomly generated circuits are typically passed to a compiler to optimize them for each specific hardware platform. The optimized circuits should be provided together with the original unoptimized circuits in order to ensure reproducibility. 
    \item The quantum volume protocol is constructed from random circuits, and as such, the effect of coherent errors in the quantum volume test may not be representative of the effect of coherent errors in certain quantum algorithms~\cite{baldwin2022re}.
\end{itemize}

\subsection{Source code}
A tutorial demonstrating quantum volume can be found at \sourceurl{circuit_execution_quality_metrics/quantum_volume}.
\metricbibliography
\end{refsegment}

\section{Mirrored circuits average polarization}
\label{sec:mirror_circuits}

\begin{refsegment}

The mirrored circuits polarization metric measures the capability of a quantum computer to run a specified type of quantum circuits successfully~\cite{proctorMeasuringCapabilitiesQuantum2022}. Mirrored circuits are based on the application of a circuit $C$, and then on the application of its inverse. Randomization layers are applied before, after and between $C$ and its inverse.
The randomization layers map the initial random circuit $C$ to a family of circuits, $M_C$. This reduces unwanted potential cancellation of systematic errors, which may occur when directly applying a circuit and its inverse. In the noiseless case, each circuit in $M_C$ results in a deterministic output state, which is easy to compute classically. 

The success probability, $p_{\mathrm{success}}$, is the probability of measuring the correct output state for the full mirrored circuit. In the case of a fully depolarized state, each possible outcome has the same probability, and so $p_{\mathrm{success}}=1/2^w$, where $w$ is the number of qubits, also denoted as the width of the circuit. To avoid having a non-zero $p_{\mathrm{success}}$ for this case, the interpretation of circuit execution quality for a given value of the metric depends on the number of qubits. The normalized difference between $p_{\mathrm{success}}$ and this fully depolarized limit, denoted as the polarization, $J$, is taken as the success metric. $J$ is given by
\begin{equation}
    \label{eq:ciruit_polarization}
    J = \frac{p_{\mathrm{success}}- \frac{1}{2^w}}{1-\frac{1}{2^w}}.
\end{equation}
A $J=0$ typically indicates a fully depolarized state. Note however that a $J=0$ can also be due to coherent errors, and that $J$ can also be negative, for example if the success probability is zero. 

If the base circuit $C$  can be generalized to different widths and numbers of layers, then this protocol can be run following the volumetric benchmarking framework (section \ref{sec:volumetric_benchmarking}). Since the ideal output can be classically computed also for large systems, the method is scalable. 

\subsection{Description}

\begin{figure}[ht]
\centering
\small{
\[
\begin{array}{c}
\Qcircuit @C=1em @R=1em {
  && \lstick{\ket{0}} & \qw &  \multigate{6}{L_0}  &  \multigate{6}{C} &  \multigate{6}{Q_0}   & \multigate{6}{\tilde{C}^{-1}}  & \multigate{6}{L_0^{-1}} & \qw & \meter   \\
  && \lstick{\ket{0}}  & \qw & \ghost{L_0}  & \ghost{C}  & \ghost{Q_0} &  \ghost{\tilde{C}^{-1}} & \ghost{L_0^{-1}} & \qw & \meter \\
  && \lstick{\ket{0}} & \qw &  \ghost{L_0}  & \ghost{C}  & \ghost{Q_0} &  \ghost{\tilde{C}^{-1}}  & \ghost{L_0^{-1}}  & \qw & \meter   \\
  && \lstick{\ket{0}}  & \qw & \ghost{L_0}  & \ghost{C}  & \ghost{Q_0} &  \ghost{\tilde{C}^{-1}}  & \ghost{L_0^{-1}} & \qw & \meter \\
  &\qvdots &  & \qvdots &       \nghost{L_0} & \ghost{C} & \ghost{Q_0} &  \ghost{\tilde{C}^{-1}} & \ghost{L_0^{-1}}   & \qvdots   &  \qvdots  \\
  && \lstick{\ket{0}}  & \qw & \ghost{L_0}   & \ghost{C} & \ghost{Q_0} &  \ghost{\tilde{C}^{-1}}  & \ghost{L_0^{-1}}  & \qw & \meter \\
  & & \lstick{\ket{0}} & \qw & \ghost{L_0}  &  \ghost{C} & \ghost{Q_0} &  \ghost{\tilde{C}^{-1}}  & \ghost{L_0^{-1}}   &\qw &  \meter}
\end{array}
\]
 \caption{Circuit diagram showing the structure of a mirrored circuit. The first layer $L_0$ is a $w$-fold tensor product of randomly selected single qubit Clifford gates, where $w$ is the number of qubits in the circuit. $C$ is the base circuit, $Q_0$ is the $w$-fold tensor product of randomly selected Pauli gates. The $\tilde{C}^{-1}$ block is the quasi inverse of $C$, though it is just as valid to be the inverse $C^{-1}$. The final block $L_0^{-1}$ is the inverse of the first layer $L_0$.}
\label{fig:mirror_circuit}
}
\end{figure}
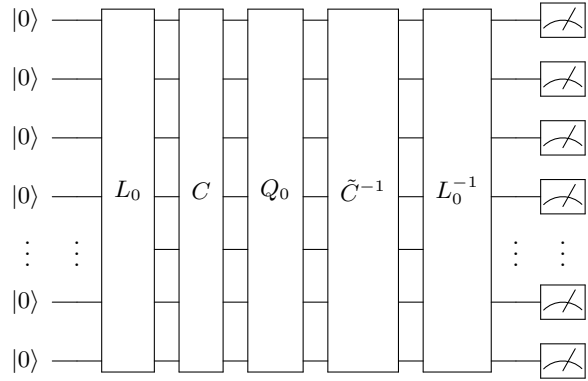

A $w$-qubit circuit $C$ is selected, which is split into $m$ layers, such that $C = L_m L_{m-1} ... L_{2}L_{1}$. $C$ can be a dedicated test circuit, or also a circuit of used within a specific application, such as a state preparation circuit within a variational quantum eigensolver. 
The mirrored circuits protocol transforms $C$ into a set of circuits $M_C$ as follows.
Given an initial $\ket{0}^{\otimes w}$ state and measurement in the computational basis, a mirrored circuit in $M_C$, as depicted in Fig.~\ref{fig:mirror_circuit}, is written in matrix form as $L_0^{-1} \tilde{C}^{-1} Q C L_0$, where $L_0$ is a random layer of single-qubit Clifford gates, $Q$ is a random Pauli string, such that $Q \in (X,Y,Z,I)^{\otimes w}$, and $\tilde{C}^{-1}$ is a quasi-inversion circuit, $\tilde{C} =  \tilde{L}_1^{-1} \tilde{L}_{2}^{-1} ... \tilde{L}_{m-1}^{-1}\tilde{L}_{m}^{-1}$. Each quasi-inversion layer satisfies $\tilde{L}_{i}^{-1}L_i = Q_i$, where $Q_i$ is a random Pauli layer. Adding this additional randomization can further reduce systematic error cancellations. The insertion of randomization layers can vary. For example, all the $Q_i$ can also be set to the identity, in which case $\tilde{C}$ is the inverse of $C$. For each circuit $C$, a family of circuits $M_C$ is generated by selecting different $Q_0$ and $L_0$. The layer $L_0$ is selected randomly from $\mathbb{C}_1^{w}$, the $w$-fold tensor product of all single qubit Clifford gates. The layer $Q_0$ is selected randomly from the $w$-fold tensor product of the Pauli gates. As part of the protocol, the detailed distribution of randomization layers needs to be specified for reproducibility of the results. 

The protocol is defined for Clifford circuits, as well as for circuits containing alternating layers of Clifford and $R_z(\theta)$ operations~\cite{proctorMeasuringCapabilitiesQuantum2022}. The circuit unitary is $ L_0^{-1} \tilde{C}^{-1} Q C L_0 = Q'$, where $Q'$ is another Pauli layer. The value of $Q'$ can be easily computed classically. This is due to the fact that if $C$ is a Clifford circuit, one can use the property that any Clifford multiplied with a Pauli returns another Pauli, so that for an arbitrary Pauli layer, $Q^{(1)}$, one obtains  $\tilde{L}_i^{-1} Q^{(1)} L_i = Q^{(2)}$, where $ Q^{(2)}$ is another Pauli layer, and which can be efficiently computed.
In the more general case, where $R_z(\theta)$ gates are included in $C$, the construction is more involved, as outlined in the supplementary material of Ref.~\cite{proctorMeasuringCapabilitiesQuantum2022}. The main reason that $Q'$ can still be efficiently calculated is that commuting a Pauli gate through a $R_z(\theta)$ gate either does nothing, or just flips the sign of the angle. 
In the noiseless case, applying the circuit $ L_0^{-1} \tilde{C}^{-1} Q C L_0=Q'$ on $\ket{0}^{\otimes w}$ returns a single deterministic output state. For example, for a single qubit the output state is $\ket{0}$ when $Q'$ is $I$ or $Z$, and $\ket{1}$ when $Q'$ is $X$ or $Y$.

For each circuit $c \in M_C$ the circuit is executed on hardware, and the polarization $J_c$ is calculated using Eq.~\ref{eq:ciruit_polarization}. The average mirrored circuits polarization is given by 
\begin{equation}
    \label{eq:average_circuit_polarization}
    J_{\text{ave}} = \frac{1}{\abs{M_C}}\sum_{ c \in M_C} J_c.
\end{equation}

 When executed as part of a volumetric benchmarking approach, $J_{\text{ave}}$ is determined for pairs of integers $(w,d_c)$, where $d_c$ is the full depth of the circuit, such that $d_c=2m + 3$. The $2m$ layers come from the base circuit and its inverse, and the remaining 3 come from $L_0$, its inverse, and $Q_0$. Within a depolarising error model, where the error is the same for all layers, the circuit polarization decays exponentially with depth~\cite{proctorMeasuringCapabilitiesQuantum2022}. Hence, the volumetric benchmarking success criteria is typically set to the threshold $J_{\text{ave}} \geq 1/e\approx 0.37$. 

\subsection{Measurement procedure}

In what follows the procedure to measure the mirrored circuits average circuit polarization for a specific circuit $C$ is presented. This mirrored circuits protocol can be used as a volumetric benchmarking instance if $C$ can be set for specified $w$ and $m$. The process then is as follows:
\begin{enumerate}
    \item Select a base circuit $C$ with width $w$ and with number of layers $m$. 
    \item Choose the number of mirrored circuits, $k$, to generate $M_C$ such that $\abs{M_C} = k$. Each element in $M_C$ is generated in the following way:
    \begin{enumerate}
        \item Select the $L_0$ layer at random. This layer consists of a random single qubit Clifford gate on each qubit.
        \item Select the Pauli layer, $Q_0$, at random. $Q_0$ consists of random single qubit Pauli gates on each qubit.
        \item Select a number $k$ of quasi-inverse circuits, as defined in the description.
    \end{enumerate}
    \item All circuits in $M_C$ have the form $M_C = \qty{L_0^{-1}\tilde{C}^{-1}Q_0 C L_0 }$. Then, for each circuit $q \in M_C$:
    \begin{enumerate}
        \item  compute the output string classically for the circuit $q$ in the noiseless case.
        \item execute $q$ on the quantum computer, with initial state $\ket{0}^{\otimes w}$.
        \item Measure the outcome probabilities and compute $J$ using Eq.~\ref{eq:ciruit_polarization}.
    \end{enumerate}
    \item Calculate the average polarization $J_{\text{ave}}$, given in  Eq.~\ref{eq:average_circuit_polarization}.
\end{enumerate}

\subsection{Assumptions and limitations}

\begin{itemize}
    \item The considered gates in the circuit need to be Clifford gates or $R_z(\theta)$ gates. 
\end{itemize}

\subsection{Source code}
A tutorial for the mirrored circuits benchmarking is provided in \sourceurl{circuit_execution_quality_metrics/mirrored_circuits}.

\metricbibliography
\end{refsegment}

\section{Algorithmic qubits}
\label{sec:algorithmic_qubits}
\begin{refsegment}

The algorithmic qubits metric, denoted as \#AQ, corresponds to the largest number of qubits that can run a set of selected quantum algorithms successfully according to a specific success criterion~\cite{AlgorithmicQubitsBetter}. It is built on the volumetric benchmarking framework (section \ref{sec:volumetric_benchmarking}).

\subsection{Description}

The \#AQ metric, which corresponds to an integer number of qubits, builds on the volumetric benchmarking framework introduced in section~\ref{sec:volumetric_benchmarking}, using a pre-defined set of circuits that are based on representative quantum computing tasks~\cite{AlgorithmicQubitsBetter}. These are six tasks selected from the the benchmarking study performed by the quantum economic development consortium (QED-C)~\cite{lubinskiApplicationOrientedPerformanceBenchmarks2021}. The single valued output \#AQ corresponds to an integer number.

For each circuit $c$ within this set, the fidelity $F_c$ of the measured output probability distribution $Q_{\text{output}}$ against the ideal probability distribution obtained by a noiseless quantum computing emulator, $Q_{\text{ideal}}$, is computed. Here the following relation is used for the fidelity of the output probabilities,
\begin{equation}
F_c(Q_{\text{ideal}},Q_{\text{output}}) = \qty(\sum_x \sqrt{p_{x, \text{output}} \, p_{x, \text{ideal}}})^2,
\end{equation}
where the sum goes over all possible measurement outcomes, $x$ is a measurement output bitstring, and  $p_{x, \text{ideal}}$ and $p_{x, \text{output}}$ are the probabilities of obtaining the output bitstring $x$ in the measurements for the ideal and experimental circuits, respectively.
Within the volumetric benchmarking framework the obtained fidelities for all circuits in the set can be plotted as function of the number of qubits, $N_{\mathrm{q}}$, and of the circuit depth, $d_c$. For the algorithmic qubits metric, the depth is defined as the total number of CNOT gates in the circuit~\cite{AlgorithmicQubitsBetter}. Note that this definition of depth also includes CNOT gates that can, in principle, be executed in parallel at the same time.

To compute the value of \#AQ one defines the success criterion for a circuit $c$ as $F_c - \epsilon_c > t$, where $t$ is a constant representing the success threshold, and is set to $t=1/e\approx 0.368$; $\epsilon_c$ is the standard error of the mean, $n_s$ is the number of shots, given by $\epsilon_c = \sqrt{(F_c(1-F_c))/n_s}$. Then the value of \#AQ is defined as
the width of the largest box that can be drawn on the 2D fidelity plot with width $N_{\mathrm{q}}$ and depth $N_{\mathrm{q}}^2$, and where all points inside the box pass the success criterion. 

\subsection{Measurement procedure}
In the measurement procedure presented here, the method and choice of circuits follows the definitions of \#AQ by IonQ~\cite{AlgorithmicQubitsBetter} in its \#AQ definition version 1.0. This in turn builds on the benchmarking study from the quantum economic development consortium (QED-C)~\cite{lubinskiApplicationOrientedPerformanceBenchmarks2021}. It specifies six tasks developed within the QED-C benchmarking study to use for the calculation of \#AQ. These tasks comprise the quantum Fourier transform, as described in metric~\ref{sec:qft}, the quantum phase estimation, the amplitude estimation, Monte Carlo sampling, the variation quantum eigensolver, as described in metric~\ref{sec:vqe}, and Hamiltonian simulation.

Any circuit transpilation applied before running this set of circuits needs to comply with the following rules:
\begin{itemize}
    \item The number of CNOT gates per circuit is calculated by transpiling the circuits to a gate set of \{$R_x$, $R_y$, $R_z$, CNOT\}, regardless of whether the hardware implements this set of native gates or not. 
    \item Any optimization to the circuits are only allowed if the quantum computer executes the same unitary operator as the submitted circuits.
    \item Potentially applied error mitigation techniques, such as randomized compiling, must be reported along with the value for $\text{\#AQ}$.
\end{itemize}
The detailed rules specified by IonQ are provided in the following folder forked from IonQ repository
\sourceurltwo{circuit_execution_quality_metrics/algorithmic_qubits/code/_doc/AQ.md}.

Then the measurement procedure is defined as:
\begin{enumerate}
    \item All the circuits for the six tasks constitute the set of circuits $C$. For each circuit $c \in C$, determine the fidelity $F_c$ of the experimental output versus the ideal results obtained on a quantum computing emulator for a noiseless quantum computer.
    \item Make a 2D plot of all the obtained fidelities, with the axes being the width and depth of circuits.
    \item On the plot, draw a box with width $N_{\mathrm{q}}$, depth $N_{\mathrm{q}}^2$, and where all fidelities inside the box pass the success criterion defined in the previous section. Then \#AQ is equal to the width of the largest such box.
\end{enumerate}

\subsection{Assumptions and limitations}

\begin{itemize}
    \item The tasks and circuits included in the benchmarking set are chosen such that they span a range of potential applications. Although the choice is well motivated, ultimately the specific list of tasks included in the set is rather arbitrary, and can be amended in future.
    \item Since the \#AQ metric requires running the selected applications on noiseless emulators of quantum computers, it can only be evaluated for limited number of qubits\cite{lubinskiApplicationOrientedPerformanceBenchmarks2021}. 
\end{itemize}

\subsection{Source Code}
A tutorial for measuring \#AQ is provided in \sourceurl{circuit_execution_quality_metrics/algorithmic_qubits}.

\metricbibliography
\end{refsegment}

\section{Upper bound on the variation distance}
\label{sec:accreditation}
\begin{refsegment}

The quantum accreditation protocol provides an upper bound on the variation distance between the probability distribution of the experimental outputs of a noisy quantum circuit and its noiseless counterparts~\cite{Ferracin2019,Ferracin2021}. 
It is therefore a measure for the correctness of the outputs of a given computation on the quantum computing hardware.

\subsection{Description}

To establish the upper bound within an accreditation protocol (AP), one considers the probability distribution of the experimental outputs of a noisy quantum circuit, 
$\{p_{x,\text{output}}\}$, and its noiseless counterpart, $\{p_{x,\text{ideal}} \}$. Here $x$
denotes the bitstrings that may be obtained as output. The quantum AP provides an upper bound, denoted as $b$, on the variation distance (VD)~\cite{Ferracin2019,Ferracin2021}, given by
\begin{equation}
\label{eq:boundVD}
\textrm{VD}:=\frac{1}{2}\sum_{x}\big|p_{x,\text{ideal}} - p_{x,\text{output}}\big|,
\end{equation}
so that
\begin{equation}
\textrm{VD} \leq b.
\end{equation}
The VD $\in [0,1]$, where the lower limit corresponds to a noiseless, ideal implementation of the quantum circuit. In general, the VD is the measure of the correctness of a noisy, error-prone computation relative to the noiseless, error-free ideal. However, measuring $\{p_{x,\text{output}}\}$ is exponentially expensive due to requiring exponentially many measurements, as is computing $\{p_{x,\text{ideal}}\}$ classically. 

The AP overcomes this challenge by estimating a bound $b$ on VD in an efficient manner.
The quantity $b$ is estimated experimentally with accuracy $\mu\in(0,1)$ and confidence $\eta\in(0,1)$ chosen by the user. Obtaining the upper bound $b$ does not require exponentially expensive classical simulation of quantum circuits. This is a significant advantage for larger systems when compared to typically considered volumetric benchmarking metrics (Sec. \ref{sec:volumetric_benchmarking}), such as for example the quantum volume (metric~\ref{sec:quantum_volume}) or algorithmic qubits (metric~\ref{sec:algorithmic_qubits}), where such exponentially scaling classical computations are needed.

The AP ascertains the correctness of a noisy quantum circuit rather than the performance of individual gates or gate sets. Thus, it includes all contributions to the overall noise obtained when the quantum circuit is executed. 
The AP is motivated by ideas of quantum verification, whose origins lie in theoretical computer science. Therein, the verifiability of quantum computations is related to the existence of an interactive proof system~\cite{gheorghiu2019verification}. A primary technique is that of blindness, whereby the computation to be verified called the target, and whose outputs are hard to obtain classically, is replaced by others, called traps, and whose outputs are easy to obtain classically. The traps have the same size and structure as the target, and the correctness of several traps is used to bound the correctness of the target.
In this metric the trap circuits are designed to be Clifford circuits.

To verify the correctness of a computation, some assumption on the trustworthiness of components of the computation must be made. Trust is thus a central notion in any verification, quantum or classical. 
In theoretical computer science studies, trust assumptions are made on the preparation or measurement of single qubits~\cite{gheorghiu2019verification}. 
Formally, the AP can be thought of as shifting the assumption to trusting single-qubit gates, in the sense it is assumed that the error in single qubit gates does  not depend strongly on the specific rotation angle applied to the qubit by the gate. This is motivated by the empirical observation that single-qubit gates are the components in quantum hardware least affected by noise when compared to multi-qubit gates.

\begin{figure}[htpb!]
     \centering
    \subfloat[][]{
        \begin{tikzpicture}[scale=1.0, every node/.style={scale=0.95}]
            
            \foreach \x in {1,...,6}
                \node at (-0.4,4.8-\x*0.8) {\footnotesize $\ket{0}_{\the\numexpr \x - 1 \relax}$};
            
            \foreach \x in {0,...,5}
                \draw (-0.0,4.0-\x*0.8) -- (3.8,4.0-\x*0.8);
            
            \foreach \x in {0,...,5}
                \draw [dashed] (3.8,4.0-\x*0.8) -- (4.8,4.0-\x*0.8);
            
            \foreach \x in {0,...,5}
                \draw (4.8,4.0-\x*0.8) -- (9.0,4.0-\x*0.8);
            
            \foreach \x in {0,...,5}
                \draw [fill=white] (0.25,4.0-\x*0.8-0.3) rectangle (1.05,4.0-\x*0.8+0.3);
            
            \foreach \x in {1,...,6}
                \node at (0.65,4.8-\x*0.8) {\scriptsize $U_{\the\numexpr \x - 1 \relax,1}$};
            
            \draw [fill=black] (1.45,0.0) circle [radius=0.07cm];
            \draw [fill=black] (1.45,1.6) circle [radius=0.07cm];
            \draw [fill=black] (1.45,2.4) circle [radius=0.07cm];
            \draw [fill=black] (1.45,3.2) circle [radius=0.07cm];
            
            \draw (1.45,0.0) -- (1.45,1.6);
            \draw (1.45,3.2) -- (1.45,2.4);

            \foreach \x in {0,...,5}
                \draw [fill=white] (0.25+1*1.6,4.0-\x*0.8-0.3) rectangle (1.05+1*1.6,4.0-\x*0.8+0.3);
            
            \node at (0.65+1*1.6,4.8-1*0.8) {\scriptsize $U_{0,2}$};
            \node at (0.65+1*1.6,4.8-2*0.8) {\scriptsize $U_{1,2}$};
            \node at (0.65+1*1.6,4.8-3*0.8) {\scriptsize $U_{2,2}$};
            \node at (0.65+1*1.6,4.8-4*0.8) {\scriptsize $U_{3,2}$};
            \node at (0.65+1*1.6,4.8-5*0.8) {\scriptsize $U_{4,2}$};
            \node at (0.65+1*1.6,4.8-6*0.8) {\scriptsize $U_{5,2}$};
            
            \draw [fill=black] (1.75+1*1.6,4.8-1*0.8) circle [radius=0.07cm];
            \draw [fill=black] (1.45+1*1.6,4.8-2*0.8) circle [radius=0.07cm];
            \draw [fill=black] (1.75+1*1.6,4.8-5*0.8) circle [radius=0.07cm];
            \draw [fill=black] (1.45+1*1.6,4.8-6*0.8) circle [radius=0.07cm];
            
            \draw (1.45+1*1.6,0.0) -- (1.45+1*1.6,3.2);
            \draw (1.75+1*1.6,4.0) -- (1.75+1*1.6,0.8);

            \foreach \x in {0,...,5}
                \draw [fill=white] (5.30,4.0-\x*0.8-0.3) rectangle (6.6,4.0-\x*0.8+0.3);
            
            \node at (5.4+0.55,4.8-1*0.8) {\scriptsize $U_{0,m-1}$};
            \node at (5.4+0.55,4.8-2*0.8) {\scriptsize $U_{1,m-1}$};
            \node at (5.4+0.55,4.8-3*0.8) {\scriptsize $U_{2,m-1}$};
            \node at (5.4+0.55,4.8-4*0.8) {\scriptsize $U_{3,m-1}$};
            \node at (5.4+0.55,4.8-5*0.8) {\scriptsize $U_{4,m-1}$};
            \node at (5.4+0.55,4.8-6*0.8) {\scriptsize $U_{5,m-1}$};
            
            \draw [fill=black] (6.95,4.8-3*0.8) circle [radius=0.07cm];
            \draw [fill=black] (6.95,4.8-4*0.8) circle [radius=0.07cm];
            
            \draw (6.95,1.6) -- (6.95,2.4);

            \foreach \x in {0,...,5}
                \draw [fill=white] (7.35,4.0-\x*0.8-0.3) rectangle (8.25,4.0-\x*0.8+0.3);
            
            \foreach \x in {1,...,6}
                \node at (7.4+0.4,4.8-\x*0.8) {\scriptsize $U_{\the\numexpr \x - 1 \relax,m}$};

            \foreach \x in {1,...,6}
                \draw (9.1,4.8-\x*0.8-0.05) -- (9.3,4.8-\x*0.8-0.05);
            \foreach \x in {1,...,6}
                \draw (9.1,4.8-\x*0.8+0.05) -- (9.3,4.8-\x*0.8+0.05);
            
            \foreach \x in {0,...,5}
                \draw [fill=white] (8.8,\x*0.8) circle [radius=0.3cm];
            \foreach \x in {1,...,6}
                \node at (8.8,4.8-\x*0.8) {\scriptsize $Z$};
        \end{tikzpicture}
        \label{fig:QAtarget}
    }
    \qquad
    \qquad
     \subfloat[][]
     {
        \begin{tikzpicture}[scale=1.0, every node/.style={scale=0.95}]

            \foreach \x in {1,...,6}
            \node at (-0.3,4.8-\x*0.8) {\footnotesize $\ket{0}_{\the\numexpr \x - 1 \relax}$};
            
            \foreach \x in {0,...,5}
            \draw (0.1,4.0-\x*0.8) -- (5.5,4.0-\x*0.8);
            
            \foreach \x in {0,...,5}
            \draw [dashed] (5.5,4.0-\x*0.8) -- (6.5,4.0-\x*0.8);
            
            \foreach \x in {0,...,5}
            \draw (11,4.0-\x*0.8) -- (6.5,4.0-\x*0.8);

            \foreach \x in {0,...,5}
            \draw [fill=white] (0.4,4.0-\x*0.8-0.3) rectangle (1.1,4.0-\x*0.8+0.3);
            
            \foreach \x in {1,...,6}
            \node at (0.75,4.8-\x*0.8) {\scriptsize $H^t$};
            
            \foreach \x in {0,...,5}
            \draw [fill=white] (1.3,4.0-\x*0.8-0.3) rectangle (2.0,4.0-\x*0.8+0.3);
            
            \node at (1.65,4.8-1*0.8) {\scriptsize $S$};
            \node at (1.65,4.8-2*0.8) {\scriptsize $S$};
            \node at (1.65,4.8-3*0.8) {\scriptsize $H$};
            \node at (1.65,4.8-4*0.8) {\scriptsize $S$};
            \node at (1.65,4.8-5*0.8) {\scriptsize $S$};
            \node at (1.65,4.8-6*0.8) {\scriptsize $H$};
            
            \draw [fill=black] (2.45,4.8-2*0.8) circle [radius=0.07cm];
            \draw [fill=black] (2.45,4.8-3*0.8) circle [radius=0.07cm];
            \draw [fill=black] (2.45,4.8-4*0.8) circle [radius=0.07cm];
            \draw [fill=black] (2.45,4.8-6*0.8) circle [radius=0.07cm];
            
            \draw (2.45,4.8-2*0.8) -- (2.45,4.8-3*0.8);
            \draw (2.45,4.8-4*0.8) -- (2.45,4.8-6*0.8);

            \foreach \x in {0,...,5}
            \draw [fill=white] (2.8,4.0-\x*0.8-0.3) rectangle (3.6,4.0-\x*0.8+0.3);
            
            \node at (3.2,4.8-1*0.8) {\scriptsize $S^\dagger$};
            \node at (3.2,4.8-2*0.8) {\scriptsize $S^\dagger$};
            \node at (3.2,4.8-3*0.8) {\scriptsize $H$};
            \node at (3.2,4.8-4*0.8) {\scriptsize $S^\dagger$};
            \node at (3.2,4.8-5*0.8) {\scriptsize $S^\dagger$};
            \node at (3.2,4.8-6*0.8) {\scriptsize $H$};
            
            \foreach \x in {0,...,5}
            \draw [fill=white] (3.8,4.0-\x*0.8-0.3) rectangle (4.6,4.0-\x*0.8+0.3);
            
            \node at (4.2,4.8-1*0.8) {\scriptsize $H$};
            \node at (4.2,4.8-2*0.8) {\scriptsize $S$};
            \node at (4.2,4.8-3*0.8) {\scriptsize $S$};
            \node at (4.2,4.8-4*0.8) {\scriptsize $S$};
            \node at (4.2,4.8-5*0.8) {\scriptsize $S$};
            \node at (4.2,4.8-6*0.8) {\scriptsize $H$};

            \draw [fill=black] (5.25,4.8-1*0.8) circle [radius=0.07cm];
            \draw [fill=black] (4.9+.15,4.8-2*0.8) circle [radius=0.07cm];
            \draw [fill=black] (5.25,4.8-5*0.8) circle [radius=0.07cm];
            \draw [fill=black] (4.9+.15,4.8-6*0.8) circle [radius=0.07cm];
            
            \draw (4.9+.15,0.0) -- (4.9+.15,3.2);
            \draw (5.25,4.0) -- (5.25,0.8);

            \foreach \x in {0,...,5}
            \draw [fill=white] (6.9,4.0-\x*0.8-0.3) rectangle (7.7,4.0-\x*0.8+0.3);
            
            \node at (7.3,4.8-1*0.8) {\scriptsize $H$};
            \node at (7.3,4.8-2*0.8) {\scriptsize $S$};
            \node at (7.3,4.8-3*0.8) {\scriptsize $S$};
            \node at (7.3,4.8-4*0.8) {\scriptsize $H$};
            \node at (7.3,4.8-5*0.8) {\scriptsize $H$};
            \node at (7.3,4.8-6*0.8) {\scriptsize $S$};
            
            \draw [fill=black] (8+.15,4.8-3*0.8) circle [radius=0.07cm];
            \draw [fill=black] (8+.15,4.8-4*0.8) circle [radius=0.07cm];
            
            \draw (8+.15,1.6) -- (8+.15,2.4);

            \foreach \x in {0,...,5}
            \draw [fill=white] (8.6,4.0-\x*0.8-0.3) rectangle (9.4,4.0-\x*0.8+0.3);
            
            \node at (9,4.8-1*0.8) {\scriptsize $H$};
            \node at (9,4.8-2*0.8) {\scriptsize $S^\dagger$};
            \node at (9,4.8-3*0.8) {\scriptsize $S^\dagger$};
            \node at (9,4.8-4*0.8) {\scriptsize $H$};
            \node at (9,4.8-5*0.8) {\scriptsize $H$};
            \node at (9,4.8-6*0.8) {\scriptsize $S^\dagger$};
            
            \foreach \x in {0,...,5}
            \draw [fill=white] (9.6,4.0-\x*0.8-0.3) rectangle (10.4,4.0-\x*0.8+0.3);
            
            \foreach \x in {0,...,5}
            \node at (10,4.0-\x*0.8) {\scriptsize $H^t$};
            
            \foreach \x in {1,...,6}
            \draw (9.1+2.2,4.8-\x*0.8-0.05) -- (9.3+2.2,4.8-\x*0.8-0.05);
            \foreach \x in {1,...,6}
            \draw (9.1+2.2,4.8-\x*0.8+0.05) -- (9.3+2.2,4.8-\x*0.8+0.05);
            
            \foreach \x in {0,...,5}
            \draw [fill=white] (11,\x*0.8) circle [radius=0.3cm];
            \foreach \x in {1,...,6}
            \node at (9.8+1.2,4.8-\x*0.8) {\scriptsize $Z$};
            
        \end{tikzpicture}
        \label{fig:QAtrap}
     
}
     \caption{(a) Example of a target circuit used in the quantum accreditation protocol. The target circuit must be compiled into $m$ cycles of one-qubit gates $U_{i,j}$ (on qubit $i$ and cycle $j$). Each cycle of $U_{i,j}$, apart from the last cycle, is followed by a cycle of controlled-$Z$ ($\mathrm{C}_Z$) gates, denoted as vertically connected dots. This gives a circuit depth $d_c=2m-1$. The qubits are initialized in the state $\ket{0}$ and measurements are in the computational basis. 
     Subfigure (b) shows an example of a trap circuit for the target circuit in (a). The trap circuit is obtained by replacing the one-qubit gates in the target circuit with one-qubit Clifford gates, such as the $H$ gate and the $S$ gate (see Sec.~\ref{sec:methodsmaths}). Neighboring cycles of one-qubit gates can be recompiled into a single cycle. Thus, the trap circuit has the same circuit depth as the target. }
\end{figure}
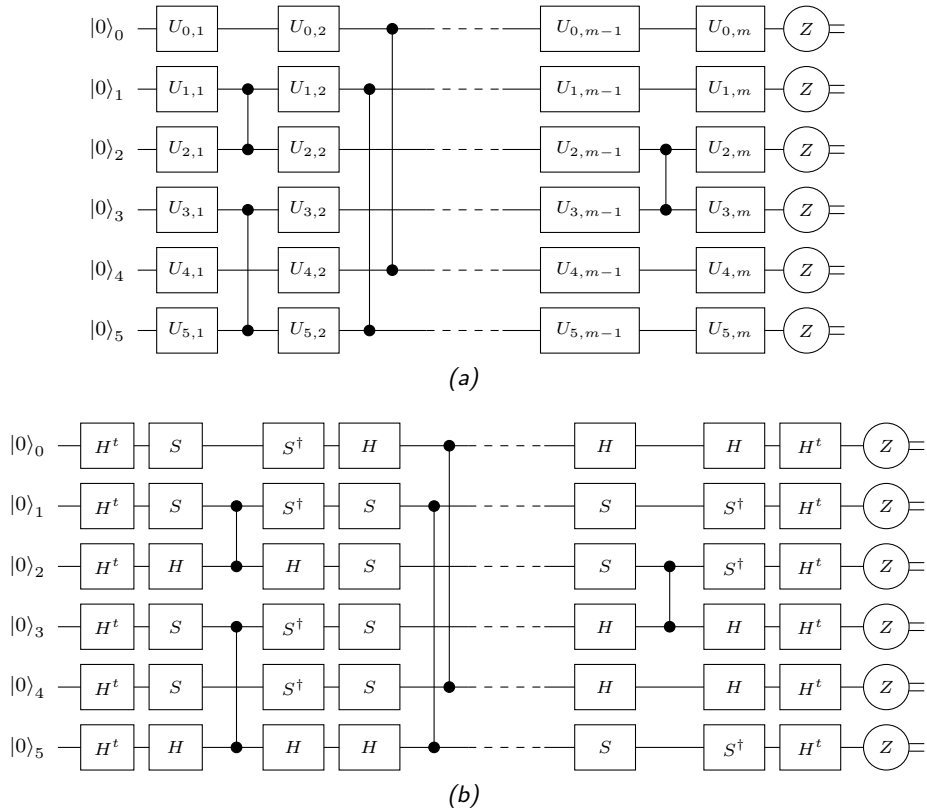

The AP~\cite{Ferracin2019,Ferracin2021} 
uses trap circuits to obtain the bound $b.$ The AP takes as input a target circuit, and two numbers, $\mu$ and $\eta\in(0,1)$, which quantify the desired accuracy and confidence on the final bound, respectively. The target circuit must:
\begin{enumerate}
    \item take as input~$N_{\mathrm{q}}$ qubits in the state $\ket{0}^{\otimes N_{\mathrm{q}}}$,
    \item contain $2m$ cycles alternating between a cycle of one-qubit gates and a cycle of two-qubit gates, which can be efficiently chosen to be $\mathrm{C}_Z$ or another two-qubit Clifford gate (see Fig.~\ref{fig:QAtarget}),
    \item end with measurements in the Pauli-$Z$ basis (see Fig.~\ref{fig:QAtarget}).
\end{enumerate} 
The trap-based AP requires executing $v$ trap circuits sequentially, where $v=\lceil2\,\textrm{ln}(2/(1-\eta))/\mu^2\rceil$, and $\lceil\cdot\rceil$ is the ceiling function. Given a target circuit, the trap circuits are obtained as in Fig.~\ref{fig:QAtrap}.

\subsection{Measurement procedure}

\begin{enumerate}
    \item Select a target circuit following the restrictions outlined in the previous section. Select two numbers, $\mu$ and $\eta\in(0,1)$, which quantify the desired accuracy and confidence on the final bound.
    \item Generate $v$ trap circuits, where $v=\lceil2\textrm{ln}(2/(1-\eta))/\mu^2\rceil$, and $\lceil\cdot\rceil$ is the ceiling function. Each trap circuit is obtained by replacing the single qubit gates in the target circuit with random one-qubit Clifford gates as follows (see Fig.~\ref{fig:QAtrap}):
\begin{enumerate}
\item For all $j\in\{1,\ldots,m-1\}$ and for all $i\in\{0,\ldots,n-1\}$:
    \begin{itemize}
        \item  If the $j$-th cycle of $\mathrm{C}_Z$ (controlled-Z) gates connects qubit $i$ to qubit~$i'$, randomly replace $U_{i,j}$ with an $S$ (phase) gate, and $U_{i^\prime,j}$ with an $H$ (Hadamard) gate, or $U_{i,j}$ with $H$ and $U_{i',j}$ with $S$.  For definitions of the Hadamard and phase gates, see section~\ref{sec:methodsmaths}.
        After the $j$-th cycle of $\mathrm{C}_Z$ gates, insert a $S^\dagger$ gate if $S$ has replaced $U_{i,j}$, or insert another $H$ if $H$ has replaced $U_{i,j}$. This is referred as undoing the $H$ and $S$ gates.
        \item If the $j$-th cycle of $\mathrm{C}_Z$ gates does not connect qubit~$i$ to any other qubit, randomly replace $U_{i,j}$ with $H$ or $S$. Undo this gate after the cycle of $\mathrm{C}_Z$ gates.
\end{itemize}
\item Initialize a random bit $t\in\{0,1\}$. If $t=0$, do nothing. If $t=1$, append a Hadamard gate on every qubit at the beginning and at the end of the circuit. 
\end{enumerate}

\item Append a cycle of random Pauli gates after every cycle of one-qubit gates, followed by a second cycle of Pauli gates before the next cycle of one-qubit gates that undoes the first. 

\item Recompile neighbouring cycles of one-qubit gates into a single cycle. 

\item Implement all the circuits on hardware, and count the number $N_\textrm{inc}$ of traps that do not return the output bitstring $(0,0, \cdots, 0).$ 

\item Set $b= 2 N_\textrm{inc}/v.$
\end{enumerate}

\subsection{Assumptions and limitations}

\begin{enumerate}
    \item The trap-based AP requires that all two-qubit gates in the target circuit are Clifford gates.
    \item The noise in state preparation, measurements and cycles can be modelled by completely-positive, trace-preserving (CPTP) maps acting on all the qubits (see section~\ref{sec:noise_effects}).
    \item An implementation of a cycle of gates $\mathcal{G}$ on a state $\rho$ at circuit depth $d_c$ returns $\Phi_{\mathcal{G},j}\mathcal{G}(\rho)$, where $\Phi_{\mathcal{G},j}$ is a CPTP map that potentially acts on the whole system and depends on both $\mathcal{G}$ and the depth~$d_c$. This is a Markovian noise model that encompasses a broad class of noise processes afflicting current platforms, such as gate-dependent noise and crosstalk between neighbouring qubits.
    \item Cycles of one-qubit gates suffer gate-independent noise, i.e. $\Phi_{\mathcal{G}_\mathrm{1q},j}=\Phi_j$ for all the cycles of one-qubit gates $\mathcal{G}_\mathrm{1q}$. However, the AP is robust to noise that depends weakly on the cycles of one-qubit gates~\cite[Appendix 2]{Ferracin2021}.
\end{enumerate}

An alternate trap-based~\cite{Ferracin2019} AP can detect even more complex noise, such as temporally correlated qubit-environment couplings, albeit at the cost of looser bounds on the VD~\cite[Appendix 3]{Ferracin2021}.

\subsection{Source code}

A tutorial for calculating the upper bound on the VD using the AP protocol is provided in \sourceurl{circuit_execution_quality_metrics/upper_bound_on_the_variation_distance}.

\metricbibliography

\end{refsegment}

\def\thechapter{M5}
\chapter{Well-studied tasks execution quality metrics}
\label{chapter:well_studied_tasks}

\section{Variational quantum eigensolver (VQE) metric}
\label{sec:vqe}
\begin{refsegment}
This metric evaluates how well a quantum computer can execute the variational quantum eigensolver (VQE) algorithm~\cite{TILLY20221}, which is a type of variational quantum algorithm (VQA)~\cite{RevModPhys.94.015004}. VQAs use parameterized quantum circuits to prepare wave functions on the quantum computer, and use a classical optimizer to find the state that minimizes a cost function. The parameters of the quantum circuit typically correspond to the parameters of single-qubit and two-qubit gates, and the cost function can be the energy in materials or quantum chemistry simulations. VQAs have some level of resilience to noise~\cite{fontana2021}, and are hence often used for demonstrations on noisy intermediate scale quantum (NISQ) hardware~\cite{TILLY20221}. The optimization of circuit parameters can become difficult as system sizes increase, where the effect of noise also becomes significant~\cite{TILLY20221}.

The VQE is a widely used VQA, which has been applied extensively to find the ground state of electronic structure problems. VQE has two components: the first component consists of the use of a quantum computer to obtain the energy expectation value of a given trial wave function for a specified Hamiltonian, while the second component is an optimization loop on a classical computer, where the parameters of the quantum circuit are adjusted to minimize this energy. 

\subsection{Description}

For an extensive review of VQE see Ref.~\cite{TILLY20221}. In this section, in order to distinguish between operators and scalar values, hats on symbols are used to denote operators. For a quantum system of interest with a Hamiltonian, $\hat{H}$, a common task is to find a wave function that minimizes the energy of the quantum system. The VQE solves this task with a parameterized quantum circuit that produces a trial wave function, $\ket{\Psi(\boldsymbol{\theta})}$, on the quantum computer, where $\boldsymbol{\theta}$ is a vector of parameters. By minimizing the expectation value of $\hat{H}$, given by
\begin{equation}
    E = \expval{\hat{H}}{\Psi(\boldsymbol{\theta})},
    \label{eq:enevqe}
\end{equation}
 with respect to the parameters $\boldsymbol{\theta}$, the energy is minimized. Thus, the two components of VQE include the method to calculate $E$, and the method for finding the optimal values for $\boldsymbol{\theta}$, such that the optimized $\ket{\Psi(\boldsymbol{\theta})}$ minimizes $E$.
 
To calculate $E$ with a quantum computer, $\hat{H}$ needs to be rewritten as a sum of Pauli strings that can be measured on quantum computers, $\hat{H}=\sum_i w_i \hat{P}_i$, where the range of terms in the sum over $i$ depends on the Hamiltonian, $w_i$ are weights and $\hat{P}_i$ are Pauli strings. Calculating $E$ for a given trial wave function can then be achieved by measuring the expectation value for each of the Pauli strings separately, and then summing the results weighted by $w_i$. The number of shots, with which the Pauli strings are measured, directly impacts the accuracy of the obtained $E$.

The structure of the parameterized quantum circuit is denoted as the ansatz. The circuit ansatz determines how the trial wave function $\ket{\Psi(\boldsymbol{\theta})}$ is constructed on a quantum computer. When choosing an ansatz, key aspects to consider are the expressibility, trainability and scalability. The expressibility indicates whether an ansatz can generate a large or only a restricted class of wave functions in the Hilbert space. Trainability indicates how easy it is to optimize the parameters $\boldsymbol{\theta}$ with classical optimization techniques. As the expressibility increases, it may generally be more difficult to train the ansatz, hence a trade-off between expressibility and trainability often needs to be made. Scalability indicates how the depth of the circuit, as well as trainability, changes for an increasing size of the system, typically related to the number of qubits. If an ansatz performs well for small systems, but has poor scalability, it may not be practically useful for large systems.

Finding the optimal $\boldsymbol{\theta}$ is achieved using a classical optimizer that iteratively updates $\boldsymbol{\theta}$ based on the value of $E$ until it converges. The final solution to the wave-function-finding problem is then given by $\ket{\Psi(\boldsymbol{\theta}_{\text{optimal}})}$, produced by the parameterized circuit with the optimal parameters. The choice of optimizer is important, as it strongly influences the speed of convergence and whether the final state is optimal. Both these aspects are relevant metrics for the performance of a VQE run. However, the aim of the metric in this section is to benchmark the quality of the quantum processor rather than that of the classical optimizer. The quality of the quantum processor determines how accurately the energy in Eq.~\ref{eq:enevqe} is computed for a given trial wave function. Therefore, the output of a quantum hardware metric based on the VQE well-studied task requires the evaluation of the fidelity of the computed energy for one or more wave functions and Hamiltonians, without optimizing the parameters. The chosen Hamiltonians needs to be representative of the classes of Hamiltonians appearing in quantum chemistry and materials science problems, and the chosen fixed wave functions for a given Hamiltonian need to be representative for the types of wave functions appearing during the optimization loop.

\subsection{Measurement procedure}
VQE performance depends on the choices of the different components outlined in the previous metric~\cite{TILLY20221}. Here a measurement procedure to evaluate the accuracy of the computed energy for a representative system is presented.

The target system chosen in this metric is the Hubbard model~\cite{doi:10.1098/rspa.1963.0204}. Specifically, the 1D Fermi-Hubbard model with $N_s$ sites is used and is the same as described in detail in metric~\ref{sec:hubbard_model}. 
Given $2N_s$ distinct fermionic sites, where there are $N_s$ physical sites each with 2 spin degrees of freedom, the Hamiltonian is given by
\begin{equation}
\label{eq:hubbard_hamiltonian_vqe}
\hat{H} = -\gamma \sum_{i=1}^{N_s-1}\sum_{\sigma \in \{\uparrow,\downarrow\}} \left(\hat{c}_{i\sigma}^\dagger\hat{c}_{i+1\sigma} +\hat{c}_{i+1\sigma}^\dagger\hat{c}_{i\sigma} \right) + U_H \sum_{i=1}^{N_s} \hat{n}_{i\uparrow}\hat{n}_{i\downarrow},
\end{equation}
where $\hat{c}^\dagger_{i\sigma}$ ($\hat{c}_{i\sigma}$) is the creation (annihilation) operator for a Fermion on site $i$ with spin $\sigma$, $\hat{n}_{i\sigma}$ is the corresponding number operator, $\gamma$ is the hopping integral, and $U_H$ is the on-site electron-electron interaction strength. 

This Hamiltonian maps to Pauli operators using the Jordan-Wigner transformation~\cite{Nielsen2005TheFC}. In order to map the system onto qubits, it is first necessary to specify the ordering of the spin sites. This methodology uses the following ordering of site and spin indices into an up-and-down chain
\begin{equation}
    \{1\uparrow, 2\uparrow, \cdots, N_s \uparrow, 1\downarrow, 2\downarrow, \cdots, N_s\downarrow \}\rightarrow \{1,2,\cdots, N_s, N_s+1, N_s+2, \cdots, 2N_s\}.
\end{equation}With this ordering and the use of the Jordan-Wigner mapping, the Hamiltonian maps onto a $2N_s$ qubit Hamiltonian that can be expressed as
\begin{align}
\hat{H}_{\mathrm{q}}= -\frac{\gamma}{2} \sum_{\substack{i=1 \\ i\neq N_s}}^{2N_s-1} \left(\hat{\sigma}_{i}^x \hat{\sigma}_{i+1}^x + \hat{\sigma}_{i}^y \hat{\sigma}_{i+1}^y\right)+ \frac{U_H}{4} \sum_{i=1}^{N_s}\hat{\sigma}_{i}^z\hat{\sigma}_{i+N_s}^z - \frac{U_H}{4}\sum_{i=1}^{2N_s} \hat{\sigma}_i^z + \frac{N_sU_H}{4},
\label{eq_vqe_hubbard_qubit_h}
\end{align}
where $\hat{\sigma}_i^\alpha$ (for $\alpha=x,y,z$) is the Pauli-$\alpha$ operator acting on qubit $i$. Here, the ratio of $U_H / \gamma$ is what determines the regime of the system, such as whether a metallic or insulating behaviour is described. As a result, $\gamma$ is set to 1 so the ratio is determined solely by the value of $U_H$.

Next, the choice of ansatz for the VQE is taken to be the Hamiltonian variational ansatz (HVA)~\cite{PhysRevA.92.042303, Reiner_2019}. The ansatz consists of parameterized gates which form a layer of circuit, and the circuit layer can be repeated $m$ times with each repetition having different parameters. Increasing the number of layers $m$ of the ansatz improves the expressibility of the quantum circuit~\cite{PhysRevResearch.4.023190}. In order to apply the HVA, an initial state, $\ket{\Psi_0}$, needs to be prepared. The state $\ket{\Psi_0}$ is chosen such that it is the ground state of the non-interacting problem where $U_H = 0$, which can be computed efficiently classically~\cite{PhysRevApplied.9.044036}. 

The following measurement procedure largely follows Ref.~\cite{PhysRevResearch.4.023190}. However, since this is a metric for a quantum computer, the classical optimization part of the VQE has been removed.
The measurement procedure then is as follows:
\begin{enumerate}
    \item Select the number of sites $N_s$, and choose the onsite energy $U_H$ for the Fermi-Hubbard model. Then convert the Hamiltonian into a weighted sum of Pauli strings using the Jordan-Wigner transformation following Eq.~\ref{eq_vqe_hubbard_qubit_h}. The number of qubits $N_{\mathrm{q}} = 2N_s$. 
    \item Classically compute $\ket{\Psi_0}$, the ground state when $U_H=0$ for the Hamiltonian.
    \item Select the number of layers of the Hamiltonian variational ansatz $m$ to apply , and choose the number of trials, $n_{\mathrm{trials}}$, for which to calculate the metric.
    \item Then, for each trial $i\in\{1,2,\cdots,n_{\mathrm{trials}}\}$:
        \begin{enumerate}
            \item Prepare the qubits in the initial state $\ket{\Psi_0}$.
            \item Randomly select the angles, $\boldsymbol{\theta}$, for the parameterized gates in all of the $m$ layers of the HVA, and then apply the $m$ layers of the HVA to the qubits with the selected angles.
            \item Compute the numerically exact energy for the wave function given by this ansatz using a classical computer, denoted as $E^{(i)}_{\text{exact}}$. 
            \item Compute the energy on the quantum computer being tested. For each Pauli string from step 1, measure the qubits with a chosen number of shots, then compute the energy using the weighted sum, denoted as $E^{(i)}$. The number of shots should be large enough so that the uncertainty in $E^{(i)}$ is significantly smaller than its difference to $E^{(i)}_{\text{exact}}$.
        \end{enumerate}
    \item Calculate $E_{\mathrm{diff}}$, the average of the difference between the quantum computer estimated energy and exact energy per-site for all of the trials 
    \begin{equation}
        E_{\mathrm{diff}} = \frac{1}{n_{\mathrm{trials}}N_s}\sum_{i=1}^{n_{\mathrm{trials}}} \lvert E^{(i)}  -E^{(i)}_{\mathrm{exact}}\rvert.
    \end{equation}
    This difference is the metric for the quality of the quantum circuit execution on the quantum computer within the VQE algorithm. We note that as  alternative or additional component of the metric one may also compute the fidelity between the wave functions obtained on the quantum computer and the numerically exact classical result, which is the metric used for the Fermi-Hubbard model simulation in metric \ref{sec:hubbard_model}. However, since the quality of the VQE optimization loop depends mainly on the accuracy of the energy, we use the energy difference as representative metric rather than the fidelity.
   
\end{enumerate}
\subsection{Assumptions and limitations}
\begin{itemize}
    \item The VQE performance depends significantly on the type of Hamiltonian, on the choice of ansatz and on the number of qubits. Hence, evaluating the VQE well-studied task for a specific system and size may not be representative of the general hardware performance. For example, the obtained performance can change significantly for increasing number of qubits~\cite{TILLY20221}. 
    \item In this metric only the quality of the energy obtained for a trial wave function is considered. The choice of classical optimizer also affects the overall performance and cost of the full VQE algorithm~\cite{TILLY20221}.
\end{itemize}
    
\subsection{Source code}
A tutorial for measuring the ability of the quantum computer to execute VQE circuits is provided in \sourceurl{well_studied_task_execution_quality_metrics/vqe}, with associated source code.
\metricbibliography
\end{refsegment}

\section{Quantum approximate optimization algorithm (QAOA) metric\label{sec:Q-Score}}
\begin{refsegment}

This metric evaluates how well a quantum computer can execute the quantum approximate optimization algorithm (QAOA) by using it to solve the so called MaxCut optimization problem. MaxCut requires one to assign the vertices in a given graph into two sets, such that the number of edges between the two sets is maximized. It is a well defined fundamental problem in combinatorial optimization~\cite{10.1145/502090.502098}. QAOA is a type of variational quantum algorithm (VQA) for solving optimization tasks by re-expressing the problem as a Hamiltonian and then finding its ground state, which corresponds to the solution of the problem~\cite{farhi2014quantum}. Note that QAOA can also be applied to other optimization problems~\cite{farhi2014quantum, PhysRevX.10.021067}.

Solving MaxCut for random graphs is NP-hard for classical computers~\cite{10.1145/502090.502098}. There are classical approximation methods that can find solutions that are close to the optimal solution up to an approximation ratio~\cite{ahn2009graph}. MaxCut can be re-expressed as an Ising Hamiltonian~\cite{RevModPhys.94.015004}, so that it can be solved on a quantum computer using QAOA. One particular implementation of a QAOA based metric to solve the MaxCut problem is the so called Q-Score~\cite{martielBenchmarkingQuantumCoprocessors2021}. It specifies the specific parameters to be used within the QAOA, and is the one that is described in the measurement procedure section of this metric~\cite{martielBenchmarkingQuantumCoprocessors2021}.

\subsection{Description}

Given a graph $G(V_G,E_G)$, where $V_G$ is the set of vertices and $E_G$ is the set of edges with each edge connecting a pair of vertices, the MaxCut problem is to partition the vertices into two sets, $A$ and $B$, such that the number of edges between $A$ and $B$ is maximized~\cite{farhi2014quantum}. 

This problem can be solved on quantum computers using QAOA, outlined in what follows~\cite{RevModPhys.94.015004, PhysRevX.10.021067}. Given a graph $G(V_G,E_G)$, define the Hamiltonians $H_0=-\sum_{1\leq i\leq n} X_i$ and $H_G=\sum_{(i,j)\in E_G}Z_i\otimes Z_j-\frac{|E_G|}{2}$, where $n=|V_G|$ is the size of $V_G$, and $X,Z$ are Pauli operators. Then implement the following unitary as a parameterized circuit on $n$ qubits: 
\begin{equation}
    U(\boldsymbol{\gamma}, \boldsymbol{\beta})=\prod_{1\leq q\leq l}e^{-i\beta_qH_0/2}e^{-i\gamma_q H_G/2},
\end{equation}
where $l$ is a parameter determining the number of circuit layers, also known as the depth parameter, $q$ is the index of the layer, and $\boldsymbol{\gamma}$ and $\boldsymbol{\beta}$ are two vectors of parameters with size $Q$. In each layer,
for each edge $(i, j)\in E_G$, the $e^{\gamma Z_i\otimes Z_j}$ terms are decomposed to two CNOT gates on qubits $(i,j)$ and a single $R_z$ rotation on qubit $j$. Then the term involving the $H_0$ is implemented with $R_x$ gates on each qubit. 

The cost function is defined as $\mathcal{L}(\boldsymbol{\gamma},\boldsymbol{\beta})=-\bra{0}U^\dagger(\boldsymbol{\gamma},\boldsymbol{\beta}) H_G U(\boldsymbol{\gamma}, \boldsymbol{\beta}) \ket{0}$.
Using a classical optimizer and the cost computed above, QAOA seeks to find the parameters $\boldsymbol{\gamma},\boldsymbol{\beta}$ that minimize the cost $\mathcal{L}(\boldsymbol{\gamma},\boldsymbol{\beta})$. For current NISQ devices, the depth parameter $l$ is restricted to be 1 or 2, since for higher depths the noise induced errors become too large.

After the parameters are optimized, the solution to MaxCut is determined by the bitstring that has the highest probability from measuring the circuit. For each digit in the bitstring, if the digit is 0, then the corresponding vertex in the graph is assigned to the set $A$, otherwise if the digit is 1, then it is assigned to $B$.

\subsection{Measurement procedure}
The measurement procedure presented here for benchmarking QAOA on a quantum computer follows from Ref.~\cite{martielBenchmarkingQuantumCoprocessors2021}, where a number of fixed choices for the used parameters are made. For example, this includes the choice of the number of shots, and of the threshold above which a computation is considered successful. The metric is named Q-Score, defined as the largest size of graph for which the hardware can solve the MaxCut problem using QAOA sufficiently accurately on average. The methodology then is as follows:
\begin{enumerate}
    \item Select the size of the problem, $N_{\mathrm{q}}$.
    \item Generate a random graph $G(V_G,E_G)$ with $N_{\mathrm{q}}$ vertices in the following way. For each possible edge between the vertices, randomly decide whether it should be included in $E_G$ with a probability of $1/2$.
    \item Select $N_{\mathrm{q}}$ qubits from the hardware that perform best and that have good connectivity, and initialize them into $\ket{0}^{\otimes N_{\mathrm{q}}}$.
    \item Select a depth $l$, usually 1 or 2 for current NISQ devices. Randomly initialize the parameters $\boldsymbol{\gamma}$ and $\boldsymbol{\beta}$, and apply the parameterized circuit $U(\boldsymbol{\gamma},\boldsymbol{\beta})$ for $l$ layers of QAOA, defined in the previous section. 
    \item Using a fixed number of shots, set to 2048, compute the cost $\mathcal{L}(\boldsymbol{\gamma},\boldsymbol{\beta})$. 
    \item Using the COBYLA optimizer~\cite{Powell1994}, find the values $(\boldsymbol{\gamma},\boldsymbol{\beta})$ that minimize the cost.
    \item Repeat steps 2-6 for 100 different graphs. The average cost of all graphs defines the value $\mathcal{L}(N_{\mathrm{q}})$, the average cost that the device can achieve. Then compute the ratio $\beta(N_{\mathrm{q}}):=\frac{\mathcal{L}(N_{\mathrm{q}})-N_{\mathrm{q}}^2/8}{\lambda N_{\mathrm{q}}^{3/2}}$, where $\lambda$ is taken to be 0.178. This number is obtained from the costs achieved by a classical exact solver~\cite{martielBenchmarkingQuantumCoprocessors2021}, so that optimal solutions give $\beta(N_{\mathrm{q}})=1$.
    \item Repeat steps 1-7 for different $N_{\mathrm{q}}$, and find the largest $N_{\mathrm{q}}$ that the ratio $\beta(n)$ exceeds the value $\beta^*=0.2$. The value of the largest $N_{\mathrm{q}}$ is the Q-score of the device. 
\end{enumerate}

\subsection{Assumptions and limitations}

\begin{itemize}
    \item A number of choices are made within Q-Score, which, while justified, are somewhat arbitrary. This includes choice of the depth parameter $l$, the choice of optimizer (COBYLA), the choice of class of graphs (random graphs with edge probability $1/2$), the number of different graphs for each $N_{\mathrm{q}}$ (100 graphs), the value of the ratio that is considered success ($\beta^*=0.2$). 
    \item Because of those choices, it is not straight forward to generalize this score for different types of hardware. In Ref.~\cite{q-score_annealing}, a generalization that applies to quantum annealers is introduced. It is conceivable that a digital quantum device could perform better for the same task using a different algorithm, or supplemented with quantum error mitigation techniques, hence a comparison with quantum annealing must be performed with care.
    \item For a small number of vertices corresponding to a small number of qubits, the $N_{\mathrm{q}}^2/8$ term in the calculation of $\beta(N_{\mathrm{q}})$ no longer corresponds to the score that would be achieved with a random solution. For example, the problem is ill-defined when there is no edge between any of the vertices in the graph, hence no cuts can be made and the MaxCut problem is not valid. In these cases, the random solution cannot achieve the expected average score of $N_{\mathrm{q}}^2/8$. This problem is particularly significant for a small number of vertices, where there is a greater chance that a random graph contains no edges. Therefore the Q-Score test is more likely to fail in these cases. This problem is most apparent for $N_{\mathrm{q}}=2$, where it is impossible to pass the Q-Score test because the pass criteria $N_{\mathrm{q}}^2/8= 0.5$ is too high. In Ref.~\cite{martielBenchmarkingQuantumCoprocessors2021} the authors evaluate the results from a problem instance with $N_{\mathrm{q}}\geq 5$.
\end{itemize}

\subsection{Source Code}
A tutorial to evaluate the Q-Score metric is provided in: \sourceurl{well_studied_task_execution_quality_metrics/qscore}.

\metricbibliography
\end{refsegment}

\section{Fermi-Hubbard model simulation (FHMS) metric} 
\label{sec:hubbard_model}
\begin{refsegment}

This metric evaluates how well a quantum computer can perform Fermi-Hubbard model simulations (FHMS). The Fermi-Hubbard model is a prototype many-body Hamiltonian for materials science simulations~\cite{hubbard1963electron,qin2022hubbard,Georges1996dynamicalmeanfieldtheory,Kotliar2006electronicstructure,paul2019applications,jamet2021krylov,jamet2022quantum,jamet2023anderson}. Depending on the quantities of interest and the method used,
the cost of the classical computation scales exponentially with the number of orbitals in the Hamiltonian, while that for quantum computation can scale polynomially. It is thus a potential testbed for demonstrating quantum advantage. A number of quantum methods have been proposed, such as those based on time-evolution of the wavefunction, which requires long quantum circuits and hence long coherence times, or  VQAs, as demonstrated on existing hardware~\cite{montanaro2020, stanisic2022, arute2020observation}. This metric discusses benchmarking methods of the former.

\subsection{Description}
In this section, in order to distinguish between operators and scalar values, hats on symbols are used to denote operators. The 1D Fermi-Hubbard model Hamiltonian is given by 
\begin{equation}
\label{eq:hubbard_hamiltonian_fhms}
\hat{H} = -\gamma \sum_{i=1}^{N_s-1}\sum_{\sigma \in \{\uparrow,\downarrow\}} \left(\hat{c}_{i\sigma}^\dagger\hat{c}_{i+1\sigma} +\hat{c}_{i+1\sigma}^\dagger\hat{c}_{i\sigma} \right) + U_H \sum_{i=1}^{N_s} \hat{n}_{i\uparrow}\hat{n}_{i\downarrow},
\end{equation}
where $\hat{c}^\dagger_{i\sigma}$ ($\hat{c}_{i\sigma}$) is the creation (annihilation) operator for a Fermion in site $i$ with spin $\sigma$, $\hat{n}_{i\sigma}$ is the corresponding number operator, $\gamma$ is the hopping integral, and $U_H$ is the on-site electron-electron interaction strength. Typically $\gamma$ is taken to be positive, whereas $U_H$ can be positive (repulsive interactions) or negative (attractive interactions).

In order to obtain a quantum circuit representation of the evolution operator for the Fermi-Hubbard model, it is first necessary to map the fermionic system onto a system of qubits. For $N_s$ physical sites each with 2 spin sites, there are a total of $2N_s$ distinct fermionic sites. Using a Jordan-Wigner mapping, this can be mapped onto a system of $2N_s$ qubits. In order to do so it is first necessary to specify the ordering of the spin sites to qubit index, for example
$$\{1\uparrow, 1\downarrow, 2\downarrow, 2\uparrow, \dots, N_s\uparrow, N_s\downarrow\} \rightarrow \{1, 2, 3, 4, \dots, 2N_s-1, 2N_s\}.$$

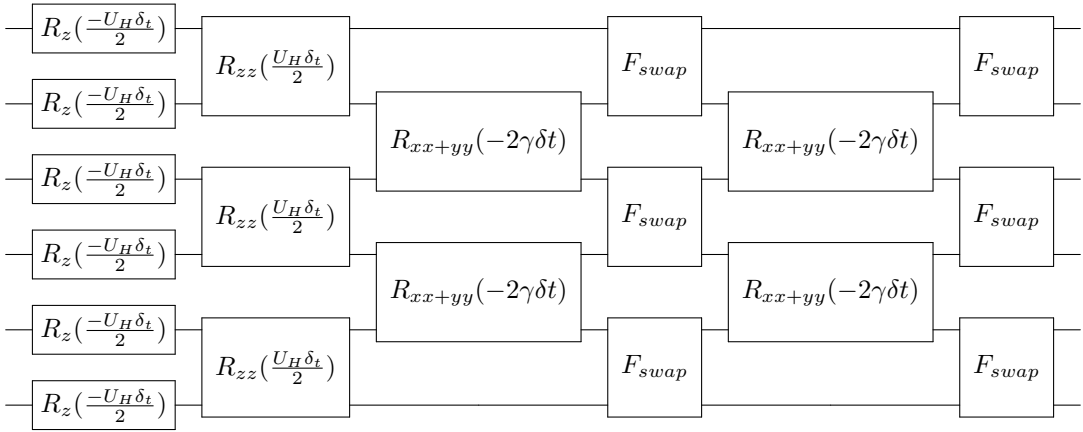
\begin{figure}[ht!]
\[
\begin{array}{c}
    \Qcircuit @C=1em @R=1em {
& \gate{R_z(\frac{-U_H \delta_t}{2})} & \multigate{1}{R_{zz}(\frac{U_H \delta_t}{2})} & \qw & \multigate{1}{F_{swap}} & \qw & \multigate{1}{F_{swap}} & \qw \\
& \gate{R_z(\frac{-U_H \delta_t}{2})} & \ghost{R_{zz}(\frac{U_H \delta_t}{2})} & \multigate{1}{R_{xx+yy}(-2\gamma \delta t)} & \ghost{F_{swap}} & \multigate{1}{R_{xx+yy}(-2\gamma \delta t)} & \ghost{F_{swap}} & \qw \\
& \gate{R_z(\frac{-U_H \delta_t}{2})} & \multigate{1}{R_{zz}(\frac{U_H \delta_t}{2})} & \ghost{R_{xx+yy}(-2\gamma \delta t)} & \multigate{1}{F_{swap}} & \ghost{R_{xx+yy}(-2\gamma \delta t)} & \multigate{1}{F_{swap}} & \qw \\
& \gate{R_z(\frac{-U_H \delta_t}{2})} & \ghost{R_{zz}(\frac{U_H \delta_t}{2})} & \multigate{1}{R_{xx+yy}(-2\gamma \delta t)} & \ghost{F_{swap}} & \multigate{1}{R_{xx+yy}(-2\gamma \delta t)} & \ghost{F_{swap}} & \qw \\
& \gate{R_z(\frac{-U_H \delta_t}{2})} & \multigate{1}{R_{zz}(\frac{U_H \delta_t}{2})} & \ghost{R_{xx+yy}(-2\gamma \delta t)} & \multigate{1}{F_{swap}} & \ghost{R_{xx+yy}(-2\gamma \delta t)} & \multigate{1}{F_{swap}} & \qw \\
& \gate{R_z(\frac{-U_H \delta_t}{2})} & \ghost{R_{zz}(\frac{U_H \delta_t}{2})} & \qw & \ghost{F_{swap}} & \qw & \ghost{F_{swap}} & \qw \\
}
\end{array}
\]
    \caption{Example 6 qubit quantum circuit diagram for each Trotter step for 1D Fermi-Hubbard model time-evolution simulation with $N_s=3$ physical sites. The gates used in this circuit are described in detail in the description of this section.}
    \label{fig:hms}
\end{figure}

With this ordering and the use of the Jordan-Wigner mapping, the fermionic Hamiltonian $\hat{H}$ maps onto a qubit Hamiltonian of the form 
\begin{align}
\nonumber \hat{H}_{\mathrm{q}}=&-\frac{\gamma}{2} \sum_{\substack{i=1 \\ i\text{ mod } 2 = 1}}^{N_s-1} \left(\hat{\sigma}_{2i-1}^x \hat{\sigma}_{2i}^z \hat{\sigma}_{2i+1}^z \hat{\sigma}_{2i+2}^x + \hat{\sigma}_{2i-1}^y \hat{\sigma}_{2i}^z \hat{\sigma}_{2i+1}^z \hat{\sigma}_{2i+2}^y\right) \\
&-\frac{\gamma}{2} \sum_{\substack{i=1 \\ i\text{ mod } 2 = 0}}^{N_s-1} \left(\hat{\sigma}_{2i-2}^x \hat{\sigma}_{2i-1}^x + \hat{\sigma}_{2i-2}^y \hat{\sigma}_{2i-1}^y\right)+ \frac{U_H}{4} \sum_{i=1}^{N_s}\hat{\sigma}_{2i-1}^z\hat{\sigma}_{2i}^z - \frac{U_H}{4}\sum_{i=1}^{2N_s} \hat{\sigma}_i^z + \frac{N_sU_H}{4},
\end{align}
where $\hat{\sigma}_i^\alpha$ (for $\alpha=x,y,z$) is the Pauli-$\alpha$ operator acting on qubit $i$. Here, the hopping terms are split into two sets of terms. The first corresponds to the case where there are two hopping sites with opposite spins in between (odd $i$). For example, hopping from $1\uparrow$ to $2\uparrow$ has $1\downarrow$ and $2\downarrow$ in between, hence $\hat{\sigma}^z$ terms arise from the Jordan-Wigner mapping. The second corresponds to the case where the two hopping sites occur on neighbouring qubits (even $i$). Note that different choices of the fermion ordering will give rise to different forms of the spin Hamiltonian.  The constant energy shift in the last term simply corresponds to an arbitrary phase rotation and so can be ignored. The dynamics associated with this model is generated by the propagator $\hat{U} = e^{-i\hat{H}t} = \left[e^{-i\hat{H}\frac{t}{N_s}}\right]^{N_s},$ working in units where $\hbar=1$. The resulting Trotterized short time propagator $\hat{U}_{\delta t}$ can be implemented with quantum circuits. An example for 3 physical sites corresponding to 6 qubits is shown in Fig. \ref{fig:hms}, where $R_{xx+yy}(\theta) = \exp\left[-i\frac{\theta}{4} \left(\hat{\sigma}_x\otimes \hat{\sigma}_x + \hat{\sigma}_y \otimes \hat{\sigma}_y \right) \right]$, $R_{zz}=\exp\left[ -i\frac{\theta}{2} \hat{\sigma}_z\otimes \hat{\sigma}_z \right]$ and $F_{swap}=\begin{bmatrix} 
    1 & 0 & 0 & 0\\
    0 & 0 & 1 & 0\\
    0 & 1 & 0 & 0\\
    0 & 0 & 0 & -1
    \end{bmatrix}$. These gates are decomposed into $R_y$, $R_z$ and $\mathrm{C}_X$ gates for implementation on quantum hardware.

Details of the derivation from the qubit Hamiltonian to the circuit implementation can be found in the source code provided, with the link in the section below. This circuit is repeated for each Trotter step. The quantum state at the end of the circuit is then the state of the time-evolved quantum system of interest.

\subsection{Measurement procedure}
\begin{enumerate}
    \item Specify the parameters for the 1D Fermi-Hubbard model Trotterized time-evolution simulation, including the number of physical sites $N_s$, the hopping integral $\gamma$, the onsite energy $U_H$, the evolution time for each Trotter step $\delta t$, the number of Trotter steps and the initial configuration of the sites, where it is chosen which qubits should be initialized into the $\ket{1}$ state.
    \item Construct the quantum circuit for time evolution according to the methods described above. Fig.~\ref{fig:hms} shows an example circuit for one Trotter step for 3 physical sites corresponding to 6 qubits. For each Trotter step, repeatedly apply the circuit.
    \item Run the circuit on hardware and measure all qubits with a certain number of shots to obtain a probability distribution of output bitstrings, denoted as $Q_{\text{output}}$. The number of shots should be chosen depending on the level of desired benchmarking precision. 
    \item Obtain an ideal probability distribution, $Q_{\text{ideal}}$. For small systems this can be done in a few different ways, including noiseless quantum circuit emulation or brute-force matrix exponential calculations.
    \item Calculate the normalized fidelity between $Q_{\text{ideal}}$ and the $Q_{\text{output}}$ in the following way. Define the fidelity between two probability distributions $Q_1$ and $Q_2$ over the same set of random events:
    \begin{equation}
        F_c(Q_1, Q_2) = \left( \sum_x \sqrt{p_{x,1}p_{x,2}} \right)^2,
    \end{equation}
    where $x$ is a random event, and $p_{x,i}$ is the probability of $x$ happening in the distribution $Q_i$. For the output of quantum computers, the random events $x$ are the bitstrings. The normalized fidelity between $Q_{\text{ideal}}$ and $Q_{\text{output}}$ is then given by~\cite{lubinskiApplicationOrientedPerformanceBenchmarks2021}:
    \begin{equation}
        F_{\mathrm{norm}}(Q_{\text{ideal}}, Q_{\text{output}}) = \text{max}\left\{\frac{F_c(Q_{\text{ideal}}, Q_{\text{output}}) - F_c(Q_{\text{ideal}}, Q_{\text{uni}})}{1 - F_c(Q_{\text{ideal}}, Q_{\text{uni}})}, 0\right\},
    \label{eq:normalised_fidelity_hms}
    \end{equation}
    where $Q_{\text{uni}}$ is the uniform distribution.

\end{enumerate}

\metricfig{
An example of the results one might obtain is shown in Fig.~\ref{fig:HMS_result}.
\begin{figure}[htpb]
    \centering
    \includegraphics{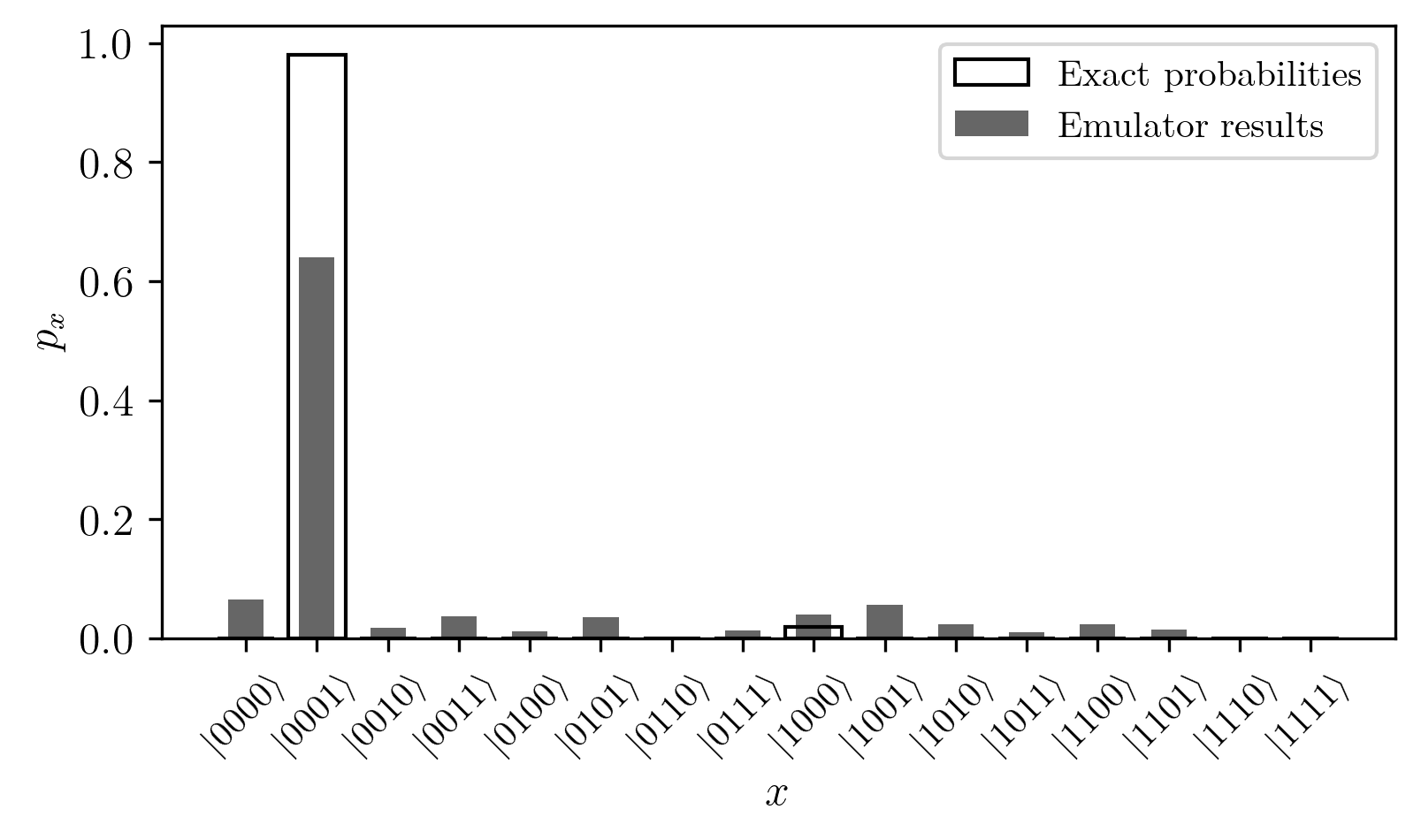}
    \caption{Results of emulator runs with the methodology outlined in this section to obtain the Fermi-Hubbard model simulation metric, where the noise model used is specified in Sec.~\ref{sec:noise_model}. The bars show the probability $p_x$ of measuring the output bitstrings. The black outlined bar shows the exact probability distribution, and the gray solid bar shows the results obtained from the emulator with the specified noise model. The normalized fidelity between the two distributions is calculated using Eq.~\ref{eq:normalised_fidelity_hms} and found to be $ F_{\mathrm{norm}}(Q_{\text{ideal}}, Q_{\text{emulator}}) = 0.572$.}
    \label{fig:HMS_result}
\end{figure}
}

\subsection{Assumptions and limitations}
\begin{itemize}
    \item This way of constructing the quantum circuits assumes that the quantum device only supports nearest-neighbour operations on qubits. If the hardware provides better connectivity between qubits, one may adjust how the circuit is constructed to tailor for the specific hardware.
    \item For large systems it can be very difficult or impossible to obtain an ideal probability distribution to benchmark against. Alternative methods of benchmarking the performance of the circuit may be used, such as quantum accreditation, which is introduced in metric~\ref{sec:accreditation}.
\end{itemize}
    
\subsection{Source Code}
A tutorial to run the 1D Fermi-Hubbard model Trotterized time-evolution simulation and calculate the fidelity is provided in: \sourceurl{well_studied_task_execution_quality_metrics/hubbard_model_simulation}.

\metricbibliography
\end{refsegment}

\section{Quantum Fourier transform (QFT) metric}
\label{sec:qft}
\begin{refsegment}
This metric evaluates how well a quantum computer can perform the quantum Fourier transform (QFT)~\cite{shor1997polynomial, nielsen2002quantum}. The QFT is widely used as a subroutine in quantum algorithms, most importantly within quantum phase estimation, which is a central component of Shor's factorization algorithm~\cite{shor1997polynomial}. Implementation of QFT has been demonstrated for small systems, such as within quantum accreditation on 4 qubits~\cite{Ferracin2021}, while for larger numbers of qubits higher quality QPUs are required, as well as QEC~\cite{mayer2024benchmarking, park2023reducing}.

\subsection{Description}
In what follows, a brief description of QFT is provided. For a thorough review on QFT, please see Ref.~\cite{nielsen2002quantum}. The QFT transforms the amplitudes of a quantum state according to the classical discrete Fourier transform (DFT). The DFT transforms a vector with $N$ elements $\boldsymbol{x}=(x_0,x_1,\cdots, x_{N-1})^T$ into a vector $\boldsymbol{y}=(y_0, y_1, \cdots, y_{N-1})^T$ according to
\begin{equation}
    y_k = \frac{1}{\sqrt{N}}\sum_{j=0}^{N-1} e^{2\pi ijk/N}x_j,
\label{eq:DFT}
\end{equation}
where $k$ is the index of the element in the transformed vector.
Applying the QFT to an arbitrary state $\sum_{j=0}^{N-1} x_j \ket{j}$ gives the following mapping,
\begin{equation}
    \sum_{j=0}^{N-1} x_j \ket{j} \rightarrow  \sum_{k=0}^{N-1} y_k \ket{k},
\label{eq:QFT}
\end{equation}
where the amplitudes $y_k$ are given by DFT following Eq.~\ref{eq:DFT}. This mapping can be implemented as a unitary transformation that can be achieved on a quantum computer.

For an $N_{\mathrm{q}}$-qubit basis state $\ket{j}$, one may write the number $j$ in its binary form as a string of $N_{\mathrm{q}}$ binary digits $j_1j_2\cdots j_n$, so that each digit $j_n$ is either $0$ or $1$, corresponding to a qubit in a state of $\ket{0}$ or $\ket{1}$, respectively. By adopting the so-called binary fraction notation where $0.j_a j_{a+1}\cdots j_b:=j_a/2+j_{a+1}/4+\cdots+j_b/2^{b-a+1}$, the QFT mapping for $\ket{j}$ is equivalent to the following mapping~\cite{nielsen2002quantum}:
\begin{equation}
\ket{j} \rightarrow \frac{
\left(\ket{0}+e^{2\pi i 0.j_{N_{\mathrm{q}}}}\ket{1}\right) \otimes
\left(\ket{0}+e^{2\pi i 0.j_{{N_{\mathrm{q}}}-1}j_{N_{\mathrm{q}}}}\ket{1}\right) \otimes \cdots \otimes
\left(\ket{0}+e^{2\pi i 0.j_1 j_2\cdots j_{N_{\mathrm{q}}}}\ket{1}\right) }
{2^{{N_{\mathrm{q}}}/2}
}.
\label{eq:QFTproduct}
\end{equation}

\begin{figure}[ht]
\centering
\small{
\[
\begin{array}{c}
\hspace*{-4.5em} 
\Qcircuit @C=0.25em @R=1em {
   & \lstick{\ket{j_1}}  & \gate{H} & \gate{R_2} & \qw & \push{\cdots} &  & \gate{R_{{N_{\mathrm{q}}}-1}} & \gate{R_{N_{\mathrm{q}}}} \qw & \qw & \qw & \qw & \qw & \qw & \qw & \qw & \qw & \qw &\qw & \qw & \qw & \qw & \rstick{\ket{0}+e^{2\pi i 0.j_1\cdots j_{N_{\mathrm{q}}}}\ket{1}}\\ 
& \lstick{\ket{j_2}} & \qw & \ctrl{-1} & \qw & \push{\cdots} & & \qw & \qw & \gate{H} & \qw & \push{\cdots} & & \gate{R_{{N_{\mathrm{q}}}-2}} & \gate{R_{{N_{\mathrm{q}}}-1}} & \qw & \push{\cdots} & & \qw & \qw & \qw & \qw & \rstick{\ket{0}+e^{2\pi i 0.j_2\cdots j_{N_{\mathrm{q}}}}\ket{1}} \\
   & \vdots  &  & \vdots  & & & & & \\   
& \lstick{\ket{j_{{N_{\mathrm{q}}}-1}}}  & \qw & \qw & \qw & \qw & \qw & \ctrl{-3} & \qw & \qw & \qw & \qw & \qw & \ctrl{-2}  & \qw & \qw & \push{\cdots} & & \gate{H} & \gate{R_2} & \qw & \qw & \rstick{\ket{0}+e^{2\pi i 0.j_{{N_{\mathrm{q}}}-1} j_{N_{\mathrm{q}}}}\ket{1}}\\
& \lstick{\ket{j_{{N_{\mathrm{q}}}}}}  & \qw & \qw & \qw & \qw & \qw & \qw & \ctrl{-4} \qw &  \qw & \qw &  \qw & \qw & \qw & \ctrl{-3}  & \qw & \push{\cdots} &  & \qw & \ctrl{-1} & \gate{H} \qw & \qw & \rstick{\ket{0}+e^{2\pi i 0.j_{N_{\mathrm{q}}}}\ket{1}} }
\end{array}
\]
 \caption{Part of the quantum circuit diagram for QFT. Normalization coefficients are omitted. The output of this circuit is a QFT-transformed state in a reversed order of qubits. Therefore, SWAP gates need to be added at the end of this circuit for each pair of qubits with index $i$ and index ${N_{\mathrm{q}}}+1-i$, so that QFT is achieved.}
    \label{fig:qft_circuit}
}
\end{figure}
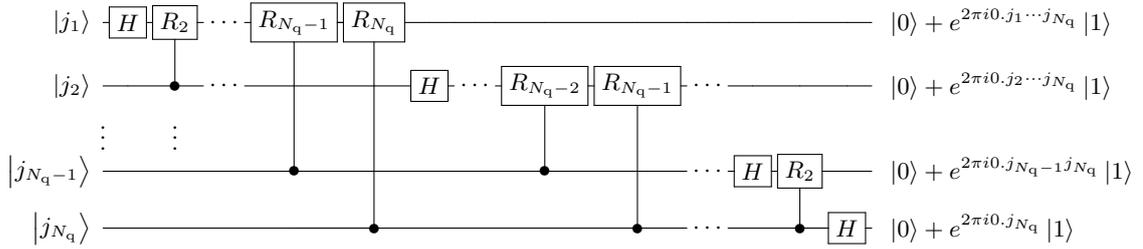

From Eq.~\ref{eq:QFTproduct}, one can then derive a quantum circuit to implement QFT within the gate-based model by defining a rotation gate $R_k$ as
\begin{equation}
    R_k = \begin{bmatrix}
        1 & 0 \\
        0 & e^{2\pi i/2^k}
    \end{bmatrix}.
\end{equation}
The first part of the QFT circuit is shown in Fig.~\ref{fig:qft_circuit}. Each qubit has a Hadamard gate applied, followed by a sequence of controlled-$R_2$ up to controlled-$R_{{N_{\mathrm{q}}}+1-i}$ gates, where $i$ is the index of the qubit, controlled sequentially by all following qubits. At the end of this circuit, the first qubit gives the last term in the tensor product in Eq.~\ref{eq:QFTproduct}, the second qubit gives the second-last term, etc. Therefore, a second part of circuit needs to be added, which consists of SWAP gates acting on each pair of qubits with index $i$ and index ${N_{\mathrm{q}}}+1-i$, so that the QFT is obtained.

\subsection{Measurement procedure}
Since QFT is not a metric in itself, one can devise a benchmarking method using QFT on a quantum computer in different ways. Here a measurement procedure to benchmark the normalized fidelity of hardware outputs and ideal results is presented. This method follows from the QED-C application oriented benchmarks~\cite{lubinskiApplicationOrientedPerformanceBenchmarks2021}, wherein it is named as method 1 within three methods of QFT benchmarks.
\begin{enumerate}
    \item Initialize $N_{\mathrm{q}}$ qubits to an arbitrarily chosen initial state $\ket{a}$, where $a\leq 2^{N_\mathrm{q}}-1 \in \mathbb{N}$.
    \item Apply the QFT circuit as outlined in the section above.
    \item For each qubit with index $i\in\{1,2,\cdots,{N_{\mathrm{q}}}\}$, apply an $R_z(\pi/2^{i-1})$ gate.
    \item Apply the inverse of the QFT circuit. 
    \item Measure all qubits with a certain number of shots to obtain a probability distribution of output bitstrings. The number of shots should be chosen based on the desired benchmarking precision. The ideal result should be a distribution where the bitstring representation of the number $a+1$ has a probability of 1, and all other bitstrings have 0 probability.
    \item Compute the normalized fidelity between the hardware output distribution, denoted as $Q_{\text{output}}$, and the ideal distribution, denoted as $Q_{\text{ideal}}$, in the following way. Define the fidelity between two probability distributions $P_1$ and $P_2$ over the same set of random events:
    \begin{equation}
        F_c(Q_1, Q_2) = \left( \sum_x \sqrt{p_{x,1}p_{x,2}} \right)^2,
    \end{equation}
    where $x$ is a random event, and $p_{x,i}$ is the probability of $x$ happening in the distribution $Q_i$. For the output of quantum computers, the random events $x$ are the bitstrings. The normalized fidelity between $Q_{\text{ideal}}$ and $Q_{\text{output}}$ is then given by~\cite{lubinskiApplicationOrientedPerformanceBenchmarks2021}:
    \begin{equation}
        F_{\mathrm{norm}}(Q_{\text{ideal}}, Q_{\text{output}}) = \text{max}\left\{\frac{F_c(Q_{\text{ideal}}, Q_{\text{output}}) - F_c(Q_{\text{ideal}}, Q_{\text{uni}})}{1 - F_c(Q_{\text{ideal}}, Q_{\text{uni}})}, 0\right\},
    \label{eq:normalised_fidelity}
    \end{equation}
    where $Q_{\text{uni}}$ is the uniform distribution.
\end{enumerate}

\subsection{Assumptions and limitations}

\begin{itemize}
    \item The controlled-$R_k$ gates need to be decomposed into the native gates of the device, which introduces multiple single and two qubit gates per controlled-$R_k$. In addition, the QFT circuit contains controlled-$R_k$ gates for every pair of qubits. Therefore, if the hardware does not support all-to-all connectivity, extra gates are needed for non-connected qubits.
\end{itemize}

\subsection{Source code}
A tutorial to run the QFT and calculate the fidelity is provided in \sourceurl{well_studied_task_execution_quality_metrics/qft}. 
The described measurement procedure follows from method 1 within the QFT benchmarks in the QED-C application oriented benchmarks suite~\cite{lubinskiApplicationOrientedPerformanceBenchmarks2021}.

\metricbibliography
\end{refsegment}

\def\thechapter{M6}
\chapter{Speed metrics}
\label{chapter:speed_metrics}

\section{Time taken to execute a general single- or multi-qubit gate}
\label{sec:1q_gate_speed}
\begin{refsegment}

This metric gives the time taken to apply a gate that executes any general single- or multi-qubit unitary operation on a quantum computer. 

\subsection{Description} 
The time taken to execute gates on a quantum computer is the determining factor for the total computation time of an algorithm. This metric quantifies the time taken to execute one general $N_\mathrm{q}$-qubit gate. These general $N_\mathrm{q}$-qubit gates can be decomposed into the native gates on the hardware. However, the size of the circuits in terms of the number of native gates grows exponentially with $N_\mathrm{q}$, and thus in practice this metric is only to be evaluated for small values of $N_\mathrm{q}$. Below, we provide the description and methodology to measure the time taken to execute a general single qubit gate, $t_{U1}$, corresponding to $N_\mathrm{q}=1$. The methodology can then be generalized to the $N_\mathrm{q}>1$ case, $t_{UN_\mathrm{q}}$ , by using appropriate decompositions of $N_\mathrm{q}$-qubit unitary operations. The decomposition of $N_\mathrm{q}$-qubit unitary operations are typically decomposed to a combination of single- and two-qubit gates, and the theoretical worst case number the required number of CNOT gates scales as $\qty(4^{N_\mathrm{q}}-3N_\mathrm{q} -1)/4$~\cite{PhysRevA.69.062321}. There are many methods to beat the worst case scaling for $N_\mathrm{q}$-qubits, such as the cosine-sine matrix decomposition~\cite{PhysRevLett.93.130502}. For 2 qubits, Cartan KAK decomposition is often used~\cite{tucci2005introduction, 10.1063/1.1723701}.

All single qubit gates perform rotations on the Bloch sphere. In order to run generic single qubit circuits, a quantum processing unit (QPU) must be able to perform arbitrary single qubit rotations. The time taken to perform such a general single qubit gate provides an estimate relevant for the use of the QPU for general purpose algorithms. Any such general single-qubit unitary operation can be written in matrix form as 
\begin{equation}
  U(\theta, \phi, \lambda) =  \begin{pmatrix}
\cos(\theta/2) & -i e^{i\lambda} \sin(\theta/2) \\
-i e^{i\phi} \sin(\theta/2) & e^{i\qty(\lambda + \phi)}\cos(\theta/2) 
\end{pmatrix},
\label{eq:U1qgeneral}
\end{equation}
where $\theta$, $\phi$, and $\lambda$ are parameters corresponding to the rotation amount, and to the angles corresponding to the axis of rotation, respectively.

The physical implementation of a general single qubit rotation is device dependent, because the type of native single qubit gates on a QPU depends on the hardware platform. For example, on many devices an $R_z$ gate is performed virtually by adding a phase to the following gate~\cite{PhysRevA.96.022330,krantz_quantum_2019}, and as such is effectively applied instantaneously. 
Given a calibrated pulse implementing a $R_x(\pi/2)$ rotation, combined with virtual $R_z$ gates for arbitrary angles, the general single qubit unitary can be decomposed into the following sequence~\cite{PhysRevA.96.022330}:
\begin{equation}
  U_{\mathrm{1q}}(\theta, \phi, \lambda) = R_z(\phi - \frac{\pi}{2}) R_x(\frac{\pi}{2}) R_z(\pi - \theta) R_x(\frac{\pi}{2}) R_z(\lambda - \frac{\pi}{2}).
  \label{eq:U1q2Rx}
\end{equation}
For such a decomposition, the total time taken to execute $U(\theta, \phi, \lambda)$ on a single qubit, $t_{U1}$, is equal to twice the times taken to execute a single $R_x$ gate, $t_{R_x}$:
\begin{equation}
    t_{U1}=2 t_{R_x}.
\end{equation}
Here $t_{R_x}$ includes possible idle gate times between two executions of an $R_x$ gate on the device.

The time taken to implement such general unitary operations depends on the hardware For example, a typical native single qubit gate in a superconducting qubit takes \SI{25}{\ns}~\cite{Arute2019}. Using Eq.~\ref{eq:U1q2Rx} yields a general single-qubit unitary time of \SI{50}{\ns}, since the $R_z$ gates are virtual and instantaneous. Trapped ion single qubit gate times have been reported to be \SI{10}{\us}~\cite{PhysRevLett.113.220501}. If virtual $R_z$ gates are used, then Eq.~\ref{eq:U1q2Rx} can again be used, and this yields a general single-qubit unitary time of \SI{20}{\us}.

\subsection{Measurement procedure}

\subsubsection{Method 1}
The native gate time is known by the hardware manufacturer, since the manufacturer sets duration of the pulses and any required idle time after the pulse needed for each gate. This is done as part of the calibration process. If the hardware manufacturer provides this information, then the measurement procedure for calculating the general single qubit gate time is as follows: 
\begin{enumerate}
    \item Hardware manufacturer is to provide the pulse level decomposition of the general single qubit unitary given in Eq. \ref{eq:U1qgeneral}.
    \item For each pulse in this decomposition, the hardware manufacturer is to provide the pulse duration, and potential idle times between pulses. If the these are $R_x$ pulses, it corresponds to providing $t_{R_x}$
    \item Summing up the times of all pulses used to decompose Eq. \ref{eq:U1qgeneral} one obtains the total gate time. If the decomposition corresponds to the one presented in Eq. \ref{eq:U1q2Rx}, then the total gate time is $t_{U1}=2 t_{R_x}$.
\end{enumerate}

For the time taken to execute a general multi-qubit gate, the same process is followed, however this time the decomposition of the multi-qubit unitary into native pulses must be provided and reported along with the time.

\subsubsection{Method 2}
If the vendor does not provide the measurement durations, but the total circuit execution time is provided by the vendor, one can run circuits with different numbers of general single qubit gates and use the execution times to estimate the time for each general single qubit gate. The measurement procedure is as follows:

\begin{enumerate}
    \item Set up a general circuit consisting of $n_U$ repetitions of a general single qubit unitary $U$, as given in Eq. \ref{eq:U1qgeneral}. The rotation angles corresponding to the unitary $U$ are chosen randomly. 
    \item For different values of $n_U$, the circuit is executed for a chosen number of shots and the circuit execution time corresponding to each $n_U$ is obtained from the vendor. The total execution time can be written as $t_{n_U}=n_U t_{U1} + t_O$, where $t_O$ includes all contributions to the total execution time that are independent of the number of gates, for example state preparation and measurement time.
    \item Perform a linear fit of the results to extract $t_{U1}$ and $t_O$. 
\end{enumerate}

To ensure that the time spent executing gates is not negligible compared to other operations such as measurement time, the values of $n_U$ must be chosen such that for the largest $n_U$, $n_U t_{U1} \gg t_O$.

To use this method to measure the time taken to execute a general multi-qubit gate, one must provide the decomposition of unitary to native gates. This decomposition must be reported along with the execution time of the gate.

\subsubsection{Method 3}
If the vendor does not provide both the measurement durations and the total circuit execution time, one can obtain an estimate of the gate duration in units of the $T_1$ time of another qubit. If it is known or assumed that the single qubit gate durations on all qubits are identical, then one can use the method to estimate the duration of a single qubit gate in units of the $T_1$ time of that qubit.
The ratio between the gate time and $T_1$ is important, since it determines how many single qubit gate operations one can implement before the $T_1$ noise dominates the circuit outcomes. 
If the vendor has provided the physical $T_1$ times, the gate duration in physical units can also be estimated.

This method is similar to Method 2, but uses another qubit as a clock. The methodology is as follows:
\begin{enumerate}
    \item Select two qubits, such that crosstalk noise is small, so that operations on one qubit do not unintentionally induce significant noise on the other qubit. 
    \item  Set up a general circuit consisting of $n_U$ repetitions of a general single qubit unitary $U$ acting on qubit 2, as given in Eq. \ref{eq:U1qgeneral}. The rotation angles corresponding to the unitary $U$ are chosen randomly. 
    \item Then for each value of $n_U$:
    \begin{enumerate}
        \item Prepare qubit 1 in the $\ket{1}$ state. 
        \item Subsequently, on qubit 2, apply the the general single qubit gate $U$ $n_U$ times.
        \item Measure qubit 1, and ensure that the measurement takes place after the gates have been run on qubit 2. One must include a form of barrier to only allow measurements on qubit 1 after all gates on qubit 2 have been applied.
        \item Repeat step (c) for a specified number of shots, then extract $p_{1}(n_U)$, the probability of measuring qubit 1 in the $\ket{1}$ state given the $n_U$ applications of $U$ on qubit 2.
    \end{enumerate}
    \item After running these circuits for differing values of $n_U$, fit a decay curve using Eq.~\ref{eq:t1_time} for the data of $p_{1}(n_U)$ versus $n_U$, where $t$, the time before measurement, is replaced by $n_U$.
    \item After fitting the decay curve to Eq.~\ref{eq:t1_time}, it gives $T_1^{\mathrm{(n1q)}}$ as the $T_1$ time of qubit $1$ in units of number of single qubit gates applied on qubit $2$. This then means that $T_1 = T_1^{\mathrm{(n1q)}} t_\mathrm{1q}$, where $T_1$ is the $T_1$ time of qubit 1 in units of seconds, and $t_\mathrm{1q}$ is the time taken to execute a general single qubit unitary on qubit 2. 
\end{enumerate}

If the $T_1$ time of qubit 2 is known, then an approximate estimate for the single qubit gate time,  $t_\mathrm{U1q} $, can be obtained using $t_\mathrm{U1q} = T_1/T_1^{\mathrm{(n1q)}}$.  

\subsection{Assumptions and limitations}
\begin{itemize}
    \item Some special single qubit gates can be natively calibrated operations that have been optimized to run on the QPU, such as $R_x(\pi/2)$. These can have shorter gate times than those of the general single qubit gate. 
    \item The time taken to execute a general multi-qubit gate depends on the method of transpilation to the native gate set~\cite{PhysRevLett.93.130502}. 
    \item Methods 2 and 3 rely on repeatedly applying a unitary. It is assumed that the user can disable circuit optimization in order to prevent the compiler combining these repeated unitary gates into a single unitary.
    \item Method 3 assumes that either the single qubit gate durations are identical for all qubits, or that the $T_1$ times for the qubits are provided by the vendor.
    \item Method 3 relies on the assumption that cross-talk between the two qubits is small. A large cross-talk noise can alter the $T_1$ time on one qubit when gates are executed on another qubit, which results in inaccurate estimates of the gate speed.  
\end{itemize}

\subsection{Source code}
A tutorial showing how to estimate the ratio of the single qubit gate time to $T_1$ is provided in \sourceurl{speed_metrics/time_taken_to_execute_a_general_single_or_multi_qubit_gate}.

\metricbibliography

\end{refsegment}

\section{Time to measure qubits}
\label{sec:timemeasure}
\begin{refsegment}
The time to measure a qubit is the time taken to perform a projective measurement in the computational basis on that qubit.

\subsection{Description}
In order to obtain the result of a quantum computation, some of the qubits in the device need to be measured depending on the quantum circuit being run.
This metric quantifies the time taken to perform such a projective measurement on a particular qubit. This does not include any quantum gates applied before the measurement to change the basis of measurement.
Measurement times for quantum computers vary from device to device even within the same hardware platform. For example, for superconducting qubits, measurement times of $\si{100\ns}$~\cite{PhysRevApplied.17.044016} and $\si{500\ns}$~\cite{google2023suppressing} have been reported. For trapped ion qubits, measurement times of $\si{11\us}$~\cite{crainHighspeedLowcrosstalkDetection2019} and $\si{46\us}$~\cite{PhysRevLett.126.010501} are values that have been reported. 

\subsection{Measurement procedure}

The measurement procedure for this metric is analogous to the procedure for the gate times. Methods 2 and 3 rely on the capability to perform mid-circuit measurements (see metric \ref{sec:capability_mid_circuit_measurements}). Method 2 further relies on the availability of the time taken for a quantum computation. Method 3 only allows obtaining the measurement time in units of single qubit gate durations.

\subsubsection{Method 1}
The information about measurement times on each qubit is obtained from the vendor device documentation.

\subsubsection{Method 2}
If the vendor does not provide the measurement durations, but the total circuit execution time is provided by the vendor, one can use method 2 described in metric~\ref{sec:1q_gate_speed}, with the general single qubit gates replaced by mid-circuit measurements.

\subsubsection{Method 3}
If the vendor does not provide both the measurement durations and the total circuit execution time, one can use method 3 described in metric~\ref{sec:1q_gate_speed}, with the general single qubit gates replaced by mid-circuit measurements.

\subsection{Assumptions and limitations}
\begin{itemize}
\item Method 2 assumes that the time taken to run the circuits on the device is available.
\item Methods 2 and 3 assume that mid-circuit measurements can be performed, since the measurements needs to be repeated inside each executed quantum circuit.
\end{itemize}

\subsection{Source code}
A tutorial for the time taken to measure qubits obtained using method 3 is provided in \sourceurl{speed_metrics/time_to_measure_qubits}.

\metricbibliography
\end{refsegment}

\section{Time to reset qubits}
\label{sec:timereset}
\begin{refsegment}
This metric quantifies the time taken to reset all qubits in a device after a measurement to make them reusable for the next circuit execution. 
Typically, quantum circuits need to be run a large number of times, also called the number of shots, in order to extract useful results~\cite{nielsen2002quantum}. The time spent by the device in-between performing these repetitions contributes to the total computation time~\cite{10.1063/1.3435463}. After qubits are measured at the end of a quantum circuit, they typically need to be reset in preparation for the next circuit. This metric quantifies the time taken to reset all the qubits in the device.

\subsection{Description}
When a quantum computer finishes running a circuit, the qubits in the device typically undergo either an active or a passive reset in preparation for running the next circuit or next repetition of the current circuit~\cite{10.1063/1.3435463, PhysRevLett.110.120501}. A passive reset occurs when the device waits for the qubit to decay to the $\ket{0}$ state~\cite{PhysRevLett.110.120501, krantz_quantum_2019}. This typically is on the order of the longest $T_1$ time of all qubits in the device. Alternatively, active reset is done when the reset time required is much lower than the $T_1$ time or the qubit environment is hot on the scale of the qubit transition frequency~\cite{PhysRevLett.110.120501}. One simple form of active reset is where after measuring the current state of the device, pulses are sent to the qubits measured to be the $\ket{1}$ state to bring them to the $\ket{0}$ state~\cite{PhysRevLett.121.060502}. There are also some strategies where no reset time is required, such as the so-called restless measurement. In that case, instead of resetting qubits, the measured states of the qubits are defined as the $\ket{0}$ states for the next circuit execution and the redefinitions are accounted for in post-processing~\cite{werninghaus2021high}.

\subsection{Measurement procedure}
The methods for measuring the reset time are similar to the ones in metrics~\ref{sec:1q_gate_speed} and \ref{sec:timemeasure}. Unlike those, however, a different qubit cannot be used a clock to estimate the reset time, because the qubit corresponding to the clock is reset between shots.

\subsubsection{Method 1}
As in metrics~\ref{sec:1q_gate_speed} and \ref{sec:timemeasure}, the first method is to obtain the details on the reset time from the vendor device documentation. 

\subsubsection{Method 2}
As in metrics~\ref{sec:1q_gate_speed} and \ref{sec:timemeasure}, if the total circuit execution time is available, it can be used to estimate the reset times.

The estimation method requires running circuits composed of repetitions of a circuit. Denote the circuit that is repeated - excluding any measurements - as $C$. The circuit to be run on the device is constructed by repeating $C$ for a number of times. If a form of barrier operation is available, it can be added between the repetitions of $C$ to ensure there is no overlap between each repetition. After the repetitions, all qubits are measured. 

The measurement procedure involves independently changing the number of shots and the number of repetitions of $C$ inside each quantum circuit. The time taken for each shot includes the time spent on running $C$, as well as the time to measure and reset the qubits. The time spent on running each quantum circuit depends on the number of repetitions of $C$. Thus, the total computation time $t_{\mathrm{tot}}$ for running $n_s$ shots of a circuit composed of repeating $C$ for $m$ times is given by
\begin{equation}
t_{\mathrm{tot}}(n_s,m) = t_o + n_s (m\, t_C + t_m + t_{\mathrm{reset}}),
\end{equation}
where $t_o$ is the overhead time which occurs independently of the circuit, $t_C$ is the actual time to run $C$, $t_m$ is the maximum of the measurements times of all the qubits, and $t_{\mathrm{reset}}$ is the time to reset the device. The method to estimate $t_{\mathrm{reset}}$ is given below. Note that $t_o$ is different to the overhead time in metric~\ref{sec:1q_gate_speed}.

To get an estimate of $t_{\mathrm{reset}}$, one can evaluate $t_{\mathrm{tot}}(n_s,m)$ for a varying number of $n_s$ and $m$ and use it to estimate $t_m + t_{\mathrm{reset}}$. Then, one can use prior knowledge of $t_m$ - obtained via methods described in metric~\ref{sec:timemeasure} - to get an estimate of $t_{\mathrm{reset}}$.

\subsection{Assumptions and limitations}
\begin{itemize}
\item Method 2 above assumes that the time taken to run the circuits on the device is available.
\item Method 2 assumes that the physical time to run a circuit composed of $r$ repetitions of $c$ takes time $r t_c$. If a barrier operation is not available, some gates in the different repetitions of $c$ might run in parallel, affecting the estimated $t_{\mathrm{reset}}$ time.
\item Method 2 assumes that the maximum measurement time of the qubits can be obtained via methods described in metric~\ref{sec:timemeasure}.
\end{itemize}

\subsection{Source code}
A tutorial for the time taken to reset qubits obtained using method 2 is provided in \sourceurl{speed_metrics/time_to_reset_qubits}.

\metricbibliography
\end{refsegment}

\section{Overall device speed on reference tasks}
\label{sec:timereferencetasks}

\begin{refsegment}
This metric gives the time taken for a quantum computer to run reference tasks, such as the well studied tasks described in the preceding sub-sections. 

\subsection{Description}
For practical evaluation of the well studied tasks on a quantum computer, the overall device speed for running a task is an important aspect when choosing which hardware platform use. Therefore, when choosing a set of reference tasks for benchmarking purposes, one needs to include widely-used quantum algorithms that are representative for real-world applications~\cite{lubinskiApplicationOrientedPerformanceBenchmarks2021}. 

When submitting a task to a remote quantum hardware platform, it is typical that the circuits in the task are first put into a waiting queue before the hardware becomes available to execute them~\cite{10.1109.9668289}. This queue time should be excluded from the benchmarking process, as well as any time overheads not related to the task, such as network communication times. Hence, the metric of overall device speed on reference tasks is measured solely by the time taken to execute the circuits on the device.

\subsection{Measurement procedure}
The measurement procedure presented here uses a set of reference tasks which include the well-studies tasks described in the preceding sub-sections, namely variational quantum eigensolver (VQE, metric~\ref{sec:vqe}), quantum approximate optimization algorithm (QAOA, metric~\ref{sec:Q-Score}), Fermi-Hubbard model simulation (FHMS, metric~\ref{sec:hubbard_model}) and quantum Fourier transform (QFT, metric~\ref{sec:qft}). Other representative sets of well studied tasks can be those used for the determination of the algorithmic qubits metric (metric \ref{sec:algorithmic_qubits}), as well as the time taken to evaluate the quantum volume (metric~\ref{sec:quantum_volume}), which is related to the definition of the circuit layer operations per second (CLOPS) proposed by IBM~\cite{wackQualitySpeedScale2021}. Note that the set of reference tasks can be extended to include other widely-used quantum algorithms as needed. The procedure then is:
\begin{enumerate}
    \item Run the chosen set of reference tasks on the hardware.
    \item For each reference task in the set, record and report the time taken to execute the circuits in seconds. 
\end{enumerate}

\subsection{Assumptions and limitations}
\begin{itemize}
\item Care should be taken depending on the hardware platform used, because this metric is dependent on how each manufacturer reports the timings. For example, some vendors may report the total time from submission of circuits to completion, which includes the queue time. In this case the queue time must be known and must be subtracted from the total time. Due to manufacturers reporting time differently, it can be difficult to compare speeds of different hardware platforms.
\item The well studied tasks can include the execution of quantum circuits and also of classical computations within a hybrid quantum-classical approach. The metric includes both the classical CPU time in addition to the time spent executing the quantum circuits.
\end{itemize}

\subsection{Source Code}
The source code for the tasks of VQE (metric~\ref{sec:vqe}), QAOA (metric~\ref{sec:Q-Score}), FHMS (metric~\ref{sec:hubbard_model}), QFT (metric~\ref{sec:qft}), and algorithmic qubits (metric~\ref{sec:algorithmic_qubits}) is given in the corresponding sub-sections. 

\metricbibliography
\end{refsegment}

\vspace*{-4\baselineskip}
\def\thechapter{M7}
\chapter{Stability metrics}
\label{chapter:stability_metrics}

\section{Standard deviation of a specified metric evaluated over a time interval}
\label{sec:stdTime}
\begin{refsegment}

The standard deviation of a metric measured multiple times over a time scale of interest quantifies the variation of that metric over time. A larger standard deviation corresponds to larger variation of the metric, and hence corresponds to reduced stability of the device. 

\subsection{Description} 
A general discussion on the time dependent changes of the metrics is given in Sec. \ref{sec:stabilityTime}. This metric based on the standard deviation quantifies the overall stability of the measured metric over a specified time scale.

\subsection{Measurement procedure}

\begin{enumerate}
    \item Determine the the time scale of interest to measure fluctuations of the specified metric, which defines a time interval, $t_0$, that spans the considered time scale. This can be on the scale of less than a second, seconds, minutes, days, weeks or months. 
    \item Determine the value of the metric a specified number of times, $N_t$, over the time scale $t_0$. The times at which the metric is evaluated are to be distributed approximately uniformly over $t_0$. A larger $N_t$ provides a better sampling of the time-dependent fluctuations of the metric.
    \item Calculate the average and the standard deviation of the metric over the $N_t$ measurements acquired over the $t_0$ time interval. 
\end{enumerate}

\subsection{Assumptions and limitations}
\begin{itemize}
    \item The smallest time scale over which the stability can be measured is bounded by the time taken to evaluate that metric. 
\end{itemize}

\subsection{Source code}
A tutorial for calculating the standard deviation of a specified metric over a time interval, with the example metric being the Clifford randomized benchmarking average gate error, is provided in \sourceurl{stability_metrics/amount_of_fluctuations_of_metrics_over_time}.

\end{refsegment}

\def\thechapter{M8}
\chapter{Metrics for quantum annealers}
\label{chapter:annealers}

\section{Single qubit control errors}
\label{sec:annealer_single_qubit_control_errors}
\begin{refsegment}
Single qubit control errors are a type of error where the device does not implement the Hamiltonian it was designed to, and a specific case is where the $h_i$ terms from Eq.~\ref{eq:H_ising} deviate from the programmed values. There are several ways to measure this, which yield different results based on the timescales they probe. Here this error is denoted as a control error (sometimes also referred to as calibration error~\cite{Pudenz2015Annealing}), although there are numerous physical origins of such errors, many of which do not stem directly from the controls.

\subsection{Description}
Not implementing the intended Hamiltonian on a quantum annealer is potentially a serious problem~\cite{Young2013wrong}, and can lead to drastically different solutions to the target problem~\cite{Albash2019errorQA}. Since the problem statement on an annealer is fundamentally analog, noise from the environment, or even just lack of precision in the controls, can lead to the statement of the problem, which the device actually solves, being different from the one the user intended. There is extensive work on error mitigation strategies which could help combat these errors, and these methods have been shown to be somewhat successful~\cite{Young2013wrong,Pudenz2014errorQA,Nishimura2016errorQA,Vinci2018errorQA,Bennett2023precision}. 

In this section three different methods to measure this noise are presented, which probe different time scales, and probe the devices under different conditions. Because single body errors can be more efficiently measured, only the single body error terms $h_i$ are discussed, rather than $J_{ij}$ from Eq.~\ref{eq:H_ising}. The error model is therefore
\begin{equation}
H_{\mathrm{error}}=\sum_i \zeta_i Z_i. \label{eq:field_noise} 
\end{equation}
The methods discussed here could likely be extended to measure two-body biases as well. The three methods discussed here have all been outlined, as well as experimentally compared, in Ref.~\cite{Chancellor2022_U_dist}. 

The most straightforward method to measure these errors is sampling. This method consists of programming a Hamiltonian $h_i=J_{ij}=0\quad \forall i,j$ and counting the number of times each qubit is seen in the $1$ or $0$ configuration. The annealing time may affect the error rate measured this way as well, since it will determine the parameters at which dynamics stop. For this reason the error should be treated as being annealing time dependent, although in practice this dependence may not be strong enough to significantly affect the results. Up to statistical fluctuations, the counts of $1$ and $0$ measurements should be equal if no noise is present. A bias between the two states can then be inferred from any difference which is seen. For example, if the device is assumed to be in thermal equilibrium at a known temperature, then the energy difference can be back-calculated from the single qubit Boltzmann distribution. A disadvantage of this method is that the quantity calculated is the average for the time period, therefore high frequency fluctuations, with a frequency much faster than the sampling time, will be missed. 

A more complicated method, which is sensitive to higher frequency noise elements, is to again program a trivial Hamiltonian $h_i=J_{ij}=0\quad \forall i,j$, but instead examine the autocorrelation effects in the individual $Z$ measurements. Again assuming the output obeys a thermal distribution at a known temperature, Taylor expanding and taking only the first non-trivial term, one has
\begin{equation}
C_{ij}=\frac{\overline{\zeta^{(i)} \zeta^{(j)}}}{T^2},
\end{equation}
where $T$ is the known temperature, and the superscripts in this equation indicate sample number rather than qubit number. Full details can be found in the supplemental material of Ref.~\cite{Chancellor2022_U_dist}. Taking the Fourier transform of this quantity, one can extract the spectral density of the noise $s_k$, then the total RMS noise can then be shown to be approximately
\begin{equation}
\sqrt{\overline{\zeta^2}}=\frac{T}{N}\sqrt{\sum_{k=1}^N|s_k|^2\exp\left[-2 \pi i \frac{k-1}{N}\right]},
\end{equation}
 where $N$ is the number of samples taken. This method is sensitive to a much broader range of frequencies, but is still limited by the rate at which the samples are taken.

Finally, the third method to measure this noise is to program the Hamiltonian for an Ising spin chain with frustrated boundaries, given by
\begin{equation}
H_{\mathrm{domain}}=J(\sum^{n-1}_{i=1}-Z_iZ_{i+1}+2(Z_1-Z_{n})) \label{eq:frust_chain}
\end{equation}
where $n$ is the number of qubits. As long as $J$ is large and positive, then the annealer will almost exclusively sample from states where the chain contains a single domain-wall somewhere on the chain, which form a degenerate manifold. This is reminiscent of the domain-wall encoding proposed in Ref.~\cite{chancellor2019domain} and also the simulation extension in Ref.~\cite{Werner2024Fermion}, but crucially contains an extra factor of $2$, which excludes the possibility of all qubits taking the same state from the degenerate manifold. The measurements are then taken by measuring the relative position of the domain wall on the chain averaging over different local gauges, which randomly change the definition of $\ket{1}$ and $\ket{0}$ on the qubits, and potentially different ways of programming the Hamiltonian onto the physical qubits of the device. It can be show mathematically that averaging Boltzmann distributions over different noise realization yields a characteristic ``U'' shape~\cite{Chancellor2022_U_dist}, which can be related to the noise strength. In practice the noise strength can be calculated by fitting a numerically calculated Boltzmann distribution to the actual distribution on the device. For flux-qubit devices, either a correction needs to be applied to the probability of finding the domain wall on a terminal site, or these points need to be excluded from the fit.

The domain-wall method has the advantage that the effective lower frequency cutoff is set by the dynamics of the device, and not by the way in which the samples are obtained, a property which could be particularly important in comparing devices where timescales related to sampling are very different. It is also the only of these methods which can measure while some of the couplers are turned on.

\subsection{Measurement procedure}
For all three methods the bias strength is extracted as a numerical ratio between the strength of the average errors when the dynamics cease to be effective (freezed out) and the temperature of the device, which is usually known. When combined with the temperature estimates using the methods described in metric~\ref{sec:annealer_samp_temp}, this could be used to calculate the single-qubit error as a fraction of total allowed $h$ values.

The procedure for Method I, the basic sampling method, is as follows:
\begin{enumerate}
    \item Turn all single and two body Hamiltonian terms off.
    \item Perform annealing multiple times to collect statistics, only to track counts rather than the order in which they were taken.
    \item Back calculate bias strength based on assumption of Boltzmann distribution.
\end{enumerate}
The procedure for Method II, the autocorrelation method, is as follows:
\begin{enumerate}
    \item Turn all single and two body Hamiltonian terms off. 
    \item Perform annealing multiple times to collect statistics, and track the data time-series, not just aggregate statistics.
    \item Calculate autocorrelation, Fourier transform and integrate over all frequency components.
\end{enumerate}
The procedure for method III, the domain wall method, is as follows:
\begin{enumerate}
    \item Program frustrated spin chain Hamiltonians (Eq. \ref{eq:frust_chain});  a length around 10 works well in practice on current devices.
    \item Perform anneals and collect statistics on domain wall positions, averaging over gauges and chain positions.
    \item If there are significant susceptibility effects, which occur when single body terms affect two-body terms that they overlap with, one needs to perform background susceptibility corrections, or remove terminal sites from statistics.
    \item Fit against numerically calculated distribution to extract bias strengths.
\end{enumerate}
\subsection{Assumptions and limitations}
\begin{itemize}
    \item All methods assume that the final distribution is a Boltzmann distribution with a known temperature. In the case of more coherent operation the methods would need to be adapted as discussed below:
    \begin{itemize}
        \item In Methods I and III, the model of system is only used for the final analysis. In a more coherent systems they could be adapted by comparing to a different classical model.
        \item In a highly coherent, low control error situation, method III would instead yield a ``particle-in-a-box'' distribution, which again could be used in fitting.
        \item Method II would be more difficult to extend to a more coherent case, but may be possible.
    \end{itemize}
    \item Method II involves taking a Taylor series of a Boltzmann distribution, and therefore assumes that the errors are small compared to the temperature.
    \item An important quantity for computation is the dimensionless ratio of control errors to the strength of the Ising terms when the annealing dynamics effectively stop (the so-called freeze time), and this quantity is difficult to know exactly.
    \item Methods I and II both are limited in the timescales they can probe by the sampling procedure:
    \begin{itemize}
       \item Method I is limited by the total time over which samples are taken. This is problematic, because it leads to a tradeoff between statistical accuracy, which increases with number of samples taken, and the range of frequencies it is sensitive to; the highest frequency which can be detected decreases with total sampling time, because the samples are averaged over this time period, and higher frequencies would cancel out.
       \item Method II is limited by the time between samples; this means that errors faster than the time between the individual samples will be missed.
       \item Method III should in principle capture all noise which can be treated as a control error, as there is no maximum or minimum frequency imposed by the measurement method.
    \end{itemize}
    \item Only method III performs measurements when some of the coupling elements are active.
\end{itemize}

\subsection{Source code}
A tutorial on calculating the single qubit control errors using the domain-wall method is provided in \sourceurl{quantum_annealers/single_qubit_control_errors}.

\metricbibliography
\end{refsegment}

\section{Size of the largest mappable fully connected problem}
\label{sec:annealer_largest_mappable_fully_connected_problem}
\begin{refsegment}
Since superconducting qubit based quantum annealers have limited connectivity in the underlying hardware graph, problems which are incompatible with this graph must be mapped in a way which maps multiple physical qubits to the same variable in the problem which is being solved. The physical qubits are strongly coupled in a way which forces them to take the same value, and thus correspond to a single binary variable. For this reason the size of problem which can be mapped with arbitrary interactions between all variables is typically smaller than the total number of qubits.

\subsection{Description}\label{sec:minorEmbedding}
Minor embedding was first proposed as a method of mapping problems which are not compatible with the hardware graph of an annealer by Choi in Ref.~\cite{Choi2008minor}. Minor embedding is a process by which multiple individual qubits are strongly coupled within a graph minor, a connected subgraph of the original graph, to form a new logical variable. In a sequel paper a prescription to minor embed to the D-Wave hardware graph available at the time was demonstrated~\cite{choi2011minor}. An alternative method of mapping problems based on parity and requiring fourth order interactions was later proposed by Lechner, Hauke, and Zoller~\cite{Lechner2015parity}, known as the LHZ method. While fourth order interactions are not directly available in most quantum annealing hardware, various mappings can be used to achieve this encoding by building higher order interactions from lower order ones~\cite{Leib2016transmonAnnealer,Rocchetto2016stabalizerAnneler,Chancellor2017circuit}. There have also been new methods proposed along these lines which do not require higher-order coupling~\cite{Palacios2024Embedding}.

The largest fully connected graph which can be minor embedded into a specified graph is given by one plus the treewidth of the graph~\cite{diestel2017graph}. In general, finding the optimal tree decomposition of a graph, necessary for computing the treewidth, is a hard computational problem, and minor embedding is also computationally hard. In practice, quantum annealers usually have an interaction graph with regular structure, making the process of computing a clique embedding or tree decomposition substantially easier. Tree decomposition is a process in which nodes are merged until the graph has a tree structure, which is one containing no loops. A useful quantity here is known as the treewidth. The idea is to look at decompositions which combine nodes to turn a graph with loops into one without (a ``tree''). The maximum number of nodes which must be combined in the best tree decomposition (minus one by convention) is known as the treewidth. The treewidth of the best tree decomposition determines the largest clique which can be embedded. For example, the embedding in Ref.~\cite{choi2011minor} was computed by hand. If some regular structure, which enables finding an embedding by hand, is not present, then software to compute graph embeddings can be used, for example the minor-miner software developed by D-Wave Systems Inc.\footnote{\url{https://github.com/dwavesystems/minorminer}} which uses the algorithm described in Ref.~\cite{Cai2014minor} for heuristic embedding. Such an embedding has no guarantee to be optimal, but can act as a lower bound of the largest fully connected problem size which can be mapped.

\subsection{Measurement procedure}

For a connectivity corresponding to a highly structured graph, as many devices including current D-Wave devices will have, an optimal embedding may be computable by hand, as was done in Ref.~\cite{choi2011minor}. This method is preferable, because one may be able to obtain a guarantee of maximum size of the graph, not simply a lower bound. Alternatively, approximate algorithms for finding minor embeddings exist, like those described in Ref.~\cite{Cai2014minor}, and can be used as a building block for embedding software such as D-Wave minor-miner\footnote{\url{https://github.com/dwavesystems/minorminer}}.

It is important to note that unlike many other metrics, an actual device does not need to be available to test the size of problem which could be mapped. For that reason this metric is ideal for devices which have not yet been fabricated to predict this aspect of their performance.

\subsection{Assumptions and limitations}

If a heuristic is used, then this metric is only a lower bound for the best possible mapping, since there is no guarantee that a better heuristic could not find a better one. However, in practice this may be close enough, since the existence of better embeddings, which cannot practically be found, will not have major implications for performance. A larger limitation is that optimization problems are not always fully connected, and may be possible to be embedded substantially more efficiently than a fully connected graph. This point becomes particularly relevant if a special-purpose annealer is developed with a hardware graph designed to match the problem structure~\cite{Roffe2019IsingDecode}. In these cases, it would be better to consider the largest problem with a relevant structure which can be mapped. Even for general purpose quantum annealers, if problems with a specific interaction structure are of specific interest to the user, then a more targeted approach using that structure may be justified. A concrete example here is the common interaction and constraint structure to all travelling salesman and quadratic assignment problems as used in Ref.~\cite{berwald2023Domain}.

\subsection{Source code}
An open source code for minor embedding (minor-miner) can be found at \url{https://github.com/dwavesystems/minorminer}.

\metricbibliography
\end{refsegment}

\section{Dimensionless sample temperature\label{sec:annealer_samp_temp}}
\begin{refsegment}
The dimensionless sample temperature is the ratio of the coupling strength to the environment temperature when the dynamics of the quantum annealer stops, known as the freeze time. The dimensionless sample temperature determines the rate of thermal errors. For optimisation applications, a sampling temperature which is as low as possible is ideal. Certain applications are based on thermal sampling and in these case a finite sampling temperature which is application dependant is desired. The freeze time, and therefore the temperature, is dependent on the problem which is programmed into an annealer. Since these devices are programmable, comparisons between devices can be made using the same problem, although for a fair comparison a problem should be chosen which is a subgraph of both hardware graphs to avoid additional mapping overheads complicating the comparison. A review of techniques to estimate temperature can be found in Ref.~\cite{Raymond2016Warming}. 

\subsection{Description}

Quantum annealers acting in a highly dissipative regime are known to produce statistical distributions, which are well approximated by thermal distributions. This fact has been used experimentally for probabilistic inference in Refs.~\cite{Chancellor2016maxEnt,Benedetti2017PGM}, which also included early estimation methods for the dimensionless temperature of the distribution. This fact has numerous applications in machine learning, including for example Boltzmann and Helmholtz machines~\cite{Adachi2015Boltzmann,Amin2018Boltzmann,Benedetti2018Helmholtz,Caldeira2019Botlzmann}.

Since quantum annealers do not perfectly sample thermal distributions, any temperature extracted will necessarily be approximate. Known methods for estimating temperature rely on comparing between the distribution, or possibly multiple distributions, returned by the annealer and a classical distribution. Different approaches are reviewed in Ref.~\cite{Raymond2016Warming}. In all approaches, with the exception of one which we discuss later, several ingredients are needed:
\begin{itemize}
    \item An objective function(s) against which to sample
    \item A classical sampling technique which operates at a known temperature and can return a good quality distribution for the given objective function(s)
    \item A measure of difference between distributions
\end{itemize}
The annealing schedule does not need to be known to estimate the sampling temperature.
The two approaches explored in Ref.~\cite{Raymond2016Warming} use dynamic programming and parallel tempering techniques to approximate the true thermal distribution with two types of randomly generated problems. The difference comes in how the distributions are compared. The first technique is to minimize the  Kullback-Leibler divergence, which statistically corresponds to finding the the most likely temperature to produce the given distribution, and is often abbreviated as KL. This quantity takes the form 
\begin{equation}
D_\mathrm{KL}\left[Q_a,Q_b\right]=\sum_xQ_a(x)\log\left(\frac{Q_a(x)}{Q_b(x)}\right),
\end{equation}
where $Q_a$ and $Q_b$ are two arbitrary probability distributions, and $D_\mathrm{KL}$ is the KL divergence.
It is important to note that this formula is not symmetric, and in particular will diverge if even a single $Q_b(x)=0$ when $Q_a(x)\neq 0$. For this reason, the samples from the physical annealer should always be taken as $Q_a(x)$, since otherwise a sample which is never seen in the experiment would cause the distribution to diverge, and the measure would be ill-defined unless the annealer exhaustively explored the solution space. Note that based on energy, the Boltzmann probability of a given state can still be estimated, even if it was not directly seen within the sample set, see Ref.~\cite{Raymond2016Warming} for details. A symmetrized version of this formula exists, where the divergence is removed, but it was not used in that study. In practice, for models of any reasonable size and temperature, the likelihood of seeing exactly the same state in two distributions is vanishingly small, so more sophisticated techniques are needed, such as binning by temperature, or comparing measured quantities~\cite{Raymond2016Warming}.

An alternative approach, which does not resort to directly comparing probabilities, is to instead compare measured quantities, which should be smaller the closer the distribution is to a true thermal distribution. In Ref. ~\cite{Raymond2016Warming} the quantity chosen was the pairwise correlation between variables, but numerous other choices could be made.  

Measurements performed within other works tend to be generalizations on the same central theme, for example Ref.~\cite{Chancellor2016maxEnt} compared the decoding of different qubits when the Ising model used within the annealer was treated as a classical error correction code, but did the fitting over multiple objective functions. 

The only estimate of the temperature that  does not require a classical sampling method is given in Ref.~\cite{Benedetti2016Temperature}. This method relies on rescaling the problem itself, and using the change in probabilities to observe different energies to calculate the difference in temperature.  The used formulation is in terms of inverse temperature. From this change, the absolute temperature can be calculated based on the known scaling. This technique  relies on the assumption that rescaling the problem statement corresponds linearly to a change in temperature. However, such assumption is often not true in real quantum annealers. This is often a bad assumption, because the freeze time of the annealer will depend on the energy scale of the problem.

\subsection{Measurement procedure}

The most common procedure for measuring temperature can be applied as follows:
\begin{enumerate}
    \item Decide on a Hamiltonian to program for the estimation, this may or may not require minor embedding (see \ref{sec:minorEmbedding}).
    \item Choose a quantity to measure and use to fit a temperature; how global or local this is will affect the temperature values measured. ~\cite{Raymond2016Warming}; this could also be a KL divergence between the annealer outputs and a classically determined distribution.
    \item Measure the chosen quantity experimentally on the annealer.
    \item Perform fitting against a classical estimate of the same quantity using methods such as Monte Carlo.
    \item The temperature value which gives the best agreement is the estimated value.
\end{enumerate}
An alternative method presented in Ref.~\cite{Benedetti2016Temperature} proceeds as follows:
\begin{enumerate}
    \item Decide on a Hamiltonian to program for the estimation.
    \item Choose two bins of energy ranges with different midpoints, such that the annealer has a non-negligible probability of sampling within each bin.
    \item Rescale the overall Hamiltonian and plot the natural logarithm of the ratio of probabilities in each bin versus the energy difference after rescaling; this plot should be a straight line.
    \item The slope of the line is the inverse temperature scaled by Boltzmann's constant.
\end{enumerate}

\subsection{Assumptions and limitations}

There are numerous challenges to overcome in this area. Key issues are:
\begin{enumerate}
    \item The distribution returned by an annealer will never exactly be a thermal distribution.
    \item Generating a thermal distribution to compare with can be a computationally challenging problem, depending on the objective function.
    \item The dimensionless temperature observed may differ for different objective functions.
    \item An interesting operating regime for annealers is the diabatic regime, where the final distribution is no longer approximately thermal.
\end{enumerate}

As a consequence of point 1, systematic effects have been observed in temperature estimations, such as the more global features of the distribution, which the estimate captures. This effect was reported in Ref.~\cite{Raymond2016Warming}, which called it global warming. This presents a considerable issue, since both global and more local sampling are likely to be important, so that multiple estimates, which give different results, can be considered equally valid. Temperature estimates can still be used as an indicative quantity.

Point 2 is a computational difficulty. Depending on the goal of the comparison, it can be circumvented by using an objective function for which it is not hard to obtain the thermal distribution classically. This approach was taken for example in Ref.~\cite{berwald2023Domain}. Alternatively, a large amount of classical computing power may be used, although at some point this may be impractical. It is worth noting that because of point 3, it may not always be suitable to choose a classically tractable objective function, since the cases of computational interest will be those where the problem is classically intractable.

With regards to point 4  this metric is only valid when the timescale of the anneal is effectively longer than the timescale associated with the coupling to its thermal bath. This regime is sometimes called the quasistatic regime, inspired by a Kibble-Zurek type approximation. Until recent technological improvements, D-Wave superconducting quantum annealers could only operate in this regime. However, recent work has demonstrated operation in a more coherent regime due to the capability to perform faster anneals~\cite{King2022Coherent}. Similarly, as more annealers operating on intrinsically less noisy platforms come into existence, experiments within this regime are likely to become more common.

In principle, one can always choose to run an annealer slowly enough that it operates in the quasistatic regime. However, this approach has the drawback that it won't give accurate information about the coherent regime, where closed system approximations are valid as opposed to a Kibble Zurek type approximation of perfect thermalisation and freezing. A comparison only in the quasistatic regime therefore may miss the most relevant features for performance. There is a need for metrics which measure a quantum annealers' performance for sampling in a coherent regime, however we are not aware of any such metrics at the time of writing.

\metricbibliography
\end{refsegment}

\def\thechapter{M9}
\chapter{Metrics for boson sampling devices}
\label{chapter:boson_sampling}
 
\begin{refsegment}

\section{Hardware characterization and model as metrics}
\label{sec:boson_sampling_hardware_characterization_and_model_as_metrics}
Hardware characterization of a boson sampling device consists in the development of a set of protocols that enable extraction of sufficient information on the device, and that allow one to model and predict its behaviour using the theory of quantum optics. Such simulations may not be scalable, but they allow predicting properties of devices at moderate sizes.

Boson sampling machines are composed of sources, linear-interferometers and detectors, so that the characterization of the device can be reduced to the characterization of its parts. As opposed to the challenging characterization of qubit based devices such as superconducting circuits, which are prone to complex cross-talks and other collective effects, characterizing boson sampling hardware is a simpler task. The reason is that the dynamics of boson sampling devices are represented by the subset of Gaussian operations that can be efficiently represented and characterized, as opposed to an $N_{\mathrm{q}}$ qubit device, where its full tomography scales exponentially with the number of qubits $N_{\mathrm{q}}$. 

In the last decades, quantum optics experimental labs have been developing  simple techniques to characterize squeezed states sources, interferometers and detectors. For the squeezing sources the main parameter to extract is the squeezing itself, the thermal noise, and potentially the distinguishability of the source~\cite{frascella2021overcoming,park2024single}. As many experiments have a single source multiplexed over the multiple input of the linear interferometer, one often only needs to characterize a unique source and the multiplexing interferometer. 

Linear-interferometers can be characterized using classical coherent light, where one extracts the transmission matrix~\cite{popoff2010measuring} and two-point coherent correlations to obtain the phases~\cite{lobino2008complete,rahimi2013direct,dhand2016accurate}.
Typical experiments need to have a coherent source to stabilize the interferometer, 
so that the tomography of the linear-optics circuit can be done using that resource.

The detectors are usually characterized using sources of single-photons or attenuated coherent states that have very small average photon numbers. This allows one to extract the efficiency (false negative) and dark counts (false positive) information~\cite{maring2024versatile}.

While the independent characterization of the different elements of a boson sampling device is the common approach in the field, there is no guarantee that, once merged into the final boson sampling device, all the properties will be preserved, or that we may have access to all the components independently. 

It is therefore preferable to have techniques that allow one to extract all information from the actual final device conceived as a unique and non-divisible entity,  specially when all the platform is integrated in a single chip. This brings some additional challenges that can be overcome. For example, a large-size interferometer tomography can also be done using single-photons sources, therefore one can easily characterize a boson sampling on an integrated chip. 

\section{Quantum advantage demonstration as metric}
\label{sec:boson_sampling_quantum_advantage_demonstration_as_metric}
The Jiuzhang 1.0 and 2.0 demonstrations by UCTS~\cite{Jiuzhang1,Jiuzhang2}, and Xanadu's implementation in 2022
\cite{Xanadu22}, use a set of similar validation tests for quantum advantage: (i) Bayesian test against adversarial distributions; (ii) high output generation test; (iii) the comparison of truncated first to k-th order correlation functions. Each technique has it own advantages and disadvantages. As we will discuss in more detail at the end of this section, it is unlikely that an experimental test that fully certifies quantum advantage for sampling problems can be found.

An important aspect for quantum advantage experiments based on random circuit or boson sampling devices is the concept of ground truth, which corresponds to the distribution the experiment should be sampling from if everything worked as it should, and also the one any spoofing strategy should try to emulate. 

Our presentation below is an adaptation of the presentation in section 3 in Ref.~\cite{martínezcifuentes2023classical}, with some additional information and discussions. 

\subsection{Bayesian Tests}
Bayesian tests compare how good the ground truth is at explaining the observed data relative to other hypotheses such as thermal states, coherent states, distinguishable squeezed states, and uniform probability distributions. Here we provide a description of a number of such tests, which in future as the field progresses may become separate metrics.
Consider a set $\bm{S} = \{\bm{s}_1,\dots,\bm{s}_L\}$
of $L$ experimental samples, each of them containing $N_c$ clicks. A click here refers to the a photon detected in a detector. The probability of obtaining one of these samples, given that it has $N_c$ clicks, under the hypothesis $\text{HYP}$ is given by
\begin{align}
    p_{(\text{HYP})}(\bm{s}_k|N_c) = \frac {p_{(\text{HYP})}(\bm{s}_k)}  {p_{(\text{HYP})}(N_c)},  
\end{align}
where $p_{(\text{HYP})}(\bm{s}_k)$ is the probability of sample $\bm{s}_k$ under hypothesis $\text{HYP}$, and $p_{(\text{HYP})}(N_c)$ is the grouped probability of obtaining $N_c$ clicks in total, again under the hypothesis $\text{HYP}$. The probability of obtaining the set of samples $\bm{S}$ under a given hypothesis $\text{HYP}$ takes the form
\begin{equation}
    p_{(\text{HYP})}(\bm{S}|N_c) = \prod_{k=1}^L p_{(\text{HYP})}(\bm{s}_k|N_c).
    \label{eq:joint_gt_probability}
\end{equation}

We define the Bayesian ratio, $r_\text{B}(N_c)$, which can be interpreted as the probability assigned to the ground truth hypothesis $\text{GTH}$ versus an alternative hypothesis $\text{ALT}$ (thermal states, coherent states, ...) for a given number of clicks, as
\begin{align}
    \begin{split}
        r_\text{B}(N_c)&=\frac{p_{(\text{GTH})}(\bm{S}|N_c)}{p_{(\text{GTH})}(\bm{S}|N_c) + p_{(\text{ALT})}(\bm{S}|N_c)}.
    \end{split}
    \label{eq:bayesian_ratio}    
\end{align}
The Bayesian test consists in checking the convergence of $r_\text{B}(N_c)$ when the number of samples is increased: if $r_B(N_c) \rightarrow 1$ for any $N_c$, we conclude that the ground truth hypothesis is more likely to describe the experimental samples. Conversely, if $r_B(N_c) \rightarrow 0$ for any $N_c$, the alternative hypothesis becomes more likely.

An alternative way to express this test is obtained by using entropies by writing 
	\begin{align}
		\begin{split}
			\Delta H(N_c) =& -\frac{1}{L}\sum_{k=1}^L\ln\left[p_{(\text{GTH})}(\bm{s}_k|N_c)\right]
			+\frac{1}{L}\sum_{k=1}^L\ln\left[p_{(\text{ALT})}(\bm{s}_k|N_c)\right],\\
			=& H_{(\text{GTH})}(N_c) - H_{(\text{ALT})}(N_c).
		\end{split}
		\label{eq:bay_cross_ent_diff}
	\end{align}
	The quantities $H_{(\text{GTH})}$ and $H_{(\text{ALT})}$ are estimators of the cross-entropy, for a given number of counts, of the ground truth and alternative distribution relative to the real probability distribution of the experimental samples. One can see that $r_\text{B}(N_c) = \left[1+\exp(L\Delta H(N_c))\right]^{-1}$ and, for a increasing number of samples, the condition $r_B(N_c) \rightarrow 1$ is equivalent to $\Delta H(N_c)<0$, while $r_B(N_c) \rightarrow 0$ is equivalent to $\Delta H(N_c)>0$. An important step for the computation of the Bayesian test is the determination of the grouped click probability distributions $p_{(\text{GTH})}(N_c)$, for which one can use the efficient method in Ref.~\cite{Drummond22}.
 
It is important to make two remarks. The first is that Bayesian tests give the degree of confidence of one hypothesis over another. Therefore, if the ground truth explains better the experimental data than a given alternative, it does not discard the possibility that there could exist another distribution that can be classically simulated efficiently, and that also describes the experimental data better. It therefore allows discarding known suspects, but not unknown ones.

In addition, the Bayesian test relies on the computation of probabilities of individual samples, which for the ground truth is a computationally hard task for patterns with a high number of clicks. Therefore, the test cannot be used to verify that a large-size Gaussian boson sampling experiment achieves quantum computational advantage. Rather, the Bayesian test is used to build up confidence in the correct functioning of the setup by ruling out possible classical hypotheses explaining the samples.

\subsection{Heavy Output Generation}
The second test looks at how well the samples generated by the experiment have so-called heavy outputs, corresponding to events
with high probability, in the ideal distribution relative to samples generated by classically efficient methods.
One defines the heavy output generation (HOG) ratio, $r_\text{HOG}(N_c)$, as
	\begin{align}
		\begin{split}
			r_\text{HOG}(N_c)&=\frac{p_{(\text{SQUE})}(\bm{S}|N_c)}{p_{(\text{SQUE})}(\bm{S}|N_c) + p_{(\text{SQUE})}(\bm{S'}|N_c)},
		\end{split}
		\label{eq:hog_ratio}    
	\end{align}
	where $\bm{S'}=\{\bm{s'}_1,\dots,\bm{s'}_L\}$ is a set of $L$ samples obtained from the squashed states distribution (each one with $N_c$ clicks). 
 $p_{(\text{GTH})}(\bm{S}|N_c)$ is computed according to Eq.~\eqref{eq:joint_gt_probability}. It is important to notice that now all the probabilities are computed from the ground truth distribution, and the difference relies on the samples resulting from the actual experimental data $\bm{S}$ and those generated from our hypothesis distribution $\bm{S'}$.
 When $r_\text{HOG}(N_c)\rightarrow 1$ we conclude that the experimental samples have higher ground truth probability, while $r_\text{HOG}(N_c)\rightarrow 0$ when the samples from the hypothesis distribution have higher ground truth probability.
As in the case of the Bayesian test, we have
\begin{align}
    \begin{split}
        \Delta E(N_c) =& -\frac{1}{L}\sum_{k=1}^L\ln\left[p_{(\text{GTH})}(\bm{s}_k|N_c)\right]
        +\frac{1}{L}\sum_{k=1}^L\ln\left[p_{(\text{GTH})}(\bm{s'}_k|N_c)\right],\\
        =& E_{(\text{GTH})}(N_c) - E'_{(\text{GTH})}(N_c).
    \end{split}
    \label{eq:hog_cross_ent_diff}  \end{align}
In terms of the cross-entropy difference $\Delta E(N_c)$~\cite{villalonga2022efficient}, $r_\text{HOG}(N_c) = \left[1 + \exp(L\Delta E(N_c))\right]^{-1}$ and, for increasing $L$, the conditions  $r_\text{HOG}(N_c)\rightarrow 1$ and $r_\text{HOG}(N_c)\rightarrow 0$ are equivalent to $\Delta E(N_c)<0$ and $\Delta E(N_c)>0$, respectively.

\subsection{Click cumulants}
As we mention before, the Bayesian test relies on the computation of probabilities of individual samples, which for the ground truth is a computationally hard task for patterns with a high number of clicks. Similar to qubits-based verification of quantum advantage, any cross-entropy approach is not scalable. To circumvent this problems a series of test that consist on certifying the correlations in the system have been proposed.

These correlation functions are given by cumulants, truncated correlation defined in terms of moments of a multidimensional random variable $\bm{V} = \left(V_1,V_2,\ldots,V_n\right)$ as
	\begin{align}
		\kappa(V_1,\dots,V_n) =\sum_\pi& (|\pi|-1)!(-1)^{|\pi|-1}
		\times\prod_{B\in\pi} \left\langle \prod_{i\in B}V_i \right\rangle,
	\end{align}
	where $\pi$ runs through the list of all partitions of $\{ 1, ..., n \}$, $B$ runs through the list of all blocks of the partition $\pi$, and $|\pi|$ is the number of parts in the partition.
	Note that the first order cumulants are simply the means $\kappa(V_i) = \braket{V_i}$, and that the second order cumulants are the covariances $\kappa(V_i, V_j) = \braket{V_i V_j} - \braket{V_i} \braket{V_j}$. It is straightforward to see that for binary outcomes, which have zero and non-zero photon number, moments of the distribution correspond to marginal probabilities, given by
	\begin{align}
		\braket{V_{i_1}\ldots V_{i_n}} = p\left(i_1=1,\ldots, i_n = 1 \right).
	\end{align}

Most experiments compare cumulants up to a constant order (fourth to sixth in general),
and compare how well the correlations in the observed
data match the correlations predicted by alternative hypothesis.
This approach suffers from the same weakness as the Bayesian method, where there could be yet to be discovered hypothesis that can be efficiently simulated on classical devices.

\subsection{Final discussion}
As discussed above, verifying that we are in the hardness regime is itself not computationally tractable for photonic devices, but it is important to note that the same holds for qubit-based experiments. Indeed, it was shown in Ref.~\cite{StilckFranca2022gameofquantum} that for random circuits under a scenario of verification based solely on classical communication, two possibilities exist: 
(i) sampling is hard to simulate for classical computers, but verification is not scalable; (ii)
an efficient verification protocol exists, but there also exists a sampling strategy that exploits the verification tests to efficiently simulate the sampling from the circuit.
This result may hold true also for boson sampling devices.

It is reasonable to conclude that focusing on problems where the solutions can be classically verified, or provide a quantifiable advantage for an end-user application, may be the only way to provide evidence for quantum advantage for boson sampling devices.

\metricbibliography
\end{refsegment}

\newpage
\vspace*{-4\baselineskip}
\def\thechapter{M10}
\chapter{Metrics for neutral atoms devices}
\label{chapter:neutral_atoms}

\section{Analogue process fidelity}
\label{sec:neutral_atoms_analog_benchmark}
\begin{refsegment}

The analogue process fidelity compares and verifies measured state and process fidelities against a classical model to enable learning of Hamiltonian control parameters and error channels whilst using finite sampling \cite{mark23}.

\subsection{Description} 

As discussed in Sec.~\ref{sec:neutralAtoms}, neutral atom platforms offer large numbers of qubits and programmable spin models suitable for analogue quantum computation targeting classical optimization problems as well as efficient quantum simulation. In the paradigm of analog optimization, it is important to evaluate the performance of the hardware accounting both for errors and ensuring the Hamiltonian parameters encoded by the user match those implemented on the hardware to avoid encoding the wrong classical optimization problem. 
Whilst there exist a wide range of metrics for benchmarking digital processors with arbitrary unitary operations, analog based processors typically offer a limited subset of control parameters such as global driving parameters, local site-dependent detunings and programmable spin-models using the specific geometric configuration of the atomic tweezers.
To overcome this challenge, in Ref.~\cite{mark23} a benchmarking protocol is proposed that enables a system to be prepared in an initial state $\rho(t)$, evolved for a time $t$ under the target Hamiltonian $\mathcal{H}$ to induce quench dynamics, and then samples $M$ measurements of the output bit-strings measured in the $\{\ket{z}\}$ basis. This results in the empirical distribution $q(z,t)$ describing the distribution of strings $z$ measured at time $t$.
This is compared against a theoretical target distribution $p(z,t)$ calculated using classical simulation, and the infinite-time average $p_\mathrm{avg}(z) = \lim_{T \rightarrow \infty} \frac{1}{T}\int_0^T p(z,t)~dt$. These are combined to obtain the rescaled outcome probabilities $~\tilde{p}(z,t) = p(z,t)/p_\mathrm{avg}(z)$ and the normalization factor $\mathcal{Z}(t) =  \sum_z p_\mathrm{avg}(z) \tilde{p}(z,t)^2$.
From these parameters one can evaluate 
\begin{equation}
\hat{F}_d(t) = \frac{2}{M} \Big[\sum_{i=1}^M \tilde{p}(z_i,t) \Big]/\mathcal{Z}(t) - 1 \approx F_d(t),
\end{equation}
which approximates the fidelity $F(t)=\bra{\Psi(t)}\rho(t)\ket{\Psi(t)}$, where $\rho(t)$ represents the state of the experimental system and $\ket{\Psi(t)}$ the calculated ideal output state expected from the evolution of $\ket{\Psi(t=0)}=\ket{\Psi_0}$ under the target Hamiltonian $\mathcal{H}$. Analysis of statistical errors shows that using this approach percent level precision is possible using finite samples sizes $M\sim10^3$.
This works under the same approach as digital benchmark methods based on deep random unitary circuits, as it can be shown that the rescaled probabilities $\tilde{p}(z)$ follow a Porter-Thomas (PT) distribution with unity mean~\cite{mark23}.

This procedure can be used to learn Hamiltonian components by scaling parameters in the computational model and finding the values which maximize the observed output state fidelity as demonstrated experimentally for extracting randomised light-shifts and global Rabi frequency on a chain of 15 atoms~\cite{choi23}. By evaluating the evolution of $F_d(t)$ with $t$ using different error models it is also possible to calibrate noise parameters~\cite{shaw24}. 
This can also be used to facilitate comparison of analog processors to what would be possible using digital hardware.

\subsection{Measurement procedure}
This procedure follows the approach outlined in Ref.~\cite{mark23}, and is as follows:
\begin{enumerate}
    \item Experimentally prepare an initial state $\rho_0$ that approximates a pure state $\ket{\Psi_0}\bra{\Psi_0}$.
    \item Evolve the system under its natural Hamiltonian $\mathcal{H}$ for a time $t$.
    \item Measure the evolved state $\rho(t)$ in a natural basis, obtaining configurations $\{zj\}^M_{j=1}$.
    \item Classically compute
    \begin{itemize}
    \item $p(z,t) \equiv \abs{\bra{z}\ket{\Psi(t)}}^2 = \abs{\bra{z}\exp(-iHt)\ket{\Psi_0}}^2$
    \item $p_\mathrm{avg}(z) \equiv \lim_{T \rightarrow \infty} \frac{1}{T}\int_0^T p(z,t)~dt$
    \item $~\tilde{p}(z,t) \equiv p(z,t)/p_\mathrm{avg}(z)$
    \item $\mathcal{Z}(t) \equiv  \sum_z p_\mathrm{avg}(z) \tilde{p}(z,t)^2$
\end{itemize}
    \item Evaluate $\hat{F}_d(t) \equiv  \frac{2}{M} \Big[\sum_{i=1}^M \tilde{p}(z_i,t) \Big]/\mathcal{Z}(t) - 1 \simeq F_d(t)$~ which approximates the fidelity $F(t)=\bra{\Psi(t)}\rho(t)\ket{\Psi(t)}$.
\end{enumerate}

\subsection{Assumptions and limitations}
\begin{itemize}
    \item The assumption in this work is that the eigenvalues $\{ E_i\}$ of $\mathcal{H}$ possess no resonant structures. This condition is expected to hold for generic ergodic Hamiltonians.
    \item Whilst this method is limited by the requirement to benchmark against classical calculations, this can be extended to extrapolate results of approximate classical simulations to provide a global fidelity estimator for an analog simulator operating outside of the classically simulable regime as demonstrated with arrays of up to $N=60$ entangled atoms~\cite{shaw24}. 
    \item While this protocol is applicable to generic programmable Hamiltonians, it may fail in special cases where are long times the system is not described by a Porter-Thomas distribution. This occurs in instances with weakly- or non-ergodic dynamics, or the presence of correlated non-local errors. See~\cite{mark23} for details.
\end{itemize}

\metricbibliography
\end{refsegment}

\section{Trap lifetime}
\label{sec:neutral_atoms_trap_lifetime}
\begin{refsegment}
For atoms trapped in optical tweezers the characteristic $1/e$~trap lifetime, $\tau$, is relevant, since for $N$ atoms the effective array lifetime is $\tau/N$, limiting the time useful computation can be performed.

\subsection{Description}
Due to the relatively shallow trap depth of neutral atom tweezers it is possible for atoms to be ejected from the traps during computation due to collisions with residual background gases~\cite{wintersperger2023neutral,ebadi2021quantum,PhysRevA.102.063107,schymik2021single,graham22}. For readout methods dependent on ejecting atoms from the trap prior to imaging, this leads to increased state preparation and measurement (SPAM) errors~\cite{wintersperger2023neutral}, whilst also introducing errors during computation due to stochastic loss of atoms from the array. For room temperature systems typically $\tau\sim10-20$~s, whilst at cryogenic temperatures lifetimes exceeding 10~minutes have been demonstrated~\cite{schymik2021single}.

\subsection{Measurement procedure}
Measurement of trap lifetime is performed as follows:
\begin{enumerate}
    \item Load atom array with $N$ atoms and determine which sites are loaded initially
    \item Hold atoms in the trap for variable hold times $t$
    \item Read out the traps to measure the number of remaining atoms
    \item Repeat steps 1-3 for a certain number of shots to calculate the survival probability $p(t) = N_\mathrm{remaining}(t)/N$
    \item Fit the survival probability vs $t$ using $p(t)=p_0e^{-t/\tau}$ to extract the characteristic trap lifetime $\tau$
\end{enumerate}

\subsection{Assumptions and limitations}
\begin{itemize}
    \item It is assumed that the losses are due to single-body collisions, as expected for single-atom loading.
    \item The metric does not take into account the case where controlled dynamics are applied on the system.
\end{itemize}

\printbibliography[heading=subbibliography,segment=\therefsegment]
\end{refsegment}

\section{Reconfigurable connectivity}
\label{sec:neutral_atoms_reconfigurable_connectivity}
\begin{refsegment}
For most quantum computing architectures the connectivity is static, and given by the hardware setup. This is captured by the pairwise connectivity hardware architectural property ~(See \ref{sec:connectivity}). For neutral atom systems with mobile tweezers it is possible to dynamically reconfigure the array geometry during computation to implement arbitrary couplings \cite{bluvstein22,bluvstein2024logical,wintersperger2023neutral}. The reconfigurable connectivity is a hardware architectural property that states whether such a capability is available in a system, and if so to what extent.

\subsection{Description}
Using atoms in mobile tweezer traps enables atoms to be transported whilst preserving coherence, allowing arbitrary qubit connectivity to be implemented, as demonstrated in Ref.~\cite{bluvstein22} for realizing topological coupling graphs.

\subsection{Measurement procedure}
 The degree of reconfigurability is limited by the available number of mobile tweezers and hardware restrictions on allowed moves, with typical implementations favoring use of parallel column or row operations. Another approach to measure re-configurable connectivity is by considering reduction in required number of two qubit gate operations compared to a static qubit connectivity when using an optimized compiler~\cite{tan23}.

\subsection{Assumptions and limitations}
\begin{itemize}
    \item Utility of reconfigurable geometries may be limited without low-level access to specify arbitrary moves. 
\end{itemize}

\metricbibliography
\end{refsegment}
\let\thechapter\theoldchapter
\newpage
\part{Conclusions}
\renewcommand{\thechapter}{\Roman{chapter}}
\setcounter{chapter}{4}
\unchapter{Conclusions}
\renewcommand{\thechapter}{\arabic{chapter}}
\setcounter{chapter}{16}
\label{part:conclusions}
\makeatletter
\renewcommand{\sectionlinesformat}[4]{%
    \@tempswafalse
        \Ifstr{#1}{section}{%
            \hspace*{#2}%
                {%
                    \parbox{\dimexpr\linewidth-2\fboxrule-2\fboxsep-#2}{%
                        \raggedsection
                        \@hangfrom{#3}{#4}%
                    }%
                }%
            }{%
        \@hangfrom{\hskip #2#3}{#4}%
    }%
}
\makeatother
\vspace{1em}
In this article a collection of metrics is provided, with a consistent format for each metric that includes its definition, a description of the methodology, the main assumptions and limitations, and a linked open source software implementing the methodology. The open source software package transparently demonstrates the methodology and can also be used in practical evaluations of the value of metrics. 
As shown in this article, the level of maturity of the different metrics varies significantly. For example, metrics such as randomized benchmarking based fidelities are well established, while approaches to benchmark full applications still allow for large variations based on the many choices of algorithms, parameters, and success criteria used to measure device performance. Research on metrics and benchmarks is actively progressing, and is particularly important for the improvement of the efficiency and scalability of metrics across all categories to benchmark quantum computers with well over a thousand qubits, for metrics on device speed and stability, as well as in benchmarks for non-circuit based approaches.

One existing problem is that, even to evaluate the well established metrics, many different approaches with slightly diverging technical details are used in practice. While all approaches provide similar overall qualitative performance values, these diverging details in the technical implementations make quantitative comparisons difficult. One of the resulting risks is that it makes it difficult to objectively determine the progress of the quantum computing field towards quantum advantage. Combined with the fact that there are many largely different hardware platforms and quantum computing approaches, this makes it important to reduce the metrics used to evaluate the performance for a given hardware platform to a minimal but representative set agreed across manufacturers, algorithm developers and end users. 
At a minimum, to fairly and objectively benchmark quantum computers across manufacturers, these benchmarks need to follow some agreed common approaches. As outlined in the Discussion and Outlook section (Sec. \ref{chapter:conclusions_outlook}), the need for such agreement has led to international standardization activities on metrics and benchmarking, and in that section a number of work items to be considered for standardization are proposed.
A collection of metrics, such as the one presented in this article, together with the associated open-source software needed to evaluate them, can contribute to the development of standardized benchmarks for quantum computers and speed up the progress of the field towards practical quantum advantage.

\section*{Acknowledgements}
This project was funded and supported by the UK National Quantum Computer Centre [NQCC200921], which is a UKRI Centre and part of the UK National Quantum Technologies Programme (NQTP). JP acknowledges support from EPSRC (Grant No. EP/T005386/1). AR acknowledges support from UKRI (Grant No. MR/T041110/1). We thank Yannic Rath, Daniel Rodrigo Albert, Wang Wong, Nikolaos Schoinas, Masum Uddin, Manognya Acharya, David Headley, Shushmi Chowdhury, and Manav Babel for useful discussions and for reviewing manuscript and software. We thank the Collaborative Computational Project Quantum Computing (ccp-qc, EPSRC Grant No. EP/T026715/2), the EPSRC Quantum Computing and Simulation Hub (EP/T001062/1), the UKRI ExCALIBUR project QEVEC (EP/W00772X/2), and the Superconducting Quantum Materials and Systems Center (SQMS, DE-AC02-07CH11359), for facilitating collaboration and discussion. 

\begingroup
\setstretch{0.8}
\setlength\bibitemsep{0.4pt}
\AtNextBibliography{\scriptsize}
\printbibliography[heading=bibintoc,title={References}]

@article{10.1063/1.1723701,
  title = {Canonical Decompositions of N-Qubit Quantum Computations and Concurrence},
  author = {Bullock, Stephen S. and Brennen, Gavin K.},
  date = {2004-06},
  journaltitle = {J. Math. Phys},
  volume = {45},
  number = {6},
  eprint = {https://pubs.aip.org/aip/jmp/article-pdf/45/6/2447/19273936/2447\_1\_online.pdf},
  pages = {2447},
  issn = {0022-2488},
  doi = {10.1063/1.1723701},
  url = {https://doi.org/10.1063/1.1723701}
}

@article{shaw24,
	author = {Shaw, Adam L. and Chen, Zhuo and Choi, Joonhee and Mark, Daniel K. and Scholl, Pascal and Finkelstein, Ran and Elben, Andreas and Choi, Soonwon and Endres, Manuel},
	journal = {Nature},
	number = {8006},
	pages = {71},
	title = {Benchmarking highly entangled states on a 60-atom analogue quantum simulator},
	volume = {628},
    doi = {10.1038/s41586-024-07173-x},
	year = {2024}}

@article{lanthaler23,
	 title = {{Rydberg-Blockade-Based Parity Quantum Optimization}},
  author = {Lanthaler, Martin and Dlaska, Clemens and Ender, Kilian and Lechner, Wolfgang},
  journal = {Phys. Rev. Lett.},
  volume = {130},
  issue = {22},
  pages = {220601},
  numpages = {6},
  year = {2023},
  month = {5},
  publisher = {American Physical Society},
  doi = {10.1103/PhysRevLett.130.220601},
  url = {https://link.aps.org/doi/10.1103/PhysRevLett.130.220601}
}

@article{petrosyan24,
  title = {Fast measurements and multiqubit gates in dual-species atomic arrays},
  author = {Petrosyan, D. and Norrell, S. and Poole, C. and Saffman, M.},
  journal = {Phys. Rev. A},
  volume = {110},
  issue = {4},
  pages = {042404},
  numpages = {9},
  year = {2024},
  month = {10},
  publisher = {American Physical Society},
  doi = {10.1103/PhysRevA.110.042404},
  url = {https://link.aps.org/doi/10.1103/PhysRevA.110.042404}
}

@article{corlett24,
      title={Speeding up quantum measurement using space-time trade-off}, 
      author={C. Corlett and I. Čepaitė and A. J. Daley and C. Gustiani and G. Pelegrí and J. D. Pritchard and N. Linden and P. Skrzypczyk},
      year={2024},
      eprint={2407.17342},
      journal={arXiv:2407.17342},
      primaryClass={quant-ph},
      url={https://arxiv.org/abs/2407.17342}, 
}

@article{deoliveira24,
      title={Demonstration of weighted graph optimization on a Rydberg atom array using local light-shifts}, 
      author={A. G. de Oliveira and E. Diamond-Hitchcock and D. M. Walker and M. T. Wells-Pestell and G. Pelegrí and C. J. Picken and G. P. A. Malcolm and A. J. Daley and J. Bass and J. D. Pritchard},
      year={2024},
      journal={arXiv:2404.02658},
      archivePrefix={arXiv},
      primaryClass={quant-ph},
      url={https://arxiv.org/abs/2404.02658}, 
}

@article{park24,
	author = {Park, Juyoung and Jeong, Seokho and Kim, Minhyuk and Kim, Kangheun and Byun, Andrew and Vignoli, Louis and Henry, Louis-Paul and Henriet, Lo\"{\i}c and Ahn, Jaewook},
	journal = {Phys. Rev. Res.},
	pages = {023241},
	title = {{Rydberg-atom experiment for the integer factorization problem}},
	volume = {6},
	year = {2024}}

@article{pichler18,
	arxivprefix = {arXiv},
	author = {H. Pichler and S.-T. Wang and L. Zhou and S. Choi and M. D. Lukin},
	journal = {arXiv:1808.10816},
	primaryclass = {quant-ph},
	title = {{Quantum Optimization for Maximum Independent Set Using Rydberg Atom Arrays}},
	year = {2018},
    url = {https://doi.org/10.48550/arXiv.1808.10816}
}

@article{choi23,
	author = {Choi, Joonhee and Shaw, Adam L. and Madjarov, Ivaylo S. and Xie, Xin and Finkelstein, Ran and Covey, Jacob P. and Cotler, Jordan S. and Mark, Daniel K. and Huang, Hsin-Yuan and Kale, Anant and Pichler, Hannes and Brand{\~a}o, Fernando G. S. L. and Choi, Soonwon and Endres, Manuel},
	journal = {Nature},
	number = {7944},
	pages = {468},
	title = {Preparing random states and benchmarking with many-body quantum chaos},
    doi = {10.1038/s41586-022-05442-1},
	volume = {613},
	year = {2023}}

@article{mark23,
  title = {Benchmarking Quantum Simulators Using Ergodic Quantum Dynamics},
  author = {Mark, Daniel K. and Choi, Joonhee and Shaw, Adam L. and Endres, Manuel and Choi, Soonwon},
  journal = {Phys. Rev. Lett.},
  volume = {131},
  issue = {11},
  pages = {110601},
  numpages = {7},
  year = {2023},
  doi = {10.1103/PhysRevLett.131.110601}
}

@article{10.1063/1.3435463,
  title = {Fast Reset and Suppressing Spontaneous Emission of a Superconducting Qubit},
  author = {Reed, M. D. and Johnson, B. R. and Houck, A. A. and DiCarlo, L. and Chow, J. M. and Schuster, D. I. and Frunzio, L. and Schoelkopf, R. J.},
  date = {2010-05},
  journaltitle = {Appl. Phys. Lett.},
  volume = {96},
  number = {20},
  eprint = {https://pubs.aip.org/aip/apl/article-pdf/doi/10.1063/1.3435463/13988301/203110\_1\_online.pdf},
  pages = {203110},
  doi = {10.1063/1.3435463},
  url = {https://doi.org/10.1063/1.3435463}
}

@article{10.1063/5.0139825,
  title = {A Compact and Versatile Cryogenic Probe Station for Quantum Device Testing},
  author = {family=Kruijf, given=Mathieu, prefix=de, useprefix=true and Geyer, Simon and Berger, Toni and Mergenthaler, Matthias and Braakman, Floris and Warburton, Richard J. and Kuhlmann, Andreas V.},
  date = {2023-05},
  journaltitle = {Rev. Sci. Instrum.},
  volume = {94},
  number = {5},
  pages = {054707},
  issn = {0034-6748},
  doi = {10.1063/5.0139825},
  url = {https://doi.org/10.1063/5.0139825}
}

@inproceedings{10.1109.9668289,
  title = {Quantum {{Computing}} in the {{Cloud}}: {{Analyzing}} Job and Machine Characteristics},
  booktitle = {2021 {{IEEE Int Symp Workload Charact}}. {{IISWC}}},
  author = {Ravi, Gokul Subramanian and Smith, Kaitlin N. and Gokhale, Pranav and Chong, Frederic T.},
  date = {2021-11},
  pages = {39},
  doi = {10.1109/IISWC53511.2021.00015},
  url = {https://doi.org/10.1109/IISWC53511.2021.00015}
}

@inproceedings{10.1145/3307650.3322273,
  title = {Full-Stack, Real-System Quantum Computer Studies: Architectural Comparisons and Design Insights},
  booktitle = {Proc 46th {{Int Symp Comput Arch}}.},
  author = {Murali, Prakash and Linke, Norbert Matthias and Martonosi, Margaret and Abhari, Ali Javadi and Nguyen, Nhung Hong and Alderete, Cinthia Huerta},
  date = {2019},
  series = {{{ISCA}} '19},
  pages = {527},
  publisher = {Association for Computing Machinery},
  location = {New York, NY, USA},
  doi = {10.1145/3307650.3322273},
  url = {https://doi.org/10.1145/3307650.3322273},
  isbn = {978-1-4503-6669-4},
  pagetotal = {14}
}

@inproceedings{10.1145/3373376.3378477,
  title = {Software {{Mitigation}} of {{Crosstalk}} on {{Noisy Intermediate-Scale Quantum Computers}}},
  booktitle = {Proc 25th {{Int Conf Arch}}. {{Support Progra Lang Oper}}. {{Syst}}},
  author = {Murali, Prakash and Mckay, David C. and Martonosi, Margaret and Javadi-Abhari, Ali},
  date = {2020},
  series = {{{ASPLOS}} '20},
  pages = {1001},
  publisher = {Association for Computing Machinery},
  location = {New York, NY, USA},
  doi = {10.1145/3373376.3378477},
  url = {https://doi.org/10.1145/3373376.3378477},
  isbn = {978-1-4503-7102-5},
}

@article{10.1145/502090.502098,
  title = {Some Optimal Inapproximability Results},
  author = {Håstad, Johan},
  date = {2001-07},
  journaltitle = {J. ACM},
  volume = {48},
  number = {4},
  pages = {798},
  publisher = {Association for Computing Machinery},
  location = {New York, NY, USA},
  issn = {0004-5411},
  doi = {10.1145/502090.502098},
  url = {https://doi.org/10.1145/502090.502098},
  issue_date = {July 2001},
  pagetotal = {62}
}

@article{A_Yu_Kitaev_1997,
  title = {Quantum Computations: Algorithms and Error Correction},
  author = {Kitaev, A Yu},
  date = {1997-12},
  journaltitle = {Russ. Math. Surv.},
  volume = {52},
  number = {6},
  pages = {1191},
  doi = {10.1070/RM1997v052n06ABEH002155},
  url = {https://dx.doi.org/10.1070/RM1997v052n06ABEH002155}
}

@article{Aaronson2010,
  title = {The {{Computational Complexity}} of {{Linear Optics}}},
  author = {Aaronson, Scott and Arkhipov, Alex},
  date = {2013},
  journaltitle = {Theory Comput.},
  volume = {9},
  number = {4},
  pages = {143},
  publisher = {Theory of Computing},
  doi = {10.4086/toc.2013.v009a004}
}

@inproceedings{aaronson2016complexity,
  title = {Complexity-{{Theoretic Foundations}} of {{Quantum Supremacy Experiments}}},
  booktitle = {Proc 32nd {{Comput Complex}}. {{Conf}}},
  author = {Aaronson, Scott and Chen, Lijie},
  date = {2016},
  pages = {22:1},
  doi = {10.4230/LIPIcs.CCC.2017.22}
}

@article{acuaviva2024benchmarking,
  title = {Benchmarking {{Quantum Computers}}: {{Towards}} a {{Standard Performance Evaluation Approach}}},
  author = {Acuaviva, Arturo and Aguirre, David and Peña, Rubén and Sanz, Mikel},
  date = {2024},
  journaltitle = {arXiv:2407.10941},
  eprint = {2407.10941},
  eprinttype = {arXiv},
  url = {https://doi.org/10.48550/arXiv.2407.10941}
}

@article{Adachi2015Boltzmann,
  title = {Application of {{Quantum Annealing}} to {{Training}} of {{Deep Neural Networks}}},
  author = {Adachi, Steven H. and Henderson, Maxwell P.},
  date = {2015},
  journaltitle = {arXiv:1510.06356},
  eprint = {1510.06356},
  eprinttype = {arXiv},
  url = {https://doi.org/10.48550/arXiv.1510.06356}
}

@article{adedoyin2018quantum,
  title = {Quantum Algorithm Implementations for Beginners},
  author = {Adedoyin, Adetokunbo and Ambrosiano, John and Anisimov, Petr and Casper, William and Chennupati, Gopinath and Coffrin, Carleton and Djidjev, Hristo and Gunter, David and Karra, Satish and Lemons, Nathan and others},
  date = {2018},
  journaltitle = {arXiv:1804.03719},
  eprint = {1804.03719},
  eprinttype = {arXiv},
  url = {https://doi.org/10.1145/3517340}
}

@article{agarwal2023modelling,
  title={Modelling non-Markovian noise in driven superconducting qubits},
  author={Agarwal, Abhishek and Lindoy, Lachlan P and Lall, Deep and Jamet, Fran{\c{c}}ois and Rungger, Ivan},
  journal={Quantum Sci. Technol.},
  volume={9},
  number={3},
  pages={035017},
  year={2024},
  publisher={IOP Publishing},
  url={https://doi.org/10.1088/2058-9565/ad3d7e}
}

@inproceedings{aharonov1997fault,
  title = {Fault-Tolerant Quantum Computation with Constant Error},
  booktitle = {Proc. 29th {{Annu}}. {{ACM Symp}}. {{Theory Comput}}.},
  author = {Aharonov, D. and Ben-Or, M.},
  date = {1997},
  series = {{{STOC}} '97},
  pages = {176},
  publisher = {Association for Computing Machinery},
  location = {New York, NY, USA},
  doi = {10.1145/258533.258579},
  url = {https://doi.org/10.1145/258533.258579},
  isbn = {0-89791-888-6},
  pagetotal = {13}
}

@inproceedings{aharonov1998quantum,
  title = {Quantum {{Circuits}} with {{Mixed States}}},
  booktitle = {Proc. {{Thirtieth Annu}}. {{ACM Symp}}. {{Theory Comput}}.},
  author = {Aharonov, Dorit and Kitaev, Alexei and Nisan, Noam},
  date = {1998},
  series = {Proc. {{Thirtieth Annu}}. {{ACM Symp}}. {{Theory Comput}}.},
  pages = {20},
  url = {https://doi.org/10.1145/276698.276708}
}

@article{Aharonov2004adiabaticUniversal,
  title = {Adiabatic {{Quantum Computation Is Equivalent}} to {{Standard Quantum Computation}}},
  author = {Aharonov, D. and family=Dam, given=W., prefix=van, useprefix=true and Kempe, J. and Landau, Z. and Lloyd, S. and Regev, O.},
  date = {2004},
  journaltitle = {45th Annu. IEEE Symp. Found. Comput. Sci.},
  pages = {42},
  doi = {10.1109/FOCS.2004.8}
}

@inproceedings{ahn2009graph,
  title = {Graph Sparsification in the Semi-Streaming Model},
  booktitle = {Int {{Colloq Autom}}. {{Lang Program}}},
  author = {Ahn, Kook Jin and Guha, Sudipto},
  date = {2009},
  pages = {328},
  publisher = {Springer},
  url = {https://doi.org/10.1007/978-3-642-02930-1_27}
}

@inproceedings{alam2019addressing,
  title = {Addressing {{Temporal Variations}} in {{Qubit Quality Metrics}} for {{Parameterized Quantum Circuits}}},
  booktitle = {2019 {{IEEEACM Int Symp Low Power Electron ISLPED}}},
  author = {Alam, Mahabubul and Ash-Saki, Abdullah and Ghosh, Swaroop},
  date = {2019},
  pages = {1},
  doi = {10.1109/ISLPED.2019.8824907},
  url = {https://doi.org/10.1109/ISLPED.2019.8824907}
}

@inproceedings{alam2020,
  title = {Circuit {{Compilation Methodologies}} for {{Quantum Approximate Optimization Algorithm}}},
  booktitle = {2020 53rd {{Annu IEEEACM Int Symp Microarchit MICRO}}},
  author = {Alam, Mahabubul and Ash-Saki, Abdullah and Ghosh, Swaroop},
  date = {2020},
  pages = {215},
  doi = {10.1109/MICRO50266.2020.00029},
  url = {https://doi.org/10.1109/MICRO50266.2020.00029}
}

@article{Albash2018AQCreview,
  title = {Adiabatic {{Quantum Computation}}},
  author = {Albash, Tameem and Lidar, Daniel A.},
  date = {2018-01},
  journaltitle = {Rev. Mod. Phys.},
  volume = {90},
  number = {1},
  pages = {015002},
  publisher = {American Physical Society},
  doi = {10.1103/RevModPhys.90.015002}
}

@article{Albash2019errorQA,
  title = {Analog {{Errors}} in {{Ising Machines}}},
  author = {Albash, Tameem and Martin-Mayor, Victor and Hen, Itay},
  date = {2019-04},
  journaltitle = {Quantum Sci. Technol.},
  volume = {4},
  number = {2},
  pages = {02LT03},
  publisher = {IOP Publishing},
  doi = {10.1088/2058-9565/ab13ea},
}

@article{AlgorithmicQubitsBetter,
  title = {Algorithmic {{Qubits}}: {{A Better Single-Number Metric}}},
  author = {{IonQ}},
  year = {2022},
  note = {[Accessed: 1-November-2023]},
  journaltitle = {IonQ Website},
  url = {https://ionq.com/resources/algorithmic-qubits-a-better-single-number-metric}
}

@inproceedings{almudever2020realizing,
  title = {Realizing Quantum Algorithms on Real Quantum Computing Devices},
  booktitle = {2020 {{Autom}}. {{Test Eur}}. {{Conf Exhib DATE}}},
  author = {Almudever, Carmen G and Lao, Lingling and Wille, Robert and Guerreschi, Gian G},
  date = {2020},
  pages = {864},
  publisher = {IEEE},
  url = {https://doi.org/10.23919/DATE48585.2020.9116240}
}

@article{altepeter2003ancilla,
  title = {Ancilla-{{Assisted Quantum Process Tomography}}},
  author = {Altepeter, J. B. and Branning, D. and Jeffrey, E. and Wei, T. C. and Kwiat, P. G. and Thew, R. T. and O'Brien, J. L. and Nielsen, M. A. and White, A. G.},
  date = {2003-05},
  journaltitle = {Phys. Rev. Lett.},
  volume = {90},
  number = {19},
  pages = {193601},
  publisher = {American Physical Society},
  doi = {10.1103/PhysRevLett.90.193601},
  url = {https://link.aps.org/doi/10.1103/PhysRevLett.90.193601},
  pagetotal = {4}
}

@article{amico2023defining,
  title = {Defining Standard Strategies for Quantum Benchmarks},
  author = {Amico, Mirko and Zhang, Helena and Jurcevic, Petar and Bishop, Lev S and Nation, Paul and Wack, Andrew and McKay, David C},
  date = {2023},
  journaltitle = {arXiv:2303.02108},
  eprint = {2303.02108},
  eprinttype = {arXiv},
  url = {https://doi.org/10.48550/arXiv.2303.02108}
}

@article{Amin2018Boltzmann,
  title = {Quantum {{Boltzmann Machine}}},
  author = {Amin, Mohammad H. and Andriyash, Evgeny and Rolfe, Jason and Kulchytskyy, Bohdan and Melko, Roger},
  date = {2018-05},
  journaltitle = {Phys. Rev. X},
  volume = {8},
  number = {2},
  pages = {021050},
  publisher = {American Physical Society},
  doi = {10.1103/PhysRevX.8.021050}
}

@article{amy2013meet,
  title = {A Meet-in-the-Middle Algorithm for Fast Synthesis of Depth-Optimal Quantum Circuits},
  author = {Amy, Matthew and Maslov, Dmitri and Mosca, Michele and Roetteler, Martin},
  date = {2013},
  journaltitle = {IEEE Trans Comput-Aided Integr Circuits Syst},
  volume = {32},
  number = {6},
  pages = {818},
  publisher = {IEEE},
  url = {https://doi.org/10.1109/TCAD.2013.2244643}
}

@article{Arrazola2018,
  title = {Using {{Gaussian Boson Sampling}} to {{Find Dense Subgraphs}}},
  author = {Arrazola, Juan Miguel and Bromley, Thomas R.},
  date = {2018-07},
  journaltitle = {Phys. Rev. Lett.},
  volume = {121},
  number = {3},
  pages = {030503},
  publisher = {American Physical Society},
  doi = {10.1103/PhysRevLett.121.030503},
  url = {https://link.aps.org/doi/10.1103/PhysRevLett.121.030503},
  pagetotal = {6}
}

@article{Arute2019,
  title = {Quantum {{Supremacy Using}} a {{Programmable Superconducting Processor}}},
  author = {Arute, Frank and Arya, Kunal and Babbush, Ryan and Bacon, Dave and Bardin, Joseph C. and Barends, Rami and Biswas, Rupak and Boixo, Sergio and Brandao, Fernando G. S. L. and Buell, David A. and Burkett, Brian and Chen, Yu and Chen, Zijun and Chiaro, Ben and Collins, Roberto and Courtney, William and Dunsworth, Andrew and Farhi, Edward and Foxen, Brooks and Fowler, Austin and Gidney, Craig and Giustina, Marissa and Graff, Rob and Guerin, Keith and Habegger, Steve and Harrigan, Matthew P. and Hartmann, Michael J. and Ho, Alan and Hoffmann, Markus and Huang, Trent and Humble, Travis S. and Isakov, Sergei V. and Jeffrey, Evan and Jiang, Zhang and Kafri, Dvir and Kechedzhi, Kostyantyn and Kelly, Julian and Klimov, Paul V. and Knysh, Sergey and Korotkov, Alexander and Kostritsa, Fedor and Landhuis, David and Lindmark, Mike and Lucero, Erik and Lyakh, Dmitry and Mandrà, Salvatore and McClean, Jarrod R. and McEwen, Matthew and Megrant, Anthony and Mi, Xiao and Michielsen, Kristel and Mohseni, Masoud and Mutus, Josh and Naaman, Ofer and Neeley, Matthew and Neill, Charles and Niu, Murphy Yuezhen and Ostby, Eric and Petukhov, Andre and Platt, John C. and Quintana, Chris and Rieffel, Eleanor G. and Roushan, Pedram and Rubin, Nicholas C. and Sank, Daniel and Satzinger, Kevin J. and Smelyanskiy, Vadim and Sung, Kevin J. and Trevithick, Matthew D. and Vainsencher, Amit and Villalonga, Benjamin and White, Theodore and Yao, Z. Jamie and Yeh, Ping and Zalcman, Adam and Neven, Hartmut and Martinis, John M.},
  date = {2019-10},
  journaltitle = {Nature},
  volume = {574},
  number = {7779},
  pages = {505},
  publisher = {{Springer Science and Business Media LLC}},
  doi = {10.1038/s41586-019-1666-5}
}

@article{arute2020observation,
  title = {Observation of Separated Dynamics of Charge and Spin in the {{Fermi-Hubbard}} Model},
  author = {Arute, Frank and Arya, Kunal and Babbush, Ryan and Bacon, Dave and Bardin, Joseph C. and Barends, Rami and Bengtsson, Andreas and Boixo, Sergio and Broughton, Michael and Buckley, Bob B. and Buell, David A. and Burkett, Brian and Bushnell, Nicholas and Chen, Yu and Chen, Zijun and Chen, Yu-An and Chiaro, Ben and Collins, Roberto and Cotton, Stephen J. and Courtney, William and Demura, Sean and Derk, Alan and Dunsworth, Andrew and Eppens, Daniel and Eckl, Thomas and Erickson, Catherine and Farhi, Edward and Fowler, Austin and Foxen, Brooks and Gidney, Craig and Giustina, Marissa and Graff, Rob and Gross, Jonathan A. and Habegger, Steve and Harrigan, Matthew P. and Ho, Alan and Hong, Sabrina and Huang, Trent and Huggins, William and Ioffe, Lev B. and Isakov, Sergei V. and Jeffrey, Evan and Jiang, Zhang and Jones, Cody and Kafri, Dvir and Kechedzhi, Kostyantyn and Kelly, Julian and Kim, Seon and Klimov, Paul V. and Korotkov, Alexander N. and Kostritsa, Fedor and Landhuis, David and Laptev, Pavel and Lindmark, Mike and Lucero, Erik and Marthaler, Michael and Martin, Orion and Martinis, John M. and Marusczyk, Anika and McArdle, Sam and McClean, Jarrod R. and McCourt, Trevor and McEwen, Matt and Megrant, Anthony and Mejuto-Zaera, Carlos and Mi, Xiao and Mohseni, Masoud and Mruczkiewicz, Wojciech and Mutus, Josh and Naaman, Ofer and Neeley, Matthew and Neill, Charles and Neven, Hartmut and Newman, Michael and Niu, Murphy Yuezhen and O'Brien, Thomas E. and Ostby, Eric and Pató, Bálint and Petukhov, Andre and Putterman, Harald and Quintana, Chris and Reiner, Jan-Michael and Roushan, Pedram and Rubin, Nicholas C. and Sank, Daniel and Satzinger, Kevin J. and Smelyanskiy, Vadim and Strain, Doug and Sung, Kevin J. and Schmitteckert, Peter and Szalay, Marco and Tubman, Norm M. and Vainsencher, Amit and White, Theodore and Vogt, Nicolas and Yao, Z. Jamie and Yeh, Ping and Zalcman, Adam and Zanker, Sebastian},
  date = {2020},
  journaltitle = {arXiv:2010.07965},
  eprint = {2010.07965},
  eprinttype = {arXiv},
  url = {https://doi.org/10.48550/arXiv.2010.07965}
}

@article{baldwin2022re,
  title = {Re-{{Examining}} the {{Quantum Volume Test}}: {{Ideal Distributions}}, {{Compiler Optimizations}}, {{Confidence Intervals}}, and {{Scalable Resource Estimations}}},
  author = {Baldwin, Charles H and Mayer, Karl and Brown, Natalie C and Ryan-Anderson, Ciarán and Hayes, David},
  date = {2022},
  journaltitle = {Quantum},
  volume = {6},
  pages = {707},
  publisher = {Verein zur Förderung des Open Access Publizierens in den Quantenwissenschaften},
  doi = {10.22331/q-2022-05-09-707}
}

@article{banaszek2013focus,
  title = {Focus on Quantum Tomography},
  author = {Banaszek, Konrad and Cramer, Marcus and Gross, David},
  date = {2013},
  journaltitle = {New J. Phys},
  volume = {15},
  number = {12},
  pages = {125020},
  publisher = {IOP Publishing},
  url = {https://doi.org/10.1088/1367-2630/15/12/125020}
}

@article{Banchi2020,
  title = {Molecular Docking with {{Gaussian Boson Sampling}}},
  author = {Banchi, Leonardo and Fingerhuth, Mark and Babej, Tomas and Ing, Christopher and Arrazola, Juan Miguel},
  date = {2020},
  journaltitle = {Sci. Adv.},
  volume = {6},
  number = {23},
  eprint = {https://www.science.org/doi/pdf/10.1126/sciadv.aax1950},
  pages = {eaax1950},
  doi = {10.1126/sciadv.aax1950},
  url = {https://www.science.org/doi/abs/10.1126/sciadv.aax1950}
}

@article{Barenco1995,
  title = {Elementary Gates for Quantum Computation},
  author = {Barenco, Adriano and Bennett, Charles H. and Cleve, Richard and DiVincenzo, David P. and Margolus, Norman and Shor, Peter and Sleator, Tycho and Smolin, John A. and Weinfurter, Harald},
  date = {1995-11},
  journaltitle = {Phys. Rev. A},
  volume = {52},
  number = {5},
  pages = {3457},
  publisher = {American Physical Society},
  doi = {10.1103/PhysRevA.52.3457},
  url = {https://link.aps.org/doi/10.1103/PhysRevA.52.3457},
  pagetotal = {0}
}

@article{barnes22,
  title = {Assembly and {{Coherent Control}} of a {{Register}} of {{Nuclear Spin Qubits}}},
  author = {Barnes, Katrina and Battaglino, Peter and Bloom, Benjamin J. and Cassella, Kayleigh and Coxe, Robin and Crisosto, Nicole and King, Jonathan P. and Kondov, Stanimir S. and Kotru, Krish and Larsen, Stuart C. and Lauigan, Joseph and Lester, Brian J. and McDonald, Mickey and Megidish, Eli and Narayanaswami, Sandeep and Nishiguchi, Ciro and Notermans, Remy and Peng, Lucas S. and Ryou, Albert and Wu, Tsung-Yao and Yarwood, Michael},
  date = {2022},
  journaltitle = {Nat. Commun.},
  volume = {13},
  number = {1},
  pages = {2779},
  doi = {10.1038/s41467-022-29977-z}
}

@article{barredo16,
  title = {An Atom-by-Atom Assembler of Defect-Free Arbitrary Two-Dimensional Atomic Arrays},
  author = {Barredo, Daniel and family=Léséleuc, given=Sylvain, prefix=de, useprefix=true and Lienhard, Vincent and Lahaye, Thierry and Browaeys, Antoine},
  date = {2016},
  journaltitle = {Science},
  volume = {354},
  number = {6315},
  pages = {1021},
  doi = {10.1126/science.aah3778}
}

@article{barredo18,
  title = {Synthetic {{Three-Dimensional Atomic Structures Assembled Atom}} by {{Atom}}},
  author = {Barredo, Daniel and Lienhard, Vincent and family=Léséleuc, given=Sylvain, prefix=de, useprefix=true and Lahaye, Thierry and Browaeys, Antoine},
  date = {2018},
  journaltitle = {Nature},
  volume = {561},
  pages = {79},
  doi = {10.1038/s41586-018-0450-2}
}

@article{bartee2024spin,
  title = {Spin {{Qubits}} with {{Integrated}} Millikelvin {{CMOS Control}}},
  author = {Bartee, Samuel K. and Gilbert, Will and Zuo, Kun and Das, Kushal and Tanttu, Tuomo and Yang, Chih Hwan and Stuyck, Nard Dumoulin and Pauka, Sebastian J. and Su, Rocky Y. and Lim, Wee Han and Serrano, Santiago and Escott, Christopher C. and Hudson, Fay E. and Itoh, Kohei M. and Laucht, Arne and Dzurak, Andrew S. and Reilly, David J.},
  date = {2024-07},
  journaltitle = {arXiv:2407.15151},
  eprint = {2407.15151},
  eprinttype = {arXiv},
  url = {https://doi.org/10.48550/arXiv.2407.15151}
}

@article{Bartolucci2023,
  title = {Fusion-Based Quantum Computation},
  author = {Bartolucci, Sara and Birchall, Patrick and Bombín, Hector and Cable, Hugo and Dawson, Chris and Gimeno-Segovia, Mercedes and Johnston, Eric and Kieling, Konrad and Nickerson, Naomi and Pant, Mihir and Pastawski, Fernando and Rudolph, Terry and Sparrow, Chris},
  date = {2023},
  journaltitle = {Nat. Commun.},
  volume = {14},
  number = {1},
  pages = {912},
  issn = {20411723},
  doi = {10.1038/s41467-023-36493-1},
  url = {https://doi.org/10.1038/s41467-023-36493-1}
}

@article{barz2012demonstration,
  title = {Demonstration of {{Blind Quantum Computing}}},
  author = {Barz, Stefanie and Kashefi, Elham and Broadbent, Anne and Fitzsimons, Joseph F and Zeilinger, Anton and Walther, Philip},
  date = {2012},
  journaltitle = {Science},
  volume = {335},
  number = {6066},
  pages = {303},
  publisher = {American Association for the Advancement of Science},
  doi = {10.1126/science.1214707}
}

@article{barz2013experimental,
  title = {Experimental {{Verification}} of {{Quantum Computation}}},
  author = {Barz, Stefanie and Fitzsimons, Joseph F and Kashefi, Elham and Walther, Philip},
  date = {2013},
  journaltitle = {Nat. Phys.},
  volume = {9},
  number = {11},
  pages = {727},
  publisher = {Nature Publishing Group UK London},
  doi = {10.1038/nphys2763}
}

@article{bendersky2008selective,
  title = {Selective and {{Efficient Estimation}} of {{Parameters}} for {{Quantum Process Tomography}}},
  author = {Bendersky, Ariel and Pastawski, Fernando and Paz, Juan Pablo},
  date = {2008-05},
  journaltitle = {Phys. Rev. Lett.},
  volume = {100},
  number = {19},
  pages = {190403},
  publisher = {American Physical Society},
  doi = {10.1103/PhysRevLett.100.190403},
  url = {https://link.aps.org/doi/10.1103/PhysRevLett.100.190403},
  pagetotal = {4}
}

@article{Benedetti2016Temperature,
  title = {Estimation of {{Effective Temperatures}} in {{Quantum Annealers}} for {{Sampling Applications}}: {{A Case Study}} with {{Possible Applications}} in {{Deep Learning}}},
  author = {Benedetti, Marcello and Realpe-Gómez, John and Biswas, Rupak and Perdomo-Ortiz, Alejandro},
  date = {2016-08},
  journaltitle = {Phys. Rev. A},
  volume = {94},
  number = {2},
  pages = {022308},
  publisher = {American Physical Society},
  doi = {10.1103/PhysRevA.94.022308}
}

@article{Benedetti2017PGM,
  title = {Quantum-{{Assisted Learning}} of {{Hardware-Embedded Probabilistic Graphical Models}}},
  author = {Benedetti, Marcello and Realpe-Gómez, John and Biswas, Rupak and Perdomo-Ortiz, Alejandro},
  date = {2017-11},
  journaltitle = {Phys. Rev. X},
  volume = {7},
  number = {4},
  pages = {041052},
  publisher = {American Physical Society},
  doi = {10.1103/PhysRevX.7.041052}
}

@article{Benedetti2018Helmholtz,
  title = {Quantum-{{Assisted Helmholtz Machines}}: {{A Quantum}}–{{Classical Deep Learning Framework}} for {{Industrial Datasets}} in near-{{Term Devices}}},
  author = {Benedetti, Marcello and Realpe-Gómez, John and Perdomo-Ortiz, Alejandro},
  date = {2018-05},
  journaltitle = {Quantum Sci. Technol.},
  volume = {3},
  number = {3},
  pages = {034007},
  publisher = {IOP Publishing},
  doi = {10.1088/2058-9565/aabd98}
}

@article{bennett2000quantum,
  title = {Quantum Information and Computation},
  author = {Bennett, Charles H and DiVincenzo, David P},
  date = {2000},
  journaltitle = {Nature},
  volume = {404},
  number = {6775},
  pages = {247},
  publisher = {Nature Publishing Group UK London},
  url = {https://doi.org/10.1038/35005001}
}

@article{Bennett2023precision,
  title = {Using Copies Can Improve Precision in Continuous-Time Quantum Computing},
  author = {Bennett, Jemma and Callison, Adam and O’Leary, Tom and West, Mia and Chancellor, Nicholas and Kendon, Viv},
  date = {2023-07},
  journaltitle = {Quantum Sci. Technol.},
  volume = {8},
  number = {3},
  pages = {035031},
  publisher = {IOP Publishing},
  doi = {10.1088/2058-9565/acdcb5},
  url = {https://dx.doi.org/10.1088/2058-9565/acdcb5}
}

@article{Callison2019walks,
title = {Finding spin glass ground states using quantum walks},
year = {2019},
publisher = {IOP Publishing},
volume = {21},
number = {12},
pages = {123022},
author = {Adam Callison and Nicholas Chancellor and Florian Mintert and Viv Kendon},
journal = {New J. Phys.},
url = {https://doi.org/10.1088/1367-2630/ab5ca2},
}

@article{berberich2024robustness,
  title = {Robustness of Quantum Algorithms against Coherent Control Errors},
  author = {Berberich, Julian and Fink, Daniel and Holm, Christian},
  date = {2024-01},
  journaltitle = {Phys. Rev. A},
  volume = {109},
  number = {1},
  pages = {012417},
  publisher = {American Physical Society},
  doi = {10.1103/PhysRevA.109.012417},
  url = {https://link.aps.org/doi/10.1103/PhysRevA.109.012417},
  pagetotal = {14}
}

@article{Bernstein1997,
  title = {Quantum {{Complexity Theory}}},
  author = {Bernstein, Ethan and Vazirani, Umesh},
  date = {1997},
  journaltitle = {SIAM J. Comput.},
  volume = {26},
  number = {5},
  pages = {1411},
  doi = {10.1137/S0097539796300921}
}

@article{Berwald2019mathematics,
  title = {The {{Mathematics}} of {{Quantum-Enabled Applications}} on the {{D-Wave Quantum Computer}}},
  author = {J. Berwald, Jesse},
  date = {2019},
  journaltitle = {Not. Am. Math. Soc.},
  volume = {66},
  number = {6},
  pages = {832},
  url = {https://www.ams.org/journals/notices/201906/rnoti-p832.pdf}
}

@article{berwald2023Domain,
  title = {Understanding {{Domain-Wall Encoding Theoretically}} and {{Experimentally}}},
  author = {Berwald, Jesse and Chancellor, Nicholas and Dridi, Raouf},
  date = {2023},
  journaltitle = {Philos. Trans. R. Soc. A: Math. Phys. Eng. Sci.},
  volume = {381},
  number = {2241},
  eprint = {https://royalsocietypublishing.org/doi/pdf/10.1098/rsta.2021.0410},
  pages = {20210410},
  doi = {10.1098/rsta.2021.0410}
}

@article{bhandari2016general,
  title = {On the General Constraints in Single Qubit Quantum Process Tomography},
  author = {Bhandari, Ramesh and Peters, Nicholas A},
  date = {2016},
  journaltitle = {Sci. Rep.},
  volume = {6},
  number = {1},
  pages = {26004},
  publisher = {Nature Publishing Group UK London},
  url = {https://doi.org/10.1038/srep26004}
}

@article{bialczak2010quantum,
  title = {Quantum Process Tomography of a Universal Entangling Gate Implemented with {{Josephson}} Phase Qubits},
  author = {Bialczak, Radoslaw C and Ansmann, Markus and Hofheinz, Max and Lucero, Erik and Neeley, Matthew and O’Connell, Aaron D and Sank, Daniel and Wang, Haohua and Wenner, James and Steffen, Matthias and others},
  date = {2010},
  journaltitle = {Nat. Phys.},
  volume = {6},
  number = {6},
  pages = {409},
  publisher = {Nature Publishing Group UK London},
  url = {https://doi.org/10.1038/nphys1639}
}

@article{Biamonte_AQC_universal_2008,
  title = {Realizable {{Hamiltonians}} for {{Universal Adiabatic Quantum Computers}}},
  author = {Biamonte, Jacob D. and Love, Peter J.},
  date = {2008-07},
  journaltitle = {Phys. Rev. A},
  volume = {78},
  number = {1},
  pages = {012352},
  publisher = {American Physical Society},
  doi = {10.1103/PhysRevA.78.012352}
}

@article{blume-kohoutVolumetricFrameworkQuantum2020,
  title = {A {{Volumetric Framework}} for {{Quantum Computer Benchmarks}}},
  author = {Blume-Kohout, Robin and Young, Kevin C.},
  date = {2020-11},
  journaltitle = {Quantum},
  volume = {4},
  eprint = {1904.05546},
  eprinttype = {arXiv},
  eprintclass = {quant-ph},
  pages = {362},
  issn = {2521-327X},
  doi = {10.22331/q-2020-11-15-362},
  langid = {english}
}

@article{blume2010optimal,
  title = {Optimal, Reliable Estimation of Quantum States},
  author = {Blume-Kohout, Robin},
  date = {2010},
  journaltitle = {New J. Phys.},
  volume = {12},
  number = {4},
  pages = {043034},
  publisher = {IOP Publishing},
  url = {https://doi.org/10.1088/1367-2630/12/4/043034}
}

@article{blume2013robust,
  title = {Robust, Self-Consistent, Closed-Form Tomography of Quantum Logic Gates on a Trapped Ion Qubit},
  author = {Blume-Kohout, Robin and Gamble, John King and Nielsen, Erik and Mizrahi, Jonathan and Sterk, Jonathan D and Maunz, Peter},
  date = {2013},
  journaltitle = {arXiv:1310.4492},
  eprint = {1310.4492},
  eprinttype = {arXiv},
  url = {https://doi.org/10.48550/arXiv.1310.4492}
}

@article{blume2017demonstration,
  title = {Demonstration of {{Qubit Operations}} below a {{Rigorous Fault Tolerance Threshold}} with {{Gate Set Tomography}}},
  author = {Blume-Kohout, Robin and Gamble, John King and Nielsen, Erik and Rudinger, Kenneth and Mizrahi, Jonathan and Fortier, Kevin and Maunz, Peter},
  date = {2017},
  journaltitle = {Nat. Commun.},
  volume = {8},
  number = {1},
  pages = {14485},
  publisher = {Nature Publishing Group UK London},
  url = {https://doi.org/10.1038/ncomms14485}
}

@article{blumoff2016implementing,
  title = {Implementing and {{Characterizing Precise Multiqubit Measurements}}},
  author = {Blumoff, J. Z. and Chou, K. and Shen, C. and Reagor, M. and Axline, C. and Brierley, R. T. and Silveri, M. P. and Wang, C. and Vlastakis, B. and Nigg, S. E. and Frunzio, L. and Devoret, M. H. and Jiang, L. and Girvin, S. M. and Schoelkopf, R. J.},
  date = {2016-09},
  journaltitle = {Phys. Rev. X},
  volume = {6},
  number = {3},
  pages = {031041},
  publisher = {American Physical Society},
  doi = {10.1103/PhysRevX.6.031041},
  url = {https://link.aps.org/doi/10.1103/PhysRevX.6.031041},
  pagetotal = {11}
}

@article{bluvstein2024logical,
  title = {Logical Quantum Processor Based on Reconfigurable Atom Arrays},
  author = {Bluvstein, Dolev and Evered, Simon J and Geim, Alexandra A and Li, Sophie H and Zhou, Hengyun and Manovitz, Tom and Ebadi, Sepehr and Cain, Madelyn and Kalinowski, Marcin and Hangleiter, Dominik and others},
  date = {2024},
  journaltitle = {Nature},
  volume = {626},
  number = {7997},
  pages = {58},
  publisher = {Nature Publishing Group UK London},
  url = {https://doi.org/10.1038/s41586-023-06927-3}
}

@article{bluvstein22,
  title = {A {{Quantum Processor Based}} on {{Coherent Transport}} of {{Entangled Atom Arrays}}},
  author = {Bluvstein, Dolev and Levine, Harry and Semeghini, Giulia and Wang, Tout T. and Ebadi, Sepehr and Kalinowski, Marcin and Keesling, Alexander and Maskara, Nishad and Pichler, Hannes and Greiner, Markus and Vuletic̀, Vladan and Lukin, Mikhail D.},
  date = {2022},
  journaltitle = {Nature},
  volume = {604},
  pages = {451},
  doi = {10.1038/s41586-022-04592-6}
}

@article{boixo16aTunneling,
  title = {Computational {{Multiqubit Tunnelling}} in {{Programmable Quantum Annealers}}},
  author = {Boixo, Sergio and Smelyanskiy, Vadim N. and Shabani, Alireza and Isakov, Sergei V. and Dykman, Mark and Denchev, Vasil S. and Amin, Mohammad H. and Smirnov, Anatoly Yu and Mohseni, Masoud and Neven, Hartmut},
  date = {2016},
  journaltitle = {Nat. Commun.},
  volume = {7},
  pages = {10327},
  url = {https://doi.org/10.1038/ncomms10327}
}

@article{Bourassa2021,
  title = {Blueprint for a Scalable Photonic Fault-Tolerant Quantum Computer},
  author = {Bourassa, J. Eli and Alexander, Rafael N. and Vasmer, Michael and Patil, Ashlesha and Tzitrin, Ilan and Matsuura, Takaya and Su, Daiqin and Baragiola, Ben Q. and Guha, Saikat and Dauphinais, Guillaume and Sabapathy, Krishna K. and Menicucci, Nicolas C. and Dhand, Ish},
  date = {2021},
  journaltitle = {Quantum},
  volume = {5},
  pages = {392},
  issn = {2521327X},
  doi = {10.22331/Q-2021-02-04-392},
  url = {https://doi.org/10.22331/q-2021-02-04-392}
}

@incollection{bouwmeester2000physics,
  title = {The Physics of Quantum Information: Basic Concepts},
  booktitle = {The Physics of Quantum Information: Quantum Cryptography, Quantum Teleportation, Quantum Computation},
  author = {Bouwmeester, Dirk and Zeilinger, Anton},
  date = {2000},
  pages = {1},
  publisher = {Springer},
  url = {https://doi.org/10.1007/978-3-662-04209-0_1}
}

@article{bowdrey2002fidelity,
  title = {Fidelity of Single Qubit Maps},
  author = {Bowdrey, Mark D and Oi, Daniel KL and Short, Anthony J and Banaszek, Konrad and Jones, Jonathan A},
  date = {2002},
  journaltitle = {Phys. Lett. A},
  volume = {294},
  number = {5-6},
  pages = {258},
  publisher = {Elsevier},
  url = {https://doi.org/10.1016/S0375-9601(02)00069-5}
}

@article{BOYKIN2000101,
  title = {A New Universal and Fault-Tolerant Quantum Basis},
  author = {Boykin, P.Oscar and Mor, Tal and Pulver, Matthew and Roychowdhury, Vwani and Vatan, Farrokh},
  date = {2000},
  journaltitle = {Inf. Process. Lett.},
  volume = {75},
  number = {3},
  pages = {101},
  issn = {0020-0190},
  doi = {10.1016/S0020-0190(00)00084-3},
  url = {https://www.sciencedirect.com/science/article/pii/S0020019000000843}
}

@article{brandt1999qubit,
  title = {Qubit Devices and the Issue of Quantum Decoherence},
  author = {Brandt, Howard E},
  date = {1999},
  journaltitle = {Prog. Quantum Electron.},
  volume = {22},
  number = {5-6},
  pages = {257},
  publisher = {Elsevier},
  url = {https://doi.org/10.1016/S0079-6727(99)00003-8}
}

@article{bravyi2021mitigating,
  title = {Mitigating Measurement Errors in Multiqubit Experiments},
  author = {Bravyi, Sergey and Sheldon, Sarah and Kandala, Abhinav and Mckay, David C. and Gambetta, Jay M.},
  date = {2021-04},
  journaltitle = {Phys. Rev. A},
  volume = {103},
  number = {4},
  pages = {042605},
  publisher = {American Physical Society},
  doi = {10.1103/PhysRevA.103.042605},
  url = {https://link.aps.org/doi/10.1103/PhysRevA.103.042605},
  pagetotal = {12}
}

@article{breuer2002theory,
  title = {The {{Theory}} of {{Open Quantum Systems}}},
  author = {Breuer, H.P. and Petruccione, F.},
  journaltitle = {Oxf. Univ. Press 2002},
  url = {https://global.oup.com/academic/product/the-theory-of-open-quantum-systems-9780198520634}
}

@article{broadbent2009universal,
  title = {Universal {{Blind Quantum Computation}}},
  author = {Broadbent, Anne and Fitzsimons, Joseph and Kashefi, Elham},
  date = {2009},
  journaltitle = {2009 50th Annu. IEEE Symp. Found. Comput. Sci.},
  pages = {517},
  publisher = {IEEE},
  url = {https://doi.org/10.1109/FOCS.2009.36}
}

@article{Bromley_2020,
  title = {Applications of {{Near-Term Photonic Quantum Computers}}: {{Software}} and {{Algorithms}}},
  author = {Bromley, Thomas R and Arrazola, Juan Miguel and Jahangiri, Soran and Izaac, Josh and family=Quesada, given=Nicolá, prefix=s, useprefix=true and Gran, Alain Delgado and Schuld, Maria and Swinarton, Jeremy and Zabaneh, Zeid and Killoran, Nathan},
  date = {2020-05},
  journaltitle = {Quantum Sci. Technol.},
  volume = {5},
  number = {3},
  pages = {034010},
  publisher = {IOP Publishing},
  doi = {10.1088/2058-9565/ab8504}
}

@article{browaeys20,
  title = {Many-{{Body Physics}} with {{Individually Controlled Rydberg Atoms}}},
  author = {Browaeys, A. and Lahaye, T.},
  date = {2020},
  journaltitle = {Nat. Phys},
  volume = {16},
  pages = {132},
  doi = {10.1038/s41567-019-0733-z}
}

@article{Browne_2007,
  title = {Generalized Flow and Determinism in Measurement-Based Quantum Computation},
  author = {Browne, Daniel E and Kashefi, Elham and Mhalla, Mehdi and Perdrix, Simon},
  date = {2007-08},
  journaltitle = {New J. Phys.},
  volume = {9},
  number = {8},
  pages = {250},
  doi = {10.1088/1367-2630/9/8/250},
  url = {https://dx.doi.org/10.1088/1367-2630/9/8/250}
}

@article{burnett2019decoherence,
  title = {Decoherence Benchmarking of Superconducting Qubits},
  author = {Burnett, Jonathan J and Bengtsson, Andreas and Scigliuzzo, Marco and Niepce, David and Kudra, Marina and Delsing, Per and Bylander, Jonas},
  date = {2019},
  journaltitle = {npj Quantum Inf.},
  volume = {5},
  number = {1},
  pages = {54},
  publisher = {Nature Publishing Group UK London},
  url = {https://doi.org/10.1038/s41534-019-0168-5}
}

@article{Cai2014minor,
  title = {A {{Practical Heuristic}} for {{Finding Graph Minors}}},
  author = {Cai, Jun and Macready, William G. and Roy, Aidan},
  date = {2014},
  journaltitle = {arXiv:1406.2741},
  eprint = {1406.2741},
  eprinttype = {arXiv},
  url = {https://doi.org/10.48550/arXiv.1406.2741}
}

@article{Caldeira2019Botlzmann,
  title = {Restricted {{Boltzmann Machines}} for {{Galaxy Morphology Classification}} with a {{Quantum Annealer}}},
  author = {Caldeira, Joao and Job, Joshua and Adachi, Steven H. and Nord, Brian and Perdue, Gabriel N.},
  date = {2019},
  journaltitle = {arXiv:1911.06259},
  eprint = {1911.06259},
  eprinttype = {arXiv},
  url = {https://doi.org/10.48550/arXiv.1911.06259}
}

@article{Callison2021diabatic,
  title = {Energetic {{Perspective}} on {{Rapid Quenches}} in {{Quantum Annealing}}},
  author = {Callison, Adam and Festenstein, Max and Chen, Jie and Nita, Laurentiu and Kendon, Viv and Chancellor, Nicholas},
  date = {2021-03},
  journaltitle = {PRX Quantum},
  volume = {2},
  number = {1},
  pages = {010338},
  publisher = {American Physical Society},
  doi = {10.1103/PRXQuantum.2.010338}
}

@article{Callison2022hybrid,
  title = {Hybrid {{Quantum-Classical Algorithms}} in the {{Noisy Intermediate-Scale Quantum Era}} and {{Beyond}}},
  author = {Callison, Adam and Chancellor, Nicholas},
  date = {2022-07},
  journaltitle = {Phys. Rev. A},
  volume = {106},
  number = {1},
  pages = {010101},
  publisher = {American Physical Society},
  doi = {10.1103/PhysRevA.106.010101}
}

@article{camenzind2022hole,
  title = {A Hole Spin Qubit in a Fin Field-Effect Transistor above 4 Kelvin},
  author = {Camenzind, Leon C and Geyer, Simon and Fuhrer, Andreas and Warburton, Richard J and Zumbühl, Dominik M and Kuhlmann, Andreas V},
  date = {2022},
  journaltitle = {Nat. Electron.},
  volume = {5},
  number = {3},
  pages = {178},
  publisher = {Nature Publishing Group UK London},
  url = {https://doi.org/10.1038/s41928-022-00722-0}
}

@article{carroll2022dynamics,
  title = {Dynamics of Superconducting Qubit Relaxation Times},
  author = {Carroll, Malcolm and Rosenblatt, Sami and Jurcevic, Petar and Lauer, Isaac and Kandala, Abhinav},
  date = {2022},
  journaltitle = {npj Quantum Inf.},
  volume = {8},
  number = {1},
  pages = {132},
  publisher = {Nature Publishing Group UK London},
  url = {https://doi.org/10.1038/s41534-022-00643-y}
}

@article{casparis2016gatemon,
  title = {Gatemon {{Benchmarking}} and {{Two-Qubit Operations}}},
  author = {Casparis, L. and Larsen, T. W. and Olsen, M. S. and Kuemmeth, F. and Krogstrup, P. and Nygård, J. and Petersson, K. D. and Marcus, C. M.},
  date = {2016-04},
  journaltitle = {Phys. Rev. Lett.},
  volume = {116},
  number = {15},
  pages = {150505},
  publisher = {American Physical Society},
  doi = {10.1103/PhysRevLett.116.150505},
  url = {https://link.aps.org/doi/10.1103/PhysRevLett.116.150505},
  pagetotal = {5}
}

@article{Chancellor2016maxEnt,
  title = {Maximum-{{Entropy Inference}} with a {{Programmable Annealer}}},
  author = {Chancellor, Nicholas and Szoke, Szilard and Vinci, Walter and Aeppli, Gabriel and Warburton, Paul A.},
  date = {2016-03},
  journaltitle = {Sci. Rep.},
  volume = {6},
  number = {1},
  pages = {22318},
  issn = {2045-2322},
  doi = {10.1038/srep22318}
}

@article{Chancellor2017circuit,
  title = {Circuit {{Design}} for {{Multi-Body Interactions}} in {{Superconducting Quantum Annealing Systems}} with {{Applications}} to a {{Scalable Architecture}}},
  author = {Chancellor, N. and Zohren, S. and Warburton, P. A.},
  date = {2017-06},
  journaltitle = {npj Quantum Inf.},
  volume = {3},
  number = {1},
  pages = {21},
  issn = {2056-6387},
  doi = {10.1038/s41534-017-0022-6},
}

@article{chancellor2019domain,
  title = {Domain {{Wall Encoding}} of {{Discrete Variables}} for {{Quantum Annealing}} and {{QAOA}}},
  author = {Chancellor, Nicholas},
  date = {2019-08},
  journaltitle = {Quantum Sci. Technol.},
  volume = {4},
  number = {4},
  pages = {045004},
  publisher = {IOP Publishing},
  doi = {10.1088/2058-9565/ab33c2}
}

@article{Chancellor2022_U_dist,
  title = {Error {{Measurements}} for a {{Quantum Annealer Using}} the {{One-Dimensional Ising Model}} with {{Twisted Boundaries}}},
  author = {Chancellor, Nicholas and Crowley, Philip J. D. and Đurić, Tanja and Vinci, Walter and Amin, Mohammad H. and Green, Andrew G. and Warburton, Paul A. and Aeppli, Gabriel},
  date = {2022-06},
  journaltitle = {npj Quantum Inf.},
  volume = {8},
  number = {1},
  pages = {73},
  issn = {2056-6387},
  doi = {10.1038/s41534-022-00580-w},
}

@article{Chancellor21SearchRange,
  title = {Experimental {{Test}} of {{Search Range}} in {{Quantum Annealing}}},
  author = {Chancellor, Nicholas and Kendon, Viv},
  date = {2021-07},
  journaltitle = {Phys. Rev. A},
  volume = {104},
  number = {1},
  pages = {012604},
  publisher = {American Physical Society},
  doi = {10.1103/PhysRevA.104.012604}
}

@article{chatterjee2021semiconductor,
  title = {Semiconductor Qubits in Practice},
  author = {Chatterjee, Anasua and Stevenson, Paul and De Franceschi, Silvano and Morello, Andrea and family=Leon, given=Nathalie P, prefix=de, useprefix=true and Kuemmeth, Ferdinand},
  date = {2021},
  journaltitle = {Nat. Rev. Phys.},
  volume = {3},
  number = {3},
  pages = {157},
  publisher = {Nature Publishing Group UK London},
  doi = {10.1038/s42254-021-00283-9},
  url = {https://doi.org/10.1038/s42254-021-00283-9}
}

@article{chen2016measuring,
  title = {Measuring and {{Suppressing Quantum State Leakage}} in a {{Superconducting Qubit}}},
  author = {Chen, Zijun and Kelly, Julian and Quintana, Chris and Barends, R. and Campbell, B. and Chen, Yu and Chiaro, B. and Dunsworth, A. and Fowler, A. G. and Lucero, E. and Jeffrey, E. and Megrant, A. and Mutus, J. and Neeley, M. and Neill, C. and O'Malley, P. J. J. and Roushan, P. and Sank, D. and Vainsencher, A. and Wenner, J. and White, T. C. and Korotkov, A. N. and Martinis, John M.},
  date = {2016-01},
  journaltitle = {Phys. Rev. Lett.},
  volume = {116},
  number = {2},
  pages = {020501},
  publisher = {American Physical Society},
  doi = {10.1103/PhysRevLett.116.020501},
  url = {https://link.aps.org/doi/10.1103/PhysRevLett.116.020501},
  pagetotal = {5}
}

@article{chen2019detector,
  title = {Detector Tomography on {{IBM}} Quantum Computers and Mitigation of an Imperfect Measurement},
  author = {Chen, Yanzhu and Farahzad, Maziar and Yoo, Shinjae and Wei, Tzu-Chieh},
  date = {2019-11},
  journaltitle = {Phys. Rev. A},
  volume = {100},
  number = {5},
  pages = {052315},
  publisher = {American Physical Society},
  doi = {10.1103/PhysRevA.100.052315},
  url = {https://link.aps.org/doi/10.1103/PhysRevA.100.052315},
  pagetotal = {17}
}

@article{chen2023benchmarking,
  title = {Benchmarking a Trapped-Ion Quantum Computer with 29 Algorithmic Qubits},
  author = {Chen, Jwo-Sy and Nielsen, Erik and Ebert, Matthew and Inlek, Volkan and Wright, Kenneth and Chaplin, Vandiver and Maksymov, Andrii and Páez, Eduardo and Poudel, Amrit and Maunz, Peter and others},
  date = {2023},
  journaltitle = {arXiv:2308.05071},
  eprint = {2308.05071},
  eprinttype = {arXiv},
  url = {https://doi.org/10.48550/arXiv.2308.05071}
}

@article{chen2023transmon,
  title = {Transmon Qubit Readout Fidelity at the Threshold for Quantum Error Correction without a Quantum-Limited Amplifier},
  author = {Chen, Liangyu and Li, Hang-Xi and Lu, Yong and Warren, Christopher W and Križan, Christian J and Kosen, Sandoko and Rommel, Marcus and Ahmed, Shahnawaz and Osman, Amr and Biznárová, Janka and others},
  date = {2023},
  journaltitle = {Npj Quantum Inf.},
  volume = {9},
  number = {1},
  pages = {26},
  publisher = {Nature Publishing Group UK London},
  doi = {10.1038/s41534-023-00689-6},
  url = {https://doi.org/10.1038/s41534-023-00689-6}
}

@article{Chen2024,
  title = {Hollow {{Core DNANF Optical Fiber}} with ¡0.11 {{dB}}/Km {{Loss}}},
  author = {Chen, Y. and Petrovich, M.N. and Fokoua, E. Numkam and Adamu, A.I. and Hassan, M.R.A. and Sakr, H. and Slavík, R. and Gorajoobi, S. Bakhtiari and Alonso, M. and Ando, R. Fatobene and Papadimopoulos, A. and Varghese, T. and Wu, D. and Ando, M. Fatobene and Wisniowski, K. and Sandoghchi, S.R. and Jasion, G.T. and Richardson, D.J. and Poletti, F.},
  date = {2024},
  journaltitle = {Opt. Fiber Commun Conf OFC},
  pages = {Th4A.8},
  publisher = {Optica Publishing Group},
  doi = {10.1364/OFC.2024.Th4A.8},
  url = {https://opg.optica.org/abstract.cfm?URI=OFC-2024-Th4A.8},
}

@article{chertkov2022holographic,
  title = {Holographic Dynamics Simulations with a Trapped-Ion Quantum Computer},
  author = {Chertkov, Eli and Bohnet, Justin and Francois, David and Gaebler, John and Gresh, Dan and Hankin, Aaron and Lee, Kenny and Hayes, David and Neyenhuis, Brian and Stutz, Russell and others},
  date = {2022},
  journaltitle = {Nat. Phys.},
  volume = {18},
  number = {9},
  pages = {1074},
  publisher = {Nature Publishing Group UK London},
  url = {https://doi.org/10.1038/s41567-022-01689-7}
}

@article{chertkov2023characterizing,
  title = {Characterizing a Non-Equilibrium Phase Transition on a Quantum Computer},
  author = {Chertkov, Eli and Cheng, Zihan and Potter, Andrew C and Gopalakrishnan, Sarang and Gatterman, Thomas M and Gerber, Justin A and Gilmore, Kevin and Gresh, Dan and Hall, Alex and Hankin, Aaron and others},
  date = {2023},
  journaltitle = {Nat. Phys.},
  volume = {19},
  number = {12},
  pages = {1799},
  publisher = {Nature Publishing Group UK London},
  url = {https://doi.org/10.1038/s41567-023-02199-w}
}

@article{childs2001realization,
  title = {Realization of Quantum Process Tomography in {{NMR}}},
  author = {Childs, Andrew M. and Chuang, Isaac L. and Leung, Debbie W.},
  date = {2001-06},
  journaltitle = {Phys. Rev. A},
  volume = {64},
  number = {1},
  pages = {012314},
  publisher = {American Physical Society},
  doi = {10.1103/PhysRevA.64.012314},
  url = {https://link.aps.org/doi/10.1103/PhysRevA.64.012314},
  pagetotal = {7}
}

@article{chirolli2008decoherence,
  title = {Decoherence in Solid-State Qubits},
  author = {Chirolli, Luca and Burkard, Guido},
  date = {2008},
  journaltitle = {Adv. Phys.},
  volume = {57},
  number = {3},
  pages = {225},
  publisher = {Taylor \& Francis},
  url = {https://doi.org/10.1080/00018730802218067}
}

@article{Choi2008minor,
  title = {Minor-{{Embedding}} in {{Adiabatic Quantum Computation}}: {{I}}. {{The Parameter Setting Problem}}},
  author = {Choi, Vicky},
  date = {2008-10},
  journaltitle = {Quantum Inf. Process.},
  volume = {7},
  number = {5},
  pages = {193},
  issn = {1573-1332},
  doi = {10.1007/s11128-008-0082-9},
}

@article{choi2011minor,
  title = {Minor-{{Embedding}} in {{Adiabatic Quantum Computation}}: {{II}}. {{Minor-universal Graph Design}}},
  author = {Choi, Vicky},
  date = {2011-06},
  journaltitle = {Quantum Inf. Process.},
  volume = {10},
  number = {3},
  pages = {343},
  publisher = {Kluwer Academic Publishers},
  location = {USA},
  issn = {1570-0755},
  doi = {10.1007/s11128-010-0200-3},
  issue_date = {June 2011}
}

@article{chow2010detecting,
  title = {Detecting Highly Entangled States with a Joint Qubit Readout},
  author = {Chow, J. M. and DiCarlo, L. and Gambetta, J. M. and Nunnenkamp, A. and Bishop, Lev S. and Frunzio, L. and Devoret, M. H. and Girvin, S. M. and Schoelkopf, R. J.},
  date = {2010-06},
  journaltitle = {Phys. Rev. A},
  volume = {81},
  number = {6},
  pages = {062325},
  publisher = {American Physical Society},
  doi = {10.1103/PhysRevA.81.062325},
  url = {https://link.aps.org/doi/10.1103/PhysRevA.81.062325},
  pagetotal = {8}
}

@article{christandl2012reliable,
  title = {Reliable {{Quantum State Tomography}}},
  author = {Christandl, Matthias and Renner, Renato},
  date = {2012-09},
  journaltitle = {Phys. Rev. Lett.},
  volume = {109},
  number = {12},
  pages = {120403},
  publisher = {American Physical Society},
  doi = {10.1103/PhysRevLett.109.120403},
  url = {https://link.aps.org/doi/10.1103/PhysRevLett.109.120403},
  pagetotal = {6}
}

@article{chuang1997prescription,
  title = {Prescription for Experimental Determination of the Dynamics of a Quantum Black Box},
  author = {Chuang, Isaac L and Nielsen, Michael A},
  date = {1997},
  journaltitle = {J. Mod. Opt.},
  volume = {44},
  number = {11-12},
  pages = {2455},
  publisher = {Taylor \& Francis},
  url = {https://doi.org/10.1080/09500349708231894}
}

@article{cirac2000scalable,
  title = {A Scalable Quantum Computer with Ions in an Array of Microtraps},
  author = {Cirac, Juan Ignacio and Zoller, Peter},
  date = {2000},
  journaltitle = {Nature},
  volume = {404},
  number = {6778},
  pages = {579},
  publisher = {Nature Publishing Group UK London},
  url = {https://doi.org/10.1038/35007021}
}

@inproceedings{Claude2022,
  title = {Secure Multi-Party Quantum Computation},
  booktitle = {Proc. 34th {{Annu}}. {{ACM Symp}}. {{Theory Comput}}.},
  author = {Crépeau, Claude and Gottesman, Daniel and Smith, Adam},
  date = {2002-05},
  pages = {643},
  publisher = {Association for Computing Machinery (ACM)},
  doi = {10.1145/509907.510000},
  url = {https://doi.org/10.1145/509907.510000}
}

@inproceedings{clifford2017classical,
  title = {The {{Classical Complexity}} of {{Boson Sampling}}},
  booktitle = {Proc. 29th {{Annu}}. {{ACM-SIAM Symp}}. {{Discrete Algorithms}}},
  author = {Clifford, Peter and Clifford, Raphaël},
  date = {2017},
  pages = {146},
  doi = {10.1137/1.9781611975031.10},
  pagetotal = {10}
}

@article{cong22,
  title = {Hardware-{{Efficient}}, {{Fault-Tolerant Quantum Computation}} with {{Rydberg Atoms}}},
  author = {Cong, Iris and Levine, Harry and Keesling, Alexander and Bluvstein, Dolev and Wang, Sheng-Tao and Lukin, Mikhail D.},
  date = {2022-06},
  journaltitle = {Phys. Rev. X},
  volume = {12},
  number = {2},
  pages = {021049},
  publisher = {American Physical Society},
  doi = {10.1103/PhysRevX.12.021049},
  url = {https://link.aps.org/doi/10.1103/PhysRevX.12.021049},
  pagetotal = {31}
}

@article{corcolesChallengesOpportunitiesNearTerm2020,
  title = {Challenges and {{Opportunities}} of {{Near-Term Quantum Computing Systems}}},
  author = {Corcoles, A. D. and Kandala, A. and Javadi-Abhari, A. and McClure, D. T. and Cross, A. W. and Temme, K. and Nation, P. D. and Steffen, M. and Gambetta, J. M.},
  date = {2020-08},
  journaltitle = {Proc. IEEE},
  volume = {108},
  number = {8},
  eprintclass = {quant-ph},
  pages = {1338},
  doi = {10.1109/JPROC.2019.2954005}
}

@article{cornelissenScalableBenchmarksGateBased2021,
  title = {Scalable {{Benchmarks}} for {{Gate-Based Quantum Computers}}},
  author = {Cornelissen, Arjan and Bausch, Johannes and Gilyén, András},
  date = {2021-04},
  journaltitle = {arXiv:2104.10698},
  eprint = {2104.10698},
  eprinttype = {arXiv},
  url = {https://doi.org/10.48550/arXiv.2104.10698}
}

@article{cowtan2019qubit,
  title = {On the Qubit Routing Problem},
  author = {Cowtan, Alexander and Dilkes, Silas and Duncan, Ross and Krajenbrink, Alexandre and Simmons, Will and Sivarajah, Seyon},
  date = {2019},
  journaltitle = {arXiv:1902.08091},
  eprint = {1902.08091},
  eprinttype = {arXiv},
  url = {https://doi.org/10.48550/arXiv.1902.08091}
}

@article{crainHighspeedLowcrosstalkDetection2019,
  title = {High-{{Speed Low-Crosstalk Detection}} of a {{171Yb}}+ {{Qubit Using Superconducting Nanowire Single Photon Detectors}}},
  author = {Crain, Stephen and Cahall, Clinton and Vrijsen, Geert and Wollman, Emma E. and Shaw, Matthew D. and Verma, Varun B. and Nam, Sae Woo and Kim, Jungsang},
  date = {2019-08},
  journaltitle = {Nat. Commun.},
  volume = {2},
  number = {1},
  pages = {1},
  publisher = {Nature Publishing Group},
  issn = {2399-3650},
  doi = {10.1038/s42005-019-0195-8},
  url = {https://doi.org/10.1038/s42005-019-0195-8}
}

@article{Crosson2021diabatic,
  title = {Prospects for {{Quantum Enhancement}} with {{Diabatic Quantum Annealing}}},
  author = {Crosson, E. J. and Lidar, D. A.},
  date = {2021-07},
  journaltitle = {Nat. Rev. Phys.},
  volume = {3},
  number = {7},
  pages = {466},
  issn = {2522-5820},
  doi = {10.1038/s42254-021-00313-6},
}

@article{crossValidatingQuantumComputers2019,
  title = {Validating {{Quantum Computers Using Randomized Model Circuits}}},
  author = {Cross, Andrew W. and Bishop, Lev S. and Sheldon, Sarah and Nation, Paul D. and Gambetta, Jay M.},
  date = {2019-09},
  journaltitle = {Phys. Rev. A},
  volume = {100},
  number = {3},
  eprint = {1811.12926},
  eprintclass = {quant-ph},
  pages = {032328},
  issn = {2469-9926, 2469-9934},
  doi = {10.1103/PhysRevA.100.032328},
  url = {https://link.aps.org/doi/10.1103/PhysRevA.100.032328}
}

@article{da2024demonstration,
  title = {Demonstration of Logical Qubits and Repeated Error Correction with Better-than-Physical Error Rates},
  author = {family=Da Silva, given=MP, given-i=MP and Ryan-Anderson, C and family=Bello-Rivas, given=JM, given-i=JM and Chernoguzov, A and family=Dreiling, given=JM, given-i=JM and Foltz, C and family=Gaebler, given=JP, given-i=JP and family=Gatterman, given=TM, given-i=TM and Hayes, D and Hewitt, N and others},
  date = {2024},
  journaltitle = {arXiv:2404.02280},
  eprint = {2404.02280},
  eprinttype = {arXiv},
  url = {https://doi.org/10.48550/arXiv.2404.02280}
}

@article{dallaire-demersApplicationBenchmarkFermionic2020,
  title = {An {{Application Benchmark}} for {{Fermionic Quantum Simulations}}},
  author = {Dallaire-Demers, Pierre-Luc and Stęchły, Michał and Gonthier, Jerome F. and Bashige, Ntwali Toussaint and Romero, Jonathan and Cao, Yudong},
  date = {2020-03},
  journaltitle = {arXiv:2003.01862},
  eprint = {2003.01862},
  eprinttype = {arXiv},
  url = {https://doi.org/10.48550/arXiv.2003.01862}
}

@article{dankert2009exact,
  title = {Exact and {{Approximate Unitary}} 2-{{Designs}} and {{Their Application}} to {{Fidelity Estimation}}},
  author = {Dankert, Christoph and Cleve, Richard and Emerson, Joseph and Livine, Etera},
  date = {2009},
  journaltitle = {Phys. Rev. A},
  volume = {80},
  number = {1},
  pages = {012304},
  publisher = {APS},
  doi = {10.1103/PhysRevA.80.012304},
  url = {https://link.aps.org/doi/10.1103/PhysRevA.80.012304}
}

@article{dariano2001quantum,
  title = {Quantum {{Tomography}} for {{Measuring Experimentally}} the {{Matrix Elements}} of an {{Arbitrary Quantum Operation}}},
  author = {D'Ariano, G. M. and Lo Presti, P.},
  date = {2001-05},
  journaltitle = {Phys. Rev. Lett.},
  volume = {86},
  number = {19},
  pages = {4195},
  publisher = {American Physical Society},
  doi = {10.1103/PhysRevLett.86.4195},
  url = {https://link.aps.org/doi/10.1103/PhysRevLett.86.4195},
  pagetotal = {0}
}

@inproceedings{dasgupta2020characterizing,
  title = {Characterizing the {{Stability}} of {{NISQ Devices}}},
  booktitle = {2020 {{IEEE Int Conf Quantum Comput Eng QCE}}},
  author = {Dasgupta, Samudra and Humble, Travis S.},
  date = {2020},
  pages = {419},
  doi = {10.1109/QCE49297.2020.00059},
  url = {https://doi.org/10.1109/QCE49297.2020.00059}
}

@article{dasgupta2021stability,
  title = {Stability of Noisy Quantum Computing Devices},
  author = {Dasgupta, Samudra and Humble, Travis S},
  date = {2021},
  journaltitle = {arXiv:2105.09472},
  eprint = {2105.09472},
  eprinttype = {arXiv},
  url = {https://doi.org/10.48550/arXiv.2105.09472}
}

@article{dasgupta2024stability,
  title = {Stability of {{Quantum Computers}}},
  author = {Dasgupta, Samudra},
  date = {2024},
  journaltitle = {arXiv:2404.19082},
  eprint = {2404.19082},
  eprinttype = {arXiv},
  url = {https://doi.org/10.48550/arXiv.2404.19082}
}

@inproceedings{dAvossa2023,
  title = {Towards the {{Quantum Internet}}: {{Entanglement Rate Analysis}} of {{High-Efficiency Electro-Optic Transducer}}},
  booktitle = {2023 {{IEEE Int Conf Quantum Comput Eng QCE}}},
  author = {family=Avossa, given=Laura, prefix=d', useprefix=true and Caleffi, Marcello and Wang, Changqing and Illiano, Jessica and Zorzetti, Silvia and Cacciapuoti, Angela Sara},
  date = {2023},
  pages = {1325},
  doi = {10.1109/QCE57702.2023.00150},
  url = {https://doi.org/10.1109/QCE57702.2023.00150}
}

@article{De_Michielis_2023,
  title = {Silicon Spin Qubits from Laboratory to Industry},
  author = {Michielis, Marco De and Ferraro, Elena and Prati, Enrico and Hutin, Louis and Bertrand, Benoit and Charbon, Edoardo and Ibberson, David J and Gonzalez-Zalba, Miguel Fernando},
  date = {2023-06},
  journaltitle = {J. Phys. D: Appl. Phys.},
  volume = {56},
  number = {36},
  pages = {363001},
  publisher = {IOP Publishing},
  doi = {10.1088/1361-6463/acd8c7},
  url = {https://dx.doi.org/10.1088/1361-6463/acd8c7}
}

@article{de2021quantifying,
  title = {Quantifying Dynamics and Interactions of Individual Spurious Low-Energy Fluctuators in Superconducting Circuits},
  author = {family=Graaf, given=S. E., prefix=de, useprefix=true and Mahashabde, S. and Kubatkin, S. E. and family=Tzalenchuk, given=A. Ya., given-i=A{{Ya}} and Danilov, A. V.},
  date = {2021-05},
  journaltitle = {Phys. Rev. B},
  volume = {103},
  number = {17},
  pages = {174103},
  publisher = {American Physical Society},
  doi = {10.1103/PhysRevB.103.174103},
  url = {https://link.aps.org/doi/10.1103/PhysRevB.103.174103},
  pagetotal = {9}
}

@article{de2022chemical,
  title = {Chemical and {{Structural Identification}} of {{Material Defects}} in {{Superconducting Quantum Circuits}}},
  author = {family=Graaf, given=S E, prefix=de, useprefix=true and Un, S and Shard, A G and Lindström, T},
  date = {2022},
  journaltitle = {Mater. Quantum Technol.},
  volume = {2},
  number = {3},
  pages = {032001},
  publisher = {IOP Publishing},
  doi = {10.1088/2633-4356/ac78ba}
}

@article{debnath2016demonstration,
  title = {Demonstration of a Small Programmable Quantum Computer with Atomic Qubits},
  author = {Debnath, Shantanu and Linke, Norbert M and Figgatt, Caroline and Landsman, Kevin A and Wright, Kevin and Monroe, Christopher},
  date = {2016},
  journaltitle = {Nature},
  volume = {536},
  number = {7614},
  pages = {63},
  publisher = {Nature Publishing Group UK London},
  url = {https://doi.org/10.1038/nature18648}
}

@article{dehollain2016optimization,
  title = {Optimization of a Solid-State Electron Spin Qubit Using Gate Set Tomography},
  author = {Dehollain, Juan P and Muhonen, Juha T and Blume-Kohout, Robin and Rudinger, Kenneth M and Gamble, John King and Nielsen, Erik and Laucht, Arne and Simmons, Stephanie and Kalra, Rachpon and Dzurak, Andrew S and others},
  date = {2016},
  journaltitle = {New J. Phys.},
  volume = {18},
  number = {10},
  pages = {103018},
  publisher = {IOP Publishing},
  url = {https://doi.org/10.1088/1367-2630/18/10/103018}
}

@article{derbyshireRandomizedBenchmarkingAnalogue2020,
  title = {Randomized {{Benchmarking}} in the {{Analogue Setting}}},
  author = {Derbyshire, Ellen and Malo, Jorge Yago and Daley, Andrew and Kashefi, Elham and Wallden, Petros},
  date = {2020-07},
  journaltitle = {Quantum Sci. Technol.},
  volume = {5},
  number = {3},
  eprint = {1909.01295},
  eprinttype = {arXiv},
  eprintclass = {quant-ph},
  pages = {034001},
  issn = {2058-9565},
  doi = {10.1088/2058-9565/ab7eec}
}

@article{Deutsch1985,
  title = {Quantum {{Theory}}, the {{Church}}–{{Turing Principle}} and the {{Universal Quantum Computer}}},
  author = {Deutsch, David},
  date = {1985-07},
  journaltitle = {Proc. R. Soc. Lond. Math. Phys. Sci.},
  volume = {400},
  number = {1818},
  pages = {97},
  publisher = {The Royal Society},
  doi = {10.1098/rspa.1985.0070}
}

@article{Deutsch1989,
  title = {Quantum {{Computational Networks}}},
  author = {Deutsch, David},
  date = {1989-09},
  journaltitle = {Proc. R. Soc. Lond. Math. Phys. Sci.},
  volume = {425},
  number = {1868},
  pages = {73},
  publisher = {The Royal Society},
  doi = {10.1098/rspa.1989.0099}
}

@article{devitt2013quantum,
  title = {Quantum Error Correction for Beginners},
  author = {Devitt, Simon J and Munro, William J and Nemoto, Kae},
  date = {2013},
  journaltitle = {Rep. Prog. Phys.},
  volume = {76},
  number = {7},
  pages = {076001},
  publisher = {IOP Publishing},
  url = {https://doi.org/10.1088/0034-4885/76/7/076001}
}

@article{dicarlo2010preparation,
  title = {Preparation and Measurement of Three-Qubit Entanglement in a Superconducting Circuit},
  author = {DiCarlo, Leonardo and Reed, Matthew D and Sun, Luyan and Johnson, Blake R and Chow, Jerry M and Gambetta, Jay M and Frunzio, Luigi and Girvin, Steven M and Devoret, Michel H and Schoelkopf, Robert J},
  date = {2010},
  journaltitle = {Nature},
  volume = {467},
  number = {7315},
  pages = {574},
  publisher = {Nature Publishing Group UK London},
  url = {https://doi.org/10.1038/nature09416}
}

@article{diestel2017graph,
  title = {Graph {{Theory}}: 5th {{Edition}}},
  author = {Diestel, R.},
  date = {2017},
  journaltitle = {Grad. Texts Math. Springer Berl. Heidelb.},
  url = {https://doi.org/10.1007/978-3-662-53622-3}
}

@article{dimatteoUnderstandingHaarMeasure2021,
  title = {Understanding the {{Haar Measure}}},
  author = {Di Matteo, Olivia},
  note = {[Accessed: 25-June-2024]},
  year= {2021},
  journaltitle = {Pennylane Website},
  publisher = {Xanadu},
  url = {https://pennylane.ai/qml/demos/tutorial_haar_measure/}
}

@article{divincenzo2000physical,
  title = {The {{Physical Implementation}} of {{Quantum Computation}}},
  author = {DiVincenzo, David P},
  date = {2000},
  journaltitle = {Fortschr. Phys.},
  volume = {48},
  number = {9-11},
  pages = {771},
  publisher = {Wiley Online Library},
  url = {https://doi.org/10.1002/1521-3978(200009)48:9/11<771::AID-PROP771>3.0.CO;2-E}
}

@article{doi:10.1080/00107514.2019.1667078,
  title = {Quantum Error Correction: An Introductory Guide},
  author = {Roffe, Joschka},
  date = {2019},
  journaltitle = {Contemp. Phys.},
  volume = {60},
  number = {3},
  eprint = {https://doi.org/10.1080/00107514.2019.1667078},
  pages = {226},
  publisher = {Taylor \& Francis},
  doi = {10.1080/00107514.2019.1667078},
  url = {https://doi.org/10.1080/00107514.2019.1667078}
}

@article{doi:10.1098/rspa.1963.0204,
  title = {Electron Correlations in Narrow Energy Bands},
  author = {Hubbard, J. and Flowers, Brian Hilton},
  date = {1963},
  journaltitle = {Proc R Soc Lond. A},
  volume = {276},
  number = {1365},
  eprint = {https://royalsocietypublishing.org/doi/pdf/10.1098/rspa.1963.0204},
  pages = {238},
  doi = {10.1098/rspa.1963.0204},
  url = {https://royalsocietypublishing.org/doi/abs/10.1098/rspa.1963.0204}
}

@article{doi:10.1126/sciadv.aar3960,
  title = {A Crossbar Network for Silicon Quantum Dot Qubits},
  author = {Li, Ruoyu and Petit, Luca and Franke, David P. and Dehollain, Juan Pablo and Helsen, Jonas and Steudtner, Mark and Thomas, Nicole K. and Yoscovits, Zachary R. and Singh, Kanwal J. and Wehner, Stephanie and Vandersypen, Lieven M. K. and Clarke, James S. and Veldhorst, Menno},
  date = {2018},
  journaltitle = {Sci. Adv.},
  volume = {4},
  number = {7},
  pages = {eaar3960},
  doi = {10.1126/sciadv.aar3960},
  url = {https://www.science.org/doi/abs/10.1126/sciadv.aar3960}
}

@article{doi:10.1126/sciadv.abn1717,
  title = {Low-Overhead Fault-Tolerant Quantum Computing Using Long-Range Connectivity},
  author = {Cohen, Lawrence Z. and Kim, Isaac H. and Bartlett, Stephen D. and Brown, Benjamin J.},
  date = {2022},
  journaltitle = {Sci. Adv.},
  volume = {8},
  number = {20},
  eprint = {https://www.science.org/doi/pdf/10.1126/sciadv.abn1717},
  pages = {eabn1717},
  doi = {10.1126/sciadv.abn1717},
  url = {https://www.science.org/doi/abs/10.1126/sciadv.abn1717}
}

@article{doi:10.1126/sciadv.abn5130,
  title = {Two-Qubit Silicon Quantum Processor with Operation Fidelity Exceeding 99\%},
  author = {Mills, Adam R. and Guinn, Charles R. and Gullans, Michael J. and Sigillito, Anthony J. and Feldman, Mayer M. and Nielsen, Erik and Petta, Jason R.},
  date = {2022},
  journaltitle = {Sci. Adv.},
  volume = {8},
  number = {14},
  pages = {eabn5130},
  doi = {10.1126/sciadv.abn5130},
  url = {https://www.science.org/doi/abs/10.1126/sciadv.abn5130}
}

@article{doi:10.1137/19M1305045,
  title = {Scalable {{Semidefinite Programming}}},
  author = {Yurtsever, Alp and Tropp, Joel A. and Fercoq, Olivier and Udell, Madeleine and Cevher, Volkan},
  date = {2021},
  journaltitle = {SIAM J. Math. Data Sci.},
  volume = {3},
  number = {1},
  eprint = {https://doi.org/10.1137/19M1305045},
  pages = {171},
  doi = {10.1137/19M1305045},
  url = {https://doi.org/10.1137/19M1305045}
}

@article{dongRandomCircuitBlockencoded2021,
  title = {Random {{Circuit Block-Encoded Matrix}} and a {{Proposal}} of {{Quantum LINPACK Benchmark}}},
  author = {Dong, Yulong and Lin, Lin},
  date = {2021-06},
  journaltitle = {Phys. Rev. A},
  volume = {103},
  number = {6},
  eprint = {2006.04010},
  eprinttype = {arXiv},
  eprintclass = {quant-ph},
  pages = {062412},
  issn = {2469-9926, 2469-9934},
  doi = {10.1103/PhysRevA.103.062412}
}

@article{donkersQPackScoresQuantitative2022,
  title = {{{QPack Scores}}: {{Quantitative Performance Metrics}} for {{Application-Oriented Quantum Computer Benchmarking}}},
  shorttitle = {{{QPack Scores}}},
  author = {Donkers, Huub and Mesman, Koen and Al-Ars, Zaid and Möller, Matthias},
  date = {2022-05},
  journaltitle = {arXiv:2205.12142},
  eprint = {2205.12142},
  eprinttype = {arXiv},
  url = {https://doi.org/10.48550/arXiv.2205.12142}
}

@article{Drummond22,
  title = {Simulating {{Complex Networks}} in {{Phase Space}}: {{Gaussian Boson Sampling}}},
  author = {Drummond, Peter D. and Opanchuk, Bogdan and Dellios, A. and Reid, M. D.},
  date = {2022-01},
  journaltitle = {Phys. Rev. A},
  volume = {105},
  number = {1},
  pages = {012427},
  publisher = {American Physical Society},
  doi = {10.1103/PhysRevA.105.012427}
}

@article{eastoe2024efficient,
  title = {Efficient System for Bulk Characterization of Cryogenic {{CMOS}} Components},
  author = {Eastoe, Jonathan and Noah, Grayson M and Dutta, Debargha and Rossi, Alessandro and Fletcher, Jonathan D and Gomez-Saiz, Alberto},
  date = {2024},
  journaltitle = {arXiv:2404.11451},
  eprint = {2404.11451},
  eprinttype = {arXiv},
  url = {https://arxiv.org/abs/2404.11451v1}
}

@article{ebadi22,
  title = {Quantum {{Optimization}} of {{Maximum Independent Set Using Rydberg Atom Arrays}}},
  author = {Ebadi, S. and Keesling, A. and Cain, M. and Wang, T. T. and Levine, H. and Bluvstein, D. and Semeghini, G. and Omran, A. and Liu, J.-G. and Samajdar, R. and Luo, X.-Z. and Nash, B. and Gao, X. and Barak, B. and Farhi, E. and Sachdev, S. and Gemelke, N. and Zhou, L. and Choi, S. and Pichler, H. and Wang, S.-T. and Greiner, M. and Vuletic̀, V. and Lukin, M. D.},
  date = {2022},
  journaltitle = {Science},
  volume = {376},
  number = {1209},
  pages = {1209},
  url = {https://doi.org/10.1126/science.abo6587}
}

@article{egan2021fault,
  title = {Fault-Tolerant Control of an Error-Corrected Qubit},
  author = {Egan, Laird and Debroy, Dripto M and Noel, Crystal and Risinger, Andrew and Zhu, Daiwei and Biswas, Debopriyo and Newman, Michael and Li, Muyuan and Brown, Kenneth R and Cetina, Marko and others},
  date = {2021},
  journaltitle = {Nature},
  volume = {598},
  number = {7880},
  pages = {281},
  publisher = {Nature Publishing Group UK London},
  url = {https://doi.org/10.1038/s41586-021-03928-y}
}

@article{eisertQuantumCertificationBenchmarking2020,
  title = {Quantum {{Certification}} and {{Benchmarking}}},
  author = {Eisert, Jens and Hangleiter, Dominik and Walk, Nathan and Roth, Ingo and Markham, Damian and Parekh, Rhea and Chabaud, Ulysse and Kashefi, Elham},
  date = {2020-07},
  journaltitle = {Nat. Rev. Phys.},
  volume = {2},
  number = {7},
  pages = {382},
  issn = {2522-5820},
  doi = {10.1038/s42254-020-0186-4}
}

@article{Elizabeth2019,
  title = {The {{Random Matrix Theory}} of the {{Classical Compact Groups}}},
  author = {Meckes, Elizabeth S.},
  journaltitle = {Camb. Univ. Press 2019},
  url = {https://doi.org/10.1017/9781108303453}
}

@article{emerson2005scalable,
  title = {Scalable {{Noise Estimation}} with {{Random Unitary Operators}}},
  author = {Emerson, Joseph and Alicki, Robert and Życzkowski, Karol},
  date = {2005},
  journaltitle = {J. Opt. B Quantum Semiclassical Opt.},
  volume = {7},
  number = {10},
  pages = {S347},
  publisher = {IOP Publishing},
  doi = {10.1088/1464-4266/7/10/021}
}

@article{emerson2007symmetrized,
  title = {Symmetrized Characterization of Noisy Quantum Processes},
  author = {Emerson, Joseph and Silva, Marcus and Moussa, Osama and Ryan, Colm and Laforest, Martin and Baugh, Jonathan and Cory, David G and Laflamme, Raymond},
  date = {2007},
  journaltitle = {Science},
  volume = {317},
  number = {5846},
  pages = {1893},
  publisher = {American Association for the Advancement of Science},
  url = {https://doi.org/10.1126/science.1145699}
}

@article{endres16,
  title = {Atom-by-Atom Assembly of Defect-Free One-Dimensional Cold Atom Arrays},
  author = {Endres, Manuel and Bernien, Hannes and Keesling, Alexander and Levine, Harry and Anschuetz, Eric R. and Krajenbrink, Alexandre and Senko, Crystal and Vuletic, Vladan and Greiner, Markus and Lukin, Mikhail D.},
  date = {2016},
  journaltitle = {Science},
  volume = {354},
  number = {6315},
  pages = {1024},
  doi = {10.1126/science.aah3752}
}

@article{erhard2021entangling,
  title = {Entangling Logical Qubits with Lattice Surgery},
  author = {Erhard, Alexander and Poulsen Nautrup, Hendrik and Meth, Michael and Postler, Lukas and Stricker, Roman and Stadler, Martin and Negnevitsky, Vlad and Ringbauer, Martin and Schindler, Philipp and Briegel, Hans J and others},
  date = {2021},
  journaltitle = {Nature},
  volume = {589},
  number = {7841},
  pages = {220},
  publisher = {Nature Publishing Group UK London},
  url = {https://doi.org/10.1038/s41586-020-03079-6}
}

@article{erhardCharacterizingLargescaleQuantum2019a,
  title = {Characterizing {{Large-Scale Quantum Computers}} via {{Cycle Benchmarking}}},
  author = {Erhard, Alexander and Wallman, Joel J. and Postler, Lukas and Meth, Michael and Stricker, Roman and Martinez, Esteban A. and Schindler, Philipp and Monz, Thomas and Emerson, Joseph and Blatt, Rainer},
  date = {2019-12},
  journaltitle = {Nat. Commun.},
  volume = {10},
  number = {1},
  pages = {5347},
  issn = {2041-1723},
  doi = {10.1038/s41467-019-13068-7},
  langid = {english}
}

@article{evered23,
  title = {High-{{Fidelity Parallel Entangling Gates}} on a {{Neutral-Atom Quantum Computer}}},
  author = {Evered, Simon J. and Bluvstein, Dolev and Kalinowski, Marcin and Ebadi, Sepehr and Manovitz, Tom and Zhou, Hengyun and Li, Sophie H. and Geim, Alexandra A. and Wang, Tout T. and Maskara, Nishad and Levine, Harry and Semeghini, Giulia and Greiner, Markus and Vuletić, Vladan and Lukin, Mikhail D.},
  date = {2023},
  journaltitle = {Nature},
  volume = {622},
  number = {7982},
  pages = {268},
  doi = {10.1038/s41586-023-06481-y}
}

@article{fabrizio2007,
  title = {Effective Method to Compute {{Franck-Condon}} Integrals for Optical Spectra of Large Molecules in Solution},
  author = {Santoro, Fabrizio and Improta, Roberto and Lami, Alessandro and Bloino, Julien and Barone, Vincenzo},
  date = {2007-02},
  journaltitle = {J. Chem. Phys},
  volume = {126},
  number = {8},
  eprint = {https://pubs.aip.org/aip/jcp/article-pdf/doi/10.1063/1.2437197/15395332/084509\_1\_online.pdf},
  pages = {084509},
  doi = {10.1063/1.2437197},
  url = {https://doi.org/10.1063/1.2437197}
}

@article{Farhi2000AQC,
  title = {Quantum {{Computation}} by {{Adiabatic Evolution}}},
  author = {Farhi, Edward and Goldstone, Jeffrey and Gutmann, Sam and Sipser, Michael},
  date = {2000-01},
  journaltitle = {arXiv:0001106},
  url = {https://doi.org/10.48550/arXiv.quant-ph/0001106}
}

@article{farhi2014quantum,
  title = {A Quantum Approximate Optimization Algorithm},
  author = {Farhi, Edward and Goldstone, Jeffrey and Gutmann, Sam},
  date = {2014},
  journaltitle = {arXiv:1411.4028},
  eprint = {1411.4028},
  eprinttype = {arXiv},
  url = {https://arxiv.org/abs/1411.4028}
}

@article{Ferracin2019,
  title = {Accrediting {{Outputs}} of {{Noisy Intermediate-Scale Quantum Computing Devices}}},
  author = {Ferracin, Samuele and Kapourniotis, Theodoros and Datta, Animesh},
  date = {2019-11},
  journaltitle = {New J. Phys.},
  volume = {21},
  pages = {113038},
  publisher = {IOP Publishing},
  doi = {10.1088/1367-2630/ab4fd6}
}

@article{Ferracin2021,
  title = {Experimental {{Accreditation}} of {{Outputs}} of {{Noisy Quantum Computers}}},
  author = {Ferracin, Samuele and Merkel, Seth T. and McKay, David and Datta, Animesh},
  date = {2021-10},
  journaltitle = {Phys. Rev. A},
  volume = {104},
  pages = {042603},
  publisher = {American Physical Society},
  doi = {10.1103/PhysRevA.104.042603}
}

@article{finilla1994quantumannealing,
  title = {Quantum {{Annealing}}: {{A New Method}} for {{Minimizing Multidimensional}} f {{Unctions}}},
  author = {Finilla, A. B. and Gomez, M. A. and Sebenik, C. and Doll, J. D.},
  date = {1994},
  journaltitle = {Chem. Phys. Lett.},
  volume = {219},
  pages = {343},
  doi = {10.1016/0009-2614(94)00117-0}
}

@inproceedings{finzgarQUARKFrameworkQuantum2022,
  title = {{{QUARK}}: {{A Framework}} for {{Quantum Computing Application Benchmarking}}},
  booktitle = {2022 {{IEEE Int Conf Quantum Comput Eng QCE}}},
  author = {Finžgar, Jernej Rudi and Ross, Philipp and Klepsch, Johannes and Luckow, Andre},
  date = {2022-03},
  pages = {226},
  url = {https://doi.ieeecomputersociety.org/10.1109/QCE53715.2022.00042}
}

@article{Fitzsimons2017,
  title = {Unconditionally Verifiable Blind Quantum Computation},
  author = {Fitzsimons, Joseph F. and Kashefi, Elham},
  date = {2017-07},
  journaltitle = {Phys. Rev. A},
  volume = {96},
  number = {1},
  pages = {012303},
  publisher = {American Physical Society},
  doi = {10.1103/PhysRevA.96.012303},
  url = {https://link.aps.org/doi/10.1103/PhysRevA.96.012303},
  pagetotal = {27}
}

@article{fiuravsek2001maximum,
  title = {Maximum-Likelihood Estimation of Quantum Measurement},
  author = {Fiuráš šek, Jaromı́r},
  date = {2001-07},
  journaltitle = {Phys. Rev. A},
  volume = {64},
  number = {2},
  pages = {024102},
  publisher = {American Physical Society},
  doi = {10.1103/PhysRevA.64.024102},
  url = {https://link.aps.org/doi/10.1103/PhysRevA.64.024102},
  pagetotal = {4}
}

@inproceedings{Floyd2023,
  title = {Long-Distance Entanglement Distribution through Satellite Intermediary Entanglement Swapping},
  booktitle = {Proc. {{SPIE Quantum Comput}}. {{Commun}}. {{Simul}}. {{III}}},
  author = {Floyd, John and Kwiat, Paul},
  date = {2023},
  volume = {12446},
  pages = {124460L},
  doi = {10.1117/12.2650193},
  url = {https://doi.org/10.1117/12.2650193}
}

@article{Fokoua2023,
  title = {Loss in Hollow-Core Optical Fibers: Mechanisms, Scaling Rules, and Limits},
  author = {Fokoua, Eric Numkam and Mousavi, Seyed Abokhamis and Jasion, Gregory T. and Richardson, David J. and Poletti, Francesco},
  date = {2023},
  journaltitle = {Adv. Opt. Photon.},
  volume = {15},
  number = {1},
  pages = {1},
  issn = {19438206},
  doi = {10.1364/aop.470592},
  url = {https://doi.org/10.1364/aop.470592}
}

@article{fontana2021,
  title = {Evaluating the Noise Resilience of Variational Quantum Algorithms},
  author = {Fontana, Enrico and Fitzpatrick, Nathan and Ramo, David Muñoz and Duncan, Ross and Rungger, Ivan},
  date = {2021-08},
  journaltitle = {Phys. Rev. A},
  volume = {104},
  number = {2},
  pages = {022403},
  publisher = {American Physical Society},
  doi = {10.1103/PhysRevA.104.022403},
  url = {https://link.aps.org/doi/10.1103/PhysRevA.104.022403},
  pagetotal = {19}
}

@article{fortunato2002implementation,
  title = {Implementation of Universal Control on a Decoherence-Free Qubit},
  author = {Fortunato, Evan M and Viola, Lorenza and Hodges, Jonathan and Teklemariam, Grum and Cory, David G},
  date = {2002},
  journaltitle = {New J. Phys.},
  volume = {4},
  number = {1},
  pages = {5},
  publisher = {IOP Publishing},
  url = {https://doi.org/10.1088/1367-2630/4/1/305}
}

@article{foss2021holographic,
  title = {Holographic Quantum Algorithms for Simulating Correlated Spin Systems},
  author = {Foss-Feig, Michael and Hayes, David and Dreiling, Joan M. and Figgatt, Caroline and Gaebler, John P. and Moses, Steven A. and Pino, Juan M. and Potter, Andrew C.},
  date = {2021-07},
  journaltitle = {Phys. Rev. Res.},
  volume = {3},
  number = {3},
  pages = {033002},
  publisher = {American Physical Society},
  doi = {10.1103/PhysRevResearch.3.033002},
  url = {https://link.aps.org/doi/10.1103/PhysRevResearch.3.033002},
  pagetotal = {12}
}

@article{gaebler2021suppression,
  title = {Suppression of Midcircuit Measurement Crosstalk Errors with Micromotion},
  author = {Gaebler, J. P. and Baldwin, C. H. and Moses, S. A. and Dreiling, J. M. and Figgatt, C. and Foss-Feig, M. and Hayes, D. and Pino, J. M.},
  date = {2021-12},
  journaltitle = {Phys. Rev. A},
  volume = {104},
  number = {6},
  pages = {062440},
  publisher = {American Physical Society},
  doi = {10.1103/PhysRevA.104.062440},
  url = {https://link.aps.org/doi/10.1103/PhysRevA.104.062440},
  pagetotal = {12}
}

@article{gambetta2020ibm,
  title = {{{IBM}}’s Roadmap for Scaling Quantum Technology},
  author = {Gambetta, Jay},
  date = {2020},
  journaltitle = {IBM Res. Blog Sept. 2020},
  url = {https://www.ibm.com/quantum/blog/ibm-quantum-roadmap}
}

@article{Gard_2015,
  title = {An {{Introduction}} to {{Boson-Sampling}}},
  author = {Gard, Bryan T. and Motes, Keith R. and Olson, Jonathan P. and Rohde, Peter P. and Dowling, Jonathan P.},
  date = {2015-06},
  journaltitle = {At. Mesoscale},
  pages = {167},
  publisher = {WORLD SCIENTIFIC},
  doi = {10.1142/9789814678704_0008}
}

@article{GBS2017,
  title = {Gaussian {{Boson Sampling}}},
  author = {Hamilton, Craig S. and Kruse, Regina and Sansoni, Linda and Barkhofen, Sonja and Silberhorn, Christine and Jex, Igor},
  date = {2017-10},
  journaltitle = {Phys. Rev. Lett.},
  volume = {119},
  number = {17},
  pages = {170501},
  publisher = {American Physical Society},
  doi = {10.1103/PhysRevLett.119.170501}
}

@article{geller2021toward,
  title = {Toward Efficient Correction of Multiqubit Measurement Errors: {{Pair}} Correlation Method},
  author = {Geller, Michael R and Sun, Mingyu},
  date = {2021},
  journaltitle = {Quantum Sci Technol},
  volume = {6},
  number = {2},
  pages = {025009},
  publisher = {IOP Publishing},
  url = {https://doi.org/10.1088/2058-9565/abd5c9}
}

@article{georgopoulosQuantumComputerBenchmarking2021,
  title = {Quantum {{Computer Benchmarking}} via {{Quantum Algorithms}}},
  author = {Georgopoulos, Konstantinos and Emary, Clive and Zuliani, Paolo},
  date = {2021-12},
  journaltitle = {arXiv:2112.09457},
  eprint = {2112.09457},
  eprinttype = {arXiv},
  url = {https://doi.org/10.48550/arXiv.2112.09457}
}

@article{gheorghiu2019verification,
  title = {Verification of {{Quantum Computation}}: {{An Overview}} of {{Existing Approaches}}},
  author = {Gheorghiu, Alexandru and Kapourniotis, Theodoros and Kashefi, Elham},
  date = {2019},
  journaltitle = {Theory Comput. Syst.},
  volume = {63},
  pages = {715},
  publisher = {Springer},
  doi = {10.1007/s00224-018-9872-3}
}

@article{gilchrist2005distance,
  title = {Distance Measures to Compare Real and Ideal Quantum Processes},
  author = {Gilchrist, Alexei and Langford, Nathan K. and Nielsen, Michael A.},
  date = {2005-06},
  journaltitle = {Phys. Rev. A},
  volume = {71},
  number = {6},
  pages = {062310},
  publisher = {American Physical Society},
  doi = {10.1103/PhysRevA.71.062310},
  pagetotal = {14}
}

@article{Giovannetti2004,
  title = {Quantum-Enhanced Measurements: {{Beating}} the Standard Quantum Limit},
  author = {Giovannetti, Vittorio and Lloyd, Seth and Maccone, Lorenzo},
  date = {2004},
  journaltitle = {Science},
  volume = {306},
  number = {5700},
  pages = {1330},
  issn = {00368075},
  doi = {10.1126/science.1104149},
  url = {https://www.science.org/doi/10.1126/science.1104149}
}

@article{gokhale2021faster,
  title = {Faster and More Reliable Quantum Swaps via Native Gates},
  author = {Gokhale, Pranav and Tomesh, Teague and Suchara, Martin and Chong, Frederic T},
  date = {2021},
  journaltitle = {arXiv:2109.13199},
  eprint = {2109.13199},
  eprinttype = {arXiv},
  url = {https://doi.org/10.48550/arXiv.2109.13199}
}

@article{google2020hartree,
  title = {Hartree-{{Fock}} on a Superconducting Qubit Quantum Computer},
  author = {{Google Quantum AI and Collaborators}},
  date = {2020},
  journaltitle = {Science},
  volume = {369},
  number = {6507},
  pages = {1084},
  publisher = {American Association for the Advancement of Science},
  url = {https://doi.org/10.1126/science.abb9811}
}

@article{google2023suppressing,
  title = {Suppressing {{Quantum Errors}} by {{Scaling}} a {{Surface Code Logical Qubit}}},
  author = {{Google Quantum AI}},
  date = {2023},
  journaltitle = {Nature},
  volume = {614},
  number = {7949},
  pages = {676},
  publisher = {Nature Publishing Group UK London},
  doi = {https://doi.org/10.1038/s41586-022-05434-1}
}

@article{gottesman1997stabilizer,
  title = {Stabilizer Codes and Quantum Error Correction},
  author = {Gottesman, Daniel},
  date = {1997},
  journaltitle = {Calif. Inst. Technol.},
  url = {https://doi.org/10.7907/rzr7-dt72}
}

@inproceedings{gottesman2010introduction,
  title = {An Introduction to Quantum Error Correction and Fault-Tolerant Quantum Computation},
  booktitle = {Quantum {{Inf}}. {{Sci}}. {{Its Contrib}}. {{Math}}. {{Proc Symp Appl Math}}},
  author = {Gottesman, Daniel},
  date = {2010},
  volume = {68},
  pages = {13},
  url = {https://doi.org/10.1090/psapm/068/2762145}
}

@article{Gottesman2012,
  title = {Longer-{{Baseline Telescopes Using Quantum Repeaters}}},
  author = {Gottesman, Daniel and Jennewein, Thomas and Croke, Sarah},
  date = {2012-08},
  journaltitle = {Phys. Rev. Lett.},
  volume = {109},
  number = {7},
  pages = {070503},
  publisher = {American Physical Society},
  doi = {10.1103/PhysRevLett.109.070503},
  url = {https://link.aps.org/doi/10.1103/PhysRevLett.109.070503},
  pagetotal = {5}
}

@article{goviaRandomizedBenchmarkingSuite2022,
  title={A randomized benchmarking suite for mid-circuit measurements},
  author={Govia, LCG and Jurcevic, Petar and Wood, CJ and Kanazawa, N and Merkel, ST and McKay, DC},
  journal={New J. Phys.},
  volume={25},
  number={12},
  pages={123016},
  year={2023},
  publisher={IOP Publishing},
  url={https://doi.org/10.1088/1367-2630/ad0e19}
}

@article{graham2022multi,
  title = {Multi-{{Qubit Entanglement}} and {{Algorithms}} on a {{Neutral-Atom Quantum Computer}}},
  author = {Graham, T. M. and Song, Y. and Scott, J. and Poole, C. and Phuttitarn, L. and Jooya, K. and Eichler, P. and Jiang, X. and Marra, A. and Grinkemeyer, B. and Kwon, M. and Ebert, M. and Cherek, J. and Lichtman, M. T. and Gillette, M. and Gilbert, J. and Bowman, D. and Ballance, T. and Campbell, C. and Dahl, E. D. and Crawford, O. and Blunt, N. S. and Rogers, B. and Noel, T. and Saffman, M.},
  date = {2022},
  journaltitle = {Nature},
  volume = {604},
  number = {7906},
  pages = {457},
  publisher = {Nature Publishing Group UK London},
  doi = {10.1038/s41586-022-04603-6}
}

@article{graham22,
  title = {Multi-qubit entanglement and algorithms on a neutral-atom quantum computer},
  author = {Graham, T. M. and Song, Y. and Scott, J. and Poole, C. and Phuttitarn, L. and Jooya, K. and Eichler, P. and Jiang, X. and Marra, A. and Grinkemeyer, B. and Kwon, M. and Ebert, {and} M. and Cherek, J. and Lichtman, M. T. and Gillette, M. and Gilbert, J. and Bowman, D. and Ballance, T. and Campbell, C. and Dahl, E. D. and Crawford, O. and Blunt, N. S. and Rogers, B. and Noel, T. and Saffman, {and} M.},
  date = {2022},
  journaltitle = {Nature},
  volume = {604},
  pages = {457},
  url = {https://doi.org/10.1038/s41586-022-04603-6}
}

@article{graham23,
  title = {Midcircuit Measurements on a Single-Species Neutral Alkali Atom Quantum Processor},
  author = {Graham, T. M. and Phuttitarn, L. and Chinnarasu, R. and Song, Y. and Poole, C. and Jooya, K. and Scott, J. and Scott, A. and Eichler, P. and Saffman, M.},
  journal = {Phys. Rev. X},
  volume = {13},
  issue = {4},
  pages = {041051},
  numpages = {22},
  year = {2023},
  month = {12},
  publisher = {American Physical Society},
  url = {https://link.aps.org/doi/10.1103/PhysRevX.13.041051}
}

@article{granade2016practical,
  title = {Practical Bayesian Tomography},
  author = {Granade, Christopher and Combes, Joshua and family=Cory, given=DG, given-i=DG},
  date = {2016},
  journaltitle = {New J. Phys.},
  volume = {18},
  number = {3},
  pages = {033024},
  publisher = {IOP Publishing},
  url = {https://doi.org/10.1088/1367-2630/18/3/033024}
}

@article{greenbaum_introduction_2015,
  title = {Introduction to {{Quantum Gate Set Tomography}}},
  author = {Greenbaum, Daniel},
  date = {2015-09},
  journaltitle = {arXiv:1509.02921},
  eprint = {1509.02921},
  eprinttype = {arXiv},
  url = {https://doi.org/10.48550/arXiv.1509.02921}
}

@article{griffiths1996semiclassical,
  title = {Semiclassical {{Fourier Transform}} for {{Quantum Computation}}},
  author = {Griffiths, Robert B. and Niu, Chi-Sheng},
  date = {1996-04},
  journaltitle = {Phys. Rev. Lett.},
  volume = {76},
  number = {17},
  pages = {3228},
  publisher = {American Physical Society},
  doi = {10.1103/PhysRevLett.76.3228},
  url = {https://link.aps.org/doi/10.1103/PhysRevLett.76.3228},
  pagetotal = {0}
}

@inproceedings{grover1996,
  title = {A {{Fast Quantum Mechanical Algorithm}} for {{Database Search}}},
  booktitle = {Proc 28th {{Annu ACM Symp Theory Comput}}},
  author = {Grover, Lov K.},
  date = {1996},
  pages = {212},
  url = {https://doi.org/10.1145/237814.237866}
}

@article{gulacsi2023smoking,
  title = {Signatures of Non-{{Markovianity}} of a Superconducting Qubit},
  author = {Gulácsi, Balázs and Burkard, Guido},
  date = {2023-05},
  journaltitle = {Phys. Rev. B},
  volume = {107},
  number = {17},
  pages = {174511},
  publisher = {American Physical Society},
  doi = {10.1103/PhysRevB.107.174511},
  url = {https://link.aps.org/doi/10.1103/PhysRevB.107.174511},
  pagetotal = {9}
}

@article{guo2020tensor,
  title = {Tensor-Network-Based Machine Learning of Non-{{Markovian}} Quantum Processes},
  author = {Guo, Chu and Modi, Kavan and Poletti, Dario},
  date = {2020-12},
  journaltitle = {Phys. Rev. A},
  volume = {102},
  number = {6},
  pages = {062414},
  publisher = {American Physical Society},
  doi = {10.1103/PhysRevA.102.062414},
  url = {https://link.aps.org/doi/10.1103/PhysRevA.102.062414},
  pagetotal = {8}
}

@inproceedings{haah2016sample,
  title = {Sample-Optimal Tomography of Quantum States},
  booktitle = {Proc. 48th {{Annu}}. {{ACM Symp}}. {{Theory Comput}}.},
  author = {Haah, Jeongwan and Harrow, Aram W and Ji, Zhengfeng and Wu, Xiaodi and Yu, Nengkun},
  date = {2016},
  pages = {913},
  url = {https://doi.org/10.1145/2897518.2897585}
}

@article{haner2018software,
  title = {A Software Methodology for Compiling Quantum Programs},
  author = {Häner, Thomas and Steiger, Damian S and Svore, Krysta and Troyer, Matthias},
  date = {2018},
  journaltitle = {Quantum Sci. Technol.},
  volume = {3},
  number = {2},
  pages = {020501},
  publisher = {IOP Publishing},
  url = {https://doi.org/10.1088/2058-9565/aaa5cc}
}

@article{harper2017estimating,
  title = {Estimating the Fidelity of {{T}} Gates Using Standard Interleaved Randomized Benchmarking},
  author = {Harper, Robin and Flammia, Steven T},
  date = {2017},
  journaltitle = {Quantum Sci Technol},
  volume = {2},
  number = {1},
  pages = {015008},
  publisher = {IOP Publishing},
  url = {https://doi.org/10.1088/2058-9565/aa5f8d}
}

@article{harrowQuantumComputationalSupremacy2017,
  title = {Quantum {{Computational Supremacy}}},
  author = {Harrow, Aram W. and Montanaro, Ashley},
  date = {2017-09},
  journaltitle = {Nature},
  volume = {549},
  number = {7671},
  pages = {203},
  issn = {0028-0836, 1476-4687},
  doi = {10.1038/nature23458},
  langid = {english}
}

@article{Hashim2021PRX,
  title = {Randomized {{Compiling}} for {{Scalable Quantum Computing}} on a {{Noisy Superconducting Quantum Processor}}},
  author = {Hashim, Akel and Naik, Ravi K. and Morvan, Alexis and Ville, Jean-Loup and Mitchell, Bradley and Kreikebaum, John Mark and Davis, Marc and Smith, Ethan and Iancu, Costin and O'Brien, Kevin P. and Hincks, Ian and Wallman, Joel J. and Emerson, Joseph and Siddiqi, Irfan},
  date = {2021-11},
  journaltitle = {Phys. Rev. X},
  volume = {11},
  number = {4},
  pages = {041039},
  publisher = {American Physical Society},
  doi = {10.1103/PhysRevX.11.041039},
  url = {https://link.aps.org/doi/10.1103/PhysRevX.11.041039},
  pagetotal = {12}
}

@article{hashim2023benchmarking,
  title = {Benchmarking Quantum Logic Operations Relative to Thresholds for Fault Tolerance},
  author = {Hashim, Akel and Seritan, Stefan and Proctor, Timothy and Rudinger, Kenneth and Goss, Noah and Naik, Ravi K and Kreikebaum, John Mark and Santiago, David I and Siddiqi, Irfan},
  date = {2023},
  journaltitle = {Npj Quantum Inf.},
  volume = {9},
  number = {1},
  pages = {109},
  publisher = {Nature Publishing Group UK London},
  doi = {10.1038/s41534-023-00764-y},
  url = {https://doi.org/10.1038/s41534-023-00764-y}
}

@article{hayes2020eliminating,
  title = {Eliminating {{Leakage Errors}} in {{Hyperfine Qubits}}},
  author = {Hayes, D. and Stack, D. and Bjork, B. and Potter, A. C. and Baldwin, C. H. and Stutz, R. P.},
  date = {2020-04},
  journaltitle = {Phys. Rev. Lett.},
  volume = {124},
  number = {17},
  pages = {170501},
  publisher = {American Physical Society},
  doi = {10.1103/PhysRevLett.124.170501},
  url = {https://link.aps.org/doi/10.1103/PhysRevLett.124.170501},
  pagetotal = {5}
}

@article{heinz2021,
  title = {Crosstalk Analysis for Single-Qubit and Two-Qubit Gates in Spin Qubit Arrays},
  author = {Heinz, Irina and Burkard, Guido},
  date = {2021-07},
  journaltitle = {Phys. Rev. B},
  volume = {104},
  number = {4},
  pages = {045420},
  publisher = {American Physical Society},
  doi = {10.1103/PhysRevB.104.045420},
  url = {https://link.aps.org/doi/10.1103/PhysRevB.104.045420},
  pagetotal = {9}
}

@article{helsen2022general,
  title = {General {{Framework}} for {{Randomized Benchmarking}}},
  author = {Helsen, Jonas and Roth, Ingo and Onorati, Emilio and Werner, Albert H and Eisert, Jens},
  date = {2022},
  journaltitle = {PRX Quantum},
  volume = {3},
  number = {2},
  pages = {020357},
  publisher = {APS},
  doi = {10.1103/PRXQuantum.3.020357},
  url = {https://link.aps.org/doi/10.1103/PRXQuantum.3.020357}
}

@article{herbert2018depth,
  title = {On the Depth Overhead Incurred When Running Quantum Algorithms on Near-Term Quantum Computers with Limited Qubit Connectivity},
  author = {Herbert, Steven},
  date = {2018},
  journaltitle = {arXiv:1805.12570},
  eprint = {1805.12570},
  eprinttype = {arXiv},
  url = {https://doi.org/10.48550/arXiv.1805.12570}
}

@article{HibatAllah2024,
  title = {A Framework for Demonstrating Practical Quantum Advantage: Comparing Quantum against Classical Generative Models},
  author = {Hibat-Allah, Mohamed and Mauri, Marta and Carrasquilla, Juan and Perdomo-Ortiz, Alejandro},
  date = {2024},
  journaltitle = {Commun. Phys.},
  volume = {7},
  number = {1},
  pages = {68},
  issn = {23993650},
  doi = {10.1038/s42005-024-01552-6},
  url = {https://doi.org/10.1038/s42005-024-01552-6}
}

@article{Holmes_2020,
  title = {Impact of Qubit Connectivity on Quantum Algorithm Performance},
  author = {Holmes, Adam and Johri, Sonika and Guerreschi, Gian Giacomo and Clarke, James S and Matsuura, A Y},
  date = {2020-03},
  journaltitle = {Quantum Sci. Technol.},
  volume = {5},
  number = {2},
  pages = {025009},
  publisher = {IOP Publishing},
  doi = {10.1088/2058-9565/ab73e0},
  url = {https://dx.doi.org/10.1088/2058-9565/ab73e0}
}

@article{horowitz2019quantum,
  title = {Quantum Computing: Progress and Prospects},
  author = {Horowitz, Mark and Grumbling, Emily},
  date = {2019},
  journaltitle = {Natl. Acad. Press},
  url = {https://doi.org/10.17226/25196}
}

@article{hradil1997quantum,
  title = {Quantum-State Estimation},
  author = {Hradil, Z.},
  date = {1997-03},
  journaltitle = {Phys. Rev. A},
  volume = {55},
  number = {3},
  pages = {R1561},
  publisher = {American Physical Society},
  doi = {10.1103/PhysRevA.55.R1561},
  url = {https://link.aps.org/doi/10.1103/PhysRevA.55.R1561},
  pagetotal = {0}
}

@article{Hu2015,
  title = {Boson {{Sampling}} for {{Molecular Vibronic Spectra}}},
  author = {Huh, Joonsuk and Guerreschi, Gian Giacomo and Peropadre, Borja and McClean, Jarrod R. and Aspuru-Guzik, Alán},
  date = {2015},
  journaltitle = {Nat. Photonics},
  volume = {9},
  number = {9},
  pages = {615},
  doi = {10.1038/nphoton.2015.153}
}

@article{huang2023near,
  title = {Near-Term Quantum Computing Techniques: {{Variational}} Quantum Algorithms, Error Mitigation, Circuit Compilation, Benchmarking and Classical Simulation},
  author = {Huang, He-Liang and Xu, Xiao-Yue and Guo, Chu and Tian, Guojing and Wei, Shi-Jie and Sun, Xiaoming and Bao, Wan-Su and Long, Gui-Lu},
  date = {2023},
  journaltitle = {Sci. China Phys. Mech. Astron.},
  volume = {66},
  number = {5},
  pages = {250302},
  publisher = {Springer},
  url = {https://doi.org/10.1007/s11433-022-2057-y}
}

@article{huangFidelityBenchmarksTwoqubit2019,
  title = {Fidelity {{Benchmarks}} for {{Two-Qubit Gates}} in {{Silicon}}},
  author = {Huang, W. and Yang, C. H. and Chan, K. W. and Tanttu, T. and Hensen, B. and Leon, R. C. C. and Fogarty, M. A. and Hwang, J. C. C. and Hudson, F. E. and Itoh, K. M. and Morello, A. and Laucht, A. and Dzurak, A. S.},
  date = {2019-05},
  journaltitle = {Nature},
  volume = {569},
  number = {7757},
  pages = {532},
  issn = {0028-0836, 1476-4687},
  doi = {10.1038/s41586-019-1197-0},
  langid = {english}
}

@article{huft22,
  title = {Simple, Passive Design for Large Optical Trap Arrays for Single Atoms},
  author = {Huft, P. and Song, Y. and Graham, T. M. and Jooya, K. and Deshpande, S. and Fang, C. and Kats, M. and Saffman, M.},
  date = {2022-06},
  journaltitle = {Phys. Rev. A},
  volume = {105},
  number = {6},
  pages = {063111},
  publisher = {American Physical Society},
  doi = {10.1103/PhysRevA.105.063111},
  url = {https://link.aps.org/doi/10.1103/PhysRevA.105.063111},
  pagetotal = {12}
}

@article{IBMUnveils400,
  title = {{{IBM Unveils}} 400 {{Qubit-Plus Quantum Processor}} and {{Next-Generation IBM Quantum System Two}}},
  author = {{IBM}},
  year = {2022},
  note = {[Accessed: 1-November-2023]},
  journaltitle = {IBM Newsroom},
  url = {https://newsroom.ibm.com/2022-11-09-IBM-Unveils-400-Qubit-Plus-Quantum-Processor-and-Next-Generation-IBM-Quantum-System-Two}
}

@article{Inagaki2016coherentIsing,
  title = {A {{Coherent Ising Machine}} for 2000-{{Node Optimization Problems}}},
  author = {Inagaki, Takahiro and Haribara, Yoshitaka and Igarashi, Koji and Sonobe, Tomohiro and Tamate, Shuhei and Honjo, Toshimori and Marandi, Alireza and McMahon, Peter L. and Umeki, Takeshi and Enbutsu, Koji and Tadanaga, Osamu and Takenouchi, Hirokazu and Aihara, Kazuyuki and Kawarabayashi, Ken-ichi and Inoue, Kyo and Utsunomiya, Shoko and {Hiroki Takesue}},
  date = {2016},
  journaltitle = {Science},
  volume = {354},
  number = {6312},
  eprint = {https://www.science.org/doi/pdf/10.1126/science.aah4243},
  pages = {603},
  doi = {10.1126/science.aah4243},
}

@article{ion:3d-nist,
  title = {Transport of Quantum States and Separation of Ions in a Dual {{RF}} Ion Trap},
  author = {Rowe, M. A. and Ben-Kish, A. and Demarco, B. and Leibfried, D. and Meyer, V. and Beall, J. and Britton, J. and Hughes, J. and Itano, W. M. and Jelenković, B. and Langer, C. and Rosenband, T. and Wineland, D. J.},
  date = {2002-06},
  journaltitle = {Quantum Info. Comput.},
  volume = {2},
  number = {4},
  pages = {257},
  publisher = {Rinton Press, Incorporated},
  location = {Paramus, NJ},
  issn = {1533-7146},
  url = {https://dl.acm.org/doi/abs/10.5555/2011477.2011478},
  issue_date = {June 2002},
  pagetotal = {15}
}

@article{ion:3d-npl,
  title = {A Monolithic Array of Three-Dimensional Ion Traps Fabricated with Conventional Semiconductor Technology},
  author = {Wilpers, Guido and See, Patrick and Gill, Patrick and Sinclair, Alastair G},
  date = {2012},
  journaltitle = {Nat. Nanotechnol.},
  volume = {7},
  pages = {572},
  doi = {10.1038/nnano.2012.126},
  url = {https://doi.org/10.1038/nnano.2012.126}
}

@article{ion:blueprint,
  title = {Blueprint for a Microwave Trapped Ion Quantum Computer},
  author = {Lekitsch, Bjoern and Weidt, Sebastian and Fowler, Austin G. and Mølmer, Klaus and Devitt, Simon J. and Wunderlich, Christof and Hensinger, Winfried K.},
  date = {2017},
  journaltitle = {Sci. Adv.},
  volume = {3},
  number = {2},
  eprint = {https://www.science.org/doi/pdf/10.1126/sciadv.1601540},
  pages = {e1601540},
  doi = {10.1126/sciadv.1601540},
  url = {https://www.science.org/doi/abs/10.1126/sciadv.1601540}
}

@article{ion:cirac-zoller,
  title = {Quantum {{Computations}} with {{Cold Trapped Ions}}},
  author = {Cirac, J. I. and Zoller, P.},
  date = {1995-05},
  journaltitle = {Phys. Rev. Lett.},
  volume = {74},
  number = {20},
  pages = {4091},
  publisher = {American Physical Society},
  doi = {10.1103/PhysRevLett.74.4091},
  url = {https://link.aps.org/doi/10.1103/PhysRevLett.74.4091},
  pagetotal = {0}
}

@article{ion:dynamic-phase-gate,
  title = {Experimental Demonstration of a Robust, High-Fidelity Geometric Two Ion-Qubit Phase Gate},
  author = {{D. Leibfried} and {B. DeMarco} and {V. Meyer} and {D. Lucas} and {M. Barrett} and {J. Britton} and {W. M. Itano} and {B. Jelenkovi} and {C. Langer} and {T. Rosenband} and Wineland, D. J.},
  date = {2003},
  journaltitle = {Nature},
  volume = {422},
  pages = {412},
  url = {https://doi.org/10.1038/nature01492}
}

@article{ion:honeywell,
  title = {Implementing Fault-Tolerant Entangling Gates on the Five-Qubit Code and the Color Code},
  author = {{C. Ryan-Anderson} and {N. C. Brown} and {M. S. Allman} and {B. Arkin} and {G. Asa-Attuah} and {C. Baldwin} and {J. Berg} and {J. G. Bohnet} and {S. Braxton} and {N. Burdick} and {J. P. Campora} and {A. Chernoguzov} and {J. Esposito} and {B. Evans} and {D. Francois} and {J. P. Gaebler} and {T. M. Gatterman} and {J. Gerber} and {K. Gilmore} and {D. Gresh} and {A. Hall} and {A. Hankin} and {J. Hostetter} and {D. Lucchetti} and {K. Mayer} and {J. Myers} and {B. Neyenhuis} and {J. Santiago} and {J. Sedlacek} and {T. Skripka} and {A. Slattery} and {R. P. Stutz} and {J. Tait} and {R. Tobey} and {G. Vittorini} and {J. Walker} and Hayes, D.},
  date = {2022},
  journaltitle = {arXiv:2208.01863},
  eprint = {2208.01863},
  eprinttype = {arXiv},
  url = {https://doi.org/10.48550/arXiv.2208.01863}
}

@article{ion:m-s,
  title = {Entanglement and Quantum Computation with Ions in Thermal Motion},
  author = {Sørensen, Anders and Mølmer, Klaus},
  date = {2000-07},
  journaltitle = {Phys. Rev. A},
  volume = {62},
  number = {2},
  pages = {022311},
  publisher = {American Physical Society},
  doi = {10.1103/PhysRevA.62.022311},
  url = {https://link.aps.org/doi/10.1103/PhysRevA.62.022311},
  pagetotal = {11}
}

@article{ion:modular,
  title = {Large-Scale Modular Quantum-Computer Architecture with Atomic Memory and Photonic Interconnects},
  author = {Monroe, C. and Raussendorf, R. and Ruthven, A. and Brown, K. R. and Maunz, P. and Duan, L.-M. and Kim, J.},
  date = {2014-02},
  journaltitle = {Phys. Rev. A},
  volume = {89},
  number = {2},
  pages = {022317},
  publisher = {American Physical Society},
  doi = {10.1103/PhysRevA.89.022317},
  url = {https://link.aps.org/doi/10.1103/PhysRevA.89.022317},
  pagetotal = {16}
}

@article{ion:omg,
  title = {Omg Blueprint for Trapped Ion Quantum Computing with Metastable States},
  author = {Allcock, D. T. C. and Campbell, W. C. and Chiaverini, J. and Chuang, I. L. and Hudson, E. R. and Moore, I. D. and Ransford, A. and Roman, C. and Sage, J. M. and Wineland, D. J.},
  date = {2021-11},
  journaltitle = {Appl. Phys. Lett.},
  volume = {119},
  number = {21},
  eprint = {https://pubs.aip.org/aip/apl/article-pdf/doi/10.1063/5.0069544/13267942/214002\_1\_online.pdf},
  pages = {214002},
  issn = {0003-6951},
  doi = {10.1063/5.0069544},
  url = {https://doi.org/10.1063/5.0069544}
}

@article{ion:osc-grad,
  title = {Trapped-{{Ion Quantum Logic Gates Based}} on {{Oscillating Magnetic Fields}}},
  author = {Ospelkaus, C. and Langer, C. E. and Amini, J. M. and Brown, K. R. and Leibfried, D. and Wineland, D. J.},
  date = {2008-08},
  journaltitle = {Phys. Rev. Lett.},
  volume = {101},
  number = {9},
  pages = {090502},
  publisher = {American Physical Society},
  doi = {10.1103/PhysRevLett.101.090502},
  url = {https://link.aps.org/doi/10.1103/PhysRevLett.101.090502},
  pagetotal = {4}
}

@article{ion:penning-array,
  title = {Scalable {{Arrays}} of {{Micro-Penning Traps}} for {{Quantum Computing}} and {{Simulation}}},
  author = {Jain, S. and Alonso, J. and Grau, M. and Home, J. P.},
  date = {2020-08},
  journaltitle = {Phys. Rev. X},
  volume = {10},
  number = {3},
  pages = {031027},
  publisher = {American Physical Society},
  doi = {10.1103/PhysRevX.10.031027},
  url = {https://link.aps.org/doi/10.1103/PhysRevX.10.031027},
  pagetotal = {22}
}

@article{ion:penning-sim,
  title = {Quantum Spin Dynamics and Entanglement Generation with Hundreds of Trapped Ions},
  author = {Bohnet, Justin G. and Sawyer, Brian C. and Britton, Joseph W. and Wall, Michael L. and Rey, Ana Maria and Foss-Feig, Michael and Bollinger, John J.},
  date = {2016},
  journaltitle = {Science},
  volume = {352},
  number = {6291},
  eprint = {https://www.science.org/doi/pdf/10.1126/science.aad9958},
  pages = {1297},
  doi = {10.1126/science.aad9958},
  url = {https://www.science.org/doi/abs/10.1126/science.aad9958}
}

@article{ion:progress-review,
  title = {Trapped-Ion Quantum Computing: {{Progress}} and Challenges},
  author = {Bruzewicz, Colin D. and Chiaverini, John and McConnell, Robert and Sage, Jeremy M.},
  date = {2019-05},
  journaltitle = {Appl. Phys. Rev.},
  volume = {6},
  number = {2},
  eprint = {https://pubs.aip.org/aip/apr/article-pdf/doi/10.1063/1.5088164/14577412/021314\_1\_online.pdf},
  pages = {021314},
  issn = {1931-9401},
  doi = {10.1063/1.5088164},
  url = {https://doi.org/10.1063/1.5088164}
}

@article{ion:readout,
  title = {High-{{Fidelity Readout}} of {{Trapped-Ion Qubits}}},
  author = {Myerson, A. H. and Szwer, D. J. and Webster, S. C. and Allcock, D. T. C. and Curtis, M. J. and Imreh, G. and Sherman, J. A. and Stacey, D. N. and Steane, A. M. and Lucas, D. M.},
  date = {2008-05},
  journaltitle = {Phys. Rev. Lett.},
  volume = {100},
  number = {20},
  pages = {200502},
  publisher = {American Physical Society},
  doi = {10.1103/PhysRevLett.100.200502},
  url = {https://link.aps.org/doi/10.1103/PhysRevLett.100.200502},
  pagetotal = {4}
}

@article{ion:static-grad,
  title = {Ion-{{Trap Quantum Logic Using Long-Wavelength Radiation}}},
  author = {Mintert, Florian and Wunderlich, Christof},
  date = {2001-11},
  journaltitle = {Phys. Rev. Lett.},
  volume = {87},
  number = {25},
  pages = {257904},
  publisher = {American Physical Society},
  doi = {10.1103/PhysRevLett.87.257904},
  url = {https://link.aps.org/doi/10.1103/PhysRevLett.87.257904},
  pagetotal = {4}
}

@article{ion:surface,
  title = {Surface-Electrode Architecture for Ion-Trap Quantum Information Processing},
  author = {Chiaverini, J. and Blakestad, R. B. and Britton, J. and Jost, J. D. and Langer, C. and Leibfried, D. and Ozeri, R. and Wineland, D. J.},
  date = {2005-09},
  journaltitle = {Quantum Info. Comput.},
  volume = {5},
  number = {6},
  pages = {419},
  publisher = {Rinton Press, Incorporated},
  location = {Paramus, NJ},
  issn = {1533-7146},
  doi = {10.5555/2011670.2011671},
  url = {https://dl.acm.org/doi/abs/10.5555/2011670.2011671},
  issue_date = {September 2005},
  pagetotal = {21}
}

@article{ion:two-qubit1,
  title = {High-{{Fidelity Quantum Logic Gates Using Trapped-Ion Hyperfine Qubits}}},
  author = {Ballance, C. J. and Harty, T. P. and Linke, N. M. and Sepiol, M. A. and Lucas, D. M.},
  date = {2016-08},
  journaltitle = {Phys. Rev. Lett.},
  volume = {117},
  number = {6},
  pages = {060504},
  publisher = {American Physical Society},
  doi = {10.1103/PhysRevLett.117.060504},
  url = {https://link.aps.org/doi/10.1103/PhysRevLett.117.060504},
  pagetotal = {6}
}

@article{ion:two-qubit2,
  title = {High-{{Fidelity Universal Gate Set}} for ${^{9}\mathrm{Be}}^{+}$ {{Ion Qubits}}},
  author = {Gaebler, J. P. and Tan, T. R. and Lin, Y. and Wan, Y. and Bowler, R. and Keith, A. C. and Glancy, S. and Coakley, K. and Knill, E. and Leibfried, D. and Wineland, D. J.},
  date = {2016-08},
  journaltitle = {Phys. Rev. Lett.},
  volume = {117},
  number = {6},
  pages = {060505},
  publisher = {American Physical Society},
  doi = {10.1103/PhysRevLett.117.060505},
  url = {https://link.aps.org/doi/10.1103/PhysRevLett.117.060505},
  pagetotal = {5}
}

@article{ion:wineland,
  title = {Experimental Issues in Coherent Quantum-State Manipulation of Trapped Atomic Ions},
  author = {{D. J. Wineland} and {C. Monroe} and {W. M. Itano} and {D. Leibfried} and {B. E. King} and Meekhof, D. M.},
  date = {1998},
  journaltitle = {J. Res. Natl. Inst. Stand. Tech.},
  volume = {103},
  pages = {259},
  doi = {10.6028/jres.103.019},
  url = {https://doi.org/10.6028/jres.103.019}
}

@article{ions,
  title = {Scalable {{Implementation}} of {{Boson Sampling}} with {{Trapped Ions}}},
  author = {Shen, C. and Zhang, Z. and Duan, L.-M.},
  date = {2014-02},
  journaltitle = {Phys. Rev. Lett.},
  volume = {112},
  number = {5},
  pages = {050504},
  publisher = {American Physical Society},
  doi = {10.1103/PhysRevLett.112.050504}
}

@article{Iverson_2020,
  title = {Coherence in Logical Quantum Channels},
  author = {Iverson, Joseph K and Preskill, John},
  date = {2020-07},
  journaltitle = {New J. Phys.},
  volume = {22},
  number = {7},
  pages = {073066},
  publisher = {IOP Publishing},
  doi = {10.1088/1367-2630/ab8e5c},
  url = {https://dx.doi.org/10.1088/1367-2630/ab8e5c}
}

@article{jaksch00,
  title = {Fast {{Quantum Gates}} for {{Neutral Atoms}}},
  author = {Jaksch, D. and Cirac, J. I. and Zoller, P. and Rolston, S. L. and Côté, R. and Lukin, M. D.},
  date = {2000-09},
  journaltitle = {Phys. Rev. Lett.},
  volume = {85},
  number = {10},
  pages = {2208},
  publisher = {American Physical Society},
  doi = {10.1103/PhysRevLett.85.2208},
  url = {https://link.aps.org/doi/10.1103/PhysRevLett.85.2208},
  pagetotal = {0}
}

@article{james2001measurement,
  title = {Measurement of Qubits},
  author = {James, Daniel F. V. and Kwiat, Paul G. and Munro, William J. and White, Andrew G.},
  date = {2001-10},
  journaltitle = {Phys. Rev. A},
  volume = {64},
  number = {5},
  pages = {052312},
  publisher = {American Physical Society},
  doi = {10.1103/PhysRevA.64.052312},
  url = {https://link.aps.org/doi/10.1103/PhysRevA.64.052312},
  pagetotal = {15}
}

@article{Jiuzhang1,
  title = {Quantum {{Computational Advantage Using Photons}}},
  author = {Zhong, Han-Sen and Wang, Hui and Deng, Yu-Hao and Chen, Ming-Cheng and Peng, Li-Chao and Luo, Yi-Han and Qin, Jian and Wu, Dian and Ding, Xing and Hu, Yi and Hu, Peng and Yang, Xiao-Yan and Zhang, Wei-Jun and Li, Hao and Li, Yuxuan and Jiang, Xiao and Gan, Lin and Yang, Guangwen and You, Lixing and Wang, Zhen and Li, Li and Liu, Nai-Le and Lu, Chao-Yang and {Jian-Wei Pan}},
  date = {2020},
  journaltitle = {Science},
  volume = {370},
  number = {6523},
  eprint = {https://www.science.org/doi/pdf/10.1126/science.abe8770},
  pages = {1460},
  doi = {10.1126/science.abe8770}
}

@article{Jiuzhang2,
  title = {Phase-{{Programmable Gaussian Boson Sampling Using Stimulated Squeezed Light}}},
  author = {Zhong, Han-Sen and Deng, Yu-Hao and Qin, Jian and Wang, Hui and Chen, Ming-Cheng and Peng, Li-Chao and Luo, Yi-Han and Wu, Dian and Gong, Si-Qiu and Su, Hao and Hu, Yi and Hu, Peng and Yang, Xiao-Yan and Zhang, Wei-Jun and Li, Hao and Li, Yuxuan and Jiang, Xiao and Gan, Lin and Yang, Guangwen and You, Lixing and Wang, Zhen and Li, Li and Liu, Nai-Le and Renema, Jelmer J. and Lu, Chao-Yang and Pan, Jian-Wei},
  date = {2021-10},
  journaltitle = {Phys. Rev. Lett.},
  volume = {127},
  number = {18},
  pages = {180502},
  publisher = {American Physical Society},
  doi = {10.1103/PhysRevLett.127.180502},
  url = {https://link.aps.org/doi/10.1103/PhysRevLett.127.180502},
  pagetotal = {9}
}

@article{Jiuzhang3,
  title = {Gaussian {{Boson Sampling}} with {{Pseudo-Photon-Number Resolving Detectors}} and {{Quantum Computational Advantage}}},
  author = {Deng, Yu-Hao and Gu, Yi-Chao and Liu, Hua-Liang and Gong, Si-Qiu and Su, Hao and Zhang, Zhi-Jiong and Tang, Hao-Yang and Jia, Meng-Hao and Xu, Jia-Min and Chen, Ming-Cheng and Qin, Jian and Peng, Li-Chao and Yan, Jiarong and Hu, Yi and Huang, Jia and Li, Hao and Li, Yuxuan and Chen, Yaojian and Jiang, Xiao and Gan, Lin and Yang, Guangwen and You, Lixing and Li, Li and Zhong, Han-Sen and Wang, Hui and Liu, Nai-Le and Renema, Jelmer J. and Lu, Chao-Yang and Pan, Jian-Wei},
  date = {2023-10},
  journaltitle = {Phys. Rev. Lett.},
  volume = {131},
  number = {15},
  pages = {150601},
  publisher = {American Physical Society},
  doi = {10.1103/PhysRevLett.131.150601},
  url = {https://link.aps.org/doi/10.1103/PhysRevLett.131.150601},
  pagetotal = {7}
}

@article{jock2022silicon,
  title = {A Silicon Singlet–Triplet Qubit Driven by Spin-Valley Coupling},
  author = {Jock, Ryan M and Jacobson, N Tobias and Rudolph, Martin and Ward, Daniel R and Carroll, Malcolm S and Luhman, Dwight R},
  date = {2022},
  journaltitle = {Nat Commun.},
  volume = {13},
  number = {1},
  pages = {641},
  url = {https://doi.org/10.1038/s41467-022-28302-y}
}

@article{johnson11aManufacturedSpins,
  title = {Quantum {{Annealing}} with {{Manufactured Spins}}},
  author = {Johnson, M. W. and Amin, M. H. S. and Gildert, S. and Lanting, T. and Hamze, F. and Dickson, N. and Harris, R. and Berkley, A. J. and Johansson, J. and Bunyk, P. and Chapple, E. M. and Enderud, C. and Hilton, J. P. and Karimi, K. and Ladizinsky, E. and Ladizinsky, N. and Oh, T. and Perminov, I. and Rich, C. and Thom, M. C. and Tolkacheva, E. and Truncik, C. J. S. and Uchaikin, S. and Wang, J. and Rose, B. Wilsonand G.},
  date = {2011},
  journaltitle = {Nature},
  volume = {473},
  pages = {194},
  doi = {doi:10.1038/nature10012}
}

@article{jozsa1994fidelity,
  title = {Fidelity for Mixed Quantum States},
  author = {Jozsa, Richard},
  date = {1994},
  journaltitle = {J. Mod. Opt.},
  volume = {41},
  number = {12},
  pages = {2315},
  publisher = {Taylor \& Francis},
  url = {https://doi.org/10.1080/09500349414552171}
}

@article{jurcevic2021demonstration,
  title = {Demonstration of Quantum Volume 64 on a Superconducting Quantum Computing System},
  author = {Jurcevic, Petar and Javadi-Abhari, Ali and Bishop, Lev S and Lauer, Isaac and Bogorin, Daniela F and Brink, Markus and Capelluto, Lauren and Günlük, Oktay and Itoko, Toshinari and Kanazawa, Naoki and others},
  date = {2021},
  journaltitle = {Quantum Sci. Technol.},
  volume = {6},
  number = {2},
  pages = {025020},
  publisher = {IOP Publishing},
  url = {https://doi.org/10.1088/2058-9565/abe519}
}

@article{Kaczmarek2018,
  title = {High-Speed Noise-Free Optical Quantum Memory},
  author = {Kaczmarek, K. T. and Ledingham, P. M. and Brecht, B. and Thomas, S. E. and Thekkadath, G. S. and Lazo-Arjona, O. and Munns, J. H. D. and Poem, E. and Feizpour, A. and Saunders, D. J. and Nunn, J. and Walmsley, I. A.},
  date = {2018-04},
  journaltitle = {Phys. Rev. A},
  volume = {97},
  number = {4},
  pages = {042316},
  publisher = {American Physical Society},
  doi = {10.1103/PhysRevA.97.042316},
  url = {https://link.aps.org/doi/10.1103/PhysRevA.97.042316},
  pagetotal = {10}
}

@article{Kadowaki1998annealing,
  title = {Quantum {{Annealing}} in the {{Transverse Ising Model}}},
  author = {Kadowaki, Tadashi and Nishimori, Hidetoshi},
  date = {1998},
  journaltitle = {Phys. Rev. E},
  volume = {58},
  number = {5},
  eprint = {9804280 [cond-mat]},
  eprinttype = {arXiv},
  pages = {5355},
  issn = {1063651X},
  doi = {10.1103/PhysRevE.58.5355},
  arxivid = {cond-mat/9804280}
}

@article{keith2018joint,
  title = {Joint Quantum-State and Measurement Tomography with Incomplete Measurements},
  author = {Keith, Adam C. and Baldwin, Charles H. and Glancy, Scott and Knill, E.},
  date = {2018-10},
  journaltitle = {Phys. Rev. A},
  volume = {98},
  number = {4},
  pages = {042318},
  publisher = {American Physical Society},
  doi = {10.1103/PhysRevA.98.042318},
  url = {https://link.aps.org/doi/10.1103/PhysRevA.98.042318},
  pagetotal = {12}
}

@article{Kempe_AQC_universal_2006,
  title = {The {{Complexity}} of the {{Local Hamiltonian Problem}}},
  author = {Kempe, Julia and Kitaev, Alexei and Regev, Oded},
  date = {2006},
  journaltitle = {SIAM J. Comput.},
  volume = {35},
  number = {5},
  eprint = {https://doi.org/10.1137/S0097539704445226},
  pages = {1070},
  doi = {10.1137/S0097539704445226},
}

@article{khatri2019quantum,
  title = {Quantum-Assisted Quantum Compiling},
  author = {Khatri, Sumeet and LaRose, Ryan and Poremba, Alexander and Cincio, Lukasz and Sornborger, Andrew T and Coles, Patrick J},
  date = {2019},
  journaltitle = {Quantum},
  volume = {3},
  pages = {140},
  publisher = {Verein zur Förderung des Open Access Publizierens in den Quantenwissenschaften},
  url = {https://doi.org/10.22331/q-2019-05-13-140}
}

@article{kielpinski2001entanglement,
  title = {Entanglement and {{Decoherence}} in a {{Trapped-ion Quantum Register}}},
  author = {Kielpinski, D.},
  date = {2001},
  journaltitle = {Univ. Colo.},
  url = {https://books.google.co.uk/books?id=x_t-0AEACAAJ}
}

@article{kielpinski2002architecture,
  title = {Architecture for a Large-Scale Ion-Trap Quantum Computer},
  author = {Kielpinski, David and Monroe, Chris and Wineland, David J},
  date = {2002},
  journaltitle = {Nature},
  volume = {417},
  number = {6890},
  pages = {709},
  publisher = {Nature Publishing Group UK London},
  url = {https://doi.org/10.1038/nature00784}
}

@article{kim2023evidence,
  title = {Evidence for the Utility of Quantum Computing before Fault Tolerance},
  author = {Kim, Youngseok and Eddins, Andrew and Anand, Sajant and Wei, Ken Xuan and Van Den Berg, Ewout and Rosenblatt, Sami and Nayfeh, Hasan and Wu, Yantao and Zaletel, Michael and Temme, Kristan and others},
  date = {2023},
  journaltitle = {Nature},
  volume = {618},
  number = {7965},
  pages = {500},
  publisher = {Nature Publishing Group UK London},
  url = {https://doi.org/10.1038/s41586-023-06096-3}
}

@article{kimmel2015robust,
  title = {Robust Calibration of a Universal Single-Qubit Gate Set via Robust Phase Estimation},
  author = {Kimmel, Shelby and Low, Guang Hao and Yoder, Theodore J.},
  date = {2015-12},
  journaltitle = {Phys. Rev. A},
  volume = {92},
  number = {6},
  pages = {062315},
  publisher = {American Physical Society},
  doi = {10.1103/PhysRevA.92.062315},
  url = {https://link.aps.org/doi/10.1103/PhysRevA.92.062315},
  pagetotal = {13}
}

@article{King2021advantageQA,
  title = {Scaling {{Advantage}} over {{Path-Integral Monte Carlo}} in {{Quantum Simulation}} of {{Geometrically Frustrated Magnets}}},
  author = {King, Andrew D. and others},
  date = {2021-02},
  journaltitle = {Nat. Commun.},
  volume = {12},
  number = {1},
  pages = {1113},
  issn = {2041-1723},
  doi = {10.1038/s41467-021-20901-5},
}

@article{King2022Coherent,
  title = {Coherent {{Quantum Annealing}} in a {{Programmable}} 2,000 Qubit {{Ising Chain}}},
  author = {King, Andrew D. and others},
  date = {2022-11},
  journaltitle = {Nat. Phys.},
  volume = {18},
  number = {11},
  pages = {1324},
  issn = {1745-2481},
  doi = {10.1038/s41567-022-01741-6}
}

@article{kitaev1995quantum,
  title = {Quantum Measurements and the {{Abelian}} Stabilizer Problem},
  author = {Kitaev, A Yu},
  date = {1995},
  journaltitle = {arXiv:quant-ph/9511026},
  eprint = {quant-ph/9511026},
  eprinttype = {arXiv},
  url = {https://doi.org/10.48550/arXiv.quant-ph/9511026}
}

@article{kjaergaard2020superconducting,
  title = {Superconducting Qubits: {{Current}} State of Play},
  author = {Kjaergaard, Morten and Schwartz, Mollie E and Braumüller, Jochen and Krantz, Philip and Wang, Joel I-J and Gustavsson, Simon and Oliver, William D},
  date = {2020},
  journaltitle = {Annu. Rev. Condens. Matter Phys.},
  volume = {11},
  number = {1},
  pages = {369},
  publisher = {Annual Reviews},
  url = {https://doi.org/10.1146/annurev-conmatphys-031119-050605}
}

@article{Klimov_2018,
  title = {Fluctuations of {{Energy-Relaxation Times}} in {{Superconducting Qubits}}},
  author = {Klimov, P. V. and Kelly, J. and Chen, Z. and Neeley, M. and Megrant, A. and Burkett, B. and Barends, R. and Arya, K. and Chiaro, B. and Chen, Yu and Dunsworth, A. and Fowler, A. and Foxen, B. and Gidney, C. and Giustina, M. and Graff, R. and Huang, T. and Jeffrey, E. and Lucero, Erik and Mutus, J. Y. and Naaman, O. and Neill, C. and Quintana, C. and Roushan, P. and Sank, Daniel and Vainsencher, A. and Wenner, J. and White, T. C. and Boixo, S. and Babbush, R. and Smelyanskiy, V. N. and Neven, H. and Martinis, John M.},
  date = {2018-08},
  journaltitle = {Phys. Rev. Lett.},
  volume = {121},
  number = {9},
  publisher = {American Physical Society (APS)},
  issn = {1079-7114},
  doi = {10.1103/physrevlett.121.090502},
  url = {http://dx.doi.org/10.1103/PhysRevLett.121.090502}
}

@article{KLM2001,
  title = {A {{Scheme}} for {{Efficient Quantum Computation}} with {{Linear Optics}}},
  author = {Knill, E. and Laflamme, R. and Milburn, G. J.},
  date = {2001},
  journaltitle = {Nature},
  volume = {409},
  number = {6816},
  pages = {46},
  doi = {10.1038/35051009}
}

@article{knill1997theory,
  title = {Theory of Quantum Error-Correcting Codes},
  author = {Knill, Emanuel and Laflamme, Raymond},
  date = {1997-02},
  journaltitle = {Phys. Rev. A},
  volume = {55},
  number = {2},
  pages = {900},
  publisher = {American Physical Society},
  doi = {10.1103/PhysRevA.55.900},
  url = {https://link.aps.org/doi/10.1103/PhysRevA.55.900},
  pagetotal = {0}
}

@article{knillAlgorithmicBenchmarkQuantum2000,
  title = {An {{Algorithmic Benchmark}} for {{Quantum Information Processing}}},
  author = {Knill, E. and Laflamme, R. and Martinez, R. and Tseng, C.-H.},
  date = {2000-03},
  journaltitle = {Nature},
  volume = {404},
  number = {6776},
  pages = {368},
  issn = {0028-0836, 1476-4687},
  doi = {10.1038/35006012},
  langid = {english}
}

@article{knillRandomizedBenchmarkingQuantum2008,
  title = {Randomized {{Benchmarking}} of {{Quantum Gates}}},
  author = {Knill, E. and Leibfried, D. and Reichle, R. and Britton, J. and Blakestad, R. B. and Jost, J. D. and Langer, C. and Ozeri, R. and Seidelin, S. and Wineland, D. J.},
  date = {2008-01},
  journaltitle = {Phys. Rev. A},
  volume = {77},
  number = {1},
  pages = {012307},
  issn = {1050-2947, 1094-1622},
  doi = {10.1103/PhysRevA.77.012307},
  langid = {english}
}

@article{Koashi2001,
  title = {Probabilistic Manipulation of Entangled Photons},
  author = {Koashi, Masato and Yamamoto, Takashi and Imoto, Nobuyuki},
  date = {2001-02},
  journaltitle = {Phys. Rev. A},
  volume = {63},
  number = {3},
  pages = {030301},
  publisher = {American Physical Society},
  doi = {10.1103/PhysRevA.63.030301},
  url = {https://link.aps.org/doi/10.1103/PhysRevA.63.030301},
  pagetotal = {4}
}

@article{kochDemonstratingNISQEra2020,
  title = {Demonstrating {{NISQ Era Challenges}} in {{Algorithm Design}} on {{IBM}}'s 20 {{Qubit Quantum Computer}}},
  author = {Koch, Daniel and Martin, Brett and Patel, Saahil and Wessing, Laura and Alsing, Paul M.},
  date = {2020-09},
  journaltitle = {AIP Adv.},
  volume = {10},
  number = {9},
  pages = {095101},
  issn = {2158-3226},
  doi = {10.1063/5.0015526},
  langid = {english}
}

@article{konik2021quantum,
  title = {Quantum Coherence Confined},
  author = {Konik, Robert},
  date = {2021},
  journaltitle = {Nat. Phys.},
  volume = {17},
  number = {6},
  pages = {669},
  publisher = {Nature Publishing Group UK London},
  doi = {10.1038/s41567-021-01211-5},
  url = {https://doi.org/10.1038/s41567-021-01211-5}
}

@article{Kozlowski2023,
  title = {{{RFC}} 9340: {{Architectural Principles}} for a {{Quantum Internet}}},
  author = {Kozlowski, W and Wehner, S and Meter, Qutech R Van and Cacciapuoti, A S and Caleffi, M and Nagayama, S},
  date = {2023},
  journaltitle = {RFC},
  pages = {9340},
  doi = {10.17487/RFC9340},
  url = {https://doi.org/10.17487/RFC9340}
}

@article{krantz_quantum_2019,
  title = {A {{Quantum Engineer}}'s {{Guide}} to {{Superconducting Qubits}}},
  author = {Krantz, Philip and Kjaergaard, Morten and Yan, Fei and Orlando, Terry P. and Gustavsson, Simon and Oliver, William D.},
  date = {2019-06},
  journaltitle = {Appl. Phys. Rev.},
  volume = {6},
  number = {2},
  eprint = {1904.06560},
  eprinttype = {arXiv},
  pages = {021318},
  issn = {1931-9401},
  doi = {10.1063/1.5089550}
}

@article{krinner2020,
  title = {Benchmarking {{Coherent Errors}} in {{Controlled-Phase Gates}} Due to {{Spectator Qubits}}},
  author = {Krinner, S. and Lazar, S. and Remm, A. and Andersen, C.K. and Lacroix, N. and Norris, G.J. and Hellings, C. and Gabureac, M. and Eichler, C. and Wallraff, A.},
  date = {2020-08},
  journaltitle = {Phys. Rev. Appl.},
  volume = {14},
  number = {2},
  pages = {024042},
  publisher = {American Physical Society},
  doi = {10.1103/PhysRevApplied.14.024042},
  url = {https://link.aps.org/doi/10.1103/PhysRevApplied.14.024042},
  pagetotal = {9}
}

@article{ladd2010quantum,
  title = {Quantum Computers},
  author = {Ladd, Thaddeus D and Jelezko, Fedor and Laflamme, Raymond and Nakamura, Yasunobu and Monroe, Christopher and O’Brien, Jeremy Lloyd},
  date = {2010},
  journaltitle = {Nature},
  volume = {464},
  number = {7285},
  pages = {45},
  publisher = {Nature Publishing Group UK London},
  url = {https://doi.org/10.1038/nature08812}
}

@article{landa2022experimental,
  title = {Experimental {{Bayesian}} Estimation of Quantum State Preparation, Measurement, and Gate Errors in Multiqubit Devices},
  author = {Landa, Haggai and Meirom, Dekel and Kanazawa, Naoki and Fitzpatrick, Mattias and Wood, Christopher J.},
  date = {2022-03},
  journaltitle = {Phys. Rev. Res.},
  volume = {4},
  number = {1},
  pages = {013199},
  publisher = {American Physical Society},
  doi = {10.1103/PhysRevResearch.4.013199},
  url = {https://link.aps.org/doi/10.1103/PhysRevResearch.4.013199},
  pagetotal = {13}
}

@article{lanting14aEntanglement,
  title = {Entanglement in a {{Quantum Annealing Processor}}},
  author = {Lanting, T. and Przybysz, A. J. and family=Smirnov, given=A. Yu., given-i=A{{Yu}} and Spedalieri, F. M. and Amin, M. H. and Berkley, A. J. and Harris, R. and Altomare, F. and Boixo, S. and Bunyk, P. and Dickson, N. and Enderud, C. and Hilton, J. P. and Hoskinson, E. and Johnson, M. W. and Ladizinsky, E. and Ladizinsky, N. and Neufeld, R. and Oh, T. and Perminov, I. and Rich, C. and Thom, M. C. and Tolkacheva, E. and Uchaikin, S. and Wilson, A. B. and Rose, G.},
  date = {2014-05},
  journaltitle = {Phys. Rev. X},
  volume = {4},
  number = {2},
  pages = {021041},
  publisher = {American Physical Society},
  doi = {10.1103/PhysRevX.4.021041}
}

@article{lazuar2023calibration,
  title = {Calibration of {{Drive Nonlinearity}} for {{Arbitrary-Angle Single-Qubit Gates Using Error Amplification}}},
  author = {Lază ăr, Stefania and Ficheux, Quentin and Herrmann, Johannes and Remm, Ants and Lacroix, Nathan and Hellings, Christoph and Swiadek, Francois and Zanuz, Dante Colao and Norris, Graham J. and Panah, Mohsen Bahrami and Flasby, Alexander and Kerschbaum, Michael and Besse, Jean-Claude and Eichler, Christopher and Wallraff, Andreas},
  date = {2023-08},
  journaltitle = {Phys. Rev. Appl.},
  volume = {20},
  number = {2},
  pages = {024036},
  publisher = {American Physical Society},
  doi = {10.1103/PhysRevApplied.20.024036},
  url = {https://link.aps.org/doi/10.1103/PhysRevApplied.20.024036},
  pagetotal = {12}
}

@article{Lechner2015parity,
  title = {A {{Quantum Annealing Architecture}} with {{All-to-All Connectivity}} from {{Local Interactions}}},
  author = {Lechner, Wolfgang and Hauke, Philipp and Zoller, Peter},
  date = {2015},
  journaltitle = {Sci. Adv.},
  volume = {1},
  number = {9},
  eprint = {https://www.science.org/doi/pdf/10.1126/sciadv.1500838},
  pages = {e1500838},
  doi = {10.1126/sciadv.1500838},
}

@article{Leib2016transmonAnnealer,
  title = {A {{Transmon Quantum Annealer}}: {{Decomposing Many-Body Ising Constraints}} into {{Pair Interactions}}},
  author = {Leib, Martin and Zoller, Peter and Lechner, Wolfgang},
  date = {2016-12},
  journaltitle = {Quantum Sci. Technol.},
  volume = {1},
  number = {1},
  pages = {015008},
  publisher = {IOP Publishing},
  doi = {10.1088/2058-9565/1/1/015008},
}

@article{leibfried1996experimental,
  title = {Experimental {{Determination}} of the {{Motional Quantum State}} of a {{Trapped Atom}}},
  author = {Leibfried, D. and Meekhof, D. M. and King, B. E. and Monroe, C. and Itano, W. M. and Wineland, D. J.},
  date = {1996-11},
  journaltitle = {Phys. Rev. Lett.},
  volume = {77},
  number = {21},
  pages = {4281},
  publisher = {American Physical Society},
  doi = {10.1103/PhysRevLett.77.4281},
  url = {https://link.aps.org/doi/10.1103/PhysRevLett.77.4281},
  pagetotal = {0}
}

@article{levi2007efficient,
  title = {Efficient {{Error Characterization}} in {{Quantum Information Processing}}},
  author = {Lévi, Benjamin and López, Cecilia C and Emerson, Joseph and Cory, David G},
  date = {2007},
  journaltitle = {Phys. Rev. A},
  volume = {75},
  number = {2},
  pages = {022314},
  publisher = {APS}
}

@article{levine2018high,
  title = {High-{{Fidelity Control}} and {{Entanglement}} of {{Rydberg-Atom Qubits}}},
  author = {Levine, Harry and Keesling, Alexander and Omran, Ahmed and Bernien, Hannes and Schwartz, Sylvain and Zibrov, Alexander S. and Endres, Manuel and Greiner, Markus and Vuletić ć, Vladan and Lukin, Mikhail D.},
  date = {2018-09},
  journaltitle = {Phys. Rev. Lett.},
  volume = {121},
  number = {12},
  pages = {123603},
  publisher = {American Physical Society},
  doi = {10.1103/PhysRevLett.121.123603},
  url = {https://link.aps.org/doi/10.1103/PhysRevLett.121.123603},
  pagetotal = {6}
}

@article{levine2019parallel,
  title = {Parallel {{Implementation}} of {{High-Fidelity Multiqubit Gates}} with {{Neutral Atoms}}},
  author = {Levine, Harry and Keesling, Alexander and Semeghini, Giulia and Omran, Ahmed and Wang, Tout T. and Ebadi, Sepehr and Bernien, Hannes and Greiner, Markus and Vuletić ć, Vladan and Pichler, Hannes and Lukin, Mikhail D.},
  date = {2019-10},
  journaltitle = {Phys. Rev. Lett.},
  volume = {123},
  number = {17},
  pages = {170503},
  publisher = {American Physical Society},
  doi = {10.1103/PhysRevLett.123.170503},
  url = {https://link.aps.org/doi/10.1103/PhysRevLett.123.170503},
  pagetotal = {6}
}

@article{levine2307demonstrating,
  title = {Demonstrating a Long-Coherence Dual-Rail Erasure Qubit Using Tunable Transmons},
  author = {Levine, H. and Haim, A. and Hung, J. S. C. and Alidoust, N. and Kalaee, M. and DeLorenzo, L. and Wollack, E. A. and Arrangoiz-Arriola, P. and Khalajhedayati, A. and Sanil, R. and Moradinejad, H. and Vaknin, Y. and Kubica, A. and Hover, D. and Aghaeimeibodi, S. and Alcid, J. A. and Baek, C. and Barnett, J. and Bawdekar, K. and Bienias, P. and Carson, H. A. and Chen, C. and Chen, L. and Chinkezian, H. and Chisholm, E. M. and Clifford, A. and Cosmic, R. and Crisosto, N. and Dalzell, A. M. and Davis, E. and D'Ewart, J. M. and Diez, S. and D'Souza, N. and Dumitrescu, P. T. and Elkhouly, E. and Fang, M. T. and Fang, Y. and Flammia, S. and Fling, M. J. and Garcia, G. and Gharzai, M. K. and Gorshkov, A. V. and Gray, M. J. and Grimberg, S. and Grimsmo, A. L. and Hann, C. T. and He, Y. and Heidel, S. and Howell, S. and Hunt, M. and Iverson, J. and Jarrige, I. and Jiang, L. and Jones, W. M. and Karabalin, R. and Karalekas, P. J. and Keller, A. J. and Lasi, D. and Lee, M. and Ly, V. and MacCabe, G. and Mahuli, N. and Marcaud, G. and Matheny, M. H. and McArdle, S. and McCabe, G. and Merton, G. and Miles, C. and Milsted, A. and Mishra, A. and Moncelsi, L. and Naghiloo, M. and Noh, K. and Oblepias, E. and Ortuno, G. and Owens, J. C. and Pagdilao, J. and Panduro, A. and Paquette, J.-P. and Patel, R. N. and Peairs, G. and Perello, D. J. and Peterson, E. C. and Ponte, S. and Putterman, H. and Refael, G. and Reinhold, P. and Resnick, R. and Reyna, O. A. and Rodriguez, R. and Rose, J. and Rubin, A. H. and Runyan, M. and Ryan, C. A. and Sahmoud, A. and Scaffidi, T. and Shah, B. and Siavoshi, S. and Sivarajah, P. and Skogland, T. and Su, C.-J. and Swenson, L. J. and Sylvia, J. and Teo, S. M. and Tomada, A. and Torlai, G. and Wistrom, M. and Zhang, K. and Zuk, I. and Clerk, A. A. and Brand\~ao, F. G. S. L. and Retzker, A. and Painter, O.},
  journal = {Phys. Rev. X},
  volume = {14},
  issue = {1},
  pages = {011051},
  numpages = {21},
  year = {2024},
  month = {3},
  publisher = {American Physical Society},
  url = {https://link.aps.org/doi/10.1103/PhysRevX.14.011051}
}

@inproceedings{li2019tackling,
  title = {Tackling the Qubit Mapping Problem for {{NISQ-era}} Quantum Devices},
  booktitle = {Proc 24th {{Int Conf Archit}}. {{Support Program Lang}}. {{Oper}}. {{Syst}}},
  author = {Li, Gushu and Ding, Yufei and Xie, Yuan},
  date = {2019},
  pages = {1001},
  url = {https://doi.org/10.1145/3297858.3304023}
}

@article{li2024non,
  title = {Non-{{Markovian}} Quantum Gate Set Tomography},
  author = {Li, Ze-Tong and Zheng, Cong-Cong and Meng, Fan-Xu and Zeng, Han and Luan, Tian and Zhang, Zai-Chen and Yu, Xu-Tao},
  date = {2024},
  journaltitle = {Quantum Sci. Technol.},
  volume = {9},
  number = {3},
  pages = {035027},
  publisher = {IOP Publishing},
  url = {https://doi.org/10.1088/2058-9565/ad3d80}
}

@article{Liang_2024,
  title = {Pulse Optimization for High-Precision Motional-Mode Characterization in Trapped-Ion Quantum Computers},
  author = {Liang, Qiyao and Kang, Mingyu and Li, Ming and Nam, Yunseong},
  date = {2024-04},
  journaltitle = {Quantum Sci. Technol.},
  volume = {9},
  number = {3},
  pages = {035007},
  publisher = {IOP Publishing},
  doi = {10.1088/2058-9565/ad3a98},
  url = {https://dx.doi.org/10.1088/2058-9565/ad3a98}
}

@article{liaoBenchmarkingQuantumProtocols2022,
  title = {Benchmarking of {{Quantum Protocols}}},
  author = {Liao, Chin-Te and Bahrani, Sima and family=Silva, given=Francisco Ferreira, prefix=da, useprefix=true and Kashefi, Elham},
  date = {2022-12},
  journaltitle = {Sci. Rep.},
  volume = {12},
  number = {1},
  pages = {5298},
  issn = {2045-2322},
  doi = {10.1038/s41598-022-08901-x}
}

@article{lidar2013quantum,
  title = {Quantum Error Correction},
  author = {Lidar, Daniel A and Brun, Todd A},
  date = {2013},
  journaltitle = {Camb. Univ. Press},
  url = {https://doi.org/10.1017/CBO9781139034807}
}

@article{lin2021independent,
  title = {Independent State and Measurement Characterization for Quantum Computers},
  author = {Lin, Junan and Wallman, Joel J. and Hincks, Ian and Laflamme, Raymond},
  date = {2021-09},
  journaltitle = {Phys. Rev. Res.},
  volume = {3},
  number = {3},
  pages = {033285},
  publisher = {American Physical Society},
  doi = {10.1103/PhysRevResearch.3.033285},
  url = {https://link.aps.org/doi/10.1103/PhysRevResearch.3.033285},
  pagetotal = {11}
}

@article{lin2022domain,
  title = {Domain-Specific Quantum Architecture Optimization},
  author = {Lin, Wan-Hsuan and Tan, Bochen and Niu, Murphy Yuezhen and Kimko, Jason and Cong, Jason},
  date = {2022},
  journaltitle = {IEEE J. Emerg. Sel. Top. Circuits Syst.},
  volume = {12},
  number = {3},
  pages = {624},
  publisher = {IEEE},
  url = {https://doi.org/10.1109/JETCAS.2022.3202870}
}

@article{linkeExperimentalComparisonTwo2017,
  title = {Experimental {{Comparison}} of {{Two Quantum Computing Architectures}}},
  author = {Linke, Norbert M. and Maslov, Dmitri and Roetteler, Martin and Debnath, Shantanu and Figgatt, Caroline and Landsman, Kevin A. and Wright, Kenneth and Monroe, Christopher},
  date = {2017-03},
  journaltitle = {Proc. Natl. Acad. Sci.},
  volume = {114},
  number = {13},
  pages = {3305},
  issn = {0027-8424, 1091-6490},
  doi = {10.1073/pnas.1618020114}
}

@article{liQASMBenchLowlevelQASM2022a,
  title = {{{QASMBench}}: {{A Low-Level Quantum Benchmark Suite}} for {{NISQ Evaluation}} and {{Simulation}}},
  author = {Li, Ang and Stein, Samuel and Krishnamoorthy, Sriram and Ang, James},
  date = {2023},
  journaltitle = {ACM Trans. Quantum Comput.},
  volume = {4},
  number = {2},
  pages = {1},
  url = {https://doi.org/10.1145/3550488}
}

@article{lis23,
  title = {Midcircuit {{Operations Using}} the Omg {{Architecture}} in {{Neutral Atom Arrays}}},
  author = {Lis, Joanna W. and Senoo, Aruku and McGrew, William F. and Rönchen, Felix and Jenkins, Alec and Kaufman, Adam M.},
  date = {2023-11},
  journaltitle = {Phys. Rev. X},
  volume = {13},
  number = {4},
  pages = {041035},
  publisher = {American Physical Society},
  doi = {10.1103/PhysRevX.13.041035},
  url = {https://link.aps.org/doi/10.1103/PhysRevX.13.041035},
  pagetotal = {22}
}

@article{liuBenchmarkingNeartermQuantum2022,
  title = {Benchmarking {{Near-Term Quantum Computers}} via {{Random Circuit Sampling}}},
  author = {Liu, Yunchao and Otten, Matthew and Bassirianjahromi, Roozbeh and Jiang, Liang and Fefferman, Bill},
  date = {2022-04},
  journaltitle = {arXiv:2105.05232},
  eprint = {2105.05232},
  eprinttype = {arXiv},
  url = {https://doi.org/10.48550/arXiv.2105.05232}
}

@article{lubinski2023optimization,
  title = {Optimization Applications as Quantum Performance Benchmarks},
  author = {Lubinski, Thomas and Coffrin, Carleton and McGeoch, Catherine and Sathe, Pratik and Apanavicius, Joshua and Bernal Neira, David and {collaboration}, Quantum Economic Development Consortium (QED-C)},
  date = {2023},
  journaltitle = {ACM Trans. Quantum Comput.},
  publisher = {ACM New York, NY},
  url = {https://doi.org/10.1145/3678184}
}

@article{lubinski2024quantum,
  title = {Quantum {{Algorithm Exploration}} Using {{Application-Oriented Performance Benchmarks}}},
  author = {Lubinski, Thomas and Goings, Joshua J and Mayer, Karl and Johri, Sonika and Reddy, Nithin and Mehta, Aman and Bhatia, Niranjan and Rappaport, Sonny and Mills, Daniel and Baldwin, Charles H and others},
  date = {2024},
  journaltitle = {arXiv:2402.08985},
  eprint = {2402.08985},
  eprinttype = {arXiv},
  url = {https://doi.org/10.48550/arXiv.2402.08985}
}
\endgroup
\end{document}